\documentclass[a4paper]{article} 

\usepackage[sort&compress,comma,square,numbers]{natbib}

{\normalfont}{\normalfont}

\usepackage{enumitem}
\usepackage{amsmath,amssymb,amsfonts,amsthm,amscd,amstext}
\usepackage[percent]{overpic}

\newcommand{\ba}[1]{\begin{array}{#1}}
	\newcommand{\ea}{\end{array}}

\usepackage{graphicx}
\usepackage{epstopdf}
\graphicspath{{images/}}
\usepackage{subfigure}
\usepackage{listings} 
\usepackage[dvipsnames]{xcolor} 

\usepackage[bookmarks,colorlinks=true,	
plainpages=false,pdfstartview=FitH,
breaklinks,citecolor=red,linkcolor=black,urlcolor=blue,filecolor=black]{hyperref}

\allowdisplaybreaks

\begin{document}
	\title{Multifield asymptotic homogenization scheme for periodic Cauchy materials in non-standard thermoelasticity}
	\author{Rosaria Del Toro$^{1}$, Maria Laura De Bellis$^{1}$, Marcello Vasta$^{1}$, Andrea Bacigalupo$^{2}$}
	\date{\small $^{1}$ University of Chieti-Pescara, Department INGEO, Viale Pindaro 42, Pescara, Italy\\
		\small $^{2}$ University of Genova, Department DICCA, via Montallegro 1, Genova, Italy}	
	\maketitle
	\vspace{0.2cm}
	\begin{abstract}
		\noindent This article presents a multifield asymptotic homogenization scheme for the analysis of Bloch wave propagation in non-standard thermoelastic periodic materials, leveraging on the Green-Linsdsay theory that accounts for two relaxation times. The procedure involves several steps. Firstly, an asymptotic expansion of the micro-fields is performed, considering the characteristic size of the microstructure.
		By utilizing the derived microscale field equations and asymptotic expansions, a series of recursive differential problems are solved within the repetitive unit cell $\mathcal{Q}$. These problems are then expressed in terms of perturbation functions, which incorporate the material's geometric, physical, and mechanical properties, as well as the microstructural heterogeneities. The down-scaling relation, which connects the microscopic and macroscopic fields along with their gradients through the perturbation functions, is then established in a consistent manner. Subsequently, the average field equations of infinite order are obtained by substituting the down-scaling relation into the microscale field equations. To solve these average field equations, an asymptotic expansion of the macroscopic fields is performed based on the microstructural size, resulting in a sequence of macroscopic recursive problems.
		To illustrate the methodology, a bi-phase layered material is introduced as an example. The dispersion curves obtained from the non-local homogenization scheme are compared with those obtained from the Floquet-Bloch theory. This analysis helps validate the effectiveness and accuracy of the proposed approach in predicting the wave propagation behavior in the considered non-standard thermoelastic periodic materials. 
	\end{abstract}
	
	\smallskip
	\noindent {\bf Keywords}:\quad periodic Cauchy materials, thermo-elastic waves, Green-Lindsay theory, asymptotic approximation, homogenized model

	\bigskip
	\section{Introduction}
	Thermoelasticity is a branch of solid mechanics that studies the coupled behavior of temperature
	and mechanical deformation in materials. It explores the relationship between temperature
	changes and resulting mechanical responses, such as stress and strain.
	The fundamental theory of thermoelasticity, grounded in Fourier's law of heat conduction, posits that thermal perturbations propagate infinitely fast in a diffusive way when governed by the coupled displacement-temperature equation, which takes the form of a parabolic-type partial differential equation \cite{carlsonlinear, zhmakin2021heat}. From a practical standpoint, this implies that a sudden change of temperature in a sample instantaneously will be felt everywhere \cite{joseph1989heat}. However, experimental observations have revealed instances where temperature behaves akin to a wave, propagating through the body with finite speed, commonly referred to as 'second sound' \cite{chandrasekharaiah1986thermoelasticity, christov2005heat}. This intriguing wave-like propagation of heat has been observed in diverse systems, such as solids, sand, processed meat and dielectric crystals \cite{coleman1982thermodynamics, coleman1983nonequilibrium}. This observation disagrees with the prevailing notion that disturbances of bounded support can only generate responses within a limited time frame and spatial extent \cite{ignaczak2009thermoelasticity}. In addition to the paradox posed by the infinite propagation speeds, the conventional dynamic thermoelasticity theory fails to provide satisfactory or accurate descriptions of a solid's response under fast transient loading, such as short laser pulses, and at low temperatures. These limitations have prompted numerous researchers to propose alternative generalized thermoelasticity theories. Building upon the works of Maxwell and Cattaneo, these theories introduce thermoelastic models featuring one or two relaxation times, models specifically tailored for low-temperature scenarios, models devoid of energy dissipation, dual-phase-lag theories, and even unconventional heat conduction described by fractional calculus \cite{landau1941theory,  lord1967generalized, green1972thermoelasticity, hetnarski2000nonclassical, povstenko2004fractional, iesan2004thermoelastic, povstenko2005stresses, fabrizio2014stability, el2014dual}. In the following, the Green-Lindsay theory (or thermoelasticity with two relaxation times) will be employed. It is a \emph{non-standard} (or \emph{non-conventional}) thermoelastic model that incorporates additional terms in the stress-strain relation to capture the nonlinear effects and the Fourier heat conduction. It provides a relatively simple and general framework to analyze the coupled behavior of
	temperature and mechanical deformation and it is widely applicable to a broad range of materials and conditions, making it a practical choice for many engineering applications \cite{green1972thermoelasticity, othman2004effect, ignaczak2009thermoelasticity, sharifi2023dynamic}.\\
	The modeling of multi-phase materials with periodic microstructures, encompassing combined phenomena of elasticity and heat transfer, holds huge significance in contemporary applications such as aerospace, structural analysis, the design of thermal protection systems, geomechanics, biomedical, and electronics engineering, \cite{fruehmann2012application, kumar2017thermoelastic, nicholson2021mapping, aliyu2023three}. Solving the governing thermoelastic partial differential equations, particularly those with one or two relaxation times, can be analytically and numerically bulky due to the periodic nature of the materials \cite{nayfeh1971thermoelastic, feyel2003multilevel}. Consequently, multi-scale asymptotic homogenization approaches, demonstrated by \cite{Bakhvalov1984, Smyshlyaev2000}, emerge as remarkable tools for establishing the responses of microscopic phases and their impact on the overall properties of composites. By supplanting a heterogeneous material with an equivalent homogenized model, which can be reshaped either as a first order (Cauchy) or as a non-local continuum, these approaches provide approximate solutions that are described by constitutive tensors unaffected by the rapidly oscillating fast variable associated with the underlying microstructure.
	It is noteworthy that various homogenization methods have been used to investigate the overall properties of multi-phase periodic materials \cite{hadjiloizi2013micromechanical, hadjiloizi2013micromechanical2, del2019characterization}. For elastic materials, they can be grouped into asymptotic \cite{Bensoussan1978, Bakhvalov1984, GambinKroner1989, meguid1994asymptotic, AndrianovBolshakov2008, panasenko2009boundary, bosco2017asymptotic, de2019characterization, fantoni2020phase}, variational-asymptotic \cite{Smyshlyaev2000, Smyshlyaev2009, bacigalupo2014homogenization, fantoni2022multifield}, and identification approaches, which include the analytical \cite{bigoni2007analytical, bacca2013amindlin, bacca2013mindlin2, bacigalupo2013multi, bacigalupo2018identification, DELTORO2023112431} and the computational techniques \cite{Forest1998, KouznetsovaGeers2004,lew2004homogenisation, scarpa2009effective, bacigalupo2011non, forest2011generalized, bacigalupo2012computational, zah2013computational, salvadori2014computational, bacigalupo2014computational, bacigalupo2015auxetic, BacigalupoDeBellis2015, de2017auxetic}. Moreover, asymptotic homogenization schemes were employed to analyze thermo-piezoelectric periodic materials, elasto-thermo-diffusive periodic materials and thermo-diffusive composites \cite{ForestAifantis2010, cook2013multiscale, boldrin2016dynamic, BacigalupoMoriniPiccolroaz2016b, fantoni2017multi,  caballero2020computation, bosco2020multi, fantoni2020wave, vega2022thermo}. Concerning with thermoelastic periodic materials, a computational method is employed in \cite{dong2018multiscale}, whereas a variational-asymptotic technique is proposed by \cite{preve2021variational}, where a first order (Cauchy) continuum is retrieved.\\
	The present paper proposes a multifield asymptotic homogenization scheme for the analysis of dispersive waves in non-standard thermoelastic periodic materials based on Green-Linsdsay theory in the framework of asymptotic methods \cite{Smyshlyaev2000, Bacigalupo2014}. Specifically, the field equations at the micro-scale governing the heterogeneous thermoelastic materials are found. The micro-displacement and the micro-temperature fields are developed as asymptotic expansions in terms of the characteristic length and their substitution into the field equations at the micro-scale determines a set of recursive differential problems defined over the periodic unit cell. 
	Then, imposing solvability conditions to the nonhomogeneous recursive cell problems enables to achieve the down-scaling relation, relating the micro-displacement and the micro-temperature fields to the macroscopic ones and their gradients through the perturbation functions. Such functions depend on the geometrical and physical-mechanical properties of the material and take into account the microstructural heterogeneities. Average field equations of infinite order are obtained by replacing the down-scale relations into the micro-field equations. Their formal solutions are given thanks to asymptotic expansions of the macro-displacement and macro-temperature and, by retaining only the terms at the zeroth order, the field equations related to the equivalent first order (Cauchy) thermoealstic continuum are recovered.\\
	Section 2 displays the field equations at the microscale. Section 3 proposes the solutions of thermo-mechanical recursive differential problems, the cell problems and the related perturbation functions, the down-scaling relation, the up-scaling relation and the average field equations of infinite order.  
	Section 4 deals with the free wave propagation through a thermoelastic material with a periodic microstructure by transposing the average field equations, via the Laplace and the Fourier transforms, into the frequency and the wave vector domain. Moreover, truncating the transformed average field equations at the second-order of $\varepsilon$ leads to an approximation of the Floquet-Bloch spectrum. In Section 5, the asymptotic homogenization scheme is applied to a bi-phase layered material with orthotropic phases and the Floquet-Bloch theory is adopted to tackle with the heterogeneous material. In such a case, the problem of wave propagation is investigated and, to assess the reliability of the asymptotic homogenization scheme, the approximate dispersion curves are compared with those obtained from the the Floquet-Bloch theory and a good agreement between the models is observed. Final remarks conclude the paper. Supplementary material displays some technical issues and it will be recalled in the main text. 
	\section{Field equations for periodic thermoelastic materials}
	\label{goveq}
	\begin{figure}[!t]
		\hspace*{-2cm} \begin{overpic}[scale=0.6]{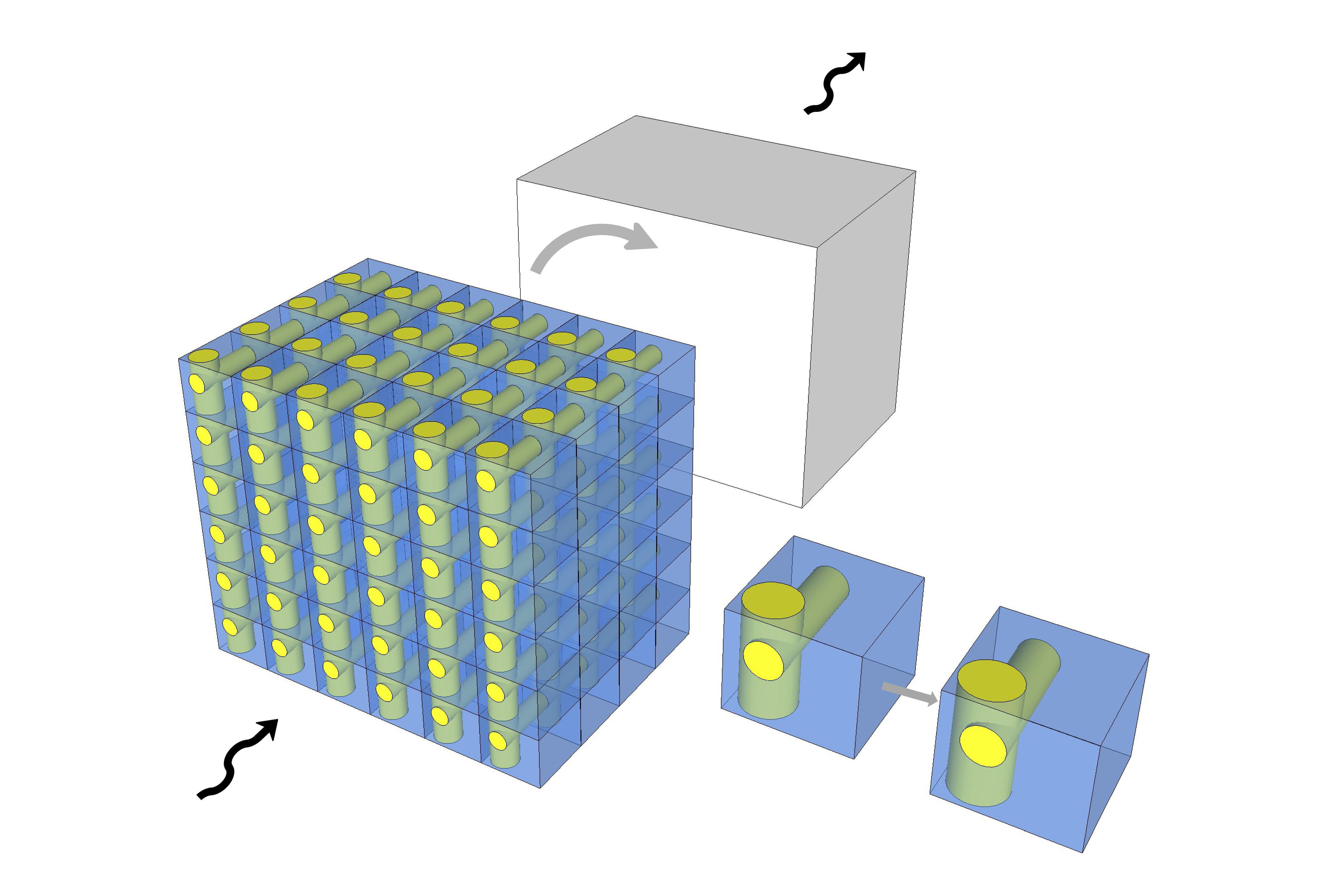}
			\put (37,6) {\colorbox{white}{\footnotesize{(a)}}}
			\put (67,34) {\footnotesize{(b)}}
			\put (61.5,8.5) {\colorbox{white}{\footnotesize{(c)}}}
			\put (79,1.5) {\colorbox{white}{\footnotesize{(d)}}}
			\put (28,50) 
			{\footnotesize{Homogenization}}
			\put (19.1,7) 
			{\footnotesize{Bloch wave}}
			\put (18,45) { $\mathcal{L}$}
			\put (54,25) { $\mathcal{A}$}
			\put (71,20) { $\mathcal{Q}$}
			\put (65.5,13) {\rotatebox{345}{$\boldsymbol{\xi} = \frac{\boldsymbol{x}}{\varepsilon}$}}
		\end{overpic}
		\caption{(a) Portion of a generic periodic thermoelastic material; (b) Corresponding portion of homogenized material (c) Detail of the periodic cell; (d) Corresponding nondimensional unit cell.}\label{uno}
		\label{figurePrima}
	\end{figure}
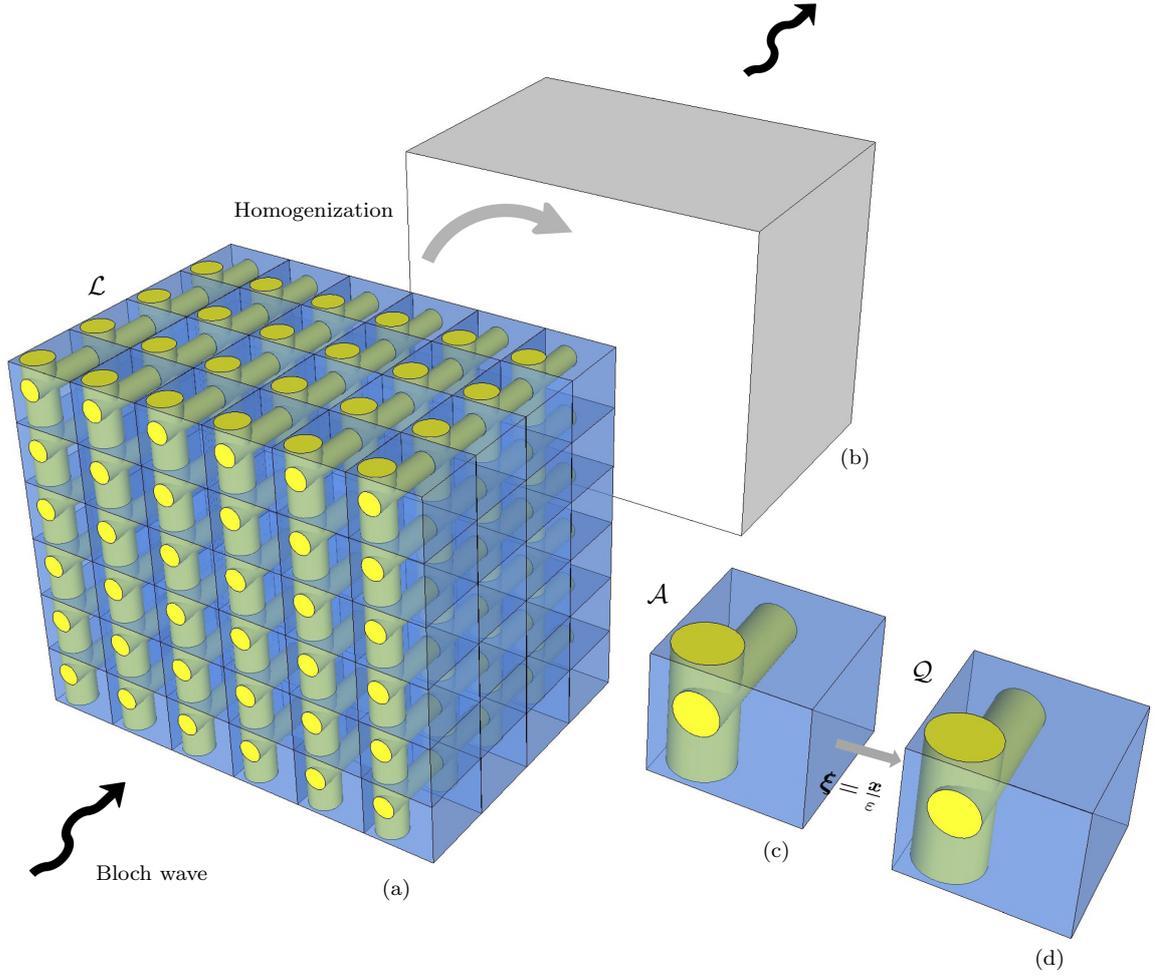
	Let $\mathcal{R}$ be a three-dimensional thermoelastic heterogeneous body endowed with a periodic microstructure. The position vector $\boldsymbol{x} = x_{1}\boldsymbol{e}_{1}+x_{2}\boldsymbol{e}_{2}+x_{3}\boldsymbol{e}_{3}$ identifies a generic point of the body after setting a system of coordinates with origin at point $O$ and orthogonal base $\{{} \boldsymbol{e}_{1},\boldsymbol{e}_{2}, \boldsymbol{e}_{3}\}$. Let $\mathcal{A}=[0,\varepsilon]\times [0,\delta \varepsilon]\times [0,\theta\varepsilon]$ be a periodic cell with characteristic size $\varepsilon$. $\mathcal{A}$ is described by three orthogonal periodicity vectors $\boldsymbol{v}_{1}$, $\boldsymbol{v}_{2}$ and $\boldsymbol{v}_{3}$ defined as $\boldsymbol{v}_{1} =  d_{1}\boldsymbol{e}_{1} = \varepsilon \boldsymbol{e}_{1} $,  $\boldsymbol{v}_{2}= d_{2}\boldsymbol{e}_{2} = \delta \varepsilon \boldsymbol{e}_{2}$ and $\boldsymbol{v}_{3} =  d_{3}\boldsymbol{e}_{3} =\theta \varepsilon \boldsymbol{e}_{3}$. The tasselation of the periodic cell $\mathcal{A}$ according to the directions of $\boldsymbol{v}_{1}$, $\boldsymbol{v}_{2}$ and $\boldsymbol{v}_{3}$ generates the material domain, (see Figure \ref{uno}).
	The constitutive equations employing the stress tensor $\boldsymbol{\sigma}(\boldsymbol{x},t)=\sigma_{ij}\boldsymbol{e}_{i}\otimes \boldsymbol{e}_{j}$, the heat flux $\boldsymbol{q}(\boldsymbol{x},t)=q_{i}\boldsymbol{e}_{i}$ and the entropy per unit of volume ${\eta}(\boldsymbol{x},t)$ are  
	\begin{subequations}
		\begin{align}
			&\boldsymbol{\sigma}(\boldsymbol{x},t)=	\mathbb{C}^{m}(\boldsymbol{x})\boldsymbol{\varepsilon}(\boldsymbol{x},t)-\boldsymbol{\alpha}^{m}(\boldsymbol{x})\upsilon(\boldsymbol{x},t)-\boldsymbol{\alpha}^{m}(\boldsymbol{x})\tau_{1}^{m}(\boldsymbol{x})\dot{\upsilon}(\boldsymbol{x},t),\label{EC1}	\\
			&{\eta}(\boldsymbol{x},t)=	  \boldsymbol{\alpha}^{m}(\boldsymbol{x})\boldsymbol{\varepsilon}(\boldsymbol{x},t)+\frac{\rho^{m}(\boldsymbol{x})C_{E}(\boldsymbol{x})}{\theta_{0}}(\upsilon(\boldsymbol{x},t)+\tau_{0}^{m}(\boldsymbol{x})\dot{\upsilon}(\boldsymbol{x},t)),\label{EC2}	\\
			&\boldsymbol{q}(\boldsymbol{x},t)=-\bar{\boldsymbol{K}}^{m}(\boldsymbol{x})\nabla \upsilon(\boldsymbol{x},t),\label{EC3}			
		\end{align}
	\end{subequations} 
	where the superscript \emph{m} stands for the microscale, $\mathbb{C}^{m}(\boldsymbol{x})=C^{m}_{ijhk}\boldsymbol{e}_{i}\otimes \boldsymbol{e}_{j}\otimes \boldsymbol{e}_{h}\otimes \boldsymbol{e}_{k}$ is the fourth-order micro elastic tensor, $\boldsymbol{\varepsilon}(\boldsymbol{x},t)=\varepsilon_{ij}\boldsymbol{e}_{i}\otimes \boldsymbol{e}_{j}$ is the second-order micro strain tensor, $\boldsymbol{\alpha}^{m}(\boldsymbol{x},t)=\alpha_{ij}^{m}\boldsymbol{e}_{i}\otimes \boldsymbol{e}_{j}$ is the symmetric second-order micro stress-temperature tensor, $\upsilon(\boldsymbol{x},t)=\theta(\boldsymbol{x},t)-\theta_{0}$ is the relative temperature field with the absolute temperature $\theta(\boldsymbol{x},t)$ and the reference stress-free temperature  $\theta_{0}$,  $\rho^{m}(\boldsymbol{x})$ is the material density, $\bar{\boldsymbol{K}}^{m}(\boldsymbol{x})= \bar{K}_{ij}^{m}\boldsymbol{e}_{i}\otimes \boldsymbol{e}_{j}$ is the symmetric second-order micro thermal conductivity tensor, $C_{E}(\boldsymbol{x})$ is the specific heat at zero strain, $\tau_{1}^{m}(\boldsymbol{x})$ and $\tau_{0}^{m}(\boldsymbol{x})$ are the relaxation times related to the Green-Lindsay theory \cite{ignaczak2009thermoelasticity, green1972thermoelasticity}. For sake of simplicity, the notations $p^{m}(\boldsymbol{x})=\frac{\rho^{m}(\boldsymbol{x})C_{E}(\boldsymbol{x})}{\theta_{0}}$, $\boldsymbol{\alpha}^{(m,1)}(\boldsymbol{x})=\boldsymbol{\alpha}^{m}(\boldsymbol{x})\tau_{1}^{m}(\boldsymbol{x})$ and $p^{(m,0)}(\boldsymbol{x})=p^{m}(\boldsymbol{x})\tau_{0}^{m}(\boldsymbol{x})$ are introduced. In addition, $t$ is the time coordinate and the superimposed dot is the time derivative. The material obeys to 
	small displacements and so the micro-strain tensor can be rewritten as
	$\boldsymbol{\varepsilon}(\boldsymbol{x},t) = \frac{1}{2} (\nabla \boldsymbol{u} (\boldsymbol{x},t)+\nabla^{T} \boldsymbol{u}(\boldsymbol{x},t))$, where $\nabla \boldsymbol{u}$ is the gradient of the micro-displacement $\boldsymbol{u}(\boldsymbol{x},t)$.  
	The balance equations are
	\begin{subequations}
		\begin{align}
			&\nabla \cdot \boldsymbol{\sigma}(\boldsymbol{x},t)+\boldsymbol{b}(\boldsymbol{x},t) = \rho^{m}(\boldsymbol{x}) {\ddot{\boldsymbol{u}}}(\boldsymbol{x},t), \label{EB1}\\
			&-\nabla \cdot \boldsymbol{q}(\boldsymbol{x},t)+\bar{r}(\boldsymbol{x},t) = \theta_{0} \dot{{\eta}}(\boldsymbol{x},t), \label{EB2}
		\end{align}
	\end{subequations} 
	where $\boldsymbol{u}(\boldsymbol{x},t)$ is the micro-displacement field, $\boldsymbol{b}(\boldsymbol{x},t)$ are the body forces and $\bar{r}(\boldsymbol{x},t)$ are the external heat sources. In order to derive a description of the thermoelastic process, replacing the relation \eqref{EC1} into the equation \eqref{EB1} and the relations \eqref{EC2}-\eqref{EC3} into the equation \eqref{EB2} leads to 
	\begin{subequations}
		\begin{align}
			&\nabla \cdot (\mathbb{C}^{m}(\boldsymbol{x})\nabla\boldsymbol{u}(\boldsymbol{x},t)-\boldsymbol{\alpha}^{m}(\boldsymbol{x})\upsilon(\boldsymbol{x},t)-\boldsymbol{\alpha}^{(m,1)}(\boldsymbol{x})\dot{\upsilon}(\boldsymbol{x},t)) +\boldsymbol{b}(\boldsymbol{x},t) = \rho^{m}(\boldsymbol{x}) {\ddot{\boldsymbol{u}}}(\boldsymbol{x},t), \label{EBTC3}\\
			&\nabla \cdot ({\boldsymbol{K}}^{m}(\boldsymbol{x})\nabla \upsilon(\boldsymbol{x},t)) -\boldsymbol{\alpha}^{m}(\boldsymbol{x})\nabla \dot{\boldsymbol{u}}(\boldsymbol{x},t)-p^{m}(\boldsymbol{x})\dot{\upsilon}(\boldsymbol{x},t)-p^{(m,0)}(\boldsymbol{x})\ddot{\upsilon}(\boldsymbol{x},t)=-{r}(\boldsymbol{x},t), \label{EBTC5}
		\end{align}
	\end{subequations} 
	where the minor symmetry property of the micro elasticity tensor $\mathbb{C}^{m}$ and the micro stress-temperature tensor $\boldsymbol{\alpha}^{m}$ are applied to the strain tensor $\boldsymbol{\varepsilon}$. For sake of simplicity, $r(\boldsymbol{x},t)=\frac{\bar{r}(\boldsymbol{x},t)}{\theta_{0}}$ and $\boldsymbol{K}^{m}(\boldsymbol{x},\boldsymbol{\xi})=\frac{\bar{\boldsymbol{K}}^{m}(\boldsymbol{x},\boldsymbol{\xi})}{\theta_{0}}$. 
	Let $[[f]]=f^{i}(\Sigma)-f^{j}(\Sigma)$ be the jump of the function values $f$ at the interface $\Sigma$ between two different phases $i$ and $j$ in the periodic cell $\mathcal{A}$, therefore the following fully-bonded interface conditions must be fulfilled
	\begin{subequations}
		\begin{align}
			&[[{\boldsymbol{u}}(\boldsymbol{x},t)]]\vert_{\boldsymbol{x} \in \Sigma}= \boldsymbol{0},	\label{icld1} \\
			&[[{\upsilon}(\boldsymbol{x},t)]]\vert_{\boldsymbol{x} \in \Sigma}= \boldsymbol{0},	\label{icld2} \\
			&[[(\mathbb{C}^{m}(\boldsymbol{x})\nabla{\boldsymbol{u}}(\boldsymbol{x},t)-\boldsymbol{\alpha}^{m}(\boldsymbol{x})\upsilon(\boldsymbol{x},t)-\boldsymbol{\alpha}^{(m,1)}(\boldsymbol{x})\dot{\upsilon}(\boldsymbol{x},t))\cdot \boldsymbol{n}]]\vert_{\boldsymbol{x} \in \Sigma}= \boldsymbol{0},	\label{icld3} \\
			&[[(\boldsymbol{K}^{m}(\boldsymbol{x})\nabla {\upsilon}(\boldsymbol{x},t)) \cdot \boldsymbol{n}]]\vert_{\boldsymbol{x} \in \Sigma}= \boldsymbol{0}	\label{icld4},
		\end{align}
	\end{subequations}
	where $\boldsymbol{n}$ represents the outward normal to the interface $\Sigma$. There is point noticing that by assuming $\boldsymbol{\alpha}^{(m,1)}=\boldsymbol{0}$ and $p^{(m,0)}=0$, the equations \eqref{EBTC3}-\eqref{EBTC5} and their interface conditions \eqref{icld1}-\eqref{icld4} can be reduced to those describing the classical thermoelasticity.   
	The $\mathcal{A}$-periodicity of the material induces the following conditions:
	\begin{subequations}
		\begin{align}
			\mathbb{C}^{m}(\boldsymbol{x}+\boldsymbol{v}_{i})&=\mathbb{C}^{m}(\boldsymbol{x}), \label{AC1} \\
			\boldsymbol{\alpha}^{(m,1)}(\boldsymbol{x}+\boldsymbol{v}_{i})&=\boldsymbol{\alpha}^{(m,1)}(\boldsymbol{x}), \label{AC2} \\
			\boldsymbol{\alpha}^{m}(\boldsymbol{x}+\boldsymbol{v}_{i})&=\boldsymbol{\alpha}^{m}(\boldsymbol{x}), \label{AC4} \\
			p^{(m,0)}(\boldsymbol{x}+\boldsymbol{v}_{i})&=p^{(m,0)}(\boldsymbol{x}), \label{AC5} \\
			p^{(m)}(\boldsymbol{x}+\boldsymbol{v}_{i})&=p^{(m)}(\boldsymbol{x}),  \label{AC11} \\
			{\boldsymbol{K}}^{m}(\boldsymbol{x}+\boldsymbol{v}_{i})&={\boldsymbol{K}}^{m}(\boldsymbol{x}),  \label{AC6} \\
			\rho^{m}(\boldsymbol{x}+\boldsymbol{v}_{i})&=\rho^{m}(\boldsymbol{x}), \quad i=1,2,3 \quad \forall \boldsymbol{x} \in \mathcal{A}. \label{AC10} 
		\end{align}
	\end{subequations}  
	The heterogeneous material undergoes to a system of $\mathcal{L}$-periodic body forces $\boldsymbol{b}(\boldsymbol{x},t)$ that are characterized by zero mean values over $\mathcal{L} = [0, L] \times [0, \delta L] \times [0, \theta L]$. The structural (or macroscopic) length L is supposed to be much greater than the microstructural length $\varepsilon$, i.e. L$>$$>$ $\varepsilon$. In such a case, the scales separation condition may take place and, as a result, $\mathcal{L}$ is a representative portion of the material. Let $\mathcal{Q} = [0,1]\times [0,\delta]\times [0,\theta]$ be the nondimensional unit cell, which can replicate the periodic microstructure of the material. $\mathcal{Q}$ is obtained by re-scaling the size of the periodic cell $\mathcal{A}$ for the characteristic length $\varepsilon$. Therefore, two variables are introduced to distinguish the two scales, namely the macroscopic (or slow) one, $\boldsymbol{x} \in \mathcal{A}$, which measures the slow fluctuations, and the microscopic (or fast) variable, $\boldsymbol{\xi} = \frac{\boldsymbol{x}}{\varepsilon} \in \mathcal{Q}$, which retains the fast propagation of the signal. After introducing the unit cell $\mathcal{Q}$, the properties \eqref{AC1}-\eqref{AC10} may be reshaped according to the microscopic variable $\boldsymbol{\xi}$ as 
	\begin{subequations}
		\begin{align}
			\mathbb{C}^{m}(\boldsymbol{x})&=\mathbb{C}^{m}(\boldsymbol{x},\boldsymbol{\xi}=\boldsymbol{x}/\varepsilon), \label{QC1}\\		\boldsymbol{\alpha}^{(m,1)}(\boldsymbol{x})&=\boldsymbol{\alpha}^{(m,1)}(\boldsymbol{x},\boldsymbol{\xi}=\boldsymbol{x}/\varepsilon), \label{QC2}\\		\boldsymbol{\alpha}^{m}(\boldsymbol{x})&=\boldsymbol{\alpha}^{m}(\boldsymbol{x},\boldsymbol{\xi}=\boldsymbol{x}/\varepsilon), \label{QC4}\\		p^{(m,0)}(\boldsymbol{x})&=p^{(m,0)}(\boldsymbol{x},\boldsymbol{\xi}=\boldsymbol{x}/\varepsilon), \label{QC5}\\
			{\boldsymbol{K}}^{m}(\boldsymbol{x})&={\boldsymbol{K}}^{m}(\boldsymbol{x},\boldsymbol{\xi}=\boldsymbol{x}/\varepsilon), \label{QC6}\\
			\rho^{m}(\boldsymbol{x})&=\rho^{m}(\boldsymbol{x},\boldsymbol{\xi}=\boldsymbol{x}/\varepsilon). \label{QC7}
		\end{align}
	\end{subequations}              
	Moreover, due to the $\mathcal{Q}$-periodicity of the micro constitutive tensors and the inertial terms and the $\mathcal{L}$-periodicity of the source terms, the micro-displacement and the micro-temperature depend on the slow variable $\boldsymbol{x}$ and the fast one $\boldsymbol{\xi}$ and they can be written as
	\begin{align}
		&\boldsymbol{u} = \boldsymbol{u}\Big(\boldsymbol{x},\frac{\boldsymbol{x}}{\varepsilon},t\Big),\quad \upsilon = \upsilon\Big(\boldsymbol{x},\frac{\boldsymbol{x}}{\varepsilon},t\Big).
	\end{align}
	It is worth noting that solving the system \eqref{EBTC3}-\eqref{EBTC5} can be both computationally and analytically demanding due to the $\mathcal{Q}$-periodic coefficients involved. Consequently, employing a non-local asymptotic homogenization technique offers a feasible approach to transform the heterogeneous material into an equivalent homogeneous one. This procedure yields equations that are equivalent to \eqref{EBTC3}-\eqref{EBTC5}, with coefficients that remain unaffected by oscillations, resulting in solutions that closely resemble those of the original equations. Furthermore, employing this technique significantly reduces the computational cost associated with handling equations \eqref{EBTC3}-\eqref{EBTC5}.
	\section{Asymptotic homogenization scheme for thermoelastic periodic materials}
	This section provides an overview of a non-local asymptotic homogenization scheme for analyzing thermoelastic heterogeneous materials with periodic microstructure. The section begins by outlining the scheme in general terms. Subsection \ref{asyexp} explores the asymptotic expansion of the micro-fields, expressing them in relation to the characteristic size $\varepsilon$ of the microstructure. Moving on to Subsection \ref{recdif}, the micro-scale field equations and asymptotic expansions will be employed to address a set of recursive differential problems defined within the periodic unit cell. Subsection \ref{cepr} displays the cell problems using perturbation functions. Subsection \ref{SCR} focuses on down-scaling and up-scaling relations, which establish connections between the microscopic and macroscopic fields, including their gradients. Finally, the infinite-order average field equations will be established and a sequence of macroscopic recursive problems will be introduced.
	\subsection{Asymptotic expansion of the field equations at the microscale}
	\label{asyexp}
	According to the asymptotic scheme exposed in \cite{Bakhvalov1984, Smyshlyaev2000, BacigalupoGambarotta2014}, the micro-displacement $\boldsymbol{u}$ and the micro-temperature $\upsilon$ may be written as asymptotic expansions with respect to the parameter $\varepsilon$ that keeps apart the slow $\boldsymbol{x}$ variable from the fast one $\boldsymbol{\xi} = \frac{\boldsymbol{x}}{\varepsilon}$ as
	\begin{subequations}
		\begin{align}
			&u_{h}\Big(\boldsymbol{x},\frac{\boldsymbol{x}}{\varepsilon},t\Big) = \sum_{l=0}^{+\infty} \varepsilon^{l} u^{(l)}_{h} = u^{(0)}_{h}\Big(\boldsymbol{x},\frac{\boldsymbol{x}}{\varepsilon},t\Big) + \varepsilon u^{(1)}_{h}\Big(\boldsymbol{x},\frac{\boldsymbol{x}}{\varepsilon},t\Big) +\varepsilon^{2}u^{(2)}_{h}\Big(\boldsymbol{x},\frac{\boldsymbol{x}}{\varepsilon},t\Big)+\textrm{O}(\varepsilon^{3}), 	\label{at1}\\
			&\upsilon\Big(\boldsymbol{x},\frac{\boldsymbol{x}}{\varepsilon},t\Big) = \sum_{l=0}^{+\infty} \varepsilon^{l} \upsilon^{(l)} = \upsilon^{(0)}\Big(\boldsymbol{x},\frac{\boldsymbol{x}}{\varepsilon},t\Big) + \varepsilon \upsilon^{(1)}\Big(\boldsymbol{x},\frac{\boldsymbol{x}}{\varepsilon},t\Big) +\varepsilon^{2}\upsilon^{(2)}\Big(\boldsymbol{x},\frac{\boldsymbol{x}}{\varepsilon},t\Big)+\textrm{O}(\varepsilon^{3}) \label{at2}.
		\end{align}
	\end{subequations} 
	Let $\frac{\partial}{\partial x_{k}} u_{h}$ and $\frac{\partial}{\partial x_{j}} {\upsilon}$ be the macroscopic derivatives of the micro-displacement and the micro-temperature. On the other hand, let $u_{h,k}$ and ${\upsilon}_{,j}$ be the microscopic derivative of the micro-displacement and the micro-temperature, respectively, which are involved in the formula
	\begin{subequations}
		\begin{align}
			&\frac{D}{Dx_{k}} {\boldsymbol{u}}\Big (\boldsymbol{x},\boldsymbol{\xi} =\frac{\boldsymbol{x}}{\varepsilon},t\Big )= \Big [\frac{\partial {u}_{h}(\boldsymbol{x},\boldsymbol{\xi},t)}{\partial x_{k}} + \frac{\partial {u}_{h}(\boldsymbol{x},\boldsymbol{\xi},t)}{\partial \xi_{k}} \frac{\partial \xi_{k}}{\partial x_{k}}\Big]\Big \vert _{\boldsymbol{\xi}= \frac{\boldsymbol{x}}{\varepsilon}}=\Big [\frac{\partial}{\partial x_{k}} {u}_{h}(\boldsymbol{x},\boldsymbol{\xi},t)+\frac{1}{\varepsilon}{u}_{h,k}\Big]\Big \vert _{\boldsymbol{\xi}= \frac{\boldsymbol{x}}{\varepsilon}}, 	\label{cr1}\\
			&\frac{D}{Dx_{j}} {\upsilon}\Big (\boldsymbol{x},\boldsymbol{\xi} =\frac{\boldsymbol{x}}{\varepsilon},t\Big)= \Big [\frac{\partial {\upsilon}(\boldsymbol{x},\boldsymbol{\xi},t)}{\partial x_{j}} + \frac{\partial {\upsilon}(\boldsymbol{x},\boldsymbol{\xi},t)}{\partial \xi_{j}} \frac{\partial \xi_{j}}{\partial x_{j}}\Big]\Big \vert _{\boldsymbol{\xi}= \frac{\boldsymbol{x}}{\varepsilon}}=\Big [\frac{\partial}{\partial x_{j}} {\upsilon}(\boldsymbol{x},\boldsymbol{\xi},t)+\frac{1}{\varepsilon}{\upsilon}_{,j}\Big]\Big \vert _{\boldsymbol{\xi}= \frac{\boldsymbol{x}}{\varepsilon}}. 	\label{cr2}
		\end{align}
	\end{subequations} 
	Applying the derivative rules \eqref{cr1}-\eqref{cr2} to the asymptotic expansions  \eqref{at1}-\eqref{at2} derives
	\begin{subequations}
		\begin{align}
			&\frac{D}{Dx_{k}} {\boldsymbol{u}}\Big (\boldsymbol{x},\boldsymbol{\xi} = \frac{\boldsymbol{x}}{\varepsilon},t\Big )=\Big [ \frac{\partial {u}^{(0)}_{h}}{\partial x_{k}} + \varepsilon \frac{\partial {u}^{(1)}_{h}}{\partial x_{k}} + \varepsilon^{2} \frac{\partial {u}^{(2)}_{h}}{\partial x_{k}}+...\Big ]+\frac{1}{\varepsilon} \Big [{u}^{(0)}_{h,k}+ \varepsilon {u}^{(1)}_{h,k} + \varepsilon^{2} {u}^{(2)}_{h,k}+...\Big ]\Big \vert _{\boldsymbol{\xi}= \frac{\boldsymbol{x}}{\varepsilon}}, 	\label{ep1}\\
			&\frac{D}{Dx_{j}} {\upsilon}\Big (\boldsymbol{x},\boldsymbol{\xi} = \frac{\boldsymbol{x}}{\varepsilon},t\Big )=\Big [ \frac{\partial {\upsilon}^{(0)}}{\partial x_{j}} + \varepsilon \frac{\partial {\upsilon}^{(1)}}{\partial x_{j}} + \varepsilon^{2} \frac{\partial {\upsilon}^{(2)}}{\partial x_{j}}+...\Big ]+\frac{1}{\varepsilon} \Big [{\upsilon}^{(0)}_{,j}+ \varepsilon {\upsilon}^{(1)}_{,j} + \varepsilon^{2} {\upsilon}^{(2)}_{,j}+...\Big ]\Big \vert _{\boldsymbol{\xi}= \frac{\boldsymbol{x}}{\varepsilon}}.\label{ep2}
		\end{align}
	\end{subequations}
	Introducing the asymptotic expansions \eqref{at1}-\eqref{at2} and the rules \eqref{ep1}-\eqref{ep2} into the field equations \eqref{EBTC3}-\eqref{EBTC5}, the regroupment of the terms with equal power $\varepsilon$ yields the asymptotic field equations
	\begin{subequations}
		\begin{align}
			&\varepsilon^{-2}\Big ({C}^{m}_{ijhk}{u}^{(0)}_{h,k} \Big )_{,j}+\varepsilon^{-1} \Big [ \Big( {C}^{m}_{ijhk}\Big ( \frac{\partial {u}^{(0)}_{h}}{\partial x_{k}}+{u}^{(1)}_{h,k} \Big) \Big)_{,j}+\frac{\partial}{\partial x_{j}} \Big ({C}^{m}_{ijhk}{u}^{(0)}_{h,k} \Big )-(\alpha_{ij}^{m}\upsilon^{(0)}+\alpha_{ij}^{(m,1)}\dot{\upsilon}^{(0)})_{,j}\Big ]+\label{fea1}\\
			&+ \varepsilon^{0} \Big [ \Big( {C}^{m}_{ijhk}\Big ( \frac{\partial {u}^{(1)}_{h}}{\partial x_{k}}+{u}^{(2)}_{h,k} \Big) \Big)_{,j}+\frac{\partial}{\partial x_{j}} \Big({C}^{m}_{ijhk}\Big(\frac{\partial{u}^{(0)}_{h}}{\partial x_{k}}+{u}^{(1)}_{h,k}\Big)\Big)-(\alpha_{ij}^{m}\upsilon^{(1)}+\alpha_{ij}^{(m,1)}\dot{\upsilon}^{(1)})_{,j}+\nonumber \\
			&-\frac{\partial}{\partial x_{j}}(\alpha_{ij}^{m}\upsilon^{(0)}+\alpha_{ij}^{(m,1)}\dot{\upsilon}^{(0)})-\rho^{m}\ddot{u}^{(0)}_{i}+b_{i}\Big ]+\nonumber\\
			&+\varepsilon \Big [ \Big( {C}^{m}_{ijhk}\Big (\frac{\partial {u}^{(2)}_{h}}{\partial x_{k}}+{u}^{(3)}_{h,k} \Big) \Big)_{,j}+\frac{\partial}{\partial x_{j}} \Big ({C}^{m}_{ijhk}\Big (\frac{\partial{u}^{(1)}_{h}}{\partial x_{k}}+{u}^{(2)}_{h,k}\Big )\Big)-(\alpha_{ij}^{m}\upsilon^{(2)}+\alpha_{ij}^{(m,1)}\dot{\upsilon}^{(2)})_{,j}+\nonumber\\	
			&-\frac{\partial}{\partial x_{j}}(\alpha_{ij}^{m}\upsilon^{(1)}+\alpha_{ij}^{(m,1)}\dot{\upsilon}^{(1)})-\rho^{m}\ddot{u}^{(1)}_{i}+\textrm{O}(\varepsilon^{2})\Big ] \Big \vert_{\boldsymbol{\xi}= \frac{\boldsymbol{x}}{\varepsilon}}=0,\nonumber\\	
			&\varepsilon^{-2}\Big ({K}^{m}_{ij}{\upsilon}^{(0)}_{,j} \Big )_{,i}+\varepsilon^{-1} \Big [ \Big( {K}^{m}_{ij}\Big ( \frac{\partial {\upsilon}^{(0)}}{\partial x_{j}}+{\upsilon}^{(1)}_{,j} \Big) \Big)_{,i}+\frac{\partial}{\partial x_{i}} \Big ({K}^{m}_{ij}{\upsilon}^{(0)}_{,j} \Big )-(\alpha_{ij}^{m}\dot{u}^{(0)})\Big ]+\label{fea2}\\
			&+ \Big [ \Big( {K}^{m}_{ij}\Big ( \frac{\partial {\upsilon}^{(1)}}{\partial x_{j}}+{\upsilon}^{(2)}_{,j} \Big) \Big)_{,i}+\frac{\partial}{\partial x_{i}} \Big({K}^{m}_{ij}\Big(\frac{\partial{\upsilon}^{(0)}}{\partial x_{j}}+{\upsilon}^{(1)}_{,j}\Big)\Big)-p^{m}\dot{\upsilon}^{(0)}-p^{(m,0)}\ddot{\upsilon}^{(0)}-\alpha_{ij}^{m}\Big(\frac{\partial \dot{u}^{(0)}_{i}}{\partial x_{j}}+\dot{u}^{(1)}_{i,j}\Big)+r\Big ]+\nonumber\\
			&+\varepsilon \Big [ \Big( {K}^{m}_{ij}\Big (\frac{\partial {\upsilon}^{(2)}}{\partial x_{j}}+{\upsilon}^{(3)}_{,j} \Big) \Big)_{,i}+\frac{\partial}{\partial x_{i}} \Big ({K}^{m}_{ij}\Big (\frac{\partial{\upsilon}^{(1)}}{\partial x_{j}}+{\upsilon}^{(2)}_{,j}\Big )\Big)-p^{m}\dot{\upsilon}^{(1)}-p^{(m,0)}\ddot{\upsilon}^{(1)}+ \nonumber\\
			&-\alpha_{ij}^{m}\Big(\frac{\partial \dot{u}^{(1)}_{i}}{\partial x_{j}}+\dot{u}^{(2)}_{i,j}\Big)+\textrm{O}(\varepsilon^{2})\Big ] \Big \vert_{\boldsymbol{\xi}= \frac{\boldsymbol{x}}{\varepsilon}}=0.\nonumber	
		\end{align}
	\end{subequations}
	Interface conditions \eqref{icld1}-\eqref{icld4} are rephrased with respect to the fast variable $\boldsymbol{\xi}$ since the micro-displacement $u_{h}(\boldsymbol{x},\boldsymbol{\xi})$ and the micro-temperature $\upsilon(\boldsymbol{x},\boldsymbol{\xi})$  are supposed to be $\mathcal{Q}-$periodic with respect to $\boldsymbol{\xi}$ and smooth in the slow variable $\boldsymbol{x}$. Denoting with $\Sigma_{1}$ the interface between two phases in the unit cell $\mathcal{Q}$ and assuming the asymptotic expansions \eqref{at1}-\eqref{at2} related to the micro-displacement and the micro-temperature, interface conditions \eqref{icld1}-\eqref{icld4} become
	\begin{subequations}
		\begin{align}
			&\Big [\Big[u^{(0)}_{h}\Big]\Big]\Big \vert_{\boldsymbol{\xi} \in \Sigma_{1}}+\varepsilon \Big[\Big[u^{(1)}_{h}\Big ]\Big]\Big \vert_{\boldsymbol{\xi} \in \Sigma_{1}}+\varepsilon^{2}\Big [\Big [u^{(2)}_{h}\Big]\Big]\Big \vert_{\boldsymbol{\xi} \in \Sigma_{1}}+\textrm{O}(\varepsilon^{3})=0,	\label{condinteq1}\\
			&\Big [\Big[\upsilon^{(0)}\Big]\Big]\Big \vert_{\boldsymbol{\xi} \in \Sigma_{1}}+\varepsilon \Big[\Big[\upsilon^{(1)}\Big ]\Big]\Big \vert_{\boldsymbol{\xi} \in \Sigma_{1}}+\varepsilon^{2}\Big [\Big [\upsilon^{(2)}\Big]\Big]\Big \vert_{\boldsymbol{\xi} \in \Sigma_{1}}+\textrm{O}(\varepsilon^{3})=0,	\label{condinteq2}\\
			&\frac{1}{\varepsilon}\Big [\Big[\Big({C}^{m}_{ijhk}{u}^{(0)}_{h,k}\Big)n_{j}\Big]\Big ]\Big \vert_{\boldsymbol{\xi} \in \Sigma_{1}}+\varepsilon^{0}\Big [\Big[ \Big ( {C}^{m}_{ijhk}\Big (\frac{\partial{u}^{(0)}_{h}}{\partial x_{k}}+{u}^{(1)}_{h,k}\Big)-\alpha_{ij}^{m}\upsilon^{(0)}-\alpha_{ij}^{(m,1)}\dot{\upsilon}^{(0)} \Big )n_{j}\Big ]\Big]\Big \vert_{\boldsymbol{\xi} \in \Sigma_{1}}+  \\
			&+\varepsilon\Big [\Big[ \Big ( {C}^{m}_{ijhk}\Big (\frac{\partial{u}^{(1)}_{h}}{\partial x_{k}}+{u}^{(2)}_{h,k}\Big)-\alpha_{ij}^{m}\upsilon^{(1)}-\alpha_{ij}^{(m,1)}\dot{\upsilon}^{(1)} \Big )n_{j}\Big ]\Big]\Big \vert_{\boldsymbol{\xi} \in \Sigma_{1}}+ \nonumber \\
			&+\varepsilon^{2}\Big [\Big[ \Big ( {C}^{m}_{ijhk}\Big (\frac{\partial{u}^{(2)}_{h}}{\partial x_{k}}+{u}^{(3)}_{h,k}\Big)-\alpha_{ij}^{m}\upsilon^{(2)}-\alpha_{ij}^{(m,1)}\dot{\upsilon}^{(2)} \Big )n_{j}\Big ]\Big]\Big \vert_{\boldsymbol{\xi} \in \Sigma_{1}}+\textrm{O}(\varepsilon^{3})=0, \nonumber \\
			&\frac{1}{\varepsilon}\Big [\Big[\Big({K}^{m}_{ij}{\upsilon}^{(0)}_{,j}\Big)n_{i}\Big]\Big ]\Big \vert_{\boldsymbol{\xi} \in \Sigma_{1}}+\varepsilon^{0}\Big [\Big[ \Big ( {K}^{m}_{ij}\Big (\frac{\partial{\upsilon}^{(0)}}{\partial x_{j}}+{\upsilon}^{(1)}_{,j}\Big) \Big )n_{i}\Big ]\Big]\Big \vert_{\boldsymbol{\xi} \in \Sigma_{1}}+\\
			&\varepsilon\Big [\Big[ \Big ( {K}^{m}_{ij}\Big (\frac{\partial{\upsilon}^{(1)}}{\partial x_{j}}+{\upsilon}^{(2)}_{,j}\Big) \Big )n_{i}\Big ]\Big]\Big \vert_{\boldsymbol{\xi} \in \Sigma_{1}}+ \varepsilon^{2}\Big [\Big[ \Big ( {K}^{m}_{ij}\Big (\frac{\partial{\upsilon}^{(2)}_{h}}{\partial x_{j}}+{\upsilon}^{(3)}_{,j}\Big)\Big )n_{i}\Big ]\Big]\Big \vert_{\boldsymbol{\xi} \in \Sigma_{1}}+\textrm{O}(\varepsilon^{3})=0. \nonumber
		\end{align}
	\end{subequations} 
	Equations \eqref{fea1}-\eqref{fea2} can be briefly written as
	\begin{subequations}
		\begin{align}
			\label{BE1}
			&\varepsilon^{-2}f_{i}^{(0)}(\boldsymbol{x},t)+\varepsilon^{-1}f_{i}^{(1)}(\boldsymbol{x},t)+\varepsilon^{0}f_{i}^{(2)}(\boldsymbol{x},t)+\varepsilon f_{i}^{(3)}(\boldsymbol{x},t)+...+\varepsilon^{l}f_{i}^{(l+2)}(\boldsymbol{x},t)+O(\varepsilon^{l+1})+b_{i}(\boldsymbol{x},t)=0,\\
			\label{BE2}
			&\varepsilon^{-2}g^{(0)}(\boldsymbol{x},t)+\varepsilon^{-1}g^{(1)}(\boldsymbol{x},t)+\varepsilon^{0}g^{(2)}(\boldsymbol{x},t)+\varepsilon g^{(3)}(\boldsymbol{x},t)+...+\varepsilon^{l}g^{(l+2)}(\boldsymbol{x},t)+O(\varepsilon^{l+1})+r(\boldsymbol{x},t)=0,	
		\end{align}
	\end{subequations} 
	where the functions $f_{i}^{(r)}(\boldsymbol{x},t)$ and $g^{(r)}(\boldsymbol{x},t)$ rely on the slow variable $\boldsymbol{x}$ and they can be determined by imposing the solvability conditions within the class of the $\mathcal{Q}-$periodic functions and $r$ is such that $r=0,1,...,l+2$ with $l\in \mathbb{N}$. 		
	\subsection{Solutions of recursive mechanical and thermal differential problems}
	\label{recdif}
	The equations \eqref{BE1}-\eqref{BE2} can identify several recursive differential problems according to a sequential order of $\varepsilon$, which enable to derive the solutions $u^{(0)}_{h}$, $u^{(1)}_{h}$,...,$\upsilon^{(0)}$, $\upsilon^{(1)}$. Specifically, at the order $\varepsilon^{-2}$, the differential problems are
	\begin{subequations}
		\begin{align}
			&\Big(C^{m}_{ijhk}u^{(0)}_{h,k}\Big)_{,j}=f^{(0)}_{i}(\boldsymbol{x}),\label{eqn:e-2}\\
			&\Big(K^{m}_{ij}\upsilon^{(0)}_{,j}\Big)_{,i}=g^{(0)}(\boldsymbol{x}),\label{eqn1:e-2}
		\end{align}
	\end{subequations}  
	with interface conditions
	\begin{subequations}
		\begin{align}
			&\Big[\Big[u^{(0)}_{h}\Big]\Big]\Big\vert_{\boldsymbol{\xi} \in \Sigma_{1}}=0, \quad\Big[\Big[\Big(C^{m}_{ijhk}u^{(0)}_{h,k}\Big)n_{j}\Big]\Big]\Big\vert_{\boldsymbol{\xi} \in \Sigma_{1}}=0,\\
			&\Big[\Big[\upsilon^{(0)}\Big]\Big]\Big\vert_{\boldsymbol{\xi} \in \Sigma_{1}}=0, \quad\Big[\Big[\Big(K^{m}_{ij}\upsilon^{(0)}_{,j}\Big)n_{i}\Big]\Big]\Big\vert_{\boldsymbol{\xi} \in \Sigma_{1}}=0.
		\end{align}
	\end{subequations}
	The solvability condition of the differential problems \eqref{eqn:e-2}-\eqref{eqn1:e-2}, within the class of $\mathcal{Q}-$periodic solutions $u^{(0)}_{h}$ and $\upsilon^{(0)}$, entails that $f^{(0)}_{i}(\boldsymbol{x})=0$ and $g^{(0)}(\boldsymbol{x})=0$. Then, the problems \eqref{eqn:e-2}-\eqref{eqn1:e-2} become
	\begin{subequations}
		\begin{align}
			&\Big(C^{m}_{ijhk}u^{(0)}_{h,k}\Big )_{,j}=0,	\label{eqn:sol}\\
			&\Big(K^{m}_{ij}\upsilon^{(0)}_{,j}\Big)_{,i}=0,\label{eqn1:sol}
		\end{align}
	\end{subequations} 
	whose solutions are
	\begin{subequations}
		\begin{align}
			&u^{(0)}_{h}(\boldsymbol{x},\boldsymbol{\xi},t) = U^{M}_{h}(\boldsymbol{x},t),	\label{eqn:so_l}\\
			&\upsilon^{(0)}(\boldsymbol{x},\boldsymbol{\xi},t) = \Upsilon^{M}(\boldsymbol{x},t),	\label{eqn1:so_l}
		\end{align}
	\end{subequations} 
	where $U^{M}_{h}(\boldsymbol{x},t)$ is the macroscopic displacement field and $\Upsilon^{M}(\boldsymbol{x},t)$ is the macroscopic temperature field that are not subject to the fast variable.\\
	Considering the solutions \eqref{eqn:so_l}-\eqref{eqn1:so_l} and the derivatives $U^{M}_{h,k}=0$, $\Upsilon^{M}_{,j}=0$, the differential problems at the order $\varepsilon^{-1}$ are
	\begin{subequations}
		\begin{align}
			&\Big (C^{m}_{ijhk}u^{(1)}_{h,k}\Big)_{,j}+C^{m}_{ijhk,j}\frac{\partial U^{M}_{h}}{\partial x_{k}}-\alpha_{ij,j}^{m}\Upsilon^{M}-\alpha_{ij,j}^{(m,1)}\dot{\Upsilon}^{M}=f^{(1)}_{i}(\boldsymbol{x}).	\label{eqn:e1n}\\
			&\Big (K^{m}_{ij}\upsilon^{(1)}_{,j}\Big)_{,i}+K^{m}_{ij,i}\frac{\partial \Upsilon^{M}}{\partial x_{j}}=g^{(1)}(\boldsymbol{x})	\label{eqn1:e1n}.
		\end{align}
	\end{subequations}  
	The interface conditions related to problems \eqref{eqn:e1n}-\eqref{eqn1:e1n} are
	\begin{subequations}
		\begin{align}
			&\Big [\Big[u^{(1)}_{h}\Big]\Big]\Big\vert_{\boldsymbol{\xi}\in \Sigma_{1}}=0, \quad \Big [\Big[ \Big ( {C}^{m}_{ijhk}\Big (\frac{\partial{U}^{M}_{h}}{\partial x_{k}}+{u}^{(1)}_{h,k}\Big)-\alpha_{ij}^{m}\Upsilon^{M}-\alpha_{ij}^{(m,1)}\dot{\Upsilon}^{M}) \Big )n_{j}\Big ]\Big]\Big \vert_{\boldsymbol{\xi} \in \Sigma_{1}}=0.\\
			&\Big [\Big[\upsilon^{(1)}\Big]\Big]\Big\vert_{\boldsymbol{\xi}\in \Sigma_{1}}=0, \quad \Big [\Big[ \Big ( {K}^{m}_{ij}\Big (\frac{\partial{\Upsilon}^{M}}{\partial x_{j}}+{\upsilon}^{(1)}_{,j}\Big) \Big )n_{i}\Big ]\Big]\Big \vert_{\boldsymbol{\xi} \in \Sigma_{1}}=0.
		\end{align}
	\end{subequations} 
	The solvability condition within the class of $\mathcal{Q}-$periodic functions implies that  
	\begin{subequations}
		\begin{align}
			&f^{(1)}_{i}(\boldsymbol{x})=\langle C^{m}_{ijhk,j} \rangle \frac{\partial {U}^{M}_{h}}{\partial x_{k}}-\langle \alpha^{m}_{ij,j} \rangle \Upsilon^{M}-\langle \alpha_{ij,j}^{(m,1)} \rangle\dot{\Upsilon}^{M},	\label{eq:e1cs}\\
			&g^{(1)}(\boldsymbol{x})=\langle K^{m}_{ij,i} \rangle \frac{\partial {\Upsilon}^{M}}{\partial x_{j}},	\label{eq1:e1cs}
		\end{align}
	\end{subequations}  
	where $\langle (\cdot) \rangle=\frac{1}{|\mathcal{Q}|} \int_{\mathcal{Q}}^{} (\cdot) d\boldsymbol{\xi}$ and  $|\mathcal{Q}|= \delta$ denotes the mean value over the unit cell $\mathcal{Q}$. Similarly, the $\mathcal{Q}$-periodicity of the components $C^{m}_{ijhk}$, $\alpha^{m}_{ij}$, $\alpha^{(m,1)}_{ij}$, $K^{m}_{ij}$ and the divergence theorem imply that $f^{(1)}_{i}(\boldsymbol{x})=0$ and $g^{(1)}(\boldsymbol{x})=0$ and so, the differential problems \eqref{eqn:e1n}-\eqref{eqn1:e1n} become
	\begin{subequations}
		\begin{align}
			&\Big (C^{m}_{ijhk}u^{(1)}_{h,k}\Big)_{,j}+C^{m}_{ijhk,j}\frac{\partial U^{M}_{h}}{\partial x_{k}}-\alpha_{ij,j}^{m}\Upsilon^{M}-\alpha_{ij,j}^{(m,1)} \dot{\Upsilon}^{M}=0, \quad \forall \frac{\partial U^{M}_{h}}{\partial x_{k}},\Upsilon^{M}	\label{eqn:e1n1}\\
			&\Big (K^{m}_{ij}\upsilon^{(1)}_{,j}\Big)_{,i}+K^{m}_{ij,i}\frac{\partial \Upsilon^{M}}{\partial x_{j}}=0, \quad \forall \frac{\partial {\Upsilon}^{M}}{\partial x_{j}}	\label{eqn1:e1n1}.
		\end{align}
	\end{subequations} 
	The solutions of the problems \eqref{eqn:e1n1}-\eqref{eqn1:e1n1} are
	\begin{subequations}
		\begin{align}
			&u^{(1)}_{h}(\boldsymbol{x},\boldsymbol{\xi},t)=N^{(1)}_{hpq_{1}}(\boldsymbol{\xi})\frac{\partial U^{M}_{p}}{\partial x_{q_{1}}}+\tilde{N}^{(1)}_{h}(\boldsymbol{\xi})\Upsilon^{M}+{\tilde{N}}^{(1,1)}_{h}(\boldsymbol{\xi})\dot{\Upsilon}^{M},	\label{eqn:solp1}\\
			&\upsilon^{(1)}(\boldsymbol{x},\boldsymbol{\xi},t)=M^{(1)}_{q_{1}}(\boldsymbol{\xi})\frac{\partial \Upsilon^{M}}{\partial x_{q_{1}}},	\label{eqn1:solp1}
		\end{align}
	\end{subequations}
	where $N^{(1)}_{hpq_{1}}$, $\tilde{N}^{(1)}_{h}$, ${\tilde{N}}^{(1,1)}_{h}$ and $M^{(1)}_{q_{1}}$ are the perturbation functions that depend on the fast variable $\boldsymbol{\xi}$. The perturbation functions have zero mean over the unit cell $\mathcal{Q}$, then $N^{(1)}_{hpq_{1}}$, $\tilde{N}^{(1)}_{h}$, ${\tilde{N}}^{(1,1)}_{h}$ and $M^{(1)}_{q_{1}}$ fulfill the normalization conditions
	\begin{subequations}
		\begin{align}
			& \langle N^{(1)}_{hpq_{1}} \rangle = \frac{1}{|\mathcal{Q}|} \int_{\mathcal{Q}} N^{(1)}_{hpq_{1}}(\boldsymbol{\xi}) d\boldsymbol{\xi}=0,\\
			& \langle \tilde{N}^{(1)}_{h} \rangle = \frac{1}{|\mathcal{Q}|} \int_{\mathcal{Q}} \tilde{N}^{(1)}_{h}(\boldsymbol{\xi}) d\boldsymbol{\xi}=0,\\
			& \langle {\tilde{N}}^{(1,1)}_{h} \rangle = \frac{1}{|\mathcal{Q}|} \int_{\mathcal{Q}} {\tilde{N}}^{(1,1)}_{h}(\boldsymbol{\xi}) d\boldsymbol{\xi}=0,\\
			& \langle M^{(1)}_{q_{1}} \rangle = \frac{1}{|\mathcal{Q}|} \int_{\mathcal{Q}} M^{(1)}_{q_{1}}(\boldsymbol{\xi}) d\boldsymbol{\xi}=0.
		\end{align}
	\end{subequations} 
	It is known that the perturbation functions are affected by the geometry and  the mechanical properties of the microstructure.
	The differential problems at the order $\varepsilon^{0}$ are 
	\begin{subequations}
		\begin{align}
			&\Big( {C}^{m}_{ijhk}\Big ( \frac{\partial {u}^{(1)}_{h}}{\partial x_{k}}+{u}^{(2)}_{h,k} \Big) \Big)_{,j}+\frac{\partial}{\partial x_{j}} \Big({C}^{m}_{ijhk}\Big(\frac{\partial{u}^{(0)}_{h}}{\partial x_{k}}+{u}^{(1)}_{h,k}\Big)\Big)-(\alpha_{ij}^{m}\upsilon^{(1)}+\alpha_{ij}^{(m,1)}\dot{\upsilon}^{(1)})_{,j}+\label{eqn1:e}\\
			&-\frac{\partial}{\partial x_{j}}(\alpha_{ij}^{m}\upsilon^{(0)}+\alpha_{ij}^{(m,1)}\dot{\upsilon}^{(0)})-\rho^{m}\ddot{u}^{(0)}_{i}=f^{(2)}_{i}(\boldsymbol{x}),\nonumber\\
			&\Big( {K}^{m}_{ij}\Big ( \frac{\partial {\upsilon}^{(1)}}{\partial x_{j}}+{\upsilon}^{(2)}_{,j} \Big) \Big)_{,i}+\frac{\partial}{\partial x_{i}} \Big({K}^{m}_{ij}\Big(\frac{\partial{\upsilon}^{(0)}}{\partial x_{j}}+{\upsilon}^{(1)}_{,j}\Big)\Big)-p^{m}\dot{\upsilon}^{(0)}-p^{(m,0)}\ddot{\upsilon}^{(0)}-\alpha_{ij}^{m}\Big(\frac{\partial \dot{u}^{(0)}_{i}}{\partial x_{j}}+\dot{u}^{(1)}_{i,j}\Big)=g^{(2)}(\boldsymbol{x}).\label{eqn2:e}
		\end{align}
	\end{subequations}
	Replacing the solutions at the orders $\varepsilon^{-1}$ and $\varepsilon^{-2}$ \eqref{eqn:so_l}, \eqref{eqn1:so_l}, \eqref{eqn:solp1}, \eqref{eqn1:solp1} into the equations \eqref{eqn1:e}-\eqref{eqn2:e} leads to
	\begin{subequations}
		\begin{align}
			&\Big (C^{m}_{ijhk}u^{(2)}_{h,k}\Big)_{,j}+\Big(\Big(C^{m}_{ijhk}N^{(1)}_{hpq_{1}}\Big)_{,j}+C^{m}_{iq_{1}pk}+\Big (C^{m}_{ikhj}N^{(1)}_{hpq_{1},j}\Big)\Big)\frac{\partial^{2}U^{M}_{p}}{\partial x_{q_{1}}\partial x_{k}}-\rho^{m}\ddot{U}^{M}_{i}+, 	\label{eqn:prob0}\\
			&+\Big(\Big(C^{m}_{ijhk}\tilde{N}^{(1)}_{h}\Big)_{,j}+C^{m}_{ikhj}\tilde{N}^{(1)}_{h,j}-\Big (\alpha^{m}_{ij}M^{(1)}_{k}\Big)_{,j}-\alpha_{ik}^{m}\Big)\frac{\partial \Upsilon^{M}}{\partial x_{k}}+\nonumber\\
			&+\Big(\Big(C^{m}_{ijhk}{\tilde{N}}^{(1,1)}_{h}\Big)_{,j}+C^{m}_{ikhj}{\tilde{N}}^{(1,1)}_{h,j}-\Big (\alpha^{(m,1)}_{ij}M^{(1)}_{k}\Big)_{,j}-\alpha_{ik}^{(m,1)}\Big)\frac{\partial \dot{\Upsilon}^{M}}{\partial x_{k}}= f^{(2)}_{i}(\boldsymbol{x}),\nonumber\\
			&\Big (K^{m}_{ij}\upsilon^{(2)}_{,j}\Big)_{,i}+\Big(\Big(K^{m}_{ij}M^{(1)}_{q_{1}}\Big)_{,i}+K^{m}_{q_{1}j}+\Big (K^{m}_{ji}M^{(1)}_{q_{1},i}\Big)\Big)\frac{\partial^{2}\Upsilon^{M}}{\partial x_{q_{1}}\partial x_{j}}-\Big(\alpha^{m}_{ij}N^{(1)}_{ipq_{1},j}+\alpha_{pq_{1}}^{m}\Big)\frac{\partial \dot{U}^{M}_{p}}{\partial x_{q_{1}}}+ 	\label{eqn1:prob0}\\
			&-(p^{m}+\alpha_{ij}^{m}\tilde{N}^{(1)}_{i,j})\dot{\Upsilon}^{M}-(p^{(m,0)}
			+\alpha_{ij}^{m}{\tilde{N}}^{(1,1)}_{i,j})\ddot{\Upsilon}^{M}=g^{(2)}(\boldsymbol{x}),\nonumber
		\end{align}
	\end{subequations} 
	whose interface conditions are
	\begin{subequations}
		\begin{align}
			&\Big[\Big[u^{(2)}_{h}\Big]\Big]\Big\vert_{\boldsymbol{\xi} \in \Sigma_{1}}=0,\\ &\Big[\Big[\upsilon^{(2)}\Big]\Big]\Big\vert_{\boldsymbol{\xi} \in \Sigma_{1}}=0,\\
			& \Big [\Big[ \Big (C^{m}_{ijhk}\Big (u^{(2)}_{h,k}+N^{(1)}_{hpq_{1}}\frac{\partial^{2} U^{M}_{p}}{\partial x_{q_{1}}\partial x_{k}}+\tilde{N}^{(1)}_{h}\frac{\partial \Upsilon^{M}}{\partial x_{k}}+{\tilde{N}}^{(1,1)}_{h}\frac{\partial \dot{\Upsilon}^{M}}{\partial x_{k}}\Big)+\nonumber\\
			& -\alpha_{ij}^{m}M^{(1)}_{k}\frac{\partial \Upsilon^{M}}{\partial x_{k}} -\alpha_{ij}^{(m,1)}M^{(1)}_{k}\frac{\partial \dot{\Upsilon}^{M}}{\partial x_{k}}\Big)n_{j}\Big ]\Big]\Big \vert_{\boldsymbol{\xi} \in \Sigma_{1}}=0,\nonumber \\
			&\Big [\Big[ \Big ( {K}^{m}_{ij}\Big (\upsilon_{,j}^{(2)}+M^{(1)}_{q_{1}}\frac{\partial^{2}{\Upsilon}^{M}}{\partial x_{q_{1}}\partial x_{j}}\Big) \Big )n_{i}\Big ]\Big]\Big \vert_{\boldsymbol{\xi} \in \Sigma_{1}}=0.
		\end{align}
	\end{subequations}
	The solvability condition of differential problems \eqref{eqn:prob0}-\eqref{eqn1:prob0} whitin the class of $\mathcal{Q}-$periodic functions and the divergence theorem enables to obtain 
	\begin{subequations}
		\begin{align}
			f^{(2)}_{i}(\boldsymbol{x})&= \langle C^{m}_{iq_{1}pk}+C^{m}_{ikhj}N^{(1)}_{hpq_{1},j}\rangle\frac{\partial^{2}U^{M}_{p}}{\partial x_{q_{1}}\partial x_{k}}+\langle C^{m}_{ikhj}\tilde{N}^{(1)}_{h,j}-\alpha_{ik}^{m}\rangle\frac{\partial \Upsilon^{M}}{\partial x_{k}}+\\
			&+\langle C^{m}_{ikhj}{\tilde{N}}^{(1,1)}_{h,j}-\alpha_{ik}^{(m,1)}\rangle\frac{\partial \dot{\Upsilon}^{M}}{\partial x_{k}}-\langle\rho^{m}\rangle \ddot{U}^{M}_{i},\nonumber \\
			g^{(2)}(\boldsymbol{x})&=\langle K^{m}_{q_{1}j}+ K^{m}_{ji}M^{(1)}_{q_{1},i}\rangle \frac{\partial^{2}\Upsilon^{M}}{\partial x_{q_{1}}\partial x_{j}}-\langle\alpha^{m}_{ij}{N}^{(1)}_{ipq_{1},j}+\alpha_{pq_{1}}^{m}\rangle\frac{\partial \dot{U}^{M}_{p}}{\partial x_{q_{1}}}+\\
			&-\langle p^{m}+\alpha_{ij}^{m}\tilde{N}^{(1)}_{i,j}\rangle \dot{\Upsilon}^{M}-\langle p^{(m,0)}+\alpha_{ij}^{m}{\tilde{N}}^{(1,1)}_{i,j}\rangle \ddot{\Upsilon}^{M}.\nonumber
		\end{align}
	\end{subequations}
	Finally, the solutions of the differential problems at the order $\varepsilon^{0}$ are
	\begin{subequations}
		\begin{align}
			\label{eqn:sol01}
			&u^{(2)}_{h}(\boldsymbol{x},\boldsymbol{\xi},t) = N^{(2)}_{hpq_{1}q_{2}}(\boldsymbol{\xi})\frac{\partial^{2}U^{M}_{p}}{\partial x_{q_{1}}\partial x_{q_{2}}}+\tilde{N}^{(2)}_{hq_{1}}(\boldsymbol{\xi})\frac{\partial \Upsilon^{M}}{\partial x_{q_{1}}}+{\tilde{N}}^{(2,1)}_{hq_{1}}(\boldsymbol{\xi})\frac{\partial \dot{\Upsilon}^{M}}{\partial x_{q_{1}}}+N^{(2,2)}_{hp}(\boldsymbol{\xi}) \ddot{U}^{M}_{p}, \\	\label{eqn:sol0}
			&\upsilon^{(2)}(\boldsymbol{x},\boldsymbol{\xi},t) = M^{(2)}_{q_{1}q_{2}}(\boldsymbol{\xi})\frac{\partial^{2}\Upsilon^{M}}{\partial x_{q_{1}}\partial x_{q_{2}}}+\tilde{M}^{(2,1)}_{pq_{1}}(\boldsymbol{\xi})\frac{\partial \dot{U}^{M}_{p}}{\partial x_{q_{1}}}+M^{(2,1)}(\boldsymbol{\xi})\dot{\Upsilon}^{M}+M^{(2,2)}(\boldsymbol{\xi}) \ddot{\Upsilon}^{M},	
		\end{align}
	\end{subequations} 
	where $N^{(2)}_{hpq_{1}q_{2}}$, $\tilde{N}^{(2)}_{hq_{1}}$, ${\tilde{N}}^{(2,1)}_{hq_{1}}$,  $N^{(2,2)}_{hp}$, $M^{(2)}_{q_{1}q_{2}}$, $\tilde{M}^{(2,1)}_{pq_{1}}$, $M^{(2,1)}$, $M^{(2,2)}$  are the second order perturbation functions.\\
	In Section A of Supplementary material there are the solutions of recursive differential problems at orders $\varepsilon$, $\varepsilon^{2}$ and $\varepsilon^{3}$.
	\subsection{Cell problems and perturbation functions}
	\label{cepr}
	The solutions $u^{(0)}_{h}$, $u^{(1)}_{h}$, $u^{(2)}_{h}$, $\upsilon^{(0)}$, $\upsilon^{(1)}$, $\upsilon^{(2)}$ obtained from the recursive differential problems discussed in Subsection \eqref{recdif} play a crucial role in establishing the cell problems. These cell problems form a set of elliptic differential problems in divergence form, which depend on the perturbation functions. Such perturbation functions are regular and exhibit periodic behavior with respect to $\mathcal{Q}$. Furthermore, the cell problems effectively capture the influence of microstructural heterogeneities and are therefore influenced by the geometric and mechanical properties of the periodic cell.
	Replacing the solutions \eqref{eqn:solp1}-\eqref{eqn1:solp1} into the recursive problems at the order $\varepsilon{^{-1}}$ \eqref{eqn:e1n1}-\eqref{eqn1:e1n1} derives the four cell problems 
	\begin{subequations}
		\begin{align}
			&\Big (C^{m}_{ijhk}N^{(1)}_{hpq_{1},k}\Big)_{,j}+{C}^{m}_{ijpq_{1},j}=0,\label{cps1}\\
			&\Big (C^{m}_{ijhk}\tilde{N}^{(1)}_{h,k}\Big)_{,j}-\alpha^{m}_{ij,j}=0,\label{cps2}\\
			&\Big (C^{m}_{ijhk}\tilde{N}^{(1,1)}_{h,k}\Big)_{,j}-\alpha^{(m,1)}_{ij,j}=0\label{cps3},\\
			&\Big (K^{m}_{ij}M^{(1)}_{q_{1},j}\Big)_{,i}+K^{(m)}_{iq_{1},i}=0,\label{cps4}
		\end{align}
	\end{subequations} 
	having the interface conditions expressed in terms of the perturbation function $N^{(1)}_{hpq_{1},k}$, $\tilde{N}^{(1)}_{h,k}$, $\tilde{N}^{(1,1)}_{h,k}$, $M^{(1)}_{q_{1},j}$ as follows    
	\begin{subequations}
		\begin{align}
			&\Big[\Big[N^{(1)}_{hpq_{1}}\Big]\Big]\Big\vert_{\boldsymbol{\xi}\in \Sigma_{1}}=0, \quad \Big [\Big [\Big ( C^{m}_{ijhk}\Big(N^{(1)}_{hpq_{1},k}+\delta_{hp}\delta_{kq_{1}}\Big )\Big )n_{j}\Big ]\Big ]\Big \vert_{\boldsymbol{\xi} \in \Sigma_{1}}=0,\label{ICCP1}\\
			&\Big[\Big[\tilde{N}^{(1)}_{h}\Big]\Big]\Big\vert_{\boldsymbol{\xi}\in \Sigma_{1}}=0, \quad\Big [\Big [\Big ( C^{m}_{ijhk}\tilde{N}^{(1)}_{h,k}-\alpha^{m}_{ij}\Big )n_{j}\Big ]\Big ]\Big \vert_{\boldsymbol{\xi} \in \Sigma_{1}}=0,\label{ICCP2}\\
			&\Big[\Big[\tilde{N}^{(1,1)}_{h}\Big]\Big]\Big\vert_{\boldsymbol{\xi}\in \Sigma_{1}}=0,\quad\Big [\Big [\Big ( C^{m}_{ijhk}\tilde{N}^{(1,1)}_{h,k}-\alpha^{(m,1)}_{ij}\Big )n_{j}\Big ]\Big ]\Big \vert_{\boldsymbol{\xi} \in \Sigma_{1}}=0,\label{ICCP3}\\
			&\Big[\Big[M^{(1)}_{q_{1}}\Big]\Big]\Big\vert_{\boldsymbol{\xi}\in \Sigma_{1}}=0,\quad\Big [\Big [\Big ( K^{m}_{ij}(M^{(1)}_{q_{1},j}-\delta_{jq_{1}})\Big )n_{i}\Big ]\Big ]\Big \vert_{\boldsymbol{\xi} \in \Sigma_{1}}=0\label{ICCP4}
		\end{align}
	\end{subequations}
	where $\delta_{hp}$, $\delta_{kq_{1}}$, $\delta_{jq_{1}}$ are the Kronecker delta functions. After determining the perturbation functions $N^{(1)}_{hpq_{1},k}$, $\tilde{N}^{(1)}_{h,k}$, $\tilde{N}^{(1,1)}_{h,k}$, $M^{(1)}_{q_{1},j}$, the differential equation \eqref{eqn:prob0} and its solution \eqref{eqn:sol01} derives the cell problems at the order $\varepsilon^{0}$ and their symmetrized form as follows 
	\begin{subequations}
		\begin{align}
			&\Big (C^{m}_{ijhk}N^{(2)}_{hpq_{1}q_{2},k}\Big )_{,j}+\frac{1}{2}\Big [\Big(C^{m}_{ikhq_{2}}N^{(1)}_{hpq_{1}}\Big)_{,k}+C^{m}_{iq_{2}pq_{1}}+C^{m}_{iq_{2}hk}N^{(1)}_{hpq_{1},k}+\Big (C^{m}_{ikhq_{1}}N^{(1)}_{hpq_{2}}\Big )_{,k}+\label{eqn:sym}\\
			&+C^{m}_{iq_{1}pq_{2}}+C^{m}_{iq_{1}hk}N^{(1)}_{hpq_{2},k}\Big ]=\frac{1}{2}\langle C^{m}_{iq_{2}hq_{1}}+C^{m}_{iq_{2}hk}N^{(1)}
			_{hpq_{1},k}+C^{m}_{iq_{1}hq_{2}}+C^{m}_{iq_{1}hk}N^{(1)}_{hpq_{2},k} \rangle, \nonumber\\
			&\Big (C^{m}_{ijhk}\tilde{N}^{(2)}_{hq_{1},k}\Big )_{,j}+\Big(C^{m}_{ijhq_{1}}\tilde{N}^{(1)}_{h}\Big)_{,j}+C^{m}_{iq_{1}hj}\tilde{N}^{(1)}_{h,j}-\Big (\alpha^{m}_{ij}M^{(1)}_{q_{1}}\Big)_{,j}-\alpha_{iq_{1}}^{m}=\langle C^{m}_{iq_{1}hj}\tilde{N}^{(1)}_{h,j}-\alpha_{iq_{1}}^{m}\rangle, \label{31b} \\
			&\Big (C^{m}_{ijhk}\tilde{N}^{(2,1)}_{hq_{1},k}\Big )_{,j}+\Big(C^{m}_{ijhq_{1}}{\tilde{N}}^{(1,1)}_{h}\Big)_{,j}+C^{m}_{iq_{1}hj}{\tilde{N}}^{(1,1)}_{h,j}-\Big (\alpha^{(m,1)}_{ij}M^{(1)}_{q_{1}}\Big)_{,j}-\alpha_{iq_{1}}^{(m,1)}=\label{32b} \\
			&=\langle C^{m}_{iq_{1}hj}{\tilde{N}}^{(1,1)}_{h,j}-\alpha_{iq_{1}}^{(m,1)}\rangle,  \nonumber \\	
			&\Big (C^{m}_{ijhk}N^{(2,2)}_{hp,k}\Big )_{,j}-\rho^{m}\delta_{ip}=-\langle\rho^{m}\rangle\delta_{ip},\label{22N}
		\end{align}
	\end{subequations}
	which are endowed with the interface conditions
	\begin{subequations}
		\begin{align}
			\label{eqn:icsym}
			&	\Big[\Big[N^{(2)}_{hpq_{1}q_{2}}\Big]\Big]\Big\vert_{\boldsymbol{\xi} \in \Sigma_{1}}=0,  \quad
			\Big [\Big[\Big (C^{m}_{ijhk}N^{(2)}_{hpq_{1}q_{2},k}+\frac{1}{2}\Big(C^{m}_{ijhq_{2}}N^{(1)}_{hpq_{1}}+C^{m}_{ijhq_{1}}N^{(1)}_{hpq_{2}}\Big ) \Big ) n_{j}\Big ] \Big ]\Big \vert_{\boldsymbol{\xi} \in \Sigma_{1}}=0,\\
			&	\Big[\Big[\tilde{N}^{(2)}_{hq_{1}}\Big]\Big]\Big\vert_{\boldsymbol{\xi} \in \Sigma_{1}}=0,\quad  \Big[\Big[\Big(C^{m}_{ijhk}(\tilde{N}^{(2)}_{hq_{1},k}+\delta_{kq_{1}}\tilde{N}^{(1)}_{h})-\alpha_{ij}^{m}M_{q_{1}}^{(1)}\Big)n_{j}\Big]\Big]\Big\vert_{\boldsymbol{\xi} \in \Sigma_{1}}=0,\\
			&	\Big[\Big[\tilde{N}^{(2,1)}_{hq_{1}}\Big]\Big]\Big\vert_{\boldsymbol{\xi} \in \Sigma_{1}}=0, \quad \Big[\Big[\Big(C^{m}_{ijhk}(\tilde{N}^{(2,1)}_{hq_{1},k}+\delta_{kq_{1}}\tilde{N}^{(1,1)}_{h})-\alpha_{ij}^{(m,1)}M_{q_{1}}^{(1)}\Big)n_{j}\Big]\Big]\Big\vert_{\boldsymbol{\xi} \in \Sigma_{1}}=0,\\
			&	\Big[\Big[N^{(2,2)}_{hp}\Big]\Big]\Big\vert_{\boldsymbol{\xi} \in \Sigma_{1}}=0,\quad \Big[\Big[\Big (C^{m}_{ijhk}N^{(2,2)}_{hp,k}\Big )n_{j}\Big]\Big]\Big\vert_{\boldsymbol{\xi} \in \Sigma_{1}}=0.
		\end{align}
	\end{subequations}
	On the other hand, replacing the solution $\eqref{eqn:sol0}$ into the problem \eqref{eqn1:prob0} allows to derive the cell problems at the order $\varepsilon^{0}$ and their symmetrized shape as follows 
	\begin{subequations}
		\begin{align}
			\label{33a}
			&\Big(K^{m}_{ij}M^{(2)}_{q_{1}q_{2},j}\Big)_{,i}+\frac{1}{2}\Big [\Big(K^{m}_{iq_{2}}M^{(1)}_{q_{1}}\Big)_{,i}+K^{m}_{q_{1}q_{2}}+K^{m}_{iq_{2}}M^{(1)}
			_{q_{1},i}+\Big(K^{m}_{iq_{1}}M^{(1)}_{q_{2}}\Big)_{,i}+K^{m}_{q_{2}q_{1}}+K^{m}_{iq_{1}}M^{(1)}
			_{q_{2},i}\Big]=\\
			&=\frac{1}{2}\langle K^{m}_{q_{1}q_{2}}+K^{m}_{iq_{2}}M^{(1)}
			_{q_{1},i}+K^{m}_{q_{2}q_{1}}+K^{m}_{iq_{1}}M^{(1)}
			_{q_{2},i}\rangle, \nonumber\\
			&\Big(K^{m}_{ij}\tilde{M}^{(2,1)}_{pq_{1},j}\Big)_{,i}-\alpha^{m}_{ij}N^{(1)}_{ipq_{1},j}-\alpha_{pq_{1}}^{m}=-\langle\alpha^{m}_{ij}{N}^{(1)}_{ipq_{1},j}+\alpha_{pq_{1}}^{m}\rangle,\label{33b} \\
			&\Big(K^{m}_{ij}M^{(2,1)}_{,j}\Big)_{,i}-(p^{m}+\alpha_{ij}^{m}\tilde{N}^{(1)}_{i,j})=-\langle p^{m}+\alpha_{ij}^{m}\tilde{N}^{(1)}_{i,j}\rangle,\label{33c} \\ 
			&\Big(K^{m}_{ij}M^{(2,2)}_{,j}\Big)_{,i}-(p^{(m,0)}+\alpha_{ij}^{m}{\tilde{N}}^{(1,1)}_{i,j})=-\langle p^{(m,0)}+\alpha_{ij}^{m}{\tilde{N}}^{(1,1)}_{i,j}\rangle,\label{33d}
		\end{align}
	\end{subequations}
	which have the interface conditions
	\begin{subequations}
		\begin{align}
			\label{BC1}
			&	\Big[\Big[M^{(2)}_{q_{1}q_{2}}\Big]\Big]\Big\vert_{\boldsymbol{\xi} \in \Sigma_{1}}=0, \quad \Big[\Big[\Big(K^{m}_{ij}M^{(2)}_{q_{1}q_{2},j}+\frac{1}{2}(K^{m}_{iq_{2}}M^{(1)}_{q_{1}}+K^{m}_{iq_{1}}M^{(1)}_{q_{2}})\Big)n_{i}\Big]\Big]\Big\vert_{\boldsymbol{\xi} \in \Sigma_{1}}=0,\\
			&	\Big[\Big[\tilde{M}^{(2,1)}_{pq_{1}}\Big]\Big]\Big\vert_{\boldsymbol{\xi} \in \Sigma_{1}}=0,\quad  \Big[\Big[\Big(K^{m}_{ij}\tilde{M}^{(2,1)}_{pq_{1},j}\Big)n_{i}\Big]\Big]\Big\vert_{\boldsymbol{\xi} \in \Sigma_{1}}=0, \label{BC2}\\
			&	\Big[\Big[M^{(2,1)}\Big]\Big]\Big\vert_{\boldsymbol{\xi} \in \Sigma_{1}}=0,\quad \Big[\Big[\Big(K^{m}_{ij}M^{(2,1)}_{,j}\Big)n_{i}\Big]\Big]\Big\vert_{\boldsymbol{\xi} \in \Sigma_{1}}=0,\\
			&	\Big[\Big[M^{(2,2)}\Big]\Big]\Big\vert_{\boldsymbol{\xi} \in \Sigma_{1}}=0,\quad \Big[\Big[(K^{m}_{ij}M^{(2,2)}_{,j}\Big)n_{i}\Big]\Big]\Big\vert_{\boldsymbol{\xi} \in \Sigma_{1}}=0.
		\end{align}
	\end{subequations}
	\subsection{Down-scaling and up-scaling relations, average field equations of infinite order and macroscopic problems}
	\label{SCR}
	The down-scaling relation related to the micro-displacement and the micro-temperature may be expressed as asymptotic expansion of powers of the microscopic length $\varepsilon$ depending on the macro-displacement $U^{M}_{h}(\boldsymbol{x},s)$, the macro-temperature $\Upsilon^{M}(\boldsymbol{x},s)$, their gradients and the $\mathcal{Q}$-periodic perturbation functions. The functions are determined by solving the cell problems that are displayed in the Subsection \eqref{recdif}. Therefore, replacing the solutions of the recursive differential problems \eqref{eqn:so_l}, \eqref{eqn:solp1}, \eqref{eqn:sol01}, \eqref{eqn1:so_l}, \eqref{eqn1:solp1} and \eqref{eqn:sol0} into the asymptotic expansions \eqref{at1}-\eqref{at2} achieves the micro-displacement $u_{h}(\boldsymbol{x},\boldsymbol{\xi},t)$ and the micro-temperature $\upsilon (\boldsymbol{x},\boldsymbol{\xi},t)$ as
	\begin{subequations}
		\begin{align}
			\label{eqn:dsr}
			u_{h}\Big(\boldsymbol{x},\frac{\boldsymbol{x}}{\varepsilon},t\Big) =&\Big [ U^{M}_{h}(\boldsymbol{x},t)+\varepsilon \Big( N^{(1)}_{hpq_{1}}(\boldsymbol{\xi})\frac{\partial U^{M}_{p}}{\partial x_{q_{1}}}+\tilde{N}^{(1)}_{h}(\boldsymbol{\xi})\Upsilon^{M}+{\tilde{N}}^{(1,1)}_{h}(\boldsymbol{\xi})\dot{\Upsilon}^{M} \Big)+\\
			&+\varepsilon^{2}\Big( N^{(2)}_{hpq_{1}q_{2}}(\boldsymbol{\xi})\frac{\partial^{2}U^{M}_{p}}{\partial x_{q_{1}}\partial x_{q_{2}}}+\tilde{N}^{(2)}_{hq_{1}}(\boldsymbol{\xi})\frac{\partial \Upsilon^{M}}{\partial x_{q_{1}}}+{\tilde{N}}^{(2,1)}_{hq_{1}}(\boldsymbol{\xi})\frac{\partial \dot{\Upsilon}^{M}}{\partial x_{q_{1}}}+N^{(2,2)}_{hp}(\boldsymbol{\xi}) \ddot{U}^{M}_{p} \Big)+\nonumber\\
			&+\textrm{O}(\varepsilon^{3})\Big]\Big\vert _{\boldsymbol{\xi}= \frac{\boldsymbol{x}}{\varepsilon}}, \nonumber \\
			\upsilon\Big(\boldsymbol{x},\frac{\boldsymbol{x}}{\varepsilon},t\Big) =&\Big [\Upsilon^{M}(\boldsymbol{x},t)+\varepsilon M^{(1)}_{q_{1}}(\boldsymbol{\xi})\frac{\partial \Upsilon^{M}}{\partial x_{q_{1}}}+\label{eqn1:dsr}\\
			&+\varepsilon^{2}\Big(M^{(2)}_{q_{1}q_{2}}(\boldsymbol{\xi})\frac{\partial^{2}\Upsilon^{M}}{\partial x_{q_{1}}\partial x_{q_{2}}}+\tilde{M}^{(2,1)}_{pq_{1}}(\boldsymbol{\xi})\frac{\partial \dot{U}^{M}_{p}}{\partial x_{q_{1}}}+M^{(2,1)}(\boldsymbol{\xi})\dot{\Upsilon}^{M}+M^{(2,2)}(\boldsymbol{\xi}) \ddot{\Upsilon}^{M}\Big)+\nonumber\\
			&+\textrm{O}(\varepsilon^{3})\Big]\Big\vert_{\boldsymbol{\xi}= \frac{\boldsymbol{x}}{\varepsilon}},\nonumber
		\end{align}
	\end{subequations} 
	where the macro-displacement $U^{M}_{h}(\boldsymbol{x},t)$ and the macro-temperature $\Upsilon^{M}(\boldsymbol{x},t)$ are $\mathcal{L}$-periodic and depend on the slow variable $\boldsymbol{x}$ and the time. In particular, the macro-displacement and the macro-temperature establish the up-scaling relations connecting the macro fields with the micro fields, which are defined as the mean value of the micro-displacement and the micro-temperature over the unit cell $\mathcal{Q}$  
	\begin{subequations}
		\begin{align}
			\label{upsca}
			&U_{h}^{M}(\boldsymbol{x},t)\dot{=} \Big \langle u_{h}\Big(\boldsymbol{x},\frac{\boldsymbol{x}}{\varepsilon}+\boldsymbol{\zeta},t\Big) \Big \rangle,\\
			&\Upsilon^{M}(\boldsymbol{x},t)\dot{=} \Big \langle \upsilon \Big(\boldsymbol{x},\frac{\boldsymbol{x}}{\varepsilon}+\boldsymbol{\zeta},t\Big) \Big \rangle, \label{upsca1}
		\end{align}
	\end{subequations}
	where the variable $\boldsymbol{\zeta} \in \mathcal{Q}$ recognizes a category of translations of the heterogeneous domain respect to the $\mathcal{L}-$periodic body forces $\boldsymbol{b}(\boldsymbol{x},t)$ \cite{Smyshlyaev2000, Bacigalupo2014}.
	Substituting the down-scaling relations \eqref{eqn:dsr}-\eqref{eqn1:dsr} into the micro-scale field equations \eqref{EBTC3}-\eqref{EBTC3} and ordering the terms with equal powers of $\varepsilon$, the average field equations of infinite order are
	\begin{subequations}
		\begin{align}
			\label{eqn:infeq}
			&n^{(2)}_{ipq_{1}q_{2}}\frac{\partial^{2}U^{M}_{p}}{\partial x_{q_{1}} \partial x_{q_{2}}}-n_{ip}^{(2,2)}\ddot{U}^{M}_{p}-\tilde{n}^{(2,1)}_{iq_{1}}\frac{\partial \dot{\Upsilon}^{M}}{\partial x_{q_{1}}}-\tilde{n}^{(2)}_{iq_{1}}\frac{\partial {\Upsilon}^{M}}{\partial x_{q_{1}}}+\sum_{n=0}^{1}\varepsilon^{n+1}\sum_{|q|=n+3}n_{ipq}^{(n+3)} \frac{\partial^{n+3}U^{M}_{p}}{\partial x_{q}}+\\
			&+\sum_{n=0}^{1}\varepsilon^{n+1}\sum_{|q|=n+1}\tilde{n}_{iq}^{(n+3)} \frac{\partial^{n+2}\Upsilon^{M}}{\partial x_{q}}+\sum_{n=0}^{1}\varepsilon^{n+1}\sum_{|q|=n+2}\tilde{n}_{iq}^{(n+3,1)} \frac{\partial^{n+2}\dot{\Upsilon}^{M}}{\partial x_{q}}+\nonumber\\
			&+\sum_{n=0}^{1}\varepsilon^{n+1}\sum_{|q|=n+1}n_{ipq}^{(n+3,2)} \frac{\partial^{n+1}\ddot{U}^{M}_{p}}{\partial x_{q}}+\sum_{n=0}^{1}\varepsilon^{n+1}\sum_{|q|=n}\tilde{\tilde{n}}_{iq}^{(n+3,2)} \frac{\partial^{n}\ddot{\Upsilon}^{M}}{\partial x_{q}}+\sum_{n=0}^{1}\varepsilon^{n+1}\sum_{|q|=n}\tilde{\tilde{n}}_{iq}^{(n+3,3)} \frac{\partial^{n}\dddot{\Upsilon}^{M}}{\partial x_{q}}+\nonumber\\
			&-\varepsilon^{2}\Big(n_{ipq_{1}q_{2}}^{(4,1)} \frac{\partial^{2}\dot{U}^{M}_{p}}{\partial x_{q_{1}}\partial x_{q_{2}}}+\tilde{n}^{(4,1)}_{iq_{1}}\frac{\partial \dot{\Upsilon}^{M}}{\partial x_{q_{1}}}+n^{(4,4)}_{ip}\ddddot{U}^{M}_{p}\Big)+O(\varepsilon^{3})+b_{i}=0,\nonumber\\
			\label{eqn:infeq2}	
			& m^{(2)}_{q_{1}q_{2}}\frac{\partial^{2}\Upsilon^{M}}{\partial x_{q_{1}}\partial x_{q_{2}}}-\tilde{m}^{(2,1)}_{pq_{1}}\frac{\partial \dot{U}^{M}_{p}}{\partial x_{q_{1}}}-m^{(2,1)}\dot{\Upsilon}^{M}-m^{(2,2)} \ddot{\Upsilon}^{M}+\\
			&+\sum_{n=0}^{1}\varepsilon^{n+1}\sum_{|q|=n+3}m_{q}^{(n+3)} \frac{\partial^{n+3} \Upsilon^{M}}{\partial x_{q}}+\sum_{n=0}^{1}\varepsilon^{n+1}\sum_{|q|=n+2}\tilde{m}_{pq}^{(n+3,1)} \frac{\partial^{n+2}\dot{U}^{M}_{p}}{\partial x_{q}}+\nonumber\\
			&+\sum_{n=0}^{1}\varepsilon^{n+1}\sum_{|q|=n+1}m_{q}^{(n+3,1)} \frac{\partial^{n+1}\dot{\Upsilon}^{M}}{\partial x_{q}}+\sum_{n=0}^{1}\varepsilon^{n+1}\sum_{|q|=n+1}m_{q}^{(n+3,2)} \frac{\partial^{n+1}\ddot{\Upsilon}^{M}}{\partial x_{q}}+\nonumber\\
			&+\sum_{n=0}^{1}\varepsilon^{n+1}\sum_{|q|=n}\tilde{m}_{pq}^{(n+3,3)} \frac{\partial^{n}\dddot{U}^{M}_{p}}{\partial x_{q}}-\varepsilon^{2}\Big(\tilde{m}_{pq_{1}}^{(4,2)} \frac{\partial \ddot{U}^{M}_{p}}{\partial x_{q_{1}}}+\tilde{\tilde{m}}^{(4,2)}\ddot{\Upsilon}^{M}+m^{(4,3)}\dddot{\Upsilon}^{M}+m^{(4,4)}\ddddot{\Upsilon}^{M}\Big)+\nonumber\\
			&+O(\varepsilon^{3})+r=0,\nonumber
		\end{align}
	\end{subequations}
	where $|q|$ is the length of the multi-index and the derivative with respect to $q$ is written as $\frac{\partial^{l}(\cdot)}{\partial x_{q}} = \frac{\partial^{l}(\cdot)}{\partial x_{q_{1}}...x_{q_{l}}}$.
	It can be observed that the coefficients of the gradients of the  macro-displacement and the macro-temperature are the known terms of the corresponding cell problems. Specifically, the coefficients related to the mechanical equation \eqref{eqn:infeq} can be identified as
	\begin{subequations}
		\begin{align}	
			n^{(2)}_{ipq_{1}q_{2}}&=\frac{1}{2}\langle C^{m}_{iq_{2}pq_{1}}+C^{m}_{iq_{2}hj}N^{(1)}
			_{hpq_{1},j}+C^{m}_{iq_{1}pq_{2}}+C^{m}_{iq_{1}hj}N^{(1)}_{hpq_{2},j} \rangle,\\
			n_{ip}^{(2,2)}&=\delta_{ip}\langle \rho^{m}\rangle,\quad \tilde{n}_{iq_{1}}^{(2)}=\langle \alpha_{iq_{1}}^{m}-C^{m}_{iq_{1}hj}\tilde{N}^{(1)}_{h,j}\rangle,\quad \tilde{n}_{iq_{1}}^{(2,1)}=\langle \alpha_{iq_{1}}^{(m,1)}-C^{m}_{iq_{1}hj}{\tilde{N}}^{(1,1)}_{h,j}\rangle, \label{CP}\\
			n_{ipq_{1}...q_{g+2}}^{(g+2)}&=\frac{1}{g+2}\sum_{P^{*}(q)}\langle C^{m}_{iq_{g+2}hj}N^{(g+1)}_{hpq_{1}...q_{g+1},j}+C^{m}_{iq_{g+1}hq_{g+2}}N^{(g)}_{hpq_{1}...q_{g}} \rangle,\\
			\tilde{n}_{iq_{1}q_{2}}^{(3)}&=\frac{1}{2}\langle C^{m}_{iq_{1}hq_{2}}\tilde{N}^{(1)}_{h}+C^{m}_{iq_{2}hq_{1}}\tilde{N}^{(1)}_{h}+C^{m}_{iq_{2}hj}N^{(2)}_{hq_{1},j}+C^{m}_{iq_{1}hj}N^{(2)}_{hq_{2},j}-\alpha^{m}_{iq_{2}}M^{(1)}_{q_{1}}-\alpha^{m}_{iq_{1}}M^{(1)}_{q_{2}}\rangle,\\
			\tilde{n}_{iq_{1}q_{2}}^{(3,1)}&=\frac{1}{2}\langle C^{m}_{iq_{1}hq_{2}}\tilde{N}^{(1,1)}_{h}+C^{m}_{iq_{2}hq_{1}}\tilde{N}^{(1,1)}_{h}+C^{m}_{iq_{2}hj}N^{(2,1)}_{hq_{1},j}+\nonumber\\
			&+C^{m}_{iq_{1}hj}N^{(2,1)}_{hq_{2},j}-\alpha^{(m,1)}_{iq_{2}}M^{(1)}_{q_{1}}-\alpha^{(m,1)}_{iq_{1}}M^{(1)}_{q_{2}}\rangle,\\
			\tilde{n}_{iq_{1}...q_{w+1}}^{(w+2)}&=\frac{1}{w+1}\sum_{P^{*}(q)}\langle C^{m}_{iq_{w}hq_{w+1}}\tilde{N}^{(w)}_{hq_{1}}+C^{m}_{iq_{w+1}hj}\tilde{N}^{(w+1)}_{hq_{1}...q_{w},j}-\alpha^{m}_{iq_{w+1}}M^{(w)}_{q_{1}...q_{w}} \rangle,\\
			\tilde{n}_{iq_{1}...q_{w+1}}^{(w+2,1)}&=\frac{1}{w+1}\sum_{P^{*}(q)}\langle C^{m}_{iq_{w}hq_{w+1}}\tilde{N}^{(w,1)}_{hq_{1}}+C^{m}_{iq_{w+1}hj}\tilde{N}^{(w+1,1)}_{hq_{1}...q_{w},j}-\alpha^{(m,1)}_{iq_{w+1}}M^{(w)}_{q_{1}...q_{w}} \rangle,\\
			n_{ipq_{1}}^{(3,2)}&=\langle C^{m}_{iq_{1}hj}N^{(2,2)}_{hp,j}- \rho^{m}N^{(1)}_{ipq_{1}}\rangle, \quad \tilde{\tilde{n}}_{iq_{1}}^{(3,2)}=-\langle \rho^{m}\tilde{N}^{(1)}_{i}\rangle, \quad \tilde{\tilde{n}}_{iq_{1}}^{(3,3)}=-\langle \rho^{m}\tilde{N}^{(1,1)}_{i}\rangle,\\
			n_{ipq_{1}q_{2}}^{(4,2)}&=\frac{1}{2}\langle C^{m}_{iq_{1}hq_{2}}N^{(2,2)}_{hp}+C^{m}_{iq_{2}hj}N^{(3,2)}_{hpq_{1},j}-\rho^{m}N^{(2)}_{ipq_{1}q_{2}}-\alpha_{iq_{2}}^{(m,1)}\tilde{M}^{(2,1)}_{pq_{1}}+\nonumber\\
			&+C^{m}_{iq_{2}hq_{1}}N^{(2,2)}_{hp}+C^{m}_{iq_{1}hj}N^{(3,2)}_{hpq_{2},j}-\rho^{m}N^{(2)}_{ipq_{2}q_{1}}-\alpha_{iq_{1}}^{(m,1)}\tilde{M}^{(2,1)}_{pq_{2}}\rangle,\\
			\tilde{\tilde{n}}_{iq_{1}}^{(4,2)}&= \langle C^{m}_{iq_{1}hj}{\tilde{\tilde{N}}}^{(3,2)}_{h,j}-\rho^{m}\tilde{N}^{(2)}_{iq_{1}}-\alpha^{m}_{iq_{1}}M^{(2,2)}-\alpha^{(m,1)}_{iq_{1}}M^{(2,1)} \rangle,\\
			\tilde{\tilde{n}}_{iq_{1}}^{(4,3)}&=\langle C^{m}_{iq_{1}hj}{\tilde{\tilde{N}}}^{(3,3)}_{h,j}-\rho^{m}\tilde{N}^{(2,1)}_{iq_{1}}-\alpha^{(m,1)}_{iq_{1}}M^{(2,2)}\rangle,\\
			n_{ipq_{1}q_{2}}^{(4,1)}&=\frac{1}{2}\langle \alpha_{iq_{2}}^{(m,1)}\tilde{M}^{(2,1)}_{pq_{1}}+\alpha_{iq_{1}}^{(m,1)}\tilde{M}^{(2,1)}_{pq_{2}}\rangle,\quad \tilde{n}^{(4,1)}_{iq_{1}}=\langle \alpha^{m}_{iq_{1}}M^{(2,1)} \rangle, \quad n^{(4,4)}_{ip}=\langle \rho^{m}N^{(2,2)}_{ip} \rangle.
		\end{align}
		On the other hand, the coefficients involved into the thermal equation \eqref{eqn:infeq2} can be written as
	\end{subequations}
	\begin{subequations}
		\begin{align}
			m^{(2)}_{q_{1}q_{2}}&=\frac{1}{2}\langle K^{m}_{q_{1}q_{2}}+K^{m}_{q_{2}i}M^{(1)}
			_{q_{1},i}+K^{m}_{q_{2}q_{1}}+K^{m}_{q_{1}i}M^{(1)}
			_{q_{2},i}\rangle,\\
			\tilde{m}^{(2,1)}_{pq_{1}}&=\langle\alpha^{m}_{ij}{N}^{(1)}_{ipq_{1},j}+\alpha_{pq_{1}}^{m}\rangle, \quad m^{(2,1)}=\langle p^{m}+\alpha_{ij}^{m}\tilde{N}^{(1)}_{i,j}\rangle, \quad m^{(2,2)}= \langle p^{(m,0)}+\alpha_{ij}^{m}{\tilde{N}}^{(1,1)}_{i,j}\rangle,\\
			\tilde{m}_{pq_{1}}^{(4,2)}&=\langle p^{m}\tilde{M}^{(2,1)}_{pj} \rangle, \quad \tilde{\tilde{m}}^{(4,2)}=\langle p^{m}M^{(2,1)}\rangle,\\
			m^{(4,3)}&=\langle p^{m}M^{(2,2)}+p^{(m,0)}M^{(2,1)}+\alpha_{ij}^{m}\tilde{\tilde{N}}^{(3,2)}_{i,j}\rangle,\\  m^{(4,4)}&=\langle p^{(m,0)}M^{(2,2)}+\alpha_{ij}^{m}\tilde{\tilde{N}}^{(3,3)}_{i,j}\rangle, \quad \tilde{m}_{p}^{(3,3)}=-\langle\alpha_{ij}^{m}\tilde{N}^{(2,2)}_{ip,j}\rangle,\\
			\tilde{m}_{pq_{1}}^{(4,3)}&=\langle K_{q_{1}i}^{m}\tilde{M}^{(3,3)}_{p,i}-p^{(m,0)}\tilde{M}^{(2,1)}_{pq_{1}}-\alpha^{m}_{iq_{1}}N^{(2,2)}_{ip}-\alpha_{ik}^{m}N^{(3,2)}_{ipq_{1},k}\rangle,\\
			\tilde{m}_{pq_{1}q_{2}}^{(3,1)}&=\frac{1}{2}\langle K^{m}_{q_{2}i}\tilde{M}^{(2,1)}_{pq_{1},i}-\alpha^{m}_{ik}N^{(2)}_{ipq_{1}q_{2},k}-\alpha^{m}_{iq_{2}}N^{(1)}_{ipq_{1}}+K^{m}_{q_{1}i}\tilde{M}^{(2,1)}_{pq_{2},i}-\alpha^{m}_{ik}N^{(2)}_{ipq_{2}q_{1},k}-\alpha^{m}_{iq_{1}}N^{(1)}_{ipq_{2}}\rangle,\\
			\tilde{m}_{pq_{1}...q_{w+1}}^{(w+2,1)}&=\frac{1}{w+1}\sum_{P^{*}(q)}\langle K^{m}_{q_{w}q_{w+1}}\tilde{M}^{(w,1)}_{hq_{1}}+K^{m}_{q_{w+1}i}\tilde{M}^{(w+1,1)}_{pq_{1}...q_{w},i}+\nonumber\\
			&-\alpha^{m}_{iq_{w+1}}N^{(w)}_{ipq_{1}...q_{w}}-\alpha^{m}_{ik}N^{(w+1)}_{ipq_{1}...q_{w+1},k} \rangle,\\
			m_{q_{1}...q_{g+2}}^{(g+2)}&=\frac{1}{g+2}\sum_{P^{*}(q)}\langle K^{m}_{q_{g+1}q_{g+2}}M^{(g)}_{q_{1}...q_{g}}+K^{m}_{q_{g+2}i}M^{(g+1)}_{q_{1}...q_{g+1},i} \rangle,\\
			m_{q_{1}}^{(3,1)}&=\langle K_{q_{1}i}^{m}{M}^{(2,1)}_{,i}-p^{m}M^{(1)}_{q_{1}} -\alpha^{m}_{iq_{1}}\tilde{N}^{(1)}_{i}-\alpha_{ik}^{m}\tilde{N}^{(2)}_{iq_{1},k}\rangle,\\
			m_{q_{1}}^{(3,2)}&=\langle K_{q_{1}i}^{m}{M}^{(2,2)}_{,i}-p^{(m,0)}M^{(1)}_{q_{1}} -\alpha^{m}_{iq_{1}}\tilde{N}^{(1,1)}_{i}-\alpha_{ik}^{m}\tilde{N}^{(2,1)}_{iq_{1},k}\rangle,\\
			m_{q_{1}q_{2}}^{(4,1)}&=\frac{1}{2}\langle K_{q_{1}q_{2}}^{m}{M}^{(2,1)}+K_{q_{2}i}^{m}{M}^{(3,1)}_{q_{1},i}-p^{m}M^{(2)}_{q_{1}q_{2}} -\alpha^{m}_{iq_{2}}\tilde{N}^{(2)}_{iq_{1}}-\alpha_{ik}^{m}\tilde{N}^{(3)}_{iq_{1}q_{2},k}+\\
			&+K_{q_{2}q_{1}}^{m}{M}^{(2,1)}+K_{q_{1}i}^{m}{M}^{(3,1)}_{q_{2},i}-p^{m}M^{(2)}_{q_{2}q_{1}} -\alpha^{m}_{iq_{1}}\tilde{N}^{(2)}_{iq_{2}}-\alpha_{ik}^{m}\tilde{N}^{(3)}_{iq_{2}q_{1},k}+\rangle, \nonumber \\
			m_{q_{1}q_{2}}^{(4,2)}&=\frac{1}{2}\langle K_{q_{1}q_{2}}^{m}{M}^{(2,2)}+K_{q_{2}i}^{m}{M}^{(3,2)}_{q_{1},i}-p^{(m,0)}M^{(2)}_{q_{1}q_{2}}-\alpha^{m}_{iq_{2}}\tilde{N}^{(2,1)}_{iq_{1}}-\alpha_{ik}^{m}\tilde{N}^{(3,1)}_{iq_{1}q_{2},k}+\\
			&+K_{q_{2}q_{1}}^{m}{M}^{(2,2)}+K_{q_{1}i}^{m}{M}^{(3,2)}_{q_{2},i}-p^{(m,0)}M^{(2)}_{q_{2}q_{1}} -\alpha^{m}_{iq_{1}}\tilde{N}^{(2,1)}_{iq_{2}}-\alpha_{ik}^{m}\tilde{N}^{(3,1)}_{iq_{2}q_{1},k} \rangle, \nonumber
		\end{align}
	\end{subequations}
	where $w=2$ and $g=1,2$, the symbol $\mathcal{P}^{*}(q)$ denotes all the possible permutations of the multi-index $q=q_{1},q_{2},...,q_{l}$ that does not show fixed indices.
	Moreover, the field equations of the corresponding first order (Cauchy) thermoelastic material can be expressed in terms of the components of the overall constitutive and inertial tensors, by exploiting the relations $n^{(2)}_{ipq_{1}q_{2}}=\frac{1}{2}(C_{pq_{1}iq_{2}}+C_{pq_{2}iq_{1}})$, $n_{ip}^{(2,2)}=\delta_{ip}\rho$, $\tilde{n}^{(2,1)}_{pq_{1}}=\alpha^{(1)}_{pq_{1}}$, $\tilde{n}^{(2)}_{iq_{1}}=\tilde{m}^{(2,1)}_{pq_{1}}=\alpha_{pq_{1}}$, $m^{(2)}_{q_{1}q_{2}}=K_{q_{1}q_{2}}$, $m^{(2,1)}=p$ and $m^{(2,2)}=p^{(0)}$ \cite{fantoni2017multi},  as
	\begin{subequations}
		\begin{align}
			&C_{iq_{1}pq_{2}}\frac{\partial^{2}U^{M}_{p}}{\partial x_{q_{1}} \partial x_{q_{2}}}-\rho\ddot{U}^{M}_{p}-\alpha^{(1)}_{iq_{1}}\frac{\partial \dot{\Upsilon}^{M}}{\partial x_{q_{1}}}-\alpha_{iq_{1}}\frac{\partial {\Upsilon}^{M}}{\partial x_{q_{1}}}+b_{i}=0,\\
			& K_{q_{1}q_{2}}\frac{\partial^{2}\Upsilon^{M}}{\partial x_{q_{1}}\partial x_{q_{2}}}-\alpha_{pq_{1}}\frac{\partial \dot{U}^{M}_{p}}{\partial x_{q_{1}}}-p\dot{\Upsilon}^{M}-p^{(0)} \ddot{\Upsilon}^{M}+r=0.
			\label{FOT}
		\end{align}
	\end{subequations}
	Managing the average field equations of infinite order \eqref{eqn:infeq}-\eqref{eqn:infeq2} may be unaffordable. Indeed, the ellipticity of the resulting differential problems could be not ensured by truncating the equations \eqref{eqn:infeq}-\eqref{eqn:infeq2} at a certain order. Therefore, several more convenient methods have been proposed to solve them, such as energetic methods or variational approaches \cite{Smyshlyaev2000, Tran2012, del2019characterization, preve2021variational}. Herein, a perturbative scheme is proposed to elevate the order of approximation and to retrieve a more accurate estimation of the solution concerning with the heterogeneous problem. Indeed, the average field equations of infinite order \eqref{eqn:infeq}-\eqref{eqn:infeq2} may be formally solved by carrying out an asymptotic expansion of the macro-displacement $U^{M}_{p}(\boldsymbol{x})$ and the macro-temperature $\Upsilon^{M}(\boldsymbol{x})$ in power of the microstructural size $\varepsilon$ leading to
	\begin{subequations}
		\begin{align}
			\label{eqn:esam}
			U^{M}_{p}(\boldsymbol{x})&=\sum_{j=0}^{+\infty}\varepsilon^{j}U^{j}_{p}(\boldsymbol{x}),\\
			\label{eqn:esam2}
			\Upsilon^{M}(\boldsymbol{x})&=\sum_{j=0}^{+\infty}\varepsilon^{j}\Upsilon^{j}(\boldsymbol{x}).
		\end{align}
	\end{subequations}
	Replacing the relations \eqref{eqn:esam}-\eqref{eqn:esam2} into the equations \eqref{eqn:infeq}-\eqref{eqn:infeq2} derives 
	\begin{subequations}
		\begin{align}
			\label{eqn:rse}
			&n^{(2)}_{ipq_{1}q_{2}}\Big(\frac{\partial^{2}U^{(0)}_{p}}{\partial x_{q_{1}} \partial x_{q_{2}}}+\varepsilon \frac{\partial^{2}U^{(2)}_{p}}{\partial x_{q_{1}} \partial x_{q_{2}}}+...\Big)-n_{ip}^{(2,2)}\Big(\ddot{U}^{(0)}_{p}+\varepsilon \ddot{U}^{(1)}_{p}+...\Big)-\tilde{n}^{(2,1)}_{iq_{1}}\Big(\frac{\partial \dot{\Upsilon}^{(0)}}{\partial x_{q_{1}}}+\varepsilon \frac{\partial \dot{\Upsilon}^{(1)}}{\partial x_{q_{1}}}+...\Big)+ \\
			&-\tilde{n}^{(2)}_{iq_{1}}\Big(\frac{\partial {\Upsilon}^{(0)}}{\partial x_{q_{1}}}+\varepsilon \frac{\partial {\Upsilon}^{(0)}}{\partial x_{q_{1}}}+...\Big)+\sum_{n=0}^{1}\varepsilon^{n+1}\sum_{|q|=n+3}n_{ipq}^{(n+3)} \Big(\frac{\partial^{n+3}U^{(0)}_{p}}{\partial x_{q}}+\varepsilon \frac{\partial^{n+3}U^{(1)}_{p}}{\partial x_{q}}+... \Big)+ \nonumber \\
			&+\sum_{n=0}^{1}\varepsilon^{n+1}\sum_{|q|=n+1}\tilde{n}_{iq}^{(n+3)} \Big(\frac{\partial^{n+2}\Upsilon^{(0)}}{\partial x_{q}}+\varepsilon \frac{\partial^{n+2}\Upsilon^{(1)}}{\partial x_{q}}+...\Big)+\nonumber \\
			&+\sum_{n=0}^{1}\varepsilon^{n+1}\sum_{|q|=n+2}\tilde{n}_{iq}^{(n+3,1)} \Big(\frac{\partial^{n+2}\dot{\Upsilon}^{(0)}}{\partial x_{q}}+\varepsilon \frac{\partial^{n+2}\dot{\Upsilon}^{(1)}}{\partial x_{q}}+...\Big)+\nonumber\\
			&+\sum_{n=0}^{1}\varepsilon^{n+1}\sum_{|q|=n+1}n_{ipq}^{(n+3,2)} \Big(\frac{\partial^{n+1}\ddot{U}^{(0)}_{p}}{\partial x_{q}}+\varepsilon \frac{\partial^{n+1}\ddot{U}^{(1)}_{p}}{\partial x_{q}}+...\Big)+\nonumber \\
			&+\sum_{n=0}^{1}\varepsilon^{n+1}\sum_{|q|=n}\tilde{\tilde{n}}_{iq}^{(n+3,2)} \Big(\frac{\partial^{n}\ddot{\Upsilon}^{(0)}}{\partial x_{q}}+\varepsilon \frac{\partial^{n}\ddot{\Upsilon}^{(1)}}{\partial x_{q}}+...\Big)+\nonumber\\
			&+\sum_{n=0}^{1}\varepsilon^{n+1}\sum_{|q|=n}\tilde{\tilde{n}}_{iq}^{(n+3,3)} \Big(\frac{\partial^{n}\dddot{\Upsilon}^{(0)}}{\partial x_{q}}+\varepsilon \frac{\partial^{n}\dddot{\Upsilon}^{(1)}}{\partial x_{q}}+...\Big)+\nonumber \\
			&-\varepsilon^{2}\Big(n_{ipq_{1}q_{2}}^{(4,1)} \Big(\frac{\partial^{2}\dot{U}^{(0)}_{p}}{\partial x_{q_{1}}\partial x_{q_{2}}}+\varepsilon \frac{\partial^{2}\dot{U}^{(1)}_{p}}{\partial x_{q_{1}}\partial x_{q_{2}}}+...\Big)+\tilde{n}^{(4,1)}_{iq_{1}}\Big(\frac{\partial \dot{\Upsilon}^{(0)}}{\partial x_{q_{1}}}+\varepsilon \frac{\partial \dot{\Upsilon}^{(1)}}{\partial x_{q_{1}}}+... \Big)+\nonumber \\
			&+n^{(4,4)}_{ip}(\ddddot{U}^{(0)}_{p}+\varepsilon \ddddot{U}^{(1)}_{p}+...) \Big)+O(\varepsilon^{3})+b_{i}=0,\nonumber
		\end{align}
	\end{subequations}
	and 
	\begin{subequations}
		\begin{align}
			\label{eqn:rse2}	
			& m^{(2)}_{q_{1}q_{2}}\Big(\frac{\partial^{2}\Upsilon^{(0)}}{\partial x_{q_{1}}\partial x_{q_{2}}}+\varepsilon \frac{\partial^{2}\Upsilon^{(1)}}{\partial x_{q_{1}}\partial x_{q_{2}}}+...\Big)-\tilde{m}^{(2,1)}_{pq_{1}}\Big(\frac{\partial \dot{U}^{(0)}_{p}}{\partial x_{q_{1}}}+\varepsilon \frac{\partial \dot{U}^{(1)}_{p}}{\partial x_{q_{1}}}+... \Big)-m^{(2,1)}(\dot{\Upsilon}^{(0)}+\varepsilon \dot{\Upsilon}^{(1)}+... )+\\
			&-m^{(2,2)} (\ddot{\Upsilon}^{(0)}+\varepsilon \ddot{\Upsilon}^{(1)}+... )+\sum_{n=0}^{1}\varepsilon^{n+1}\sum_{|q|=n+3}m_{q}^{(n+3)} \Big(\frac{\partial^{n+3} \Upsilon^{(0)}}{\partial x_{q}}+\varepsilon \frac{\partial^{n+3} \Upsilon^{(1)}}{\partial x_{q}}+...\Big)+ \nonumber \\
			&+\sum_{n=0}^{1}\varepsilon^{n+1}\sum_{|q|=n+2}\tilde{m}_{pq}^{(n+3,1)} \Big(\frac{\partial^{n+2}\dot{U}^{(0)}_{p}}{\partial x_{q}}+\varepsilon \frac{\partial^{n+2}\dot{U}^{(1)}_{p}}{\partial x_{q}}+... \Big)+\nonumber\\
			&+\sum_{n=0}^{1}\varepsilon^{n+1}\sum_{|q|=n+1}m_{q}^{(n+3,1)} \Big(\frac{\partial^{n+1}\dot{\Upsilon}^{(0)}}{\partial x_{q}}+\varepsilon \frac{\partial^{n+1}\dot{\Upsilon}^{(1)}}{\partial x_{q}}+... \Big)+\nonumber \\
			&+\sum_{n=0}^{1}\varepsilon^{n+1}\sum_{|q|=n+1}m_{q}^{(n+3,2)} \Big(\frac{\partial^{n+1}\ddot{\Upsilon}^{(0)}}{\partial x_{q}}+\varepsilon \frac{\partial^{n+1}\ddot{\Upsilon}^{(1)}}{\partial x_{q}}+... )+\nonumber\\
			&+\sum_{n=0}^{1}\varepsilon^{n+1}\sum_{|q|=n}\tilde{m}_{pq}^{(n+3,3)} \Big(\frac{\partial^{n}\dddot{U}^{(0)}_{p}}{\partial x_{q}}+\varepsilon \frac{\partial^{n}\dddot{U}^{(1)}_{p}}{\partial x_{q}}+...\Big)+\nonumber \\
			&-\varepsilon^{2}\Big(\tilde{m}_{pq_{1}}^{(4,2)} \Big(\frac{\partial \ddot{U}^{(0)}_{p}}{\partial x_{q_{1}}}+\varepsilon \frac{\partial \ddot{U}^{(1)}_{p}}{\partial x_{q_{1}}}+... \Big)+\tilde{\tilde{m}}^{(4,2)}(\ddot{\Upsilon}^{(0)}+\varepsilon \ddot{\Upsilon}^{(1)}+... )+\nonumber\\
			&+m^{(4,3)}(\dddot{\Upsilon}^{(0)}+\varepsilon \dddot{\Upsilon}^{(1)}+... )+m^{(4,4)}(\ddddot{\Upsilon}^{(0)}+\varepsilon \ddddot{\Upsilon}^{(1)}+... )\Big)+O(\varepsilon^{3})+r=0.\nonumber
		\end{align}
	\end{subequations}
	Collecting the terms of the equations \eqref{eqn:rse}-\eqref{eqn:rse2} for different orders of $\varepsilon$ provides a sequence of macroscopic recursive problems. For instance, at the order $\varepsilon^{0}$ the problems are
	\begin{subequations}
		\begin{align}
			&n^{(2)}_{ipq_{1}q_{2}}\frac{\partial^{2}U^{(0)}_{p}}{\partial x_{q_{1}} \partial x_{q_{2}}}-n_{ip}^{(2,2)}\ddot{U}^{(0)}_{p}-\tilde{n}^{(2,1)}_{iq_{1}}\frac{\partial \dot{\Upsilon}^{(0)}}{\partial x_{q_{1}}}-\tilde{n}^{(2)}_{iq_{1}}\frac{\partial {\Upsilon}^{(0)}}{\partial x_{q_{1}}}+b_{i}=0,\\
			& m^{(2)}_{q_{1}q_{2}}\frac{\partial^{2}\Upsilon^{(0)}}{\partial x_{q_{1}}\partial x_{q_{2}}}-\tilde{m}^{(2,1)}_{pq_{1}}\frac{\partial \dot{U}^{(0)}_{p}}{\partial x_{q_{1}}}-m^{(2,1)}\dot{\Upsilon}^{(0)}-m^{(2,2)} \ddot{\Upsilon}^{(0)}+r=0.
		\end{align}
	\end{subequations}
	At the order $\varepsilon$, the recursive problems are
	\begin{subequations}
		\begin{align}
			&n^{(2)}_{ipq_{1}q_{2}}\frac{\partial^{2}U^{(1)}_{p}}{\partial x_{q_{1}} \partial x_{q_{2}}}-n_{ip}^{(2,2)}\ddot{U}^{(1)}_{p}-\tilde{n}^{(2,1)}_{iq_{1}}\frac{\partial \dot{\Upsilon}^{(1)}}{\partial x_{q_{1}}}-\tilde{n}^{(2)}_{iq_{1}}\frac{\partial {\Upsilon}^{(1)}}{\partial x_{q_{1}}}+b_{i}^{(1)}=0,\\		
			& m^{(2)}_{q_{1}q_{2}}\frac{\partial^{2}\Upsilon^{(1)}}{\partial x_{q_{1}}\partial x_{q_{2}}}-\tilde{m}^{(2,1)}_{pq_{1}}\frac{\partial \dot{U}^{(1)}_{p}}{\partial x_{q_{1}}}-m^{(2,1)}\dot{\Upsilon}^{(1)}-m^{(2,2)} \ddot{\Upsilon}^{(1)}+r^{(1)}=0,
		\end{align}
	\end{subequations}
	where $b_{i}^{(1)}$ and $r^{(1)}$ are the known $\mathcal{L}$-periodic source terms that remark the non locality that arises within the average field equations of infinite order because they contain the non-local constitutive tensors. In particular, the source terms are detailed as follows
	\begin{subequations}
		\begin{align}
			\label{b1}
			b_{i}^{(1)}&=n_{ipq_{1}q_{2}q_{3}}^{(3)} \frac{\partial^{3}U^{(0)}_{p}}{\partial x_{q_{1}}...\partial x_{q_{3}}}+\tilde{n}_{iq_{1}q_{2}}^{(3)} \frac{\partial^{2}\Upsilon^{(0)}}{\partial x_{q_{1}} \partial x_{q_{2}}}+\tilde{n}_{iq_{1}q_{2}}^{(3,1)} \frac{\partial^{2}\dot{\Upsilon}^{(0)}}{\partial x_{q_{1}} \partial x_{q_{2}}}+n_{ipq_{1}}^{(3,2)} \frac{\partial \ddot{U}^{(0)}_{p}}{\partial x_{q_{1}}}+\\
			&+\tilde{\tilde{n}}_{iq_{1}}^{(3,2)} \frac{\partial\ddot{\Upsilon}^{(0)}}{\partial x_{q_{1}}}+\tilde{\tilde{n}}_{iq_{1}}^{(3,3)} \frac{\partial\dddot{\Upsilon}^{(0)}}{\partial x_{q_{1}}},\nonumber\\
			r^{(1)}&=m_{q_{1}q_{2}q_{3}}^{(3)} \frac{\partial^{3} \Upsilon^{(0)}}{\partial x_{q_{1}}...\partial x_{q_{3}}}+\tilde{m}_{pq_{1}q_{2}}^{(3,1)} \frac{\partial^{2}\dot{U}^{(0)}_{p}}{\partial x_{q_{1}}\partial x_{q_{2}}}+m_{q_{1}}^{(3,1)} \frac{\partial \dot{\Upsilon}^{(0)}}{\partial x_{q_{1}}}+m_{q_{1}}^{(3,2)} \frac{\partial \ddot{\Upsilon}^{(0)}}{\partial x_{q_{1}}}+\tilde{m}_{p}^{(3,3)} \dddot{U}^{(0)}_{p}.
		\end{align}
	\end{subequations}
	At the order $\varepsilon^{2}$, the recursive problems are
	\begin{subequations}
		\begin{align}
			\label{ep21}
			&n^{(2)}_{ipq_{1}q_{2}}\frac{\partial^{2}U^{(2)}_{p}}{\partial x_{q_{1}} \partial x_{q_{2}}}-n_{ip}^{(2,2)}\ddot{U}^{(2)}_{p}-\tilde{n}^{(2,1)}_{iq_{1}}\frac{\partial \dot{\Upsilon}^{(2)}}{\partial x_{q_{1}}}-\tilde{n}^{(2)}_{iq_{1}}\frac{\partial {\Upsilon}^{(2)}}{\partial x_{q_{1}}}+b_{i}^{(2)}=0,\\	
			\label{ep22}	
			& m^{(2)}_{q_{1}q_{2}}\frac{\partial^{2}\Upsilon^{(2)}}{\partial x_{q_{1}}\partial x_{q_{2}}}-\tilde{m}^{(2,1)}_{pq_{1}}\frac{\partial \dot{U}^{(2)}_{p}}{\partial x_{q_{1}}}-m^{(2,1)}\dot{\Upsilon}^{(2)}-m^{(2,2)} \ddot{\Upsilon}^{(2)}+r^{(2)}=0,
		\end{align}
	\end{subequations}
	where the known $\mathcal{L}$-periodic source terms $b_{i}^{(2)}$ and $r^{(2)}$ can be written as follows
	\begin{subequations}
		\begin{align}
			\label{b2}
			b_{i}^{(2)}&=n_{ipq_{1}q_{2}q_{3}q_{4}}^{(4)} \frac{\partial^{4}U^{(1)}_{p}}{\partial x_{q_{1}}...\partial x_{q_{4}}}+\tilde{n}_{iq_{1}q_{2}q_{3}}^{(4)} \frac{\partial^{3}\Upsilon^{(1)}}{\partial x_{q_{1}}... \partial x_{q_{3}}}-n_{ipq_{1}q_{2}}^{(4,1)} \frac{\partial^{2}\dot{U}^{(0)}_{p}}{\partial x_{q_{1}}\partial x_{q_{2}}}-\tilde{n}^{(4,1)}_{iq_{1}}\frac{\partial \dot{\Upsilon}^{(0)}}{\partial x_{q_{1}}}-n^{(4,4)}_{ip}\ddddot{U}^{(0)}_{p}+\\
			&+\tilde{n}_{iq_{1}q_{2}q_{3}}^{(4,1)} \frac{\partial^{3}\dot{\Upsilon}^{(1)}}{\partial x_{q_{1}}... \partial x_{q_{3}}}+n_{ipq_{1}q_{2}}^{(4,2)} \frac{\partial^{2} \ddot{U}^{(1)}_{p}}{\partial x_{q_{1}}\partial x_{q_{2}}}+\tilde{\tilde{n}}_{iq_{1}}^{(4,2)} \frac{\partial^{2}\ddot{\Upsilon}^{(1)}}{\partial x_{q_{1}}\partial x_{q_{2}}}+\tilde{\tilde{n}}_{iq_{1}}^{(4,3)} \frac{\partial^{2}\dddot{\Upsilon}^{(1)}}{\partial x_{q_{1}}\partial x_{q_{2}}}, \nonumber\\
			\label{r2}
			r^{(2)}&=m_{q_{1}q_{2}q_{3}q_{4}}^{(4)} \frac{\partial^{4} \Upsilon^{(1)}}{\partial x_{q_{1}}...\partial x_{q_{4}}}-\Big(\tilde{m}_{pq_{1}}^{(4,2)} \frac{\partial \ddot{U}^{(0)}_{p}}{\partial x_{q_{1}}}+\tilde{\tilde{m}}^{(4,2)}\ddot{\Upsilon}^{(0)}+m^{(4,3)}\dddot{\Upsilon}^{M}+m^{(4,4)}\ddddot{\Upsilon}^{(0)}\Big)+\\
			&+\tilde{m}_{pq_{1}q_{2}q_{3}}^{(4,1)} \frac{\partial^{3}\dot{U}^{(1)}_{p}}{\partial x_{q_{1}}...\partial x_{q_{3}}} +m_{q_{1}q_{2}}^{(4,1)} \frac{\partial^{2} \dot{\Upsilon}^{(1)}}{\partial x_{q_{1}}\partial x_{q_{2}}}+m_{q_{1}q_{2}}^{(4,2)} \frac{\partial^{2} \ddot{\Upsilon}^{(1)}}{\partial x_{q_{1}}\partial x_{q_{2}}}+\tilde{m}_{pq_{1}}^{(4,3)} \frac{\partial \dddot{U}^{(1)}_{p}}{\partial x_{q_{1}}}.\nonumber
		\end{align}   
	\end{subequations}
	\begin{figure}
		\centering
		\begin{minipage}[c][\width]{0.46\textwidth}
			\hspace{3pt}
			{\begin{overpic}[width=\textwidth]{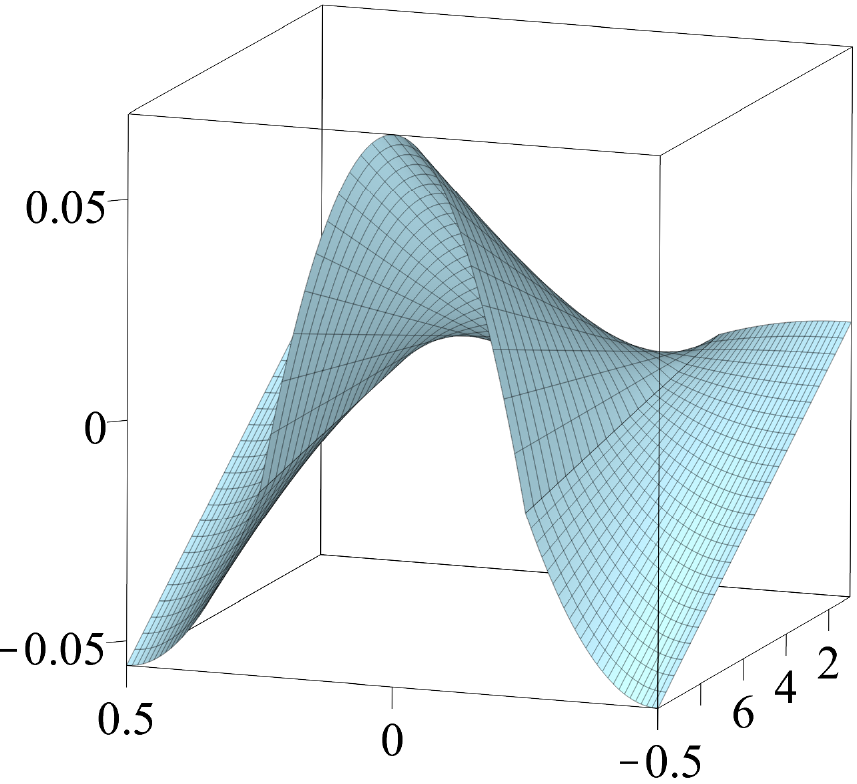}
					\put(06,83){\normalsize (a)}
					\put(00,53){\normalsize $\bar{\tilde{M}}^{(2,1)}_{22}$}
					\put(45,-02){\normalsize $\xi_{2}$ }
					\put(90,03){\normalsize $\frac{\alpha^{2}_{22}}{\alpha^{1}_{22}}$}
			\end{overpic}}
		\end{minipage}	\qquad
		\begin{minipage}[c][\width]{0.46\textwidth}
			\hspace{2pt} 
			{\begin{overpic}[width=\textwidth]{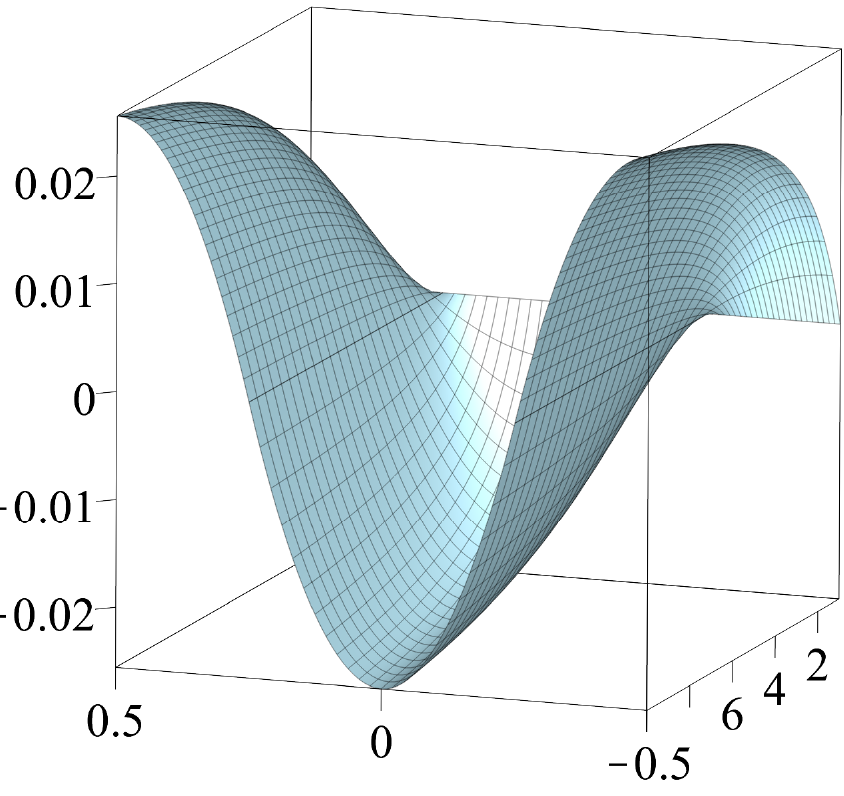}
					\put(06,83){\normalsize (b)}
					\put(02,51){\normalsize ${M}^{(2)}_{22}$}
					\put(90,04){\normalsize $\frac{K^{2}_{22}}{K^{1}_{22}}$}
					\put(45,-02){\normalsize $\xi_{2}$}
			\end{overpic}}
		\end{minipage}
		\caption{non-dimensionalized second-order perturbation function $\bar{\tilde{M}}^{(2,1)}_{22}$ represented as a function of the ratio $\frac{\alpha^{2}_{22}}{\alpha^{1}_{22}}$ and the coordinate $\xi_{2}$. The values of $\frac{K^{2}_{22}}{K^{1}_{22}}=3$ and $\frac{C^{2}_{2222}}{C^{1}_{2222}}=2$ are fixed, (a). Non-dimensionalized second-order perturbation function ${M}^{(2)}_{22}$ depicted while varying the ratio $\frac{K^{2}_{22}}{K^{1}_{22}}$. The constitutive parameters $\tilde{\nu}_{1}=\tilde{\nu}_{2}=0.3$ and $\eta=1$ remain constant, (b).}
		\label{0B}
	\end{figure}
	\begin{figure}
		\centering
		\begin{minipage}[c][\width]{0.46\textwidth}
			\hspace{3pt}
			{\begin{overpic}[width=\textwidth]{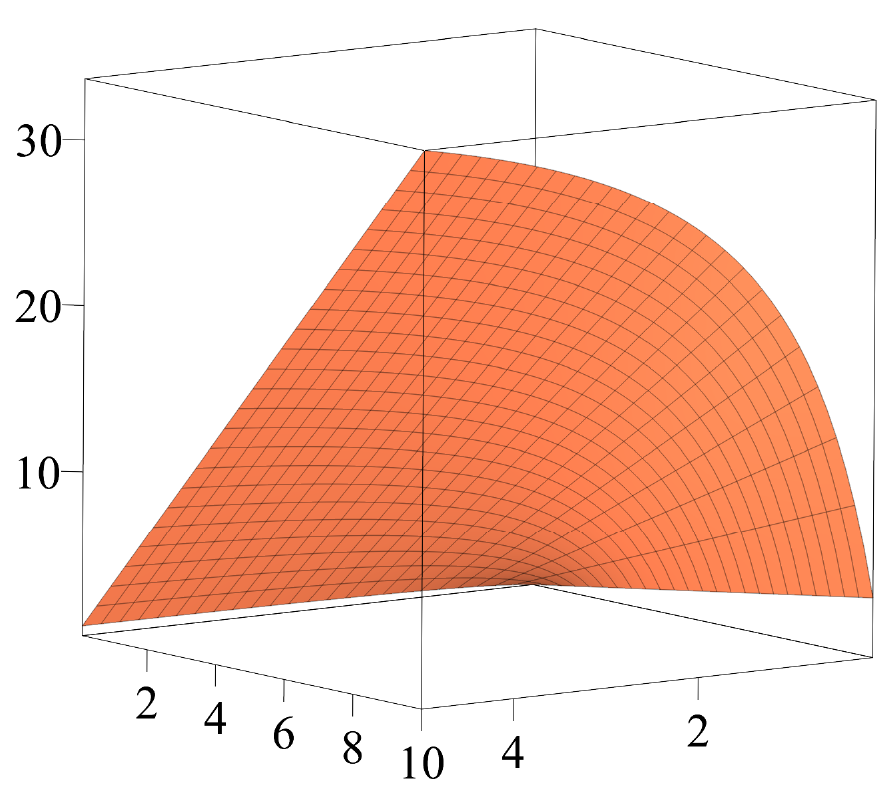}
					\put(02,85){\normalsize (a)}
					\put(10,57){\normalsize $\frac{\tilde{n}^{(2)}_{22}\theta_{0}}{\varepsilon \sqrt{(C^{1}_{2222}\rho^{1})}}$}
					\put(27,-02){\normalsize $\frac{\tau^{2}_{1}}{\tau^{1}_{1}}$}
					\put(76,-02){\normalsize $\frac{C^{2}_{2222}}{C^{1}_{2222}}$}
			\end{overpic}}
		\end{minipage}	\qquad
		\begin{minipage}[c][\width]{0.46\textwidth}
			\hspace{3pt} 
			{\begin{overpic}[width=\textwidth]{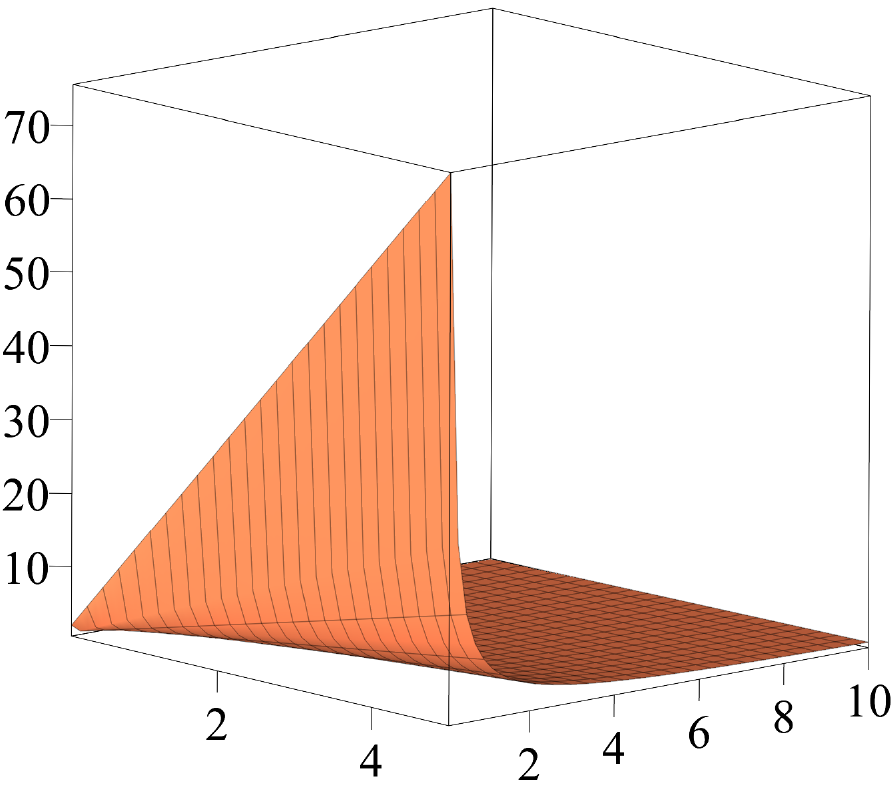}
					\put(02,85){\normalsize (b)}
					\put(10,68){\normalsize $\frac{p(\theta_{0})^{2}}{C^{1}_{2222}}$}
					\put(27,-02){\normalsize $\frac{C^{2}_{2222}}{C^{1}_{2222}}$}
					\put(76,-02){\normalsize $\frac{K^{2}_{22}}{K^{1}_{22}}$}
			\end{overpic}}
		\end{minipage}
		\caption{the behavior of the non-dimensionalized overall tensor components is shown. (a) The component $\frac{\tilde{n}^{(2)}_{22}\theta_{0}}{\varepsilon \sqrt{(C^{1}_{2222}\rho^{1})}}$ is presented. (b) The component $\frac{p(\theta_{0})^{2}}{C^{1}_{2222}}$ is displayed. These components are evaluated for fixed non-null constitutive parameters: $\frac{p^{2}}{p^{1}}=3$, $\frac{\rho^{2}}{\rho^{1}}=3$, $\tilde{\nu}_{1}=\tilde{\nu}_{2}=0.2$, $\frac{\alpha^{1}_{22}\theta_{0}}{C^{1}_{2222}}=\frac{1}{100}$, $\frac{\alpha^{2}_{22}\theta_{0}}{C^{2}_{2222}}=\frac{1}{10}$, $\frac{\alpha^{1}_{22}\eta \sqrt{C^{1}_{2222}/\rho^{1}}}{\bar{K}^{1}_{22}}=\frac{1}{100}$, $\frac{\alpha^{2}_{22}\eta \sqrt{C^{1}_{2222}/\rho^{1}}}{\bar{K}^{2}_{22}}=\frac{1}{10}$, $\frac{p^{1}\theta_{0}\eta \sqrt{C^{1}_{2222}/\rho^{1}}}{\bar{K}^{1}_{22}}=1$, $\frac{\tau^{1}_{1} \sqrt{C^{1}_{2222}/\rho^{1}}}{\varepsilon}=3$ and $\eta=1$.}
		\label{0C}
	\end{figure}
	\begin{figure*}[htbp]
		\centering
		\begin{minipage}[c][\width]{0.40\textwidth}
			\hspace{-10pt}
			{\begin{overpic}[width=\textwidth]{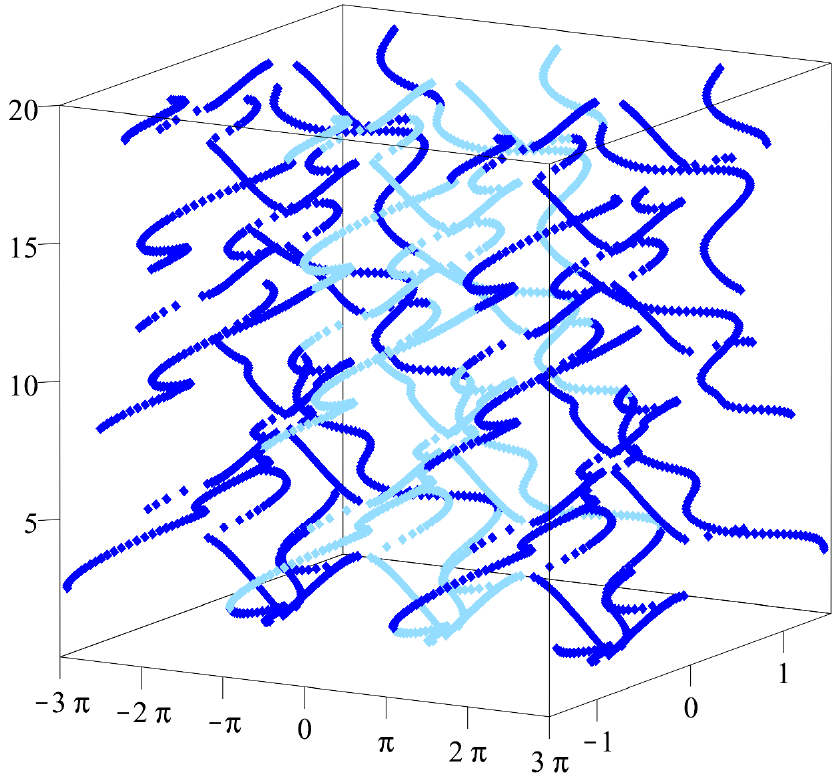}
					\put(02,92){\normalsize (a)}
					\put(-05,55){\normalsize $\bar\omega$}
					\put(27,-02){\footnotesize ${\mathcal{R}}e(\bar{k}_{2})$}
					\put(74,-01){\footnotesize ${\mathcal{I}}m(\bar{k}_{2})$}
			\end{overpic}}
		\end{minipage}	\qquad
		\begin{minipage}[c][\width]{0.40\textwidth}
			\hspace{-05pt} 
			{\begin{overpic}[width=\textwidth]{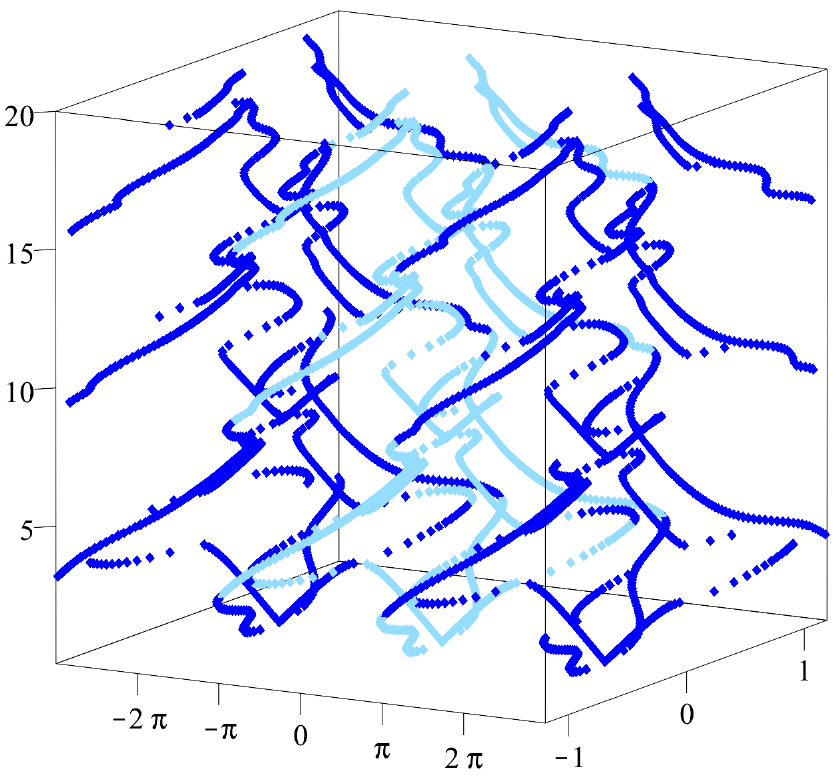}
					\put(02,92){\normalsize (b)}
					\put(-05,55){\normalsize $\bar\omega$}
					\put(27,-02){\footnotesize ${\mathcal{R}}e(\bar{k}_{2})$}
					\put(74,-01){\footnotesize ${\mathcal{I}}m(\bar{k}_{2})$}
			\end{overpic}}
		\end{minipage}\\[10pt]
		\begin{minipage}[c][1\width]{0.39\textwidth}
			\hspace{-16pt}\vspace{3pt}
			{\begin{overpic}[width=\textwidth]{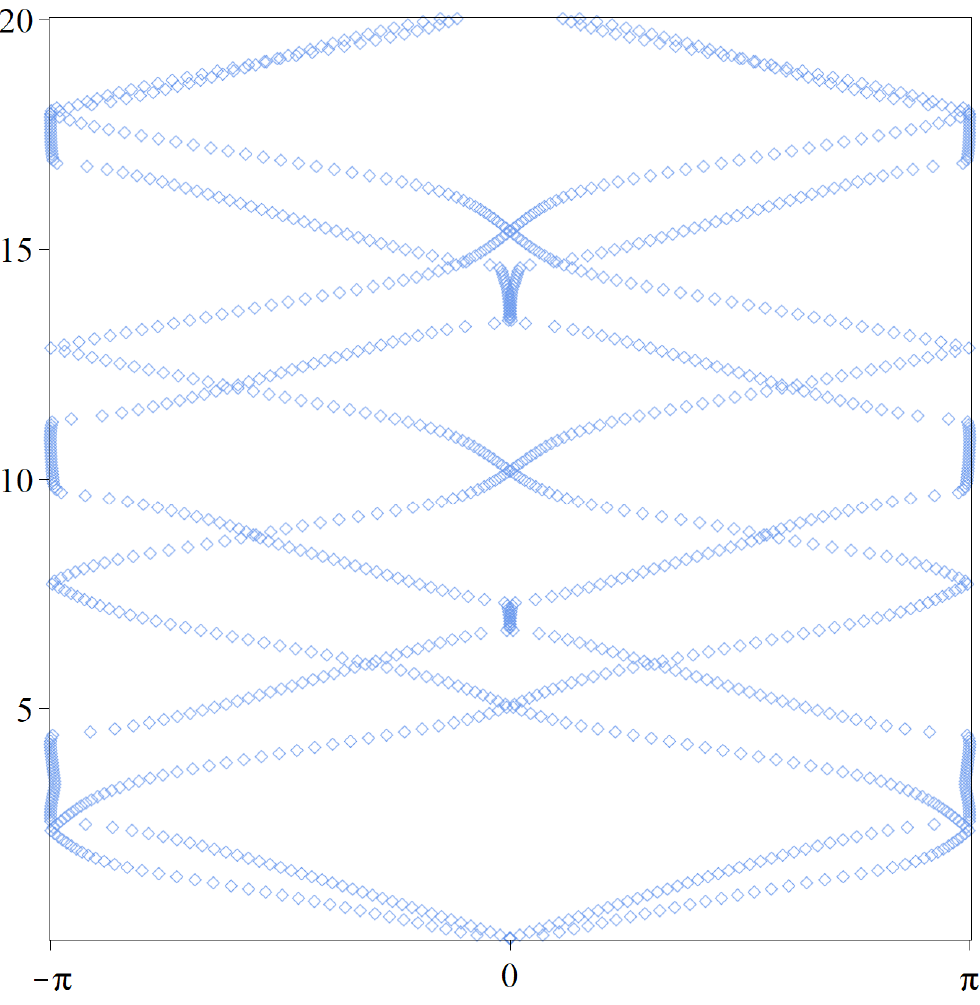}
					\put(-05,90){\normalsize (c)}
					\put(-05,65){\normalsize $\bar\omega$}
					\put(40,-05){\footnotesize ${\mathcal{R}}e(\bar{k}_{2})$}
			\end{overpic}}
		\end{minipage}	
		\begin{minipage}[c][1\width]{0.39\textwidth}
			\hspace{9pt}\vspace{3pt} 
			{\begin{overpic}[width=\textwidth]{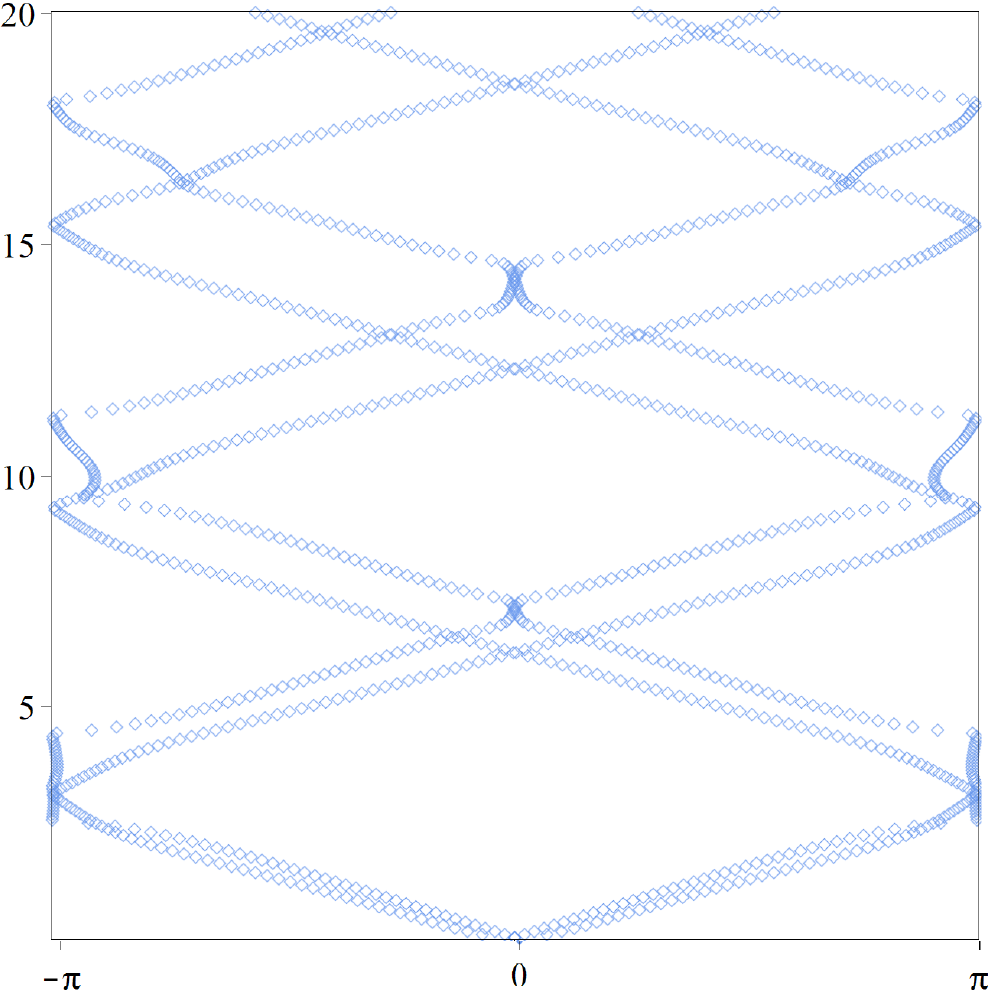}	
					\put(-05,90){\normalsize (d)}
					\put(-05,65){\normalsize $\bar\omega$}
					\put(40,-05){\footnotesize ${\mathcal{R}}e(\bar{k}_{2})$}
			\end{overpic}}
		\end{minipage}\\[15pt]
		\begin{minipage}[c][1\width]{0.39\textwidth}
			\hspace{-16pt}\vspace{3pt}
			{\begin{overpic}[width=\textwidth]{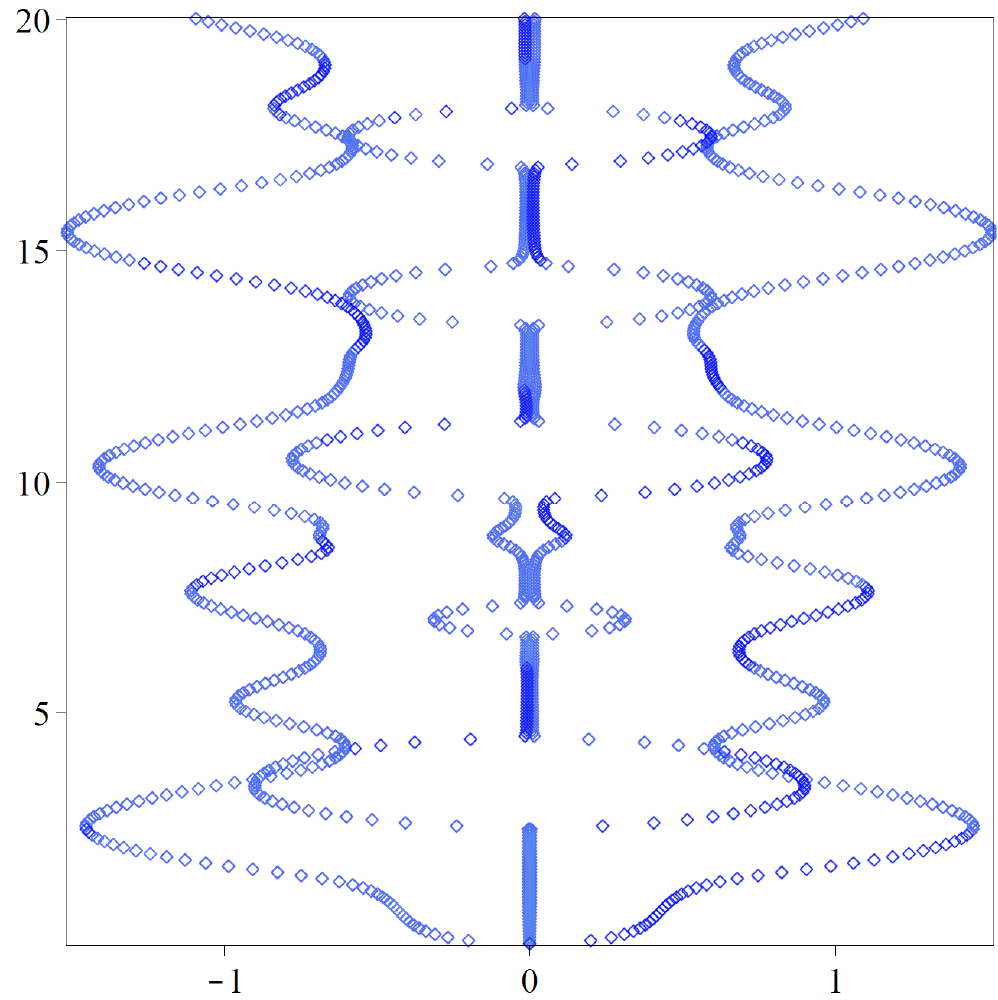}
					\put(-05,90){\normalsize (e)}
					\put(-05,65){\normalsize $\bar\omega$}
					\put(60,-02){\footnotesize ${\mathcal{I}}m(\bar{k}_{2})$}
			\end{overpic}}
		\end{minipage}	
		\begin{minipage}[c][1\width]{0.39\textwidth}
			\hspace{8pt} \vspace{3pt}
			{\begin{overpic}[width=\textwidth]{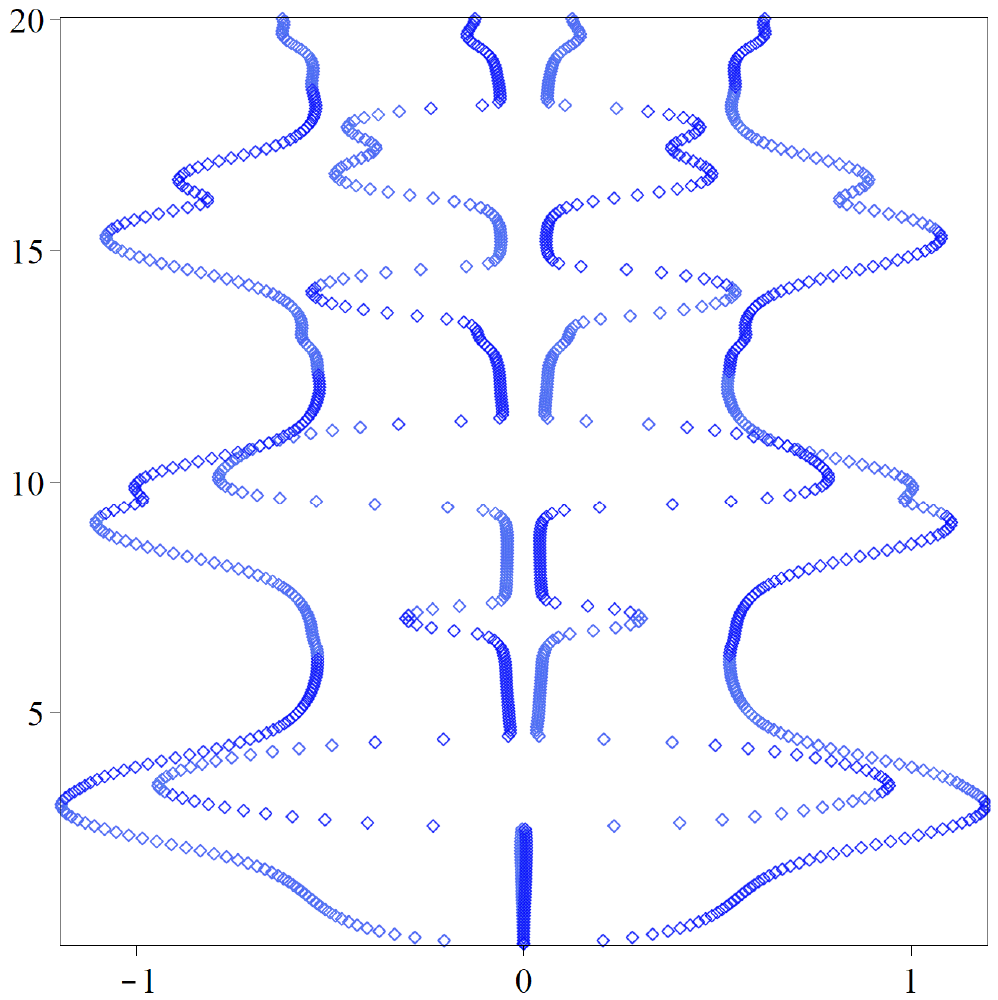}	
					\put(-05,90){\normalsize (f)}
					\put(-05,65){\normalsize $\bar\omega$}
					\put(60,-02){\footnotesize ${\mathcal{I}}m(\bar{k}_{2})$}
			\end{overpic}}
		\end{minipage}
		\vspace{0pt}
		\caption{Floquet-Bloch complex spectra and band structure associated to compressional-thermal waves with $k_{1}=0$ for fixed not-null constitutive parameters $\frac{p^{2}}{p^{1}}=3$, $\frac{C^{2}_{2222}}{C^{1}_{2222}}=2$, $\frac{\rho^{2}}{\rho^{1}}=3$, $\tilde{\nu}_{1}=\tilde{\nu}_{2}=0.2$, $\frac{\bar{K}^{2}_{22}}{\bar{K}^{1}_{22}}=3$, $\frac{\alpha^{1}_{22}\theta_{0}}{C^{1}_{2222}}=\frac{1}{100}$, $\frac{\alpha^{2}_{22}\theta_{0}}{C^{2}_{2222}}=\frac{1}{10}$, $\frac{\alpha^{1}_{22}\eta \sqrt{C^{1}_{2222}/\rho^{1}}}{\bar{K}^{1}_{22}}=\frac{1}{100}$, $\frac{\alpha^{2}_{22}\eta \sqrt{C^{1}_{2222}/\rho^{1}}}{\bar{K}^{2}_{22}}=\frac{1}{10}$, $\frac{p^{1}\theta_{0}\eta \sqrt{C^{1}_{2222}/\rho^{1}}}{\bar{K}^{1}_{22}}=1$, $\frac{\tau^{1}_{0} \sqrt{C^{1}_{2222}/\rho^{1}}}{\varepsilon}=1$, $\frac{\tau^{1}_{1} \sqrt{C^{1}_{2222}/\rho^{1}}}{\varepsilon}=3$ and $\eta=1$, by varying the ratios between the relaxation times $\tau^{m}_{0}$ and $\tau^{m}_{1}$ as $\frac{\tau^{2}_{0}}{\tau^{1}_{0}}=\frac{\tau^{2}_{1}}{\tau^{1}_{1}}=2$ in (a), (c), (e) and $\frac{\tau^{2}_{0}}{\tau^{1}_{0}}=\frac{\tau^{2}_{1}}{\tau^{1}_{1}}=1$ in (b), (d), (e). $(\bar{\omega},{\mathcal{R}}e(\bar{k}_{2}))-$plane, (c), (d). $(\bar{\omega},{\mathcal{I}}m(\bar{k}_{2}))-$plane, (e), (f).}
		\label{000}
		\vspace{3pt}
	\end{figure*}
	\section{Wave propagation in thermoelastic periodic materials}
	\begin{figure*}[htbp]
		\centering
		\begin{minipage}[c][\width]{0.40\textwidth}
			\hspace{-10pt}
			{\begin{overpic}[width=\textwidth]{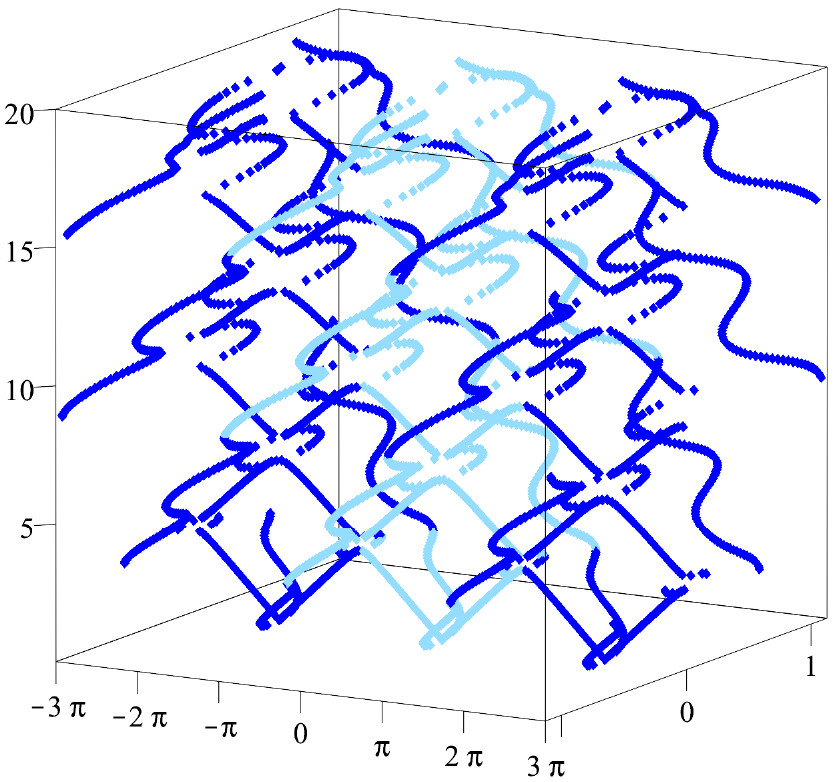}
					\put(02,92){\normalsize (a)}
					\put(-05,55){\normalsize $\bar\omega$}
					\put(27,-02){\footnotesize ${\mathcal{R}}e(\bar{k}_{2})$}
					\put(74,-01){\footnotesize ${\mathcal{I}}m(\bar{k}_{2})$}
			\end{overpic}}
		\end{minipage}	\qquad
		\begin{minipage}[c][\width]{0.40\textwidth}
			\hspace{-05pt} 
			{\begin{overpic}[width=\textwidth]{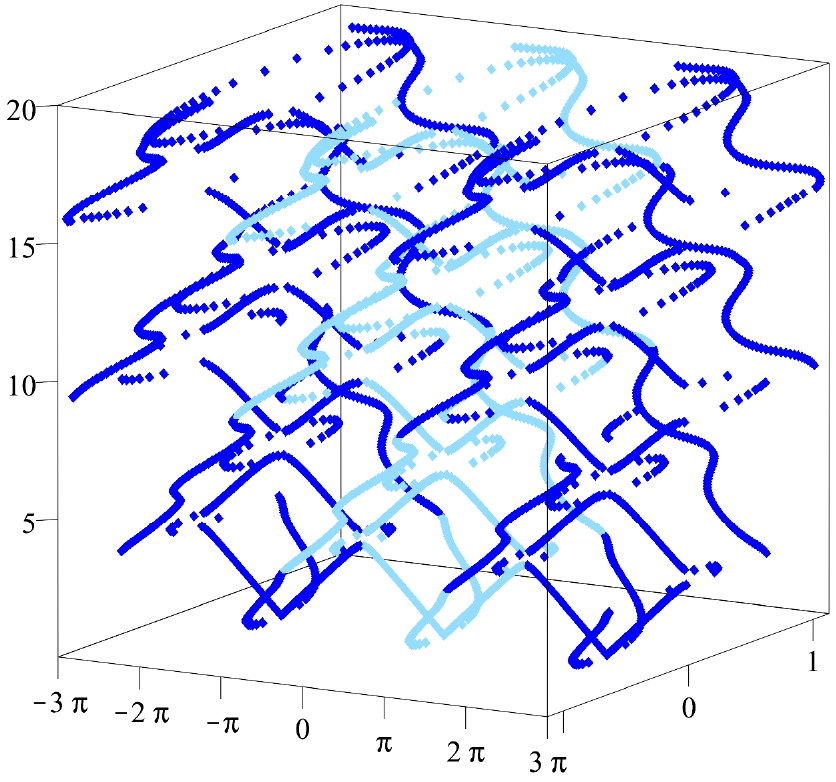}
					\put(02,92){\normalsize (b)}
					\put(-05,55){\normalsize $\bar\omega$}
					\put(27,-02){\footnotesize ${\mathcal{R}}e(\bar{k}_{2})$}
					\put(74,-01){\footnotesize ${\mathcal{I}}m(\bar{k}_{2})$}
			\end{overpic}}
		\end{minipage}\\[10pt]
		\begin{minipage}[c][1\width]{0.39\textwidth}
			\hspace{-16pt}\vspace{3pt}
			{\begin{overpic}[width=\textwidth]{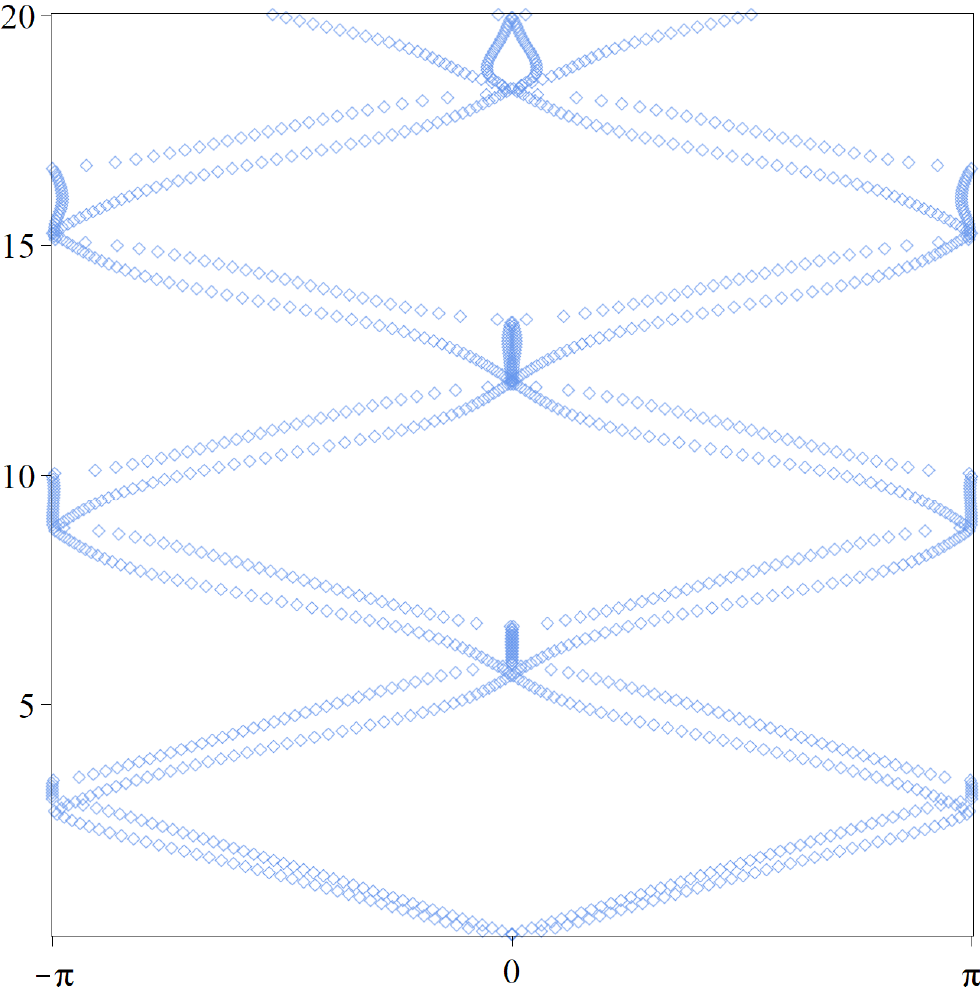}
					\put(-05,90){\normalsize (c)}
					\put(-05,65){\normalsize $\bar\omega$}
					\put(40,-05){\footnotesize ${\mathcal{R}}e(\bar{k}_{2})$}
			\end{overpic}}
		\end{minipage}	
		\begin{minipage}[c][1\width]{0.39\textwidth}
			\hspace{8pt}\vspace{3pt} 
			{\begin{overpic}[width=\textwidth]{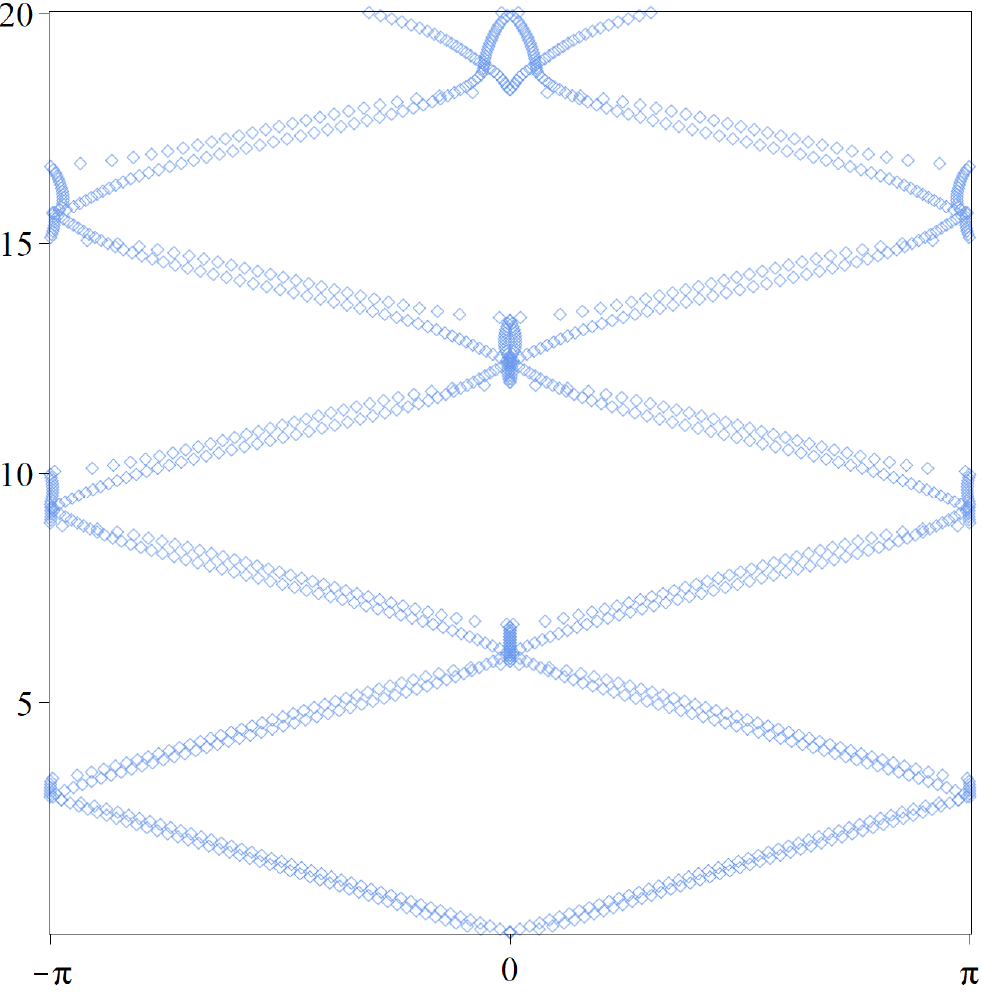}	
					\put(-05,90){\normalsize (d)}
					\put(-05,65){\normalsize $\bar\omega$}
					\put(40,-05){\footnotesize ${\mathcal{R}}e(\bar{k}_{2})$}
			\end{overpic}}
		\end{minipage}\\[20pt]
		\begin{minipage}[c][1\width]{0.39\textwidth}
			\hspace{-16pt}\vspace{3pt}
			{\begin{overpic}[width=\textwidth]{Spettro_Im1_cropped.pdf}
					\put(-05,90){\normalsize (e)}
					\put(-05,65){\normalsize $\bar\omega$}
					\put(60,-02){\footnotesize ${\mathcal{I}}m(\bar{k}_{2})$}
			\end{overpic}}
		\end{minipage}	
		\begin{minipage}[c][1\width]{0.39\textwidth}
			\hspace{8pt} \vspace{3pt}
			{\begin{overpic}[width=\textwidth]{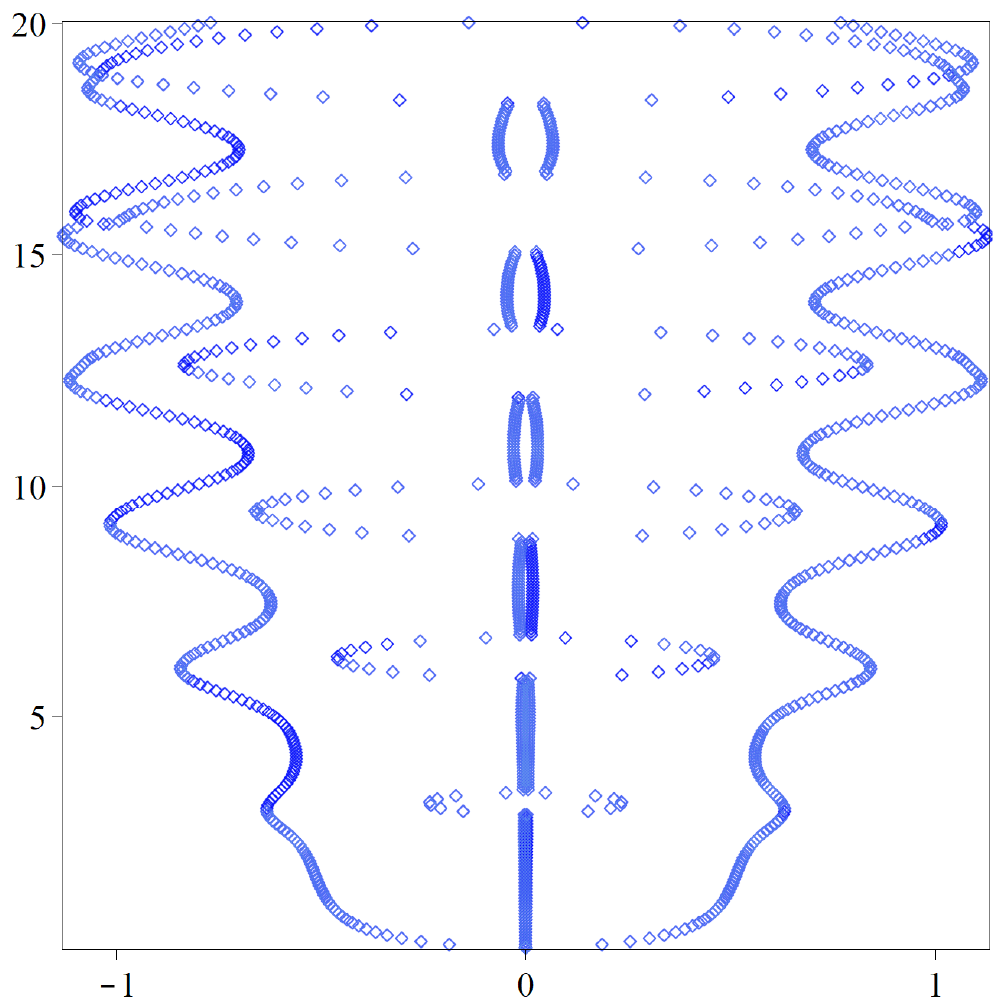}	
					\put(-05,90){\normalsize (f)}
					\put(-05,65){\normalsize $\bar\omega$}
					\put(60,-02){\footnotesize ${\mathcal{I}}m(\bar{k}_{2})$}
			\end{overpic}}
		\end{minipage}
		\vspace{0pt}
		\caption{Floquet-Bloch complex spectra and band structure associated to compressional-thermal waves with $k_{1}=0$ for fixed not-null constitutive parameters $\frac{p^{2}}{p^{1}}=3$, $\frac{C^{2}_{2222}}{C^{1}_{2222}}=2$, $\frac{\rho^{2}}{\rho^{1}}=3$, $\tilde{\nu}_{1}=\tilde{\nu}_{2}=0.2$, $\frac{\bar{K}^{2}_{22}}{\bar{K}^{1}_{22}}=3$, $\frac{\alpha^{1}_{22}\theta_{0}}{C^{1}_{2222}}=\frac{1}{100}$, $\frac{\alpha^{2}_{22}\theta_{0}}{C^{2}_{2222}}=\frac{1}{10}$, $\frac{\alpha^{1}_{22}\eta \sqrt{C^{1}_{2222}/\rho^{1}}}{\bar{K}^{1}_{22}}=\frac{1}{100}$, $\frac{\alpha^{2}_{22}\eta \sqrt{C^{1}_{2222}/\rho^{1}}}{\bar{K}^{2}_{22}}=\frac{1}{10}$, $\frac{p^{1}\theta_{0}\eta \sqrt{C^{1}_{2222}/\rho^{1}}}{\bar{K}^{1}_{22}}=1$, $\frac{\tau^{1}_{0} \sqrt{C^{1}_{2222}/\rho^{1}}}{\varepsilon}=1$, $\frac{\tau^{1}_{1} \sqrt{C^{1}_{2222}/\rho^{1}}}{\varepsilon}=3$ and $\eta=10$, by varying the ratios between the relaxation times $\tau^{m}_{0}$ and $\tau^{m}_{1}$ as $\frac{\tau^{2}_{0}}{\tau^{1}_{0}}=\frac{\tau^{2}_{1}}{\tau^{1}_{1}}=2$ in (a), (c), (e) and $\frac{\tau^{2}_{0}}{\tau^{1}_{0}}=\frac{\tau^{2}_{1}}{\tau^{1}_{1}}=1$ in (b), (d), (e). $(\bar{\omega},{\mathcal{R}}e(\bar{k}_{2}))-$plane, (c), (d). $(\bar{\omega},{\mathcal{I}}m(\bar{k}_{2}))-$plane, (e), (f).}
		\label{002}
		\vspace{3pt}
	\end{figure*}
	\label{waveprop}
	In this Section, the wave propagation through thermoelastic periodic material will be performed by carrying out the bilateral Laplace and the Fourier transforms to the macroscopic fields related to equations \eqref{eqn:infeq}-\eqref{eqn:infeq2}.
	The time bilateral Laplace transform of a real valued function (i.e. $f:\mathbb{R} \to \mathbb{R}$ ) is defined as
	\begin{equation}
		\label{eqn:lap}
		\mathcal{L}(f(t))= \hat{f}(s) = \int_{-\infty}^{+\infty} f(t) e^{-st} dt, \quad s\in \mathbb{C}, 
	\end{equation} 
	where, the Laplace argument, $s$, and the Laplace transform, $\hat{f}$, are complex valued (i.e. $\hat{f}:\mathbb{C} \to \mathbb{C}$ ) \cite{paley1934fourier}. The derivative rule for the Laplace transform is given by $\label{eqn:der}
	\mathcal{L}\Big (\frac{\partial^{n}f(t)}{\partial t^{n}}\Big) = s^{n}\hat{f}(s).$
	On the other hand, the complex space Fourier transform of an arbitrary function $f$ is defined as
	\begin{equation}
		\label{eqn:fou}
		\mathcal{F}(f(\boldsymbol{x}))= \check{f}(\boldsymbol{k}) = \int_{\mathbb{R}^{3}}^{} f(\boldsymbol{x}) e^{-\iota \boldsymbol{k} \cdot \boldsymbol{x}} d\boldsymbol{x} ,\quad \boldsymbol{k}\in \mathbb{C}^{3},
	\end{equation} 
	where $\iota$ is the imaginary unit such that $\iota^{2}=1$ and $\check{f}:\mathbb{C}^{3} \to \mathbb{C}$, whereas its derivative rule results to be
	$\label{eqn:derfou}
	\mathcal{F}\Big (\frac{\partial^{n}f(\boldsymbol{x})}{\partial x^{n}_{j}}\Big) = (\iota k_{j})^{n}\check{f}(\boldsymbol{k})$ \cite{paley1934fourier}. 
	Applying the Laplace and the Fourier transforms to the average infinite order equations \eqref{eqn:infeq}-\eqref{eqn:infeq2} with respect to the time $t$ and to the slow variable $\boldsymbol{x}$, respectively, derives the field equations at the macroscale within the frequency and the wave vector domain. Indeed it results
	\begin{subequations}
		\begin{align}
			\label{epho21wk}
			&\Big\{-k_{q_{1}}k_{q_{2}}n^{(2)}_{iq_{1}pq_{2}}-s^{2}n^{(2,2)}_{ip}-\varepsilon(\iota k_{r_{1}}k_{r_{2}}k_{q_{1}}n^{(3)}_{ir_{1}r_{2}pq_{1}}+\iota s^{2}k_{q_{1}}n^{(3,2)}_{iq_{1}})+\varepsilon^{2}(s k_{q_{1}}k_{q_{2}}n^{(4,1)}_{iq_{1}pq_{2}}-s^{4}n^{(4,4)}_{ip}+\\
			&+k_{q_{1}}k_{q_{2}}k_{r_{1}}k_{r_{2}}n^{(4)}_{iq_{1}q_{2}pr_{1}r_{2}}-s^{2}k_{q_{1}}k_{q_{2}}n^{(4,2)}_{iq_{1}pq_{2}})+O(\varepsilon^{3})\Big\} \check{\hat{U}}^{M}_{p}+\Big\{-\iota s k_{q_{1}} \tilde{n}^{(2,1)}_{iq_{1}}-\iota k_{q_{1}} \tilde{n}^{(2)}_{iq_{1}}+\nonumber \\
			&-\varepsilon(k_{q_{1}}k_{q_{2}}\tilde{n}^{(3)}_{iq_{1}q_{2}} 
			+sk_{q_{1}}k_{q_{2}}\tilde{n}^{(3,1)}_{iq_{1}q_{2}}-s^{2}\tilde{\tilde{n}}^{(3,2)}_{i}-s^{3} \tilde{\tilde{n}}^{(3,3)}_{i})+\varepsilon^{2}(-\iota s k_{q_{1}} \tilde{n}^{(4,1)}_{iq_{1}}-\iota k_{r_{1}}k_{r_{2}}k_{q_{1}}\tilde{n}^{(4)}_{ir_{1}r_{2}q_{1}}+\nonumber \\
			&-\iota s k_{r_{1}}k_{r_{2}}k_{q_{1}}\tilde{n}^{(4,1)}_{ir_{1}r_{2}q_{1}}+\iota s^{2} k_{q_{1}} \tilde{\tilde{n}}^{(4,2)}_{iq_{1}}+\iota s^{3} k_{q_{1}} \tilde{\tilde{n}}^{(4,3)}_{iq_{1}})+O(\varepsilon^{3}) \Big\}\check{\hat{\Upsilon}}^{M}+\check{\hat{b}}_{i}=0,\nonumber \\	
			\label{epho22wk}	
			&\Big\{-\iota s k_{q_{1}} \tilde{m}^{(2,1)}_{pq_{1}}+\varepsilon(s^{3}\tilde{m}^{(3,3)}_{p}-s k_{q_{1}}k_{q_{2}} \tilde{m}^{(3,1)}_{pq_{1}q_{2}} )+\varepsilon^{2}(-\iota s k_{r_{1}}k_{r_{2}}k_{q_{1}}\tilde{m}^{(4,1)}_{pr_{1}r_{2}q_{1}}-\iota s^{2}k_{q_{1}} \tilde{m}^{(4,2)}_{pq_{1}}+\\
			&+\iota s^{3}k_{q_{1}} \tilde{m}^{(4,3)}_{pq_{1}}) +O(\varepsilon^{3})\Big\}\check{\hat{U}}^{M}_{p}+\Big\{-m^{(2)}_{q_{1}q_{2}} k_{q_{1}}k_{q_{2}}-s m^{(2,1)}-s^{2}m^{(2,2)}+\varepsilon(-\iota k_{r_{1}}k_{r_{2}}k_{q_{1}}m^{(3)}_{r_{1}r_{2}q_{1}}+\nonumber\\
			&\iota s^{2} k_{q_{1}} m^{(3,2)}_{q_{1}}+\iota s k_{q_{1}} m^{(3,1)}_{q_{1}})+\varepsilon^{2}(k_{q_{1}}k_{q_{2}}k_{r_{1}}k_{r_{1}}m^{(4)}_{q_{1}q_{2}r_{1}r_{2}}-s k_{q_{1}}k_{q_{2}} m^{(4,1)}_{q_{1}q_{2}}-s^{2} k_{q_{1}}k_{q_{2}} m^{(4,2)}_{q_{1}q_{2}}\nonumber\\
			&-s^{2}\tilde{\tilde{m}}^{(4,2)}-s^{3}m^{(4,3)}-s^{4}m^{(4,4)})+O(\varepsilon^{3})\Big\}\check{\hat{\Upsilon}}^{M}+\check{\hat{r}}=0.\nonumber	
		\end{align}
	\end{subequations}
	Collecting the terms of equations \eqref{epho21wk}-\eqref{epho22wk} for increasing powers of $\varepsilon$ achieves the matricial system
	\begin{equation}
		\label{MMM}
		(\boldsymbol{A}^{(0)}+\varepsilon \boldsymbol{A}^{(1)}+ \varepsilon^{2} \boldsymbol{A}^{(2)}+O(\varepsilon^{3}))\check{\hat{\boldsymbol{P}}}(\boldsymbol{k},s)=\check{\hat{\boldsymbol{f}}}(\boldsymbol{k},s),
	\end{equation} 
	where the vector $\check{\hat{\boldsymbol{P}}}(\boldsymbol{k},s)$ gathers the transformed macro-displacement and the transformed macro-temperature such that $\check{\hat{\boldsymbol{P}}}(\boldsymbol{k},s)=(\check{\hat{\boldsymbol{U}}}^{M}(\boldsymbol{k},s) \quad \check{\hat{\Upsilon}}^{M}(\boldsymbol{k},s))^{\top}$, the vector $\check{\hat{\boldsymbol{f}}}(\boldsymbol{k},s)=(\check{\hat{\boldsymbol{b}}}(\boldsymbol{k},s) \quad \check{\hat{r}}(\boldsymbol{k},s))^{\top}$ and the $2 \times 2$ matrices $\boldsymbol{A}^{(0)}$, $\boldsymbol{A}^{(1)}$ and $\boldsymbol{A}^{(2)}$ can be written as
	\begin{subequations} 
		\begin{align}
			\label{matriceA}
			\boldsymbol{A}^{(0)}&=
			\begin{bmatrix}
				-k_{q_{1}}k_{q_{2}}n^{(2)}_{iq_{1}pq_{2}}-s^{2}n^{(2,2)}_{ip}	 & -\iota s k_{q_{1}} \tilde{n}^{(2,1)}_{iq_{1}}-\iota k_{q_{1}} \tilde{n}^{(2)}_{iq_{1}}  \\
				-\iota s k_{q_{1}} \tilde{m}^{(2,1)}_{pq_{1}}   & -m^{(2)}_{q_{1}q_{2}} k_{q_{1}}k_{q_{2}}-s m^{(2,1)}-s^{2}m^{(2,2)}\\
			\end{bmatrix},
			\\
			\label{matriceB}
			\boldsymbol{A}^{(1)}&=
			\begin{bmatrix}
				-\iota k_{r_{1}}k_{r_{2}}k_{q_{1}}n^{(3)}_{ir_{1}r_{2}pq_{1}}-\iota s^{2}k_{q_{1}}n^{(3,2)}_{iq_{1}}	 & -k_{q_{1}}k_{q_{2}}\tilde{n}^{(3)}_{iq_{1}q_{2}}-s k_{q_{1}}k_{q_{2}}\tilde{n}^{(3,1)}_{iq_{1}q_{2}}+s^{2}\tilde{\tilde{n}}^{(3,2)}_{i}+s^{3} \tilde{\tilde{n}}^{(3,3)}_{i}  \\
				s^{3}\tilde{m}^{(3,3)}_{p}-s k_{q_{1}}k_{q_{2}} \tilde{m}^{(3,1)}_{pq_{1}q_{2}}   & -\iota k_{r_{1}}k_{r_{2}}k_{q_{1}}m^{(3)}_{r_{1}r_{2}q_{1}}+\iota s^{2} k_{q_{1}} m^{(3,2)}_{q_{1}}+\iota s k_{q_{1}} m^{(3,1)}_{q_{1}}\\
			\end{bmatrix},
			\\
			\label{matriceC}
			\boldsymbol{A}^{(2)}&=
			\begin{bmatrix}
				\begin{pmatrix}sk_{q_{1}}k_{q_{2}}n^{(4,1)}_{iq_{1}pq_{2}}-s^{4}n^{(4,4)}_{ip}+\\
					+k_{q_{1}}k_{q_{2}}k_{r_{1}}k_{r_{2}}n^{(4)}_{iq_{1}q_{2}pr_{1}r_{2}}+\\
					-s^{2}k_{q_{1}}k_{q_{2}}n^{(4,2)}_{iq_{1}pq_{2}}\end{pmatrix}	 & \begin{pmatrix} -\iota s k_{q_{1}} \tilde{n}^{(4,1)}_{iq_{1}}-\iota k_{r_{1}}k_{r_{2}}k_{q_{1}}\tilde{n}^{(4)}_{ir_{1}r_{2}q_{1}}+\\
					-\iota s k_{r_{1}}k_{r_{2}}k_{q_{1}}\tilde{n}^{(4,1)}_{ir_{1}r_{2}q_{1}}+\iota s^{2} k_{q_{1}} \tilde{\tilde{n}}^{(4,2)}_{iq_{1}}+\\
					+\iota s^{3} k_{q_{1}} \tilde{\tilde{n}}^{(4,3)}_{iq_{1}} \end{pmatrix} \\
				\begin{pmatrix}
					-\iota s k_{r_{1}}k_{r_{2}}k_{q_{1}}\tilde{m}^{(4,1)}_{pr_{1}r_{2}q_{1}}-\iota s^{2}k_{q_{1}} \tilde{m}^{(4,2)}_{pq_{1}}+\\
					+\iota s^{3}k_{q_{1}} \tilde{m}^{(4,3)}_{pq_{1}}
				\end{pmatrix}	 & \begin{pmatrix}
					k_{q_{1}}k_{q_{2}}k_{r_{1}}k_{r_{1}}m^{(4)}_{q_{1}q_{2}r_{1}r_{2}}-s k_{q_{1}}k_{q_{2}} m^{(4,1)}_{q_{1}q_{2}}+\\
					-s^{2} k_{q_{1}}k_{q_{2}} m^{(4,2)}_{q_{1}q_{2}}-s^{2}\tilde{\tilde{m}}^{(4,2)}+\\
					-s^{3}m^{(4,3)}-s^{4}m^{(4,4)}
				\end{pmatrix}\\
			\end{bmatrix}.
		\end{align}
	\end{subequations} 
	The complex spectrum of a thermoelastic periodic material can be accomplished by considering $\boldsymbol{\check{\hat{f}}}=\boldsymbol{0}$ and by determining the roots of the characteristic equation that is provided by the implicit dispersion relation as
	\begin{equation}
		\label{PC}
		\mathcal{T}(\boldsymbol{k},s)=\det\Big(\boldsymbol{A}^{(0)}+\varepsilon \boldsymbol{A}^{(1)}+ \varepsilon^{2} \boldsymbol{A}^{(2)}+O(\varepsilon^{3})\Big)=0,
	\end{equation}
	which depends on the wavevector $\boldsymbol{k} \in \mathbb{C}^{3}$ and the complex angular frequency $s$, namely $\boldsymbol{k}=\mathcal{R}e(\boldsymbol{k})+\iota \mathcal{I}m(\boldsymbol{k})$ and $s=\mathcal{R}e(s)+\iota \mathcal{I}m(s)$. Alternatively, if the characteristic equation \eqref{PC} is expressed by means of its real part and its imaginary part $\mathcal{T}(\boldsymbol{k},s)=\mathcal{R}e(\mathcal{T}(\boldsymbol{k},s))+\iota \mathcal{I}m(\mathcal{T}(\boldsymbol{k},s))$, then the spectrum is given by the intersection of the hyper-surfaces
	\begin{equation}
		\begin{cases}
			\mathcal{R}e(\mathcal{T}(\mathcal{R}e(\boldsymbol{k}),\mathcal{I}m(\boldsymbol{k}),\mathcal{R}e(s), \mathcal{I}m(s))=0\\
			\mathcal{I}m(\mathcal{T}(\mathcal{R}e(\boldsymbol{k}),\mathcal{I}m(\boldsymbol{k}),\mathcal{R}e(s), \mathcal{I}m(s))=0
		\end{cases}.
		\label{hyp}
	\end{equation}
	Furthermore, the complex wavevector $\boldsymbol{k}$ is assumed to be  $\boldsymbol{k}=||\mathcal{R}e(\boldsymbol{k})||\boldsymbol{v}_{r}+\iota ||\mathcal{I}m(\boldsymbol{k})||\boldsymbol{v}_{i}$ in terms of the versors $\boldsymbol{v}_{r}$, $\boldsymbol{v}_{i} \in \mathbb{R}^{3}$, which indicate the direction of the normal to planes of constant phase and planes of constant amplitude of the propagating wave, respectively. As detailed in \cite{caviglia1992inhomogeneous, carcione2007wave, fantoni2021generalized, diana2023thermodinamically}, a plane wave is said to be homogeneous if $\boldsymbol{v}_{r}=\boldsymbol{v}_{i}=\boldsymbol{v}$, yielding $\boldsymbol{k}=(||\mathcal{R}e(\boldsymbol{k})||+\iota ||\mathcal{I}m(\boldsymbol{k})||)\boldsymbol{v}=\chi \boldsymbol{v}$, with $\chi=\mathcal{R}e(\chi)+\iota \mathcal{I}m(\chi)$. Therefore, in case of an homogeneous plane wave, the relation \eqref{hyp} becomes 
	\begin{equation}
		\begin{cases}
			\mathcal{R}e(\mathcal{T}(\mathcal{R}e(\chi),\mathcal{I}m(\chi),\mathcal{R}e(s),\mathcal{I}m(s))=0\\
			\mathcal{I}m(\mathcal{T}(\mathcal{R}e(\chi),\mathcal{I}m(\chi),\mathcal{R}e(s), \mathcal{I}m(s))=0
		\end{cases}.
		\label{hyp2}
	\end{equation}
	Moreover, to investigate the wave propagation with spatial damping, the complex angular frequency is considered to be $s=\iota \omega$, with $\omega \in \mathbb{R}$, then the relation \eqref{hyp2} can be rewritten as
	\begin{equation}
		\begin{cases}
			\mathcal{R}e(\mathcal{R}e(\chi),\mathcal{I}m(\chi),\omega)=0\\
			\mathcal{I}m(\mathcal{R}e(\chi),\mathcal{I}m(\chi),\omega)=0
		\end{cases}.
		\label{hyp1}
	\end{equation}
	Another way to achieve the dispersion spectrum is to reshape the matricial system \eqref{MMM}, for homogeneous waves and $\boldsymbol{\check{\hat{f}}}=\boldsymbol{0}$, as
	\begin{equation}
		\label{MMM2}
		(\boldsymbol{\Gamma}^{(0)}(\omega)+\chi \boldsymbol{\Gamma}^{(1)}(\omega)+ \chi^{2} \boldsymbol{\Gamma}^{(2)}(\omega)+\chi^{3}\boldsymbol{\Gamma}^{(3)}(\omega)+\chi^{4}\boldsymbol{\Gamma}^{(4)}(\omega)+O(\chi^{5}))\check{\hat{\boldsymbol{P}}}(\chi,\omega)=\boldsymbol{0},
	\end{equation}  
	which results to be an eigenproblem in terms of the wavenumber $\chi$ and the angular frequency $\omega$. Specifically, the wavenumber $\chi$ plays the role of the eigenvalue and $\check{\hat{\boldsymbol{P}}}(\chi,\omega)$ is the eigenvector. The $2 \times 2$ matrices $\boldsymbol{\Gamma}^{(0)}$, $\boldsymbol{\Gamma}^{(1)}$, $\boldsymbol{\Gamma}^{(2)}$, $\boldsymbol{\Gamma}^{(3)}$ and $\boldsymbol{\Gamma}^{(4)}$ can be written as
	\begin{align}
		\label{M} 	&\boldsymbol{\Gamma}^{(0)}=\boldsymbol{\Gamma}^{(0,0)}+\varepsilon \boldsymbol{\Gamma}^{(0,1)}+\varepsilon^{2}\boldsymbol{\Gamma}^{(0,2)}+O(\varepsilon^{3}),\\ &\boldsymbol{\Gamma}^{(1)}=\boldsymbol{\Gamma}^{(1,0)}+\varepsilon \boldsymbol{\Gamma}^{(1,1)}+\varepsilon^{2}\boldsymbol{\Gamma}^{(1,2)}+O(\varepsilon^{3}),\nonumber \\ &\boldsymbol{\Gamma}^{(2)}=\boldsymbol{\Gamma}^{(2,0)}+\varepsilon \boldsymbol{\Gamma}^{(2,1)}+\varepsilon^{2}\boldsymbol{\Gamma}^{(2,2)}+O(\varepsilon^{3}),\nonumber\\ &\boldsymbol{\Gamma}^{(3)}=\varepsilon \boldsymbol{\Gamma}^{(3,1)}+\varepsilon^{2}\boldsymbol{\Gamma}^{(3,2)}+O(\varepsilon^{3}),\nonumber\\	&\boldsymbol{\Gamma}^{(4)}=\varepsilon^{2}\boldsymbol{\Gamma}^{(4,2)}+O(\varepsilon^{3}),\nonumber
	\end{align}
	which depend on the components of the versor $\boldsymbol{v}$ and they are reported on Section B of Supplementary material.
	\subsection{Free waves propagation via a zeroeth-order approximation of Floquet-Bloch spectrum}
	\label{FWPFOH}
	To analyze the dispersion properties in the real-valued frequency domain and complex-valued wavenumber domain, the equations \eqref{epho21wk}-\eqref{epho22wk} may be truncated at the zeroeth order of $\varepsilon$ and the source terms are supposed to equal to zero ($\boldsymbol{\check{\hat{f}}}=\boldsymbol{0}$). As a result, the matricial system \eqref{MMM} is transformed into:
	\begin{equation}
		\label{SMZ}	\boldsymbol{A}^{(0)}\check{\hat{\boldsymbol{P}}}(\boldsymbol{k},s)=\boldsymbol{0}.
	\end{equation}
	The zeroeth order approximate complex spectrum of a thermoelastic periodic material can be achieved by solving the characteristic equation as follows
	\begin{equation}
		\label{PCZO}
		\mathcal{T}_{0}(\boldsymbol{k},s)=\det(\boldsymbol{A}^{(0)})=0,
	\end{equation}
	which depends on the wavevector $\boldsymbol{k} \in \mathbb{C}^{3}$ and the complex angular frequency $s$. It is worth highlighting that the zeroth order approximate complex spectrum of a thermoelastic periodic material precisely matches the complex spectrum of a first-order homogenized thermoelastic material. 
	To study the wave propagation through a thermoelastic periodic material with spatial damping, the complex angular frequency is supposed to be $s=\iota \omega$, with $\omega \in \mathbb{R}$, and, for homogeneous waves, bearing in mind that $\boldsymbol{k}=\chi \boldsymbol{v}$ the eigenproblem \eqref{SMZ} can be modified in terms of powers of $\chi$ and $\omega$ as follows
	\begin{equation}
		\label{MMM2ZO}
		(\boldsymbol{\Gamma}^{(0,0)}(\omega)+\chi \boldsymbol{\Gamma}^{(1,0)}(\omega)+ \chi^{2} \boldsymbol{\Gamma}^{(2,0)}(\omega))\check{\hat{\boldsymbol{P}}}(\chi,\omega)=\boldsymbol{0},
	\end{equation}  
	where the wavenumber $\chi$ is the eigenvalue and $\check{\hat{\boldsymbol{P}}}(\chi,\omega)$ is the eigenvector. The eigenproblem \eqref{MMM2ZO} can be linearized as
	\begin{equation}
		\label{MMM0ZO}
		(\boldsymbol{N}^{(0)}- \chi  \boldsymbol{N}^{(1)})\check{\hat{\boldsymbol{R}}}(\chi,\omega)=\boldsymbol{0},
	\end{equation}
	where the eigenvector $\check{\hat{\boldsymbol{R}}}$ is built as $\check{\hat{\boldsymbol{R}}}(\chi,\omega)=(\chi \check{\hat{\boldsymbol{P}}}(\chi,\omega) \quad \check{\hat{\boldsymbol{P}}}(\chi,\omega))^{\top}$ and the $2\times2$ matrices $\boldsymbol{N}^{(0)}$, $\boldsymbol{N}^{(1)}$ are
	\begin{align}
		\label{matrixformdue1ZO}
		\boldsymbol{N}^{(0)}&=
		\begin{pmatrix}
			\boldsymbol{\Gamma}^{(1,0)}	 & \boldsymbol{\Gamma}^{(0,0)} \\
			\boldsymbol{-1}   &\boldsymbol{0}   \\  
		\end{pmatrix},\quad 
		\boldsymbol{N}^{(1)}=
		\begin{pmatrix}
			-\boldsymbol{\Gamma}^{(2,0)}	 & \boldsymbol{0} \\
			\boldsymbol{0}	 &-\boldsymbol{1}	   \\
		\end{pmatrix}.
	\end{align}
	It can be noticed that $\boldsymbol{1}$ is the identity tensor. The diagonal matrix $\boldsymbol{N}^{(1)}$ can be inverted, enabling to rewrite the eigenproblem \eqref{MMM0ZO} as the standard form as follows
	\begin{equation}
		\label{MMM01ZO}
		(\boldsymbol{S}-\chi \boldsymbol{1}  )\check{\hat{\boldsymbol{R}}}(\chi,\omega)=\boldsymbol{0},
	\end{equation}
	where the $2\times2$ matrix $\boldsymbol{S}$ is
	\begin{align}
		\label{FSZO}
		\boldsymbol{S}&=
		\begin{pmatrix}
			-\boldsymbol{\Gamma}^{(1,0)}(\boldsymbol{\Gamma}^{(2,0)})^{-1}	 & -\boldsymbol{\Gamma}^{(0,0)} \\
			(\boldsymbol{\Gamma}^{(2,0)})^{-1}   &\boldsymbol{0}   \\  
		\end{pmatrix}.
	\end{align}  
	The characteristic polynomial stemming from the eigenproblem \eqref{MMM01ZO} is recast with respect to the invariant coefficients as 
	\begin{equation}
		\label{decseninv4ZO}
		\mathcal{M}_{0}(\chi,\omega)=\det(\boldsymbol{S}-\chi \boldsymbol{1})=\sum_{n=0}^{4} II_{n}(\omega)\chi^{n},
	\end{equation}
	where the invariant coefficients $II_{n}(\omega)$ are computed via the Faddeev-LeVerrier recursive formula \cite{helmberg1993faddeev} and they are reported on Section C.1 of Supplementary material.
	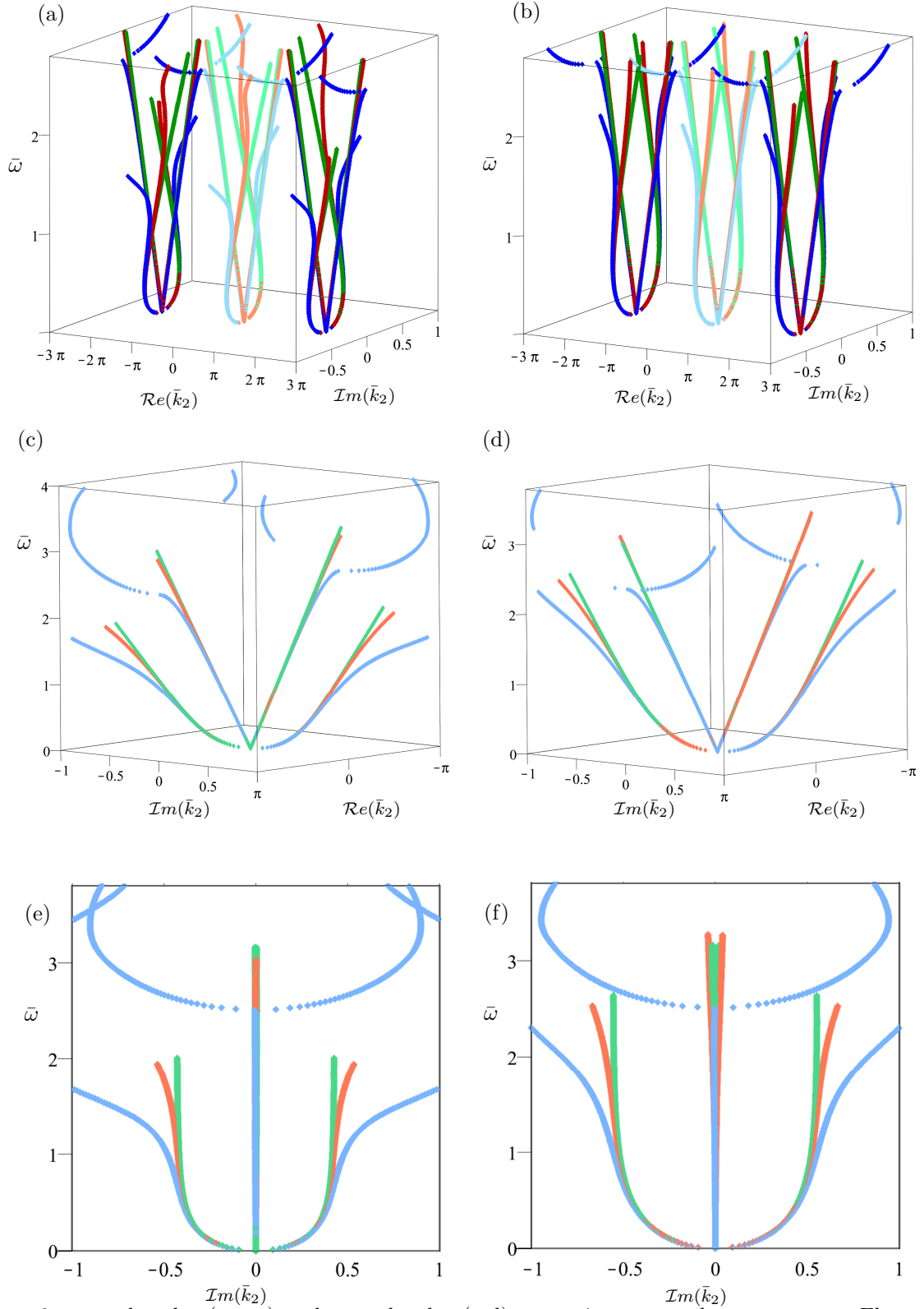
\begin{figure*}[htbp]
		\centering
		\begin{minipage}[c][\width]{0.40\textwidth}
			\hspace{-10pt}
			{\begin{overpic}[width=\textwidth]{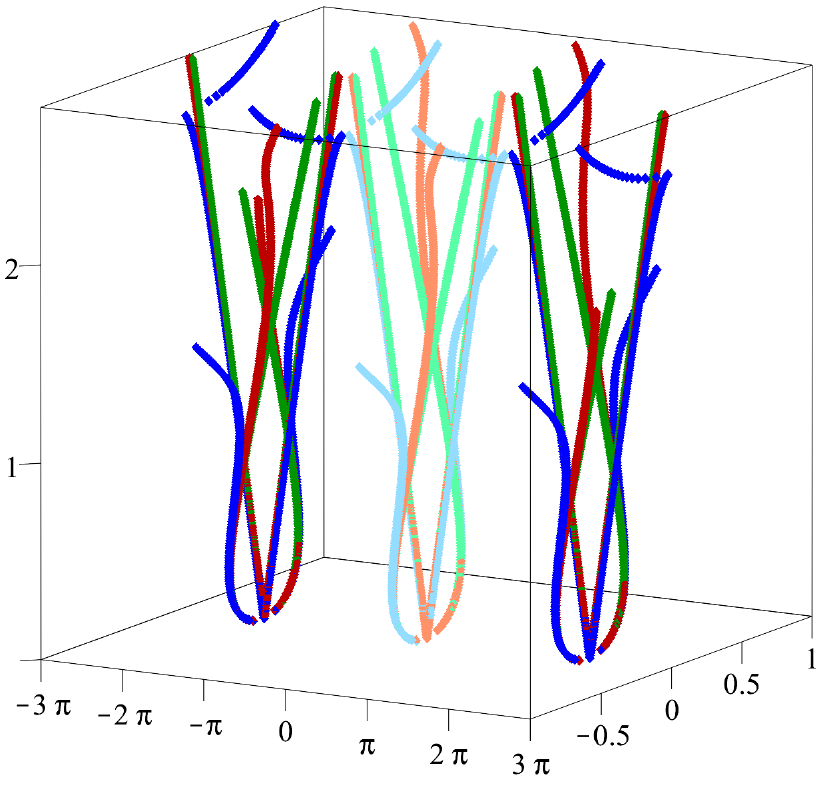}
					\put(02,92){\normalsize (a)}
					\put(-05,55){\normalsize $\bar\omega$}
					\put(27,-02){\footnotesize ${\mathcal{R}}e(\bar{k}_{2})$}
					\put(74,-01){\footnotesize ${\mathcal{I}}m(\bar{k}_{2})$}
			\end{overpic}}
		\end{minipage}	\qquad
		\begin{minipage}[c][\width]{0.40\textwidth}
			\hspace{-05pt} 
			{\begin{overpic}[width=\textwidth]{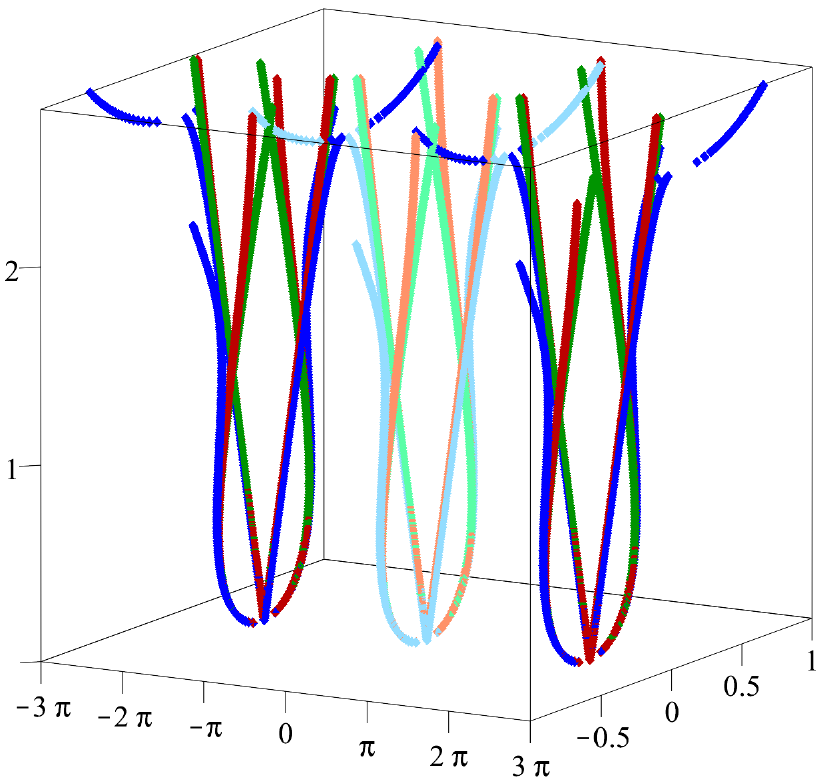}
					\put(02,92){\normalsize (b)}
					\put(-05,55){\normalsize $\bar\omega$}
					\put(27,-02){\footnotesize ${\mathcal{R}}e(\bar{k}_{2})$}
					\put(74,-01){\footnotesize ${\mathcal{I}}m(\bar{k}_{2})$}
			\end{overpic}}
		\end{minipage}\\[10pt]
		\begin{minipage}[c][1\width]{0.40\textwidth}
			\hspace{-16pt}\vspace{3pt}
			{\begin{overpic}[width=\textwidth]{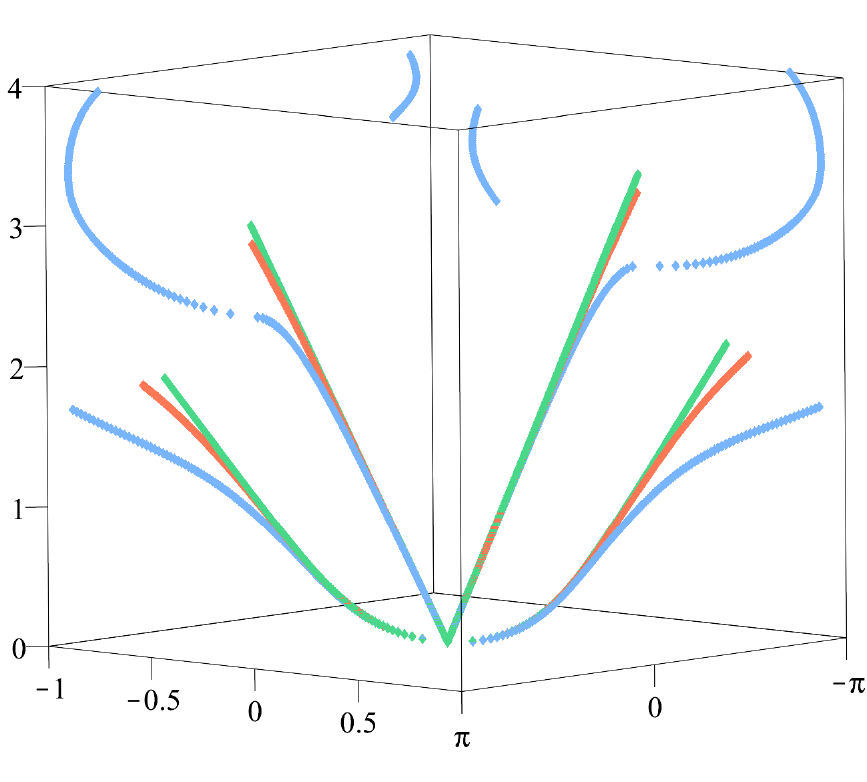}
					\put(-05,90){\normalsize (c)}
					\put(-05,65){\normalsize $\bar\omega$}
					\put(27,00){\footnotesize ${\mathcal{I}}m(\bar{k}_{2})$}
					\put(74,00){\footnotesize ${\mathcal{R}}e(\bar{k}_{2})$}
			\end{overpic}}
		\end{minipage}	
		\begin{minipage}[c][1\width]{0.40\textwidth}
			\hspace{8pt}\vspace{3pt} 
			{\begin{overpic}[width=\textwidth]{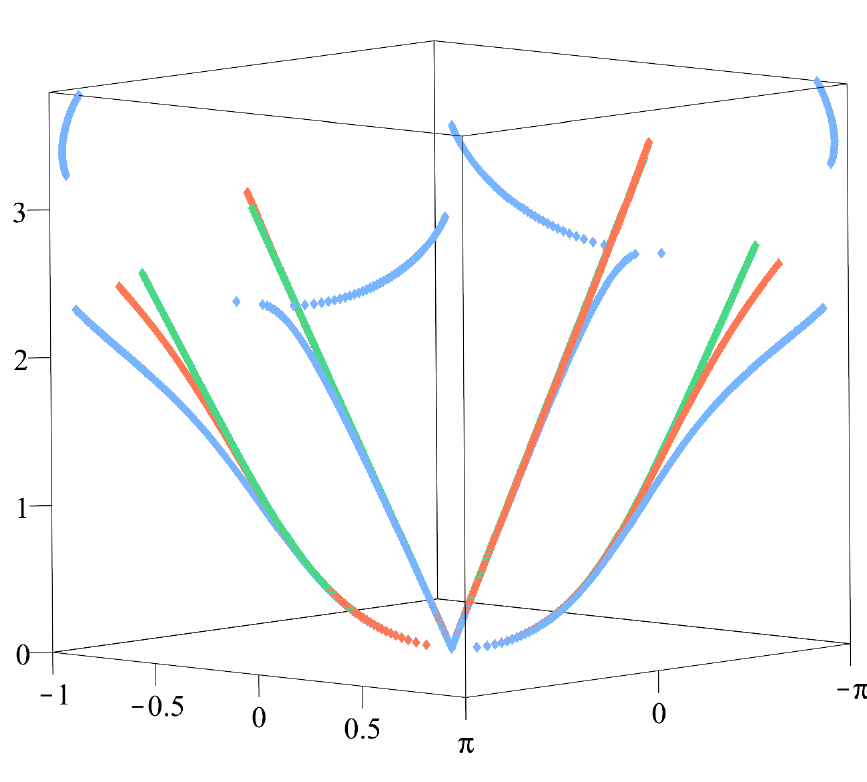}	
					\put(-05,90){\normalsize (d)}
					\put(-05,65){\normalsize $\bar\omega$}
					\put(27,00){\footnotesize ${\mathcal{I}}m(\bar{k}_{2})$}
					\put(74,00){\footnotesize ${\mathcal{R}}e(\bar{k}_{2})$}
			\end{overpic}}
		\end{minipage}\\[20pt]
		\begin{minipage}[c][1\width]{0.39\textwidth}
			\hspace{-16pt}\vspace{3pt}
			{\begin{overpic}[width=\textwidth]{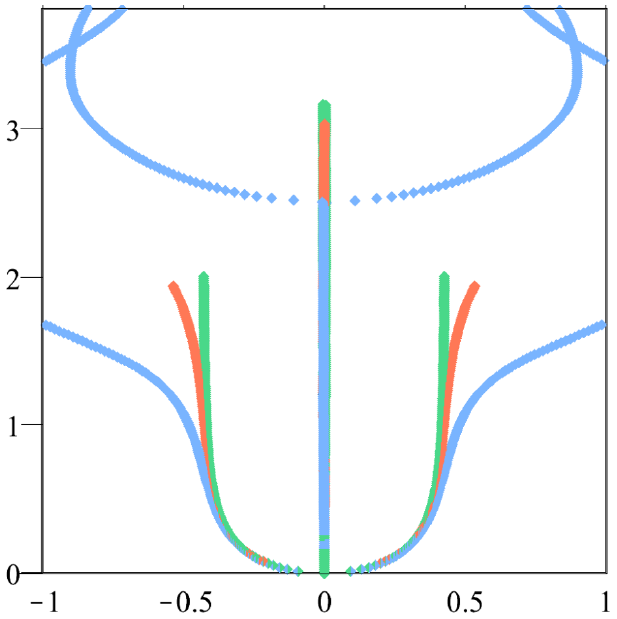}
					\put(-05,90){\normalsize (e)}
					\put(-05,65){\normalsize $\bar\omega$}
					\put(40,-05){\footnotesize ${\mathcal{I}}m(\bar{k}_{2})$}
			\end{overpic}}
		\end{minipage}	
		\begin{minipage}[c][1\width]{0.39\textwidth}
			\hspace{8pt} \vspace{3pt}
			{\begin{overpic}[width=\textwidth]{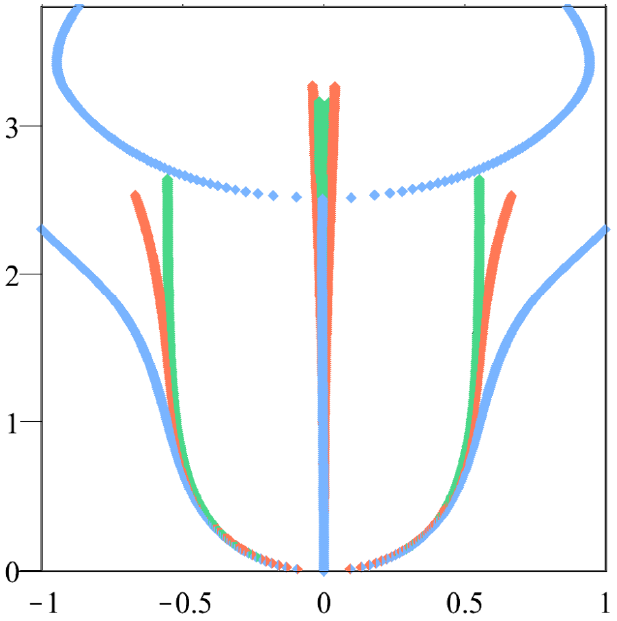}	
					\put(-05,90){\normalsize (f)}
					\put(-05,65){\normalsize $\bar\omega$}
					\put(40,-05){\footnotesize ${\mathcal{I}}m(\bar{k}_{2})$}
			\end{overpic}}
		\end{minipage}
		\vspace{0pt}
		\caption{zeroeth-order (green) and second-order (red) approximate complex spectra vs. Floquet-Bloch (blue) spectra associated with compressional-thermal waves at $k_{1}=0$. The spectra are evaluated for:  $\frac{p^{2}}{p^{1}}=3$, $\frac{C^{2}_{2222}}{C^{1}_{2222}}=2$, $\frac{\rho^{2}}{\rho^{1}}=3$, $\tilde{\nu}_{1}=\tilde{\nu}_{2}=0.2$, $\frac{\bar{K}^{2}_{22}}{\bar{K}^{1}_{22}}=3$, $\frac{\alpha^{1}_{22}\theta_{0}}{C^{1}_{2222}}=\frac{1}{100}$, $\frac{\alpha^{2}_{22}\theta_{0}}{C^{2}_{2222}}=\frac{1}{10}$, $\frac{\alpha^{1}_{22}\eta \sqrt{C^{1}_{2222}/\rho^{1}}}{\bar{K}^{1}_{22}}=\frac{1}{100}$, $\frac{\alpha^{2}_{22}\eta \sqrt{C^{1}_{2222}/\rho^{1}}}{\bar{K}^{2}_{22}}=\frac{1}{10}$, $\frac{p^{1}\theta_{0}\eta \sqrt{C^{1}_{2222}/\rho^{1}}}{\bar{K}^{1}_{22}}=1$, $\frac{\tau^{1}_{0} \sqrt{C^{1}_{2222}/\rho^{1}}}{\varepsilon}=1$, $\frac{\tau^{1}_{1} \sqrt{C^{1}_{2222}/\rho^{1}}}{\varepsilon}=3$ and $\eta=1$. The ratios between the relaxation times phases $\tau^{m}_{0}$ and $\tau^{m}_{1}$ are varied as $\frac{\tau^{2}_{0}}{\tau^{1}_{0}}=\frac{\tau^{2}_{1}}{\tau^{1}_{1}}=2$ in (a), (c), (e) and $\frac{\tau^{2}_{0}}{\tau^{1}_{0}}=\frac{\tau^{2}_{1}}{\tau^{1}_{1}}=1$ in (b), (d), (e).}
		\label{004}
		\vspace{3pt}
	\end{figure*}
	\subsection{Free waves propagation via a second-order approximation of Floquet-Bloch spectrum}
	\label{FWPFSH}
	A second-order approximation of the Floquet-Bloch spectrum related to a thermoelastic periodic material will be herein analyzed via the equations \eqref{epho21wk}-\eqref{epho22wk} truncated at the second order of $\varepsilon$ and by supposing that $\boldsymbol{\check{\hat{f}}}=\boldsymbol{0}$. Therefore, the matricial system \eqref{MMM} can be reshaped as
	\begin{equation}
		\label{T2CPt}
		(\boldsymbol{A}^{(0)}+\varepsilon \boldsymbol{A}^{(1)}+ \varepsilon^{2} \boldsymbol{A}^{(2)})\check{\hat{\boldsymbol{P}}}(\boldsymbol{k},s)=\boldsymbol{0}.
	\end{equation} 
	The second-order approximate complex spectrum of a thermoelastic material can be found by solving the roots of the characteristic equation that is provided as
	\begin{equation}
		\label{PCt}
		\mathcal{T}_{2}(\boldsymbol{k},s)=\det\Big(\boldsymbol{A}^{(0)}+\varepsilon \boldsymbol{A}^{(1)}+ \varepsilon^{2} \boldsymbol{A}^{(2)}\Big)=0,
	\end{equation}
	which relies on the wavevector $\boldsymbol{k} \in \mathbb{C}^{3}$ and the complex angular frequency $s$. The current second-order approximation scheme allows to identify a second-order thermoelastic continuum. This continuum can be obtained by imposing a second-order truncation of the transformed energy-like functional and assuming its first variation to be zero, as described in \cite{del2019characterization}.
	When considering homogeneous waves, the wavevector is assumed to be $\boldsymbol{k}=\chi \boldsymbol{v}$, where $\chi\in\mathbb{C}$. To investigate spatial damping, the complex angular frequency is denoted as $s=\iota \omega$, with $\omega \in \mathbb{R}$. Consequently, the dispersion spectrum is obtained by solving the following eigenproblem
	\begin{equation}
		\label{MMM2t}
		(\boldsymbol{\Gamma}^{(0)}(\omega)+\chi \boldsymbol{\Gamma}^{(1)}(\omega)+ \chi^{2} \boldsymbol{\Gamma}^{(2)}(\omega)+\chi^{3}\boldsymbol{\Gamma}^{(3)}(\omega)+\chi^{4}\boldsymbol{\Gamma}^{(4)}(\omega))\check{\hat{\boldsymbol{P}}}(\chi,\omega)=\boldsymbol{0},
	\end{equation}  
	where the wavenumber $\chi$ plays the role of the eigenvalue and $\check{\hat{\boldsymbol{P}}}(\chi,\omega)$ is the eigenvector. 
	The eigenproblem \eqref{MMM2t} can be linearized as
	\begin{equation}
		\label{MMM0}
		(\boldsymbol{L}^{(0)}- \chi  \boldsymbol{L}^{(1)})\check{\hat{\boldsymbol{V}}}(\chi,\omega)=\boldsymbol{0},
	\end{equation}
	where the eigenvector $\check{\hat{\boldsymbol{V}}}$ can be written as $\check{\hat{\boldsymbol{V}}}(\chi,\omega)=(\chi^{3} \check{\hat{\boldsymbol{P}}}(\chi,\omega) \quad ...\quad \check{\hat{\boldsymbol{P}}}(\chi,\omega))^{\top}$ and the $8\times8$ matrices $\boldsymbol{L}^{(0)}$, $\boldsymbol{L}^{(1)}$ can be built as
	\begin{align}
		\label{LL0}
		&\boldsymbol{L}^{(0)}=\boldsymbol{L}^{(0,0)}+\varepsilon \boldsymbol{L}^{(0,1)}+\varepsilon^{2}\boldsymbol{L}^{(0,2)},\\	
		&\boldsymbol{L}^{(1)}=\varepsilon^{2}\boldsymbol{L}^{(1,2)}.\nonumber
	\end{align}
	The submatrices that appear in \eqref{LL0} are
	\begin{subequations} 
		\begin{align}
			\label{matrixformdue1}
			\boldsymbol{L}^{(0,0)}&=
			\begin{pmatrix}
				\boldsymbol{0}	 & \boldsymbol{\Gamma}^{(2,0)} & \boldsymbol{\Gamma}^{(1,0)}& \boldsymbol{\Gamma}^{(0,0)} \\
				\boldsymbol{0}   &\boldsymbol{0} & \boldsymbol{0}&\boldsymbol{0}  \\
				\boldsymbol{0}   &\boldsymbol{0} & \boldsymbol{0}&\boldsymbol{0}  \\
				\boldsymbol{0}   &\boldsymbol{0} & \boldsymbol{0}&\boldsymbol{0}  
			\end{pmatrix}, \quad \boldsymbol{L}^{(0,1)}=
			\begin{pmatrix}
				\boldsymbol{\Gamma}^{(3,1)}	 & \boldsymbol{\Gamma}^{(2,1)} & \boldsymbol{\Gamma}^{(1,1)}& \boldsymbol{\Gamma}^{(0,1)} \\
				\boldsymbol{0}   &\boldsymbol{0} & \boldsymbol{0}&\boldsymbol{0}  \\
				\boldsymbol{0}   &\boldsymbol{0} & \boldsymbol{0}&\boldsymbol{0}  \\
				\boldsymbol{0}   &\boldsymbol{0} & \boldsymbol{0}&\boldsymbol{0}  
			\end{pmatrix}, \\ 
			\boldsymbol{L}^{(0,2)}&=
			\begin{pmatrix}
				\boldsymbol{\Gamma}^{(3,2)}	 & \boldsymbol{\Gamma}^{(2,2)} & \boldsymbol{\Gamma}^{(1,2)}& \boldsymbol{\Gamma}^{(0,2)} \\
				-\boldsymbol{1}   &\boldsymbol{0} & \boldsymbol{0}&\boldsymbol{0}  \\
				\boldsymbol{0}   &-\boldsymbol{1} & \boldsymbol{0}&\boldsymbol{0}  \\
				\boldsymbol{0}   &\boldsymbol{0} & -\boldsymbol{1}&\boldsymbol{0}  
			\end{pmatrix}, \quad
			\boldsymbol{L}^{(1,2)}=
			\begin{pmatrix}
				-\boldsymbol{\Gamma}^{(4,2)}	 & \boldsymbol{0} & \boldsymbol{0} & \boldsymbol{0}   \\
				\boldsymbol{0}	 &-\boldsymbol{1}	   & 	\boldsymbol{0} &\boldsymbol{0}\\
				\boldsymbol{0}	 &\boldsymbol{0}	   & 	-\boldsymbol{1} &\boldsymbol{0}\\
				\boldsymbol{0}	 &\boldsymbol{0}	   & 	\boldsymbol{0} &-\boldsymbol{1}
			\end{pmatrix}.
		\end{align}
	\end{subequations} 
	The invertible diagonal matrix $\boldsymbol{L}^{(1,2)}$ reduces the eigenproblem \eqref{MMM0} to the standard form as 
	\begin{equation}
		\label{MMM01}
		(\boldsymbol{D}-\varphi \boldsymbol{1}  )\check{\hat{\boldsymbol{V}}}(\chi,\omega)=\boldsymbol{0},
	\end{equation} 
	in terms of the eigenvalue $\varphi=\varepsilon^{2}\chi$ with $\boldsymbol{D}=\boldsymbol{D}^{(0)}+\varepsilon \boldsymbol{D}^{(1)}+\varepsilon^{2}\boldsymbol{D}^{(2)}$, where the matrices $\boldsymbol{D}^{(0)}$, $\boldsymbol{D}^{(1)}$ and $\boldsymbol{D}^{(2)}$ can be written as 
	\begin{subequations} 
		\begin{align}
			\label{matrixformdue2}
			\boldsymbol{D}^{(0)}&=
			\begin{pmatrix}
				\boldsymbol{0}	 & -\boldsymbol{\Gamma}^{(2,0)} & -\boldsymbol{\Gamma}^{(1,0)}& -\boldsymbol{\Gamma}^{(0,0)} \\
				\boldsymbol{0} &\boldsymbol{0} & \boldsymbol{0}&\boldsymbol{0}  \\
				\boldsymbol{0}   &\boldsymbol{1} & \boldsymbol{0}&\boldsymbol{0}  \\
				\boldsymbol{0}   &\boldsymbol{0} & \boldsymbol{1}&\boldsymbol{0}  
			\end{pmatrix},\\ \boldsymbol{D}^{(1)}&=
			\begin{pmatrix}
				-\boldsymbol{\Gamma}^{(3,1)}(\boldsymbol{\Gamma}^{(4,2)})^{-1}	 & -\boldsymbol{\Gamma}^{(2,1)} & -\boldsymbol{\Gamma}^{(1,1)}& -\boldsymbol{\Gamma}^{(0,1)} \\
				\boldsymbol{0}   &\boldsymbol{0} & \boldsymbol{0}&\boldsymbol{0}  \\
				\boldsymbol{0}   &\boldsymbol{0} & \boldsymbol{0}&\boldsymbol{0}  \\
				\boldsymbol{0}   &\boldsymbol{0} & \boldsymbol{0}&\boldsymbol{0}  
			\end{pmatrix}, \\
			\boldsymbol{D}^{(2)}&=
			\begin{pmatrix}
				-\boldsymbol{\Gamma}^{(3,2)}	(\boldsymbol{\Gamma}^{(4,2)})^{-1} & -\boldsymbol{\Gamma}^{(2,2)} & -\boldsymbol{\Gamma}^{(1,2)}& -\boldsymbol{\Gamma}^{(0,2)} \\
				(\boldsymbol{\Gamma}^{(4,2)})^{-1}   &\boldsymbol{0} & \boldsymbol{0}&\boldsymbol{0}  \\
				\boldsymbol{0}   &\boldsymbol{1} & \boldsymbol{0}&\boldsymbol{0}  \\
				\boldsymbol{0}   &\boldsymbol{0} & \boldsymbol{1}&\boldsymbol{0}  
			\end{pmatrix}.
		\end{align}
	\end{subequations} 
	The characteristic polynomial that derives from the eigenproblem \eqref{MMM01} is expressed in terms of the invariant coefficients as 
	\begin{equation}
		\label{decseninv4}
		\mathcal{M}_{2}(\varphi,\omega)=\det(\boldsymbol{D}-\varphi \boldsymbol{1})=\sum_{n=0}^{8} I_{n}(\omega,\varepsilon)\varphi^{n},
	\end{equation}
	\begin{figure*}[htbp]
		\centering
		\begin{minipage}[c][\width]{0.45\textwidth}
			\hspace{-10pt}
			{\begin{overpic}[width=\textwidth]{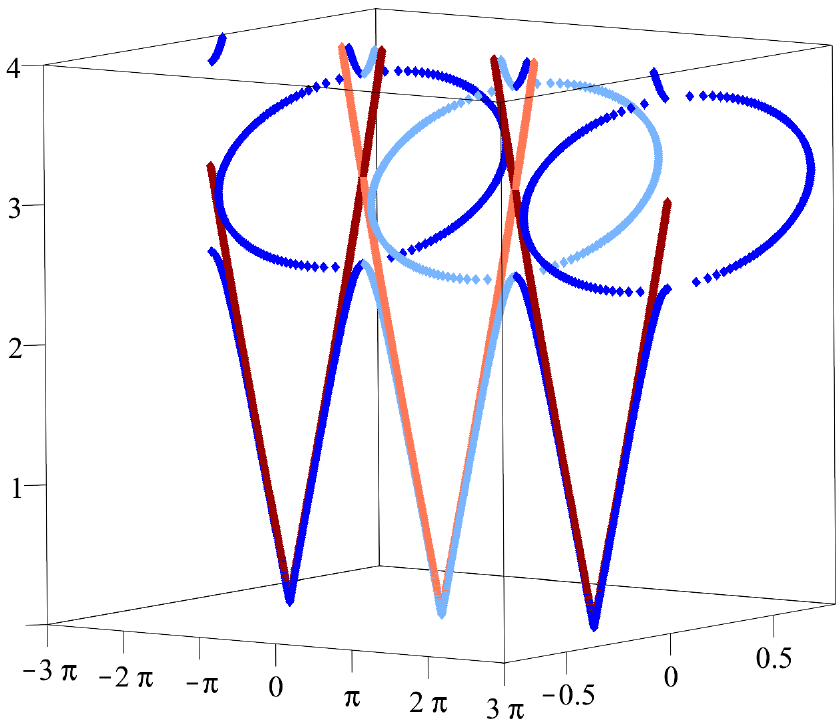}
					\put(02,88){\normalsize (a)}
					\put(-05,55){\normalsize $\bar\omega$}
					\put(27,-02){\footnotesize ${\mathcal{R}}e(\bar{k}_{2})$}
					\put(74,-01){\footnotesize ${\mathcal{I}}m(\bar{k}_{2})$}
			\end{overpic}}
		\end{minipage}	\qquad
		\begin{minipage}[c][\width]{0.40\textwidth}
			\hspace{-05pt} 
			{\begin{overpic}[width=\textwidth]{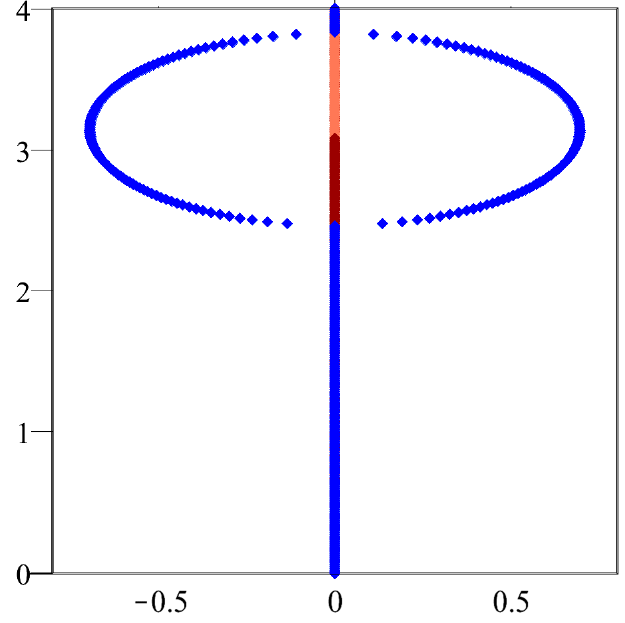}
					\put(-03,88){\normalsize (b)}
					\put(-03,58){\normalsize $\bar\omega$}
					\put(60,00){\footnotesize ${\mathcal{I}}m(\bar{k}_{2})$}
			\end{overpic}}
		\end{minipage}
		\vspace{0pt}
		\caption{comparison of second-order (red) approximate complex spectra with Floquet-Bloch (blue) spectra associated with shear waves at $k_{1}=0$. The spectra are evaluated for fixed non-null constitutive parameters $\frac{C^{2}_{1212}}{C^{1}_{1212}}=2$, $\frac{\rho^{2}}{\rho^{1}}=2$, $\eta=1$, and $\tilde{\nu}_{1}=\tilde{\nu}_{2}=0.2$. (a) Represents the translated complex spectra, while (b) displays the complex spectra represented in the $(\bar\omega, \mathcal{I}m(\bar{k}_{2}))-$plane.}
		\label{005}
		\vspace{0pt}
	\end{figure*}
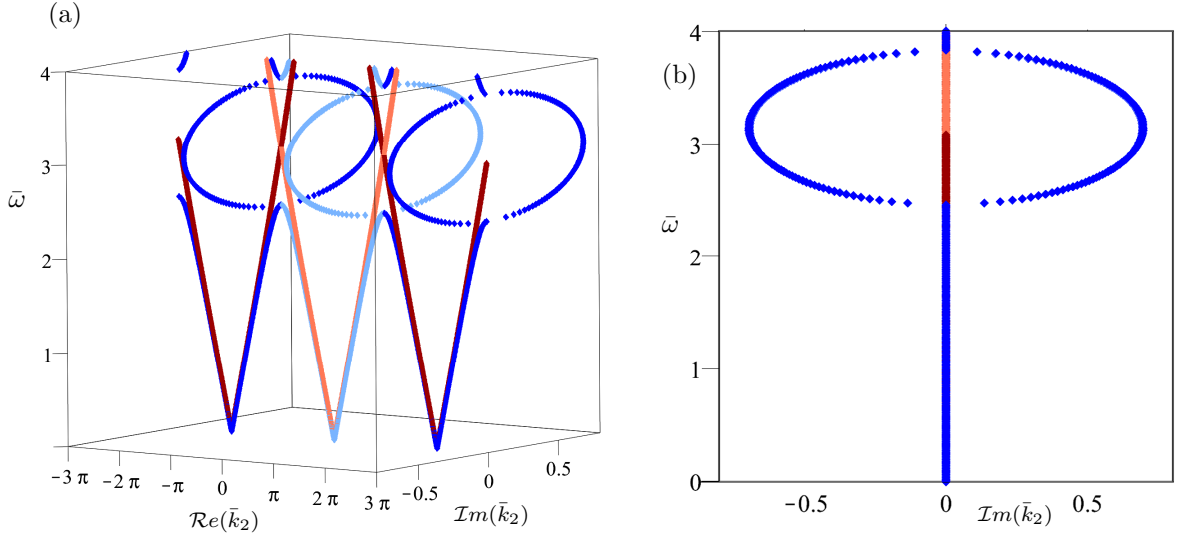
	where the invariants are reported on Section C.2 of Supplementary material.   
	\section{Illustrative examples}
	A thermoelastic periodic layered material is herein employed as an example. The material is composed of two layers, with thickness $s_{1}$ and $s_{2}$ ($\varepsilon=s_{1}+s_{2}$) and subject to $\mathcal{L}$-periodic body forces $\boldsymbol{b}(\boldsymbol{x},t)$. The material exhibits orthotropic phases and the orthotropic axis is supposed to be parallel to the direction $\boldsymbol{e}_{1}$ and the wavenumber $k_{1}=0$. In the following, the transfer matrix method is used together
	with the Floquet-Bloch theory to determine an eigenproblem governing
	the frequency dispersion spectrum of a thermoelastic periodic layered heterogeneous material. Then, the approximate dispersion curves, obtained via the scheme proposed in Section 4, will be compared with those related to the heterogeneous material via the Floquet-Bloch theory.
	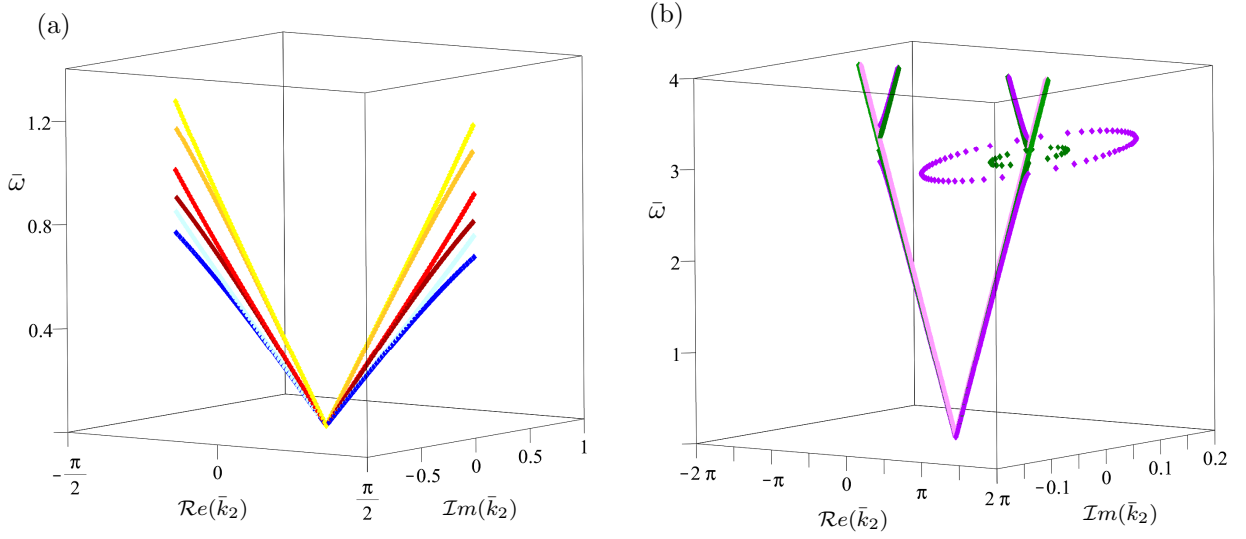
\begin{figure*}[htbp]
		\centering
		\begin{minipage}[c][\width]{0.45\textwidth}
			\hspace{-10pt}
			{\begin{overpic}[width=\textwidth]{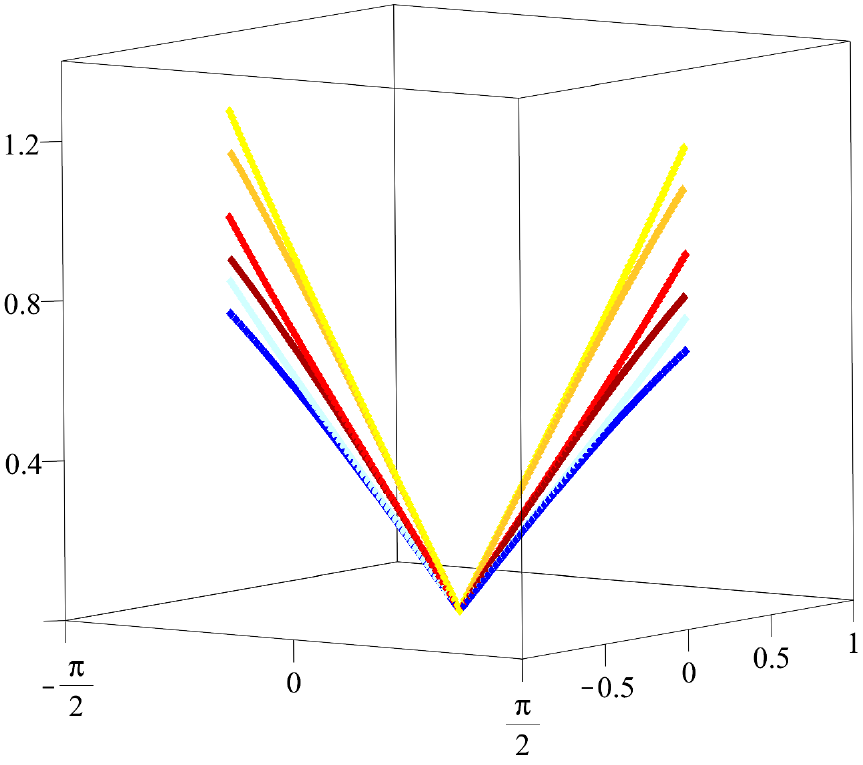}
					\put(02,88){\normalsize (a)}
					\put(-03,60){\normalsize $\bar\omega$}
					\put(27,02){\footnotesize ${\mathcal{R}}e(\bar{k}_{2})$}
					\put(74,02){\footnotesize ${\mathcal{I}}m(\bar{k}_{2})$}
			\end{overpic}}
		\end{minipage}	\qquad
		\begin{minipage}[c][\width]{0.45\textwidth}
			\hspace{-05pt} 
			{\begin{overpic}[width=\textwidth]{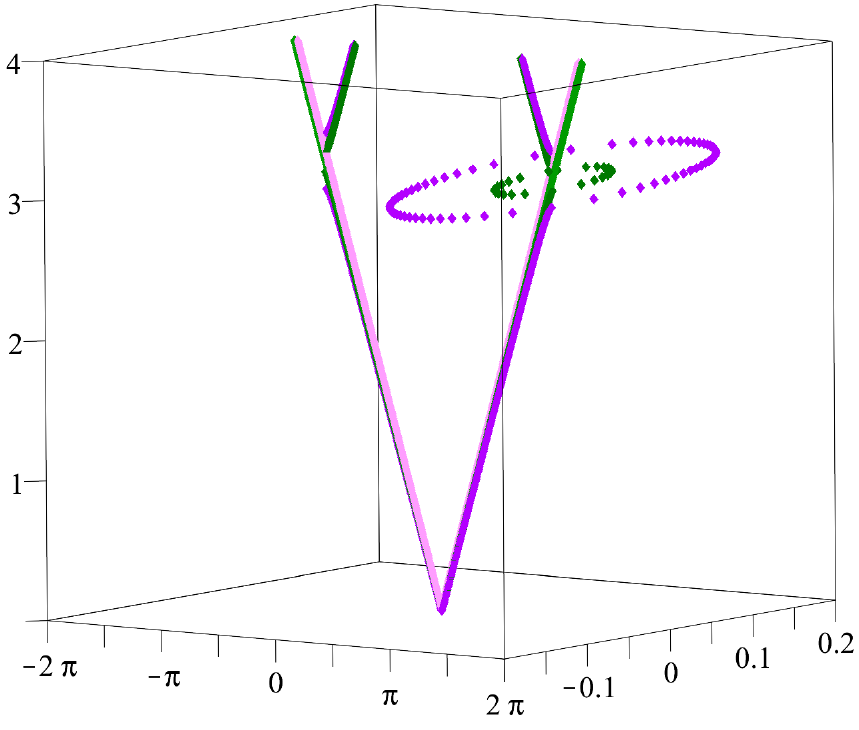}
					\put(-03,88){\normalsize (b)}
					\put(-03,53){\normalsize $\bar\omega$}
					\put(27,-02){\footnotesize ${\mathcal{R}}e(\bar{k}_{2})$}
					\put(74,-01){\footnotesize ${\mathcal{I}}m(\bar{k}_{2})$}
			\end{overpic}}
		\end{minipage}
		\vspace{0pt}
		\caption{comparison of second-order (light) approximate complex spectra with Floquet-Bloch (dark) spectra associated with shear waves at $k_{2}=0$. The spectra are evaluated under different conditions: (a) varying $\frac{C^{2}_{1212}}{C^{1}_{1212}}=\frac{\rho^{2}}{\rho^{1}}$ as $15$ (yellow), $10$ (red), and $5$ (blue), while maintaining fixed non-null constitutive parameters $\eta=1$ and $\tilde{\nu}_{1}=\tilde{\nu}_{2}=0.2$; (b) varying $\eta=10$ (green) and $\eta=30$ (violet), with fixed non-null constitutive parameters $\frac{C^{2}_{1212}}{C^{1}_{1212}}=3$, $\frac{\rho^{2}}{\rho^{1}}=2$, and $\tilde{\nu}_{1}=\tilde{\nu}_{2}=0.2$.}
		\label{006}
		\vspace{0pt}
	\end{figure*}
	\subsection{Free wave propagation for thermoelastic periodic layered heterogeneous materials}
	\label{FWP}
	In order to employ the Floquet-Bloch theory, the wave solution of field equatios \eqref{EBTC3}-\eqref{EBTC5} is expressed as
	\begin{align}
		\boldsymbol{g}(x_{2},t)=(\tilde{u}(x_{2},t) \quad \tilde{\upsilon}(x_{2},t))^{\top}=\boldsymbol{w}(x_{2}) e^{[\iota(\boldsymbol{k}\cdot \boldsymbol{x}-\omega t)]}, \label{FBBC}
	\end{align} 
	where $\omega$ is the angular frequency and the vector $\boldsymbol{w}(x_{2})=(\hat{u}(x_{2}) \quad \hat{\upsilon}(x_{2}))^{\top}$ gathers the periodic $x_{2}$-dependent Bloch amplitudes of the
	displacement and the temperature. The wave vector \eqref{FBBC} is replaced into the field equations \eqref{EBTC3}-\eqref{EBTC5} that, for a single layer $j$ (with $j=\{1,2\}$), are specialized as 
	\begin{subequations}
		\begin{align}
			&C^{j}_{1212}\hat{u}_{1,22}+2\iota k_{2} C^{j}_{1212}\hat{u}_{1,2}-(k^{2}_{2}C^{j}_{1212}+s^{2}\rho^{j})\hat{u}_{1}=0, \label{SP1}\\
			&C^{j}_{2222}\hat{u}_{2,22}+2\iota k_{2} C^{j}_{2222}\hat{u}_{2,2}-(k^{2}_{2}C^{j}_{2222}+s^{2}\rho^{j})\hat{u}_{2}-(\alpha^{j}_{22}+s\alpha^{(1,j)}_{22})\hat{\upsilon}_{,2}-\iota k_{2}(\alpha^{j}_{22}+s\alpha^{(1,j)}_{22})\hat{\upsilon}=0,\label{SP2} \\
			&K^{j}_{22}\hat{\upsilon}_{,22}+2\iota k_{2} K^{j}_{22}\hat{\upsilon}_{,2}-(k^{2}_{2}K^{j}_{22}+sp^{j}+s^{2}p^{(0,j)})\hat{v}_{2}-s\alpha^{j}_{22}(\hat{u}_{2,2}+\hat{u}_{2}\iota k_{2})=0, \label{SP3}
		\end{align}  
	\end{subequations}
	where the Bloch amplitudes are the unknown variables. The procedure to obtain the transfer matrix, as described in \cite{fantoni2021generalized, bacigalupo2022design, del2022dispersive}, is briefly outlined here. First, the constitutive relations \eqref{EC1} and \eqref{EC3} are transformed using the Floquet-Bloch decomposition. This transformation yields the transformed stress components $\hat{\sigma}_{12}$, $\hat{\sigma}_{22}$, and the transformed heat flux $\hat{q}_{2}$. Next, the vector $\boldsymbol{y}=(\hat{u}_{1} \quad \hat{u}_{2}\quad \hat{\upsilon}\quad \hat{\sigma}_{12}\quad \hat{\sigma}_{22}\quad \hat{q}_{2})^{\top}$ is evaluated at the upper $(+)$ and lower $(-)$ boundary surfaces of the $j$-th layer. Since the layers are perfectly bonded, the continuity condition $\boldsymbol{y}^{+}_{j}=\boldsymbol{y}^{-}_{j+1}$ holds at the interface between any pair of adjacent layers $j$ and $j+1$. Thus, for a periodic cell consisting of two layers, the relation connecting the generalized vector $\boldsymbol{y}^{+}_{2}$ at the upper boundary of the second layer to the generalized vector $\boldsymbol{y}^{-}_{1}$ at the lower boundary of the first layer is given by $\boldsymbol{y}^{+}_{2}=\boldsymbol{T}(\omega)\boldsymbol{y}^{-}_{1}$, where $\boldsymbol{T}(\omega)$ represents the real-valued frequency-dependent transfer matrix of the periodic cell. By imposing the Floquet-Bloch boundary condition $\boldsymbol{y}^{+}_{2}=\exp[\iota k_{2} \varepsilon]\boldsymbol{y}^{-}_{1}$, which accounts for the spatial periodicity of the cell, the eigenproblem can be formulated as
	\begin{equation}
		\label{eigen1}
		\big(\boldsymbol{T}(\omega)-\varphi \boldsymbol{I}\big)\boldsymbol{y}^{-}_{1}=\boldsymbol{0},
	\end{equation}
	where the complex-valued eigenvalue $\varphi = \exp [\iota k_{2} \varepsilon]$ plays the same role as the Floquet multipler. The characteristic polynomial that derives from the eigenproblem \eqref{eigen1} is
	\begin{equation}
		\label{eigen2}
		\mathcal{H} (\varphi,\omega)=\det(\boldsymbol{T}(\omega)-\varphi \boldsymbol{I}).
	\end{equation}
	The matrix $\boldsymbol{T}(\omega)$ possesses a unimodular property, where its determinant remains independent of $\omega$. The characteristic polynomial $\mathcal{H} (\varphi,\omega)$ of $\boldsymbol{T}(\omega)$ exhibits palindromic symmetry. As a result, both $\varphi$ and its reciprocal $1/\varphi$ must be eigenvalues of $\boldsymbol{T}(\omega)$. Furthermore, since the characteristic polynomial has real-valued coefficients, both $\varphi$ and its complex conjugate $\varphi^{*}$ are also eigenvalues. The palindromic
	characteristic polynomial $\eqref{eigen2}$ is expressed in terms of invariant coefficients as
	\begin{equation}
		\label{eigen3}
		\mathcal{H} (\varphi,\omega)=\det(\boldsymbol{T}(\omega)-\varphi \boldsymbol{I})=\sum_{n=0}^{6} III_{n}(\omega)\varphi^{n},
	\end{equation}
	where the invariants are reported on Section C.3 of Supplementary material. The equation \eqref{eigen2} represents the implicit dispersion relation of plane wave oscillations in thermoelastic periodic layered materials featured by an elementary cell made of two homogeneous layers. 
	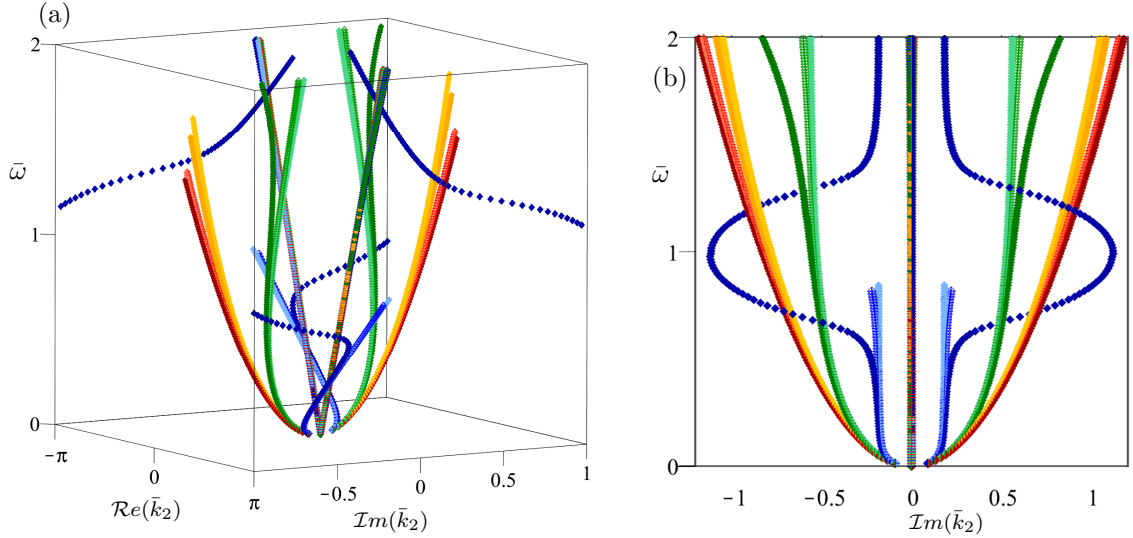
\begin{figure*}[htbp]
		\centering
		\begin{minipage}[c][\width]{0.45\textwidth}
			\hspace{-10pt}
			{\begin{overpic}[width=\textwidth]{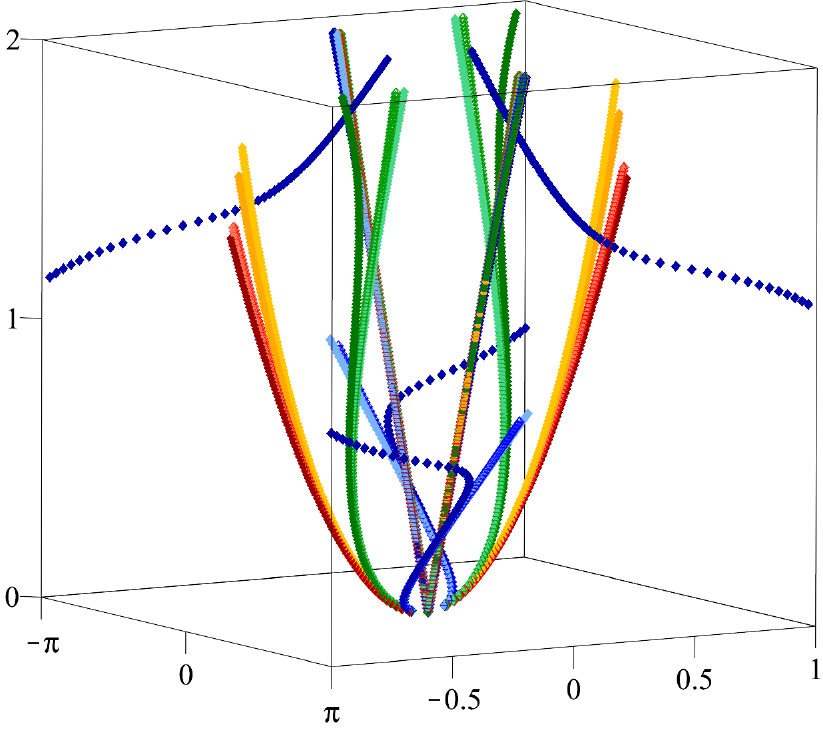}
					\put(02,88){\normalsize (a)}
					\put(-03,60){\normalsize $\bar\omega$}
					\put(15,00){\footnotesize ${\mathcal{R}}e(\bar{k}_{2})$}
					\put(58,-02){\footnotesize ${\mathcal{I}}m(\bar{k}_{2})$}
			\end{overpic}}
		\end{minipage}	\qquad
		\begin{minipage}[c][\width]{0.37\textwidth}
			\hspace{-05pt} 
			{\begin{overpic}[width=\textwidth]{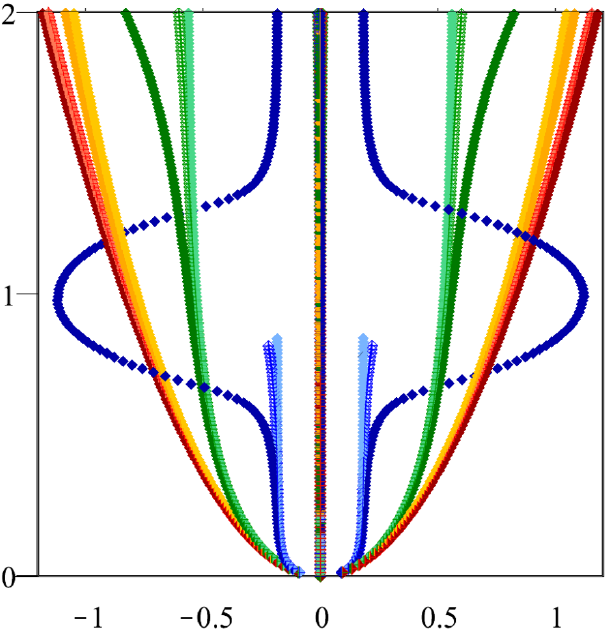}
					\put(-03,88){\normalsize (b)}
					\put(-03,68){\normalsize $\bar\omega$}
					\put(50,-04){\footnotesize ${\mathcal{I}}m(\bar{k}_{2})$}
			\end{overpic}}
		\end{minipage}
		\vspace{0pt}
		\caption{comparison of second-order (light and dyamond) approximate, zeroeth-order (light) approximate and exact (dark) complex spectra with Floquet-Bloch spectra associated with compressional-thermal waves at $k_{1}=0$. The spectra are obtained for: $\frac{p^{2}}{p^{1}}=3$, $\frac{C^{2}_{2222}}{C^{1}_{2222}}=2$, $\frac{\rho^{2}}{\rho^{1}}=3$, $\tilde{\nu}_{1}=\tilde{\nu}_{2}=0.2$, $\frac{\bar{K}^{2}_{22}}{\bar{K}^{1}_{22}}=3$, $\frac{\alpha^{1}_{22}\theta_{0}}{C^{1}_{2222}}=\frac{1}{100}$, $\frac{\alpha^{2}_{22}\theta_{0}}{C^{2}_{2222}}=\frac{1}{10}$, $\frac{\alpha^{1}_{22}\eta \sqrt{C^{1}_{2222}/\rho^{1}}}{\bar{K}^{1}_{22}}=\frac{1}{100}$, $\frac{\alpha^{2}_{22}\eta \sqrt{C^{1}_{2222}/\rho^{1}}}{\bar{K}^{2}_{22}}=\frac{1}{10}$, $\frac{p^{1}\theta_{0}\eta \sqrt{C^{1}_{2222}/\rho^{1}}}{\bar{K}^{1}_{22}}=1$, $\frac{\tau^{2}_{0}}{\tau^{1}_{0}}=\frac{\tau^{2}_{1}}{\tau^{1}_{1}}=1$ and $\eta=1$ with $\frac{\tau^{1}_{0} \sqrt{C^{1}_{2222}/\rho^{1}}}{\varepsilon}=\frac{\tau^{1}_{1} \sqrt{C^{1}_{2222}/\rho^{1}}}{\varepsilon}=0$ (red),  $\frac{\tau^{1}_{0} \sqrt{C^{1}_{2222}/\rho^{1}}}{\varepsilon}=\frac{\tau^{1}_{1} \sqrt{C^{1}_{2222}/\rho^{1}}}{\varepsilon}=1/10$ (yellow),  $\frac{\tau^{1}_{0} \sqrt{C^{1}_{2222}/\rho^{1}}}{\varepsilon}=\frac{\tau^{1}_{1} \sqrt{C^{1}_{2222}/\rho^{1}}}{\varepsilon}=1$ (green) and $\frac{\tau^{1}_{0} \sqrt{C^{1}_{2222}/\rho^{1}}}{\varepsilon}=\frac{\tau^{1}_{1} \sqrt{C^{1}_{2222}/\rho^{1}}}{\varepsilon}=10$ (blue). (a) represents the three-dimensional complex spectra, while (b) displays the complex spectra in the $(\bar\omega, \mathcal{I}m(\bar{k}_{2}))-$plane.}
		\label{007}
		\vspace{0pt}
	\end{figure*}
	\subsection{Benchmark test: heterogeneous material vs. homogenized scheme}
	In this Subsection, the Floquet-Bloch complex spectra will be analyzed and a comparative study will be conducted between the results obtained from the Floquet-Bloch theory (Subsection \ref{FWP}) and those based on the multifield asymptotic homogenization scheme (Subsections \ref{FWPFOH} and \ref{FWPFSH}) for a bi-phase layered material. In case of isotropic phases and plane-stress state, it results that $\tilde{E}=E$ and $\tilde{\nu}=\nu$, whereas the plane-strain state implies that $\tilde{E}=\frac{E}{1-\nu^{2}}$ and $\tilde{\nu}=\frac{\nu}{1-\nu}$, where $E$ is the Young's modulus and $\nu$ is the Poisson's ratio. To simplify the representation, the components of the elastic tensor are defined as $C_{1111}^{j}=C_{2222}^{j}=\frac{\tilde{E}}{1-\tilde{\nu}^{2}}, \quad C_{1122}^{j}=\frac{\tilde{E}\tilde{\nu}}{1-\tilde{\nu}^{2}},$ and $C_{1212}^{j}=\frac{\tilde{E}}{2(1+\tilde{\nu})}$, where the superscript $j=\{1,2\}$ stands for the phase 1 and the phase 2.\\
	The perturbation function, denoted as $\tilde{M}^{(2,1)}_{22}$, is computed analytically by solving the cell problem stated in equation \eqref{33b}, along with the interface conditions described in equation \eqref{BC2}. The computation is carried out with respect to phase 1 and phase 2 and the structure of $\tilde{M}^{(2,1)}_{22}$ is presented in terms of the geometric and thermo-mechanical properties of the periodic domain, as expressed in equation (55) of Supplementary material. This function depends on the fast variable $\xi_{2}$, which is perpendicular to the direction $\boldsymbol{e}_{1}$.
	Figure \ref{0B}-(a) illustrates the behaviour of the non-dimensionalized perturbation function  $\bar{\tilde{M}}^{(2,1)}_{22}=\frac{\tilde{M}^{(2,1)}_{22}{K}^{1}_{22}\theta_{0}}{C^{1}_{2222}}$ along the vertical coordinate $\xi_{2}$ and in terms of the ratio $\frac{\alpha^{2}_{22}}{\alpha^{1}_{22}}$.
	To determine the non-dimensionalized perturbation function $M^{(2)}_{22}$, the cell problem specified in equation \eqref{33a}, along with the interface conditions provided in equation \eqref{BC1}, is solved. The formulation for $M^{(2)}_{22}$ is explicitly expressed in equation (66) of Supplementary material, emphasizing its consideration of the effect of microstructural heterogeneities within the domain.
	Figure \ref{0B}-(b) shows the behaviour of the perturbation function $M^{(2)}_{22}$ by varying the vertical coordinate $\xi_{2}$ and the ratio $\frac{K^{2}_{22}}{K^{1}_{22}}$.
	In all the aforementioned figures, it is evident that the perturbation functions, namely $\bar{\tilde{M}}^{(2,1)}_{22}$ and $M^{(2)}_{22}$ are $\mathcal{Q}$-periodic. Furthermore, they possess vanishing mean values over the unit cell $\mathcal{Q}$ and exhibit smooth behavior along the boundaries of $\mathcal{Q}$.
	For the considered scenario, both phases are assumed to have equal Poisson ratios ($\tilde{\nu}_{1}=\tilde{\nu}_{2}=0.3$) and the thickness ratio between the phases is $\eta=1$. Additional first and second-order perturbation functions are detailed in Section D and E of the Supplementary material, respectively. Furthermore, the third-order perturbation functions can be found in \cite{de2019characterization}. Figure \ref{0C}-(a) depicts the behavior of the non-dimensionlized constitutive tensor component $\frac{\tilde{n}^{(2)}_{22}\theta_{0}}{\varepsilon \sqrt{(C^{1}_{2222}\rho^{1})}}$ with increasing values of the ratio between the relaxation time of the phases $\tau_{1}$ and $\frac{C^{2}_{2222}}{C^{1}_{2222}}$. The formulation for the component $\tilde{n}^{(2)}_{22}$ is referred to equation \eqref{CP}. Figure \ref{0C}-(b) displays the non-dimensionlized constitutive tensor component $\frac{p(\theta_{0})^{2}}{C^{1}_{2222}}$ with increasing values of the ratios $\frac{K^{2}_{22}}{K^{1}_{22}}$ and $\frac{C^{2}_{2222}}{C^{1}_{2222}}$. The  component $p$ is computed by means of equation \eqref{FOT},  for fixed not-null constitutive parameters $\frac{p^{2}}{p^{1}}=3$, $\frac{\rho^{2}}{\rho^{1}}=3$, $\tilde{\nu}_{1}=\tilde{\nu}_{2}=0.3$, $\frac{\alpha^{1}_{22}\theta_{0}}{C^{1}_{2222}}=\frac{1}{100}$, $\frac{\alpha^{2}_{22}\theta_{0}}{C^{2}_{2222}}=\frac{1}{10}$, $\frac{\alpha^{1}_{22}\eta \sqrt{C^{1}_{2222}/\rho^{1}}}{\bar{K}^{1}_{22}}=\frac{1}{100}$, $\frac{\alpha^{2}_{22}\eta \sqrt{C^{1}_{2222}/\rho^{1}}}{\bar{K}^{2}_{22}}=\frac{1}{10}$, $\frac{p^{1}\theta_{0}\eta \sqrt{C^{1}_{2222}/\rho^{1}}}{\bar{K}^{1}_{22}}=1$, $\frac{\tau^{1}_{1} \sqrt{C^{1}_{2222}/\rho^{1}}}{\varepsilon}=3$ and $\eta=1$.\\
	Assuming that the wavenumber $k_1$ is zero, the compressional-thermal wave function of the heterogeneous continuum is determined using the Floquet-Bloch theory in Subsection \ref{FWP}. Figures \ref{000} show the complex spectra obtained by determining the roots of the characteristic polynomial \eqref{eigen2}. The dimensionless parameters were carefully selected, namely $\frac{p^{2}}{p^{1}}=3$, $\frac{C^{2}_{2222}}{C^{1}_{2222}}=2$, $\frac{\rho^{2}}{\rho^{1}}=3$, $\tilde{\nu}{1}=\tilde{\nu}{2}=0.2$, $\frac{\bar{K}^{2}_{22}}{\bar{K}^{1}_{22}}=3$, $\frac{\alpha^{1}_{22}\theta{0}}{C^{1}_{2222}}=\frac{1}{100}$, $\frac{\alpha^{2}_{22}\theta_{0}}{C^{2}_{2222}}=\frac{1}{10}$, $\frac{\alpha^{1}_{22}\eta \sqrt{C^{1}_{2222}/\rho^{1}}}{\bar{K}^{1}_{22}}=\frac{1}{100}$, $\frac{\alpha^{2}_{22}\eta \sqrt{C^{1}_{2222}/\rho^{1}}}{\bar{K}^{2}_{22}}=\frac{1}{10}$, $\frac{p^{1}\theta_{0}\eta \sqrt{C^{1}_{2222}/\rho^{1}}}{\bar{K}^{1}_{22}}=1$, $\frac{\tau^{1}_{0} \sqrt{C^{1}_{2222}/\rho^{1}}}{\varepsilon}=1$, $\frac{\tau^{1}_{1} \sqrt{C^{1}_{2222}/\rho^{1}}}{\varepsilon}=3$, and $\eta=1$. The ratios between the relaxation times phases $\tau^{m}_{0}$ and $\tau^{m}_{1}$ vary, specifically, they are chosen as $\frac{\tau^{2}_{0}}{\tau^{1}_{0}}=\frac{\tau^{2}_{1}}{\tau^{1}_{1}}=2$ in (a), (c), (e) and $\frac{\tau^{2}_{0}}{\tau^{1}_{0}}=\frac{\tau^{2}_{1}}{\tau^{1}_{1}}=1$ in (b), (d), (f). Figures \ref{000}-(a) and (b) show the Floquet-Bloch complex spectra in the selected non-dimensionalized angular frequency range $\bar{\omega}=\omega \varepsilon \sqrt{\frac{\rho^{1}}{C^{1}_{2222}}}$ vs. the real and the imaginary parts of the non-dimensionalized wavenumber $\bar{k}_{2}=k_{2}\varepsilon$. Figures \ref{000}-(a) and (b) emphasize the translated complex spectra (dark curves) along ${\mathcal{R}}e(\bar{k}_{2})\in[-3\pi, -\pi], [\pi, 3\pi]$ due to the periodicity of the microstructure, whereas the light curves represent the spectra within the Brillouin zone.
	Figures \ref{000}-(c) and (d) show a representation of figures \ref{000}-(a) and (b) in the $(\bar{\omega},{\mathcal{R}}e(\bar{k}_{2}))-$plane, where ${\mathcal{R}}e(\bar{k}_{2})$ belongs to the Brillouin zone. They show the structure of stop-stop bands that correspond to wave attenuation. Figures \ref{000}-(e) and (f) depict a representation of figures \ref{000}-(a) and (b) in the $(\bar{\omega},{\mathcal{I}}m(\bar{k}_{2}))-$plane, where the opening of several stop-stop gaps, referred to the coupled compressional mechanics and thermal waves, can be observed. \\
	Figure \ref{002} illustrates the complex spectra and band structure derived by solving the characteristic polynomial \eqref{eigen2} with the same dimensionless parameters as in Figure \ref{002} except for the thickness ratio $\eta=10$. In Figure \ref{002}-(a) and (b), the complex spectra are translated along the $\bar{k}_{2}-$axis. Figures \ref{002}-(c) and (d) illustrate the corresponding representations in the $(\bar{\omega},{\mathcal{R}}e(\bar{k}_{2}))-$plane with ${\mathcal{R}}e(\bar{k}_{2})\in [-\pi,\pi]$. It may be remarked that, increasing $\eta$ and setting the same value of the ratio between the relaxation times of the phases, $\tau_{0}^{2}=\tau_{0}^{1}$ and $\tau_{1}^{2}=\tau_{1}^{1}$, the dispersione curves exhibit a quasi-elastic wave beaviour. In contrast, the structural characteristics of the material's stop-stop bands remain largely unaltered, as it can be observed in Figures \ref{002}-(e) and (f), where the complex spectra are depicted in the $(\bar{\omega},{\mathcal{I}}m(\bar{k}_{2}))-$plane.\\
	The approximate zeroth-order and second-order complex spectra can be obtained by solving the characteristic polynomials \eqref{decseninv4ZO} and \eqref{decseninv4}, respectively.
	Figure \ref{004} depicts the graphical representation of the complex spectra associated with compressional-thermal waves. The Floquet-Bloch theory yields the blue spectra, while the approximate zeroth-order and second-order complex spectra are shown in green and red, respectively. The frequency range is non-dimensionalized as $\bar{\omega}\in[0,4]$.
	Figures \ref{004}-(a) and (b) illustrate the translated complex spectra (dark curves) resulting from the periodicity of the material. The light curves represent the spectra within the Brillouin zone. In Figures \ref{004}-(c) and (d), both the exact complex spectra (blue) and the approximate complex spectra (green and red) are displayed within the first Brillouin zone. Additionally, Figures \ref{004}-(e) and (f) showcase the exact (blue) and approximate (green and red) complex spectra in the $(\bar{\omega},{\mathcal{I}}m(\bar{k}_{2}))-$plane.
	It is evident that as the truncation of the field equations \eqref{epho21wk}-\eqref{epho22wk} increases, a progressively more accurate estimation of the exact complex spectra can be achieved.\\
	Figure \eqref{005} presents plots illustrating the shear waves. Specifically, Figure \eqref{005}-(a) displays the translated complex spectra (dark curves) associated with shear waves within the range ${\mathcal{R}}e(\bar{k}_{2})\in[-3\pi, -\pi], [\pi, 3\pi]$. Additionally, the light curves represent the spectra related to the Brillouin zone. The blue curves represent the spectra obtained from the Floquet-Bloch theory, while the red curves correspond to the spectra related to the second-order approximation, for fixed not-null constitutive parameters $\frac{C^{2}_{1212}}{C^{1}_{1212}}=2$, $\frac{\rho^{2}}{\rho^{1}}=2$, $\eta=1$, $\tilde{\nu}_{1}=\tilde{\nu}_{2}=0.2$ and in the selected non-dimensionalized angular frequency range $\bar{\omega}=\omega \varepsilon \sqrt{\frac{\rho^{1}}{C^{1}_{1212}}}$. In Figure \ref{005}-(b), the complex spectra are depicted in the $(\bar{\omega},{\mathcal{I}}m(\bar{k}_{2}))-$plane. Notably, the second-order approximation scheme demonstrates its effectiveness in accurately capturing the shear wave behavior, as it exhibits excellent agreement with the heterogeneous continuum derived from the Floquet-Bloch theory.\\
	Figure \eqref{006} presents a comparison between the second-order (light) approximate complex spectra and those derived from the Floquet-Bloch theory (dark) associated with shear waves. The analysis is conducted by varying the parameter $\frac{C^{2}_{1212}}{C^{1}_{1212}}=\frac{\rho^{2}}{\rho^{1}}$, which takes on values of 15 (yellow), 10 (red), and 5 (blue), while the other constitutive parameters $\eta=1$ and $\tilde{\nu}_{1}=\tilde{\nu}_{2}=0.2$ remain fixed (Figure (a)).
	It can be observed that numerically increasing the values of the non-dimensional parameters $\frac{C^{2}_{1212}}{C^{1}_{1212}}=\frac{\rho^{2}}{\rho^{1}}$ leads to a reduction in the curvatures of the shear waves.
	Additionally, the effect of varying $\eta$, represented by values of 10 (green) and 30 (violet), is explored for fixed constitutive parameters $\frac{C^{2}_{1212}}{C^{1}_{1212}}=3$, $\frac{\rho^{2}}{\rho^{1}}=2$, and $\tilde{\nu}_{1}=\tilde{\nu}_{2}=0.2$ (Figure (b)). As previously observed in \cite{preve2021variational}, Figure \eqref{006} provides further evidence of the precise estimation of shear wave propagation between the two developed models. This accuracy is demonstrated by considering the range of ${\mathcal{R}}e(\bar{k}_{2})\in[-3\pi, 3\pi]$. When the dimensionless relaxation times of phase 1 are zero $\Big(\frac{\tau^{1}_{0} \sqrt{C^{1}_{2222}/\rho^{1}}}{\varepsilon}=0$, $\frac{\tau^{1}_{1} \sqrt{C^{1}_{2222}/\rho^{1}}}{\varepsilon}=0\Big)$ and the parameters representing the ratios between the relaxation times are one $\Big(\frac{\tau^{2}_{0}}{\tau^{1}_{0}}=\frac{\tau^{2}_{1}}{\tau^{1}_{1}}=1\Big)$ an interesting observation can be made. Indeed, this corresponds to the specific case known as classical thermoelasticity. In this particular situation, the relaxation times of both phases become zero ($\tau_{1}^{m}, \tau_{0}^{m}$ = 0), and the field equations \eqref{EBTC3}-\eqref{EBTC5} governing the periodic material revert back to the equations of the conventional thermoelastic problem. This circumstance is depicted in Figure \eqref{007}-(a), along with multiple compressional-thermal waves corresponding to three distinct values of the dimensionless relaxation times $\frac{\tau^{1}_{0} \sqrt{C^{1}_{2222}/\rho^{1}}}{\varepsilon}$ and $\frac{\tau^{1}_{1} \sqrt{C^{1}_{2222}/\rho^{1}}}{\varepsilon}$ for phase 1 of the layered material. In this comparison, the light curves represent the zeroeth-order approximate scheme, the light curves with diamonds represent the second-order approximate scheme, while the dark curves illustrate the waves characterized by the heterogeneous continuum with $\frac{\tau^{1}_{0} \sqrt{C^{1}_{2222}/\rho^{1}}}{\varepsilon}=\frac{\tau^{1}_{1} \sqrt{C^{1}_{2222}/\rho^{1}}}{\varepsilon}=0$ (red), $\frac{\tau^{1}_{0} \sqrt{C^{1}_{2222}/\rho^{1}}}{\varepsilon}=\frac{\tau^{1}_{1} \sqrt{C^{1}_{2222}/\rho^{1}}}{\varepsilon}=1/10$ (yellow), $\frac{\tau^{1}_{0} \sqrt{C^{1}_{2222}/\rho^{1}}}{\varepsilon}=\frac{\tau^{1}{1} \sqrt{C^{1}_{2222}/\rho^{1}}}{\varepsilon}=1$ (green), and $\frac{\tau^{1}_{0} \sqrt{C^{1}_{2222}/\rho^{1}}}{\varepsilon}=\frac{\tau^{1}{1} \sqrt{C^{1}_{2222}/\rho^{1}}}{\varepsilon}=10$ (blue) with $\frac{\tau^{2}_{0}}{\tau^{1}_{0}}=\frac{\tau^{2}_{1}}{\tau^{1}_{1}}=1$. The red curves ($\frac{\tau^{1}_{0} \sqrt{C^{1}_{2222}/\rho^{1}}}{\varepsilon}=\frac{\tau^{1}_{1} \sqrt{C^{1}_{2222}/\rho^{1}}}{\varepsilon}=0$) exhibit lesser curvature compared to the others. The curvature also goes up as the relaxation times expand. Consequently, for low frequencies, the dispersion curves associated with the quasi-thermal waves have an imaginary part of the dimensionless wavenumber $\bar{k}_{2}$ that tends to decrease in magnitude as the dimensionless relaxation times increase (Figure \eqref{007}-(b)) and modifying the dimensionless relaxation times of phase 1 reveals several frameworks in the frequency band structure of the layered material. Moreover, the dispersion curves obtained from the second-order homogenized scheme demonstrate high accuracy in this analysis with respect to the zeroeth-order ones.
	\section{Final remarks} 
	This paper has dealt with the propagation of dispersive waves in thermoelastic materials with periodic microstructures using an asymptotic homogenization scheme. The chosen framework incorporates the Green-Lindsay theory, which accounts for two relaxation times and enables the coupling of mechanical and thermal fields without the classical paradox of infinite thermal signal propagation speeds.
	Within this mathematical framework, the governing equations at the micro-scale are derived. The down-scaling relation connects the micro-displacement and micro-temperature fields to the macro-displacement, macro-temperature field, and their gradients through perturbation functions. These perturbation functions, which are solutions of cell problems defined over the unit cell $\mathcal{Q}$, are $\mathcal{Q}$-periodic and have zero mean values over the unit cell. Additionally, the up-scaling relation imposes that the macro-displacement and macro-temperature fields are the mean values of the corresponding micro-fields over the unit cell $\mathcal{Q}$. By replacing the down-scaling relation into the governing equations at the micro-scale, the average field equations of infinite order are obtained. These equations are formally solved by expanding the macro-displacement and macro-temperature fields in powers of the microstructural size and solving a cascade of macroscopic recursive problems.
	To study free wave propagation in a thermoelastic material with a periodic multi-phase microstructure, the transformed average field equations are expressed in the frequency and wave vector domains using Laplace and Fourier transforms. The transformed equations are truncated at the zeroth-order of $\varepsilon$ to derive the field equations for a homogeneous first-order (Cauchy) thermoelastic material. The resulting governing equations at the macro-scale are formulated in terms of overall constitutive tensors for the equivalent first-order homogenized material. Truncation at the second-order of $\varepsilon$ yields an approximation of the Floquet-Bloch spectrum.\\
	As an illustrative example, the study has focused on a thermoelastic periodic layered material with orthotropic phases and an orthotropy axis parallel to the layer direction. The governing equations are specialized for this case, employing the Floquet-Bloch decomposition and obtaining the closed-form uni-modular transfer matrix for the heterogeneous layered cell. An eigenproblem is then solved to determine the imaginary and real implicit dispersion functions, whose intersection identifies the frequency spectrum. The dispersion curves obtained from the homogenized models show good agreement with those derived using the Floquet-Bloch approach, indicating a high level of consistency. Notably, the second-order approximation provides a superior approximation, highlighting the significance of the additional terms accounted for in this approach. These second-order terms play a crucial role in capturing nonlinear relationships and coupling effects that are not adequately represented in the zeroth-order approximation. Applying the second-order approximation to peculiar problems of engineering interest can open opportunities for optimizing design and performance in thermoelastic systems. As future developments, one possibility is to explore higher-order approximations beyond the second order and continualization schemes. This can introduce additional terms in the asymptotic expansion or incorporate more complex mathematical techniques to capture finer details of the thermoelastic behavior, namely describing accurately the frequency stop-bands of shear waves propagating perpendicular to the layering direction. Moreover, including more detailed microstructural features and their influence on the macroscopic behavior can lead to strengthen accuracy in modeling thermoelastic materials. More complex microstructures (i.e. composites, porous materials), and their effects on the overall behavior, can be explored. Thermoelastic materials often interact with other physical phenomena, namely electromagnetic fields or chemical reactions. Finally, investigating the coupling of thermoelasticity with these fields can lead to a more comprehensive understanding of real-world scenarios and model complex multi-physics phenomena.
	\section*{Acknowledgement}
	The authors gratefully acknowledge financial support from National Group of Mathematical Physics, Italy (GNFM-IN$\delta$AM),  from University of Chieti-Pescara project Search for Excellence Ud’A 2019 and from University of Trento, project UNMASKED 2020.
	\bibliographystyle{myplainnat}
	\bibliography{hom}
\end{document}


\title{Supplementary material of \\ {Multifield asymptotic homogenization scheme for periodic Cauchy materials in non-conventional thermoelasticity}}
	\author{Rosaria Del Toro$^{1}$, Maria Laura De Bellis$^{1}$, Marcello Vasta$^{1}$, Andrea Bacigalupo$^{2}$}
	\date{\small $^{1}$ University of Chieti-Pescara, Department INGEO, Viale Pindaro 42, Pescara, Italy\\
		\small $^{2}$ University of Genova, Department DICCA, via Montallegro 1, Genova, Italy}	
	\maketitle
\section*{Section A \quad Recursive differential thermo-mechanical problems}
Section A shows the solutions of recursive differential problems stemming from equations (11a)-(11b) at the orders $\varepsilon$, $\varepsilon^{2}$ and $\varepsilon^{3}$ of Subsection 3.1 in the main text. The differential problems at the order $\varepsilon$ are 
\begin{subequations}
	\begin{align}
		&\Big( {C}^{m}_{ijhk}\Big ( \frac{\partial {u}^{(2)}_{h}}{\partial x_{k}}+{u}^{(3)}_{h,k} \Big) \Big)_{,j}+\frac{\partial}{\partial x_{j}} \Big({C}^{m}_{ijhk}\Big(\frac{\partial{u}^{(1)}_{h}}{\partial x_{k}}+{u}^{(2)}_{h,k}\Big)\Big)-(\alpha_{ij}^{m}\upsilon^{(2)}+\alpha_{ij}^{(m,1)}\dot{\upsilon}^{(2)})_{,j}+\label{eqn1:ep1}\\
		&-\frac{\partial}{\partial x_{j}}(\alpha_{ij}^{m}\upsilon^{(1)}+\alpha_{ij}^{(m,1)}\dot{\upsilon}^{(1)})-\rho^{m}\ddot{u}^{(1)}_{i}=f^{(3)}_{i}(\boldsymbol{x}),\nonumber\\
		&\Big( {K}^{m}_{ij}\Big ( \frac{\partial {\upsilon}^{(2)}}{\partial x_{j}}+{\upsilon}^{(3)}_{,j} \Big) \Big)_{,i}+\frac{\partial}{\partial x_{i}} \Big({K}^{m}_{ij}\Big(\frac{\partial{\upsilon}^{(1)}}{\partial x_{j}}+{\upsilon}^{(2)}_{,j}\Big)\Big)-p^{m}\dot{\upsilon}^{(1)}-p^{(m,0)}\ddot{\upsilon}^{(1)}-\alpha_{ij}^{m}\Big(\frac{\partial \dot{u}^{(1)}_{i}}{\partial x_{j}}+\dot{u}^{(2)}_{i,j}\Big)=g^{(3)}(\boldsymbol{x}),\label{eqn2:ep1}
	\end{align}
\end{subequations}
with the interface conditions
\begin{subequations}
	\begin{align}
		&\Big[\Big[u^{(3)}_{h}\Big]\Big]\Big\vert_{\boldsymbol{\xi} \in \Sigma_{1}}=0 \quad \Big [\Big[ \Big ( C^{m}_{ijhk}\Big (\frac{\partial u^{(2)}_{h}}{\partial x_{k}}+u^{(3)}_{h,k}\Big)-(\alpha_{ij}^{m}\upsilon^{(2)}+\alpha_{ij}^{(m,1)}\dot{\upsilon}^{(2)}) \Big )n_{j}\Big ]\Big]\Big \vert_{\boldsymbol{\xi} \in \Sigma_{1}}=0,\\
		&\Big[\Big[\upsilon^{(3)}\Big]\Big]\Big\vert_{\boldsymbol{\xi} \in \Sigma_{1}}=0 \quad \Big [\Big[ \Big ( K^{m}_{ij}\Big (\frac{\partial \upsilon^{(2)}}{\partial x_{j}}+\upsilon^{(3)}_{,j}\Big) \Big )n_{i}\Big ]\Big]\Big \vert_{\boldsymbol{\xi} \in \Sigma_{1}}=0.
	\end{align}
\end{subequations}
Replacing the solutions (21a)-(21b) obtained at the order $\varepsilon^{-1}$ and the solutions (27a)-(27b) referred to the order $\varepsilon^{0}$ into the equations \eqref{eqn1:ep1}-\eqref{eqn2:ep1} and bearing in mind the solvability conditions in the class of $\mathcal{Q}-$periodic functions result
\begin{subequations}
	\begin{align}
		f^{(3)}_{i}(\boldsymbol{x})&= \langle C^{m}_{iq_{2}hk}N^{(1)}_{hpq_{1}}+C^{m}_{ikhj}N^{(2)}_{hpq_{1}q_{2},j}\rangle\frac{\partial^{3}U^{M}_{p}}{\partial x_{q_{1}} \partial x_{q_{2}}\partial x_{k}}+\\
		&+\langle C^{m}_{iq_{1}hk}\tilde{N}^{(1)}_{h}+ C^{m}_{ikhj}{\tilde{N}}^{(2)}_{hq_{1},j}-\alpha_{ik}^{(m)}M^{(1)}_{q_{1}}\rangle\frac{\partial^{2} {\Upsilon}^{M}}{\partial x_{q_{1}}\partial x_{k}}+\nonumber \\
		&+\langle C^{m}_{iq_{1}hk}\tilde{N}^{(1,1)}_{h}+ C^{m}_{ikhj}{\tilde{N}}^{(2,1)}_{hq_{1},j}-\alpha_{ik}^{(m,1)}M^{(1)}_{q_{1}}\rangle\frac{\partial^{2} \dot{\Upsilon}^{M}}{\partial x_{q_{1}}\partial x_{k}}+\nonumber\\
		&+\langle C^{m}_{ikhj}N^{(2,2)}_{hp,j}-\rho^{m}N^{(1)}_{ipk}\rangle\frac{\partial \ddot{U}^{M}_{p}}{\partial x_{k}}-\langle \rho^{m}\tilde{N}^{(1)}_{i} \rangle \ddot{\Upsilon}^{M}-\langle \rho^{m}\tilde{N}^{(1,1)}_{i} \rangle \dddot{\Upsilon}^{M},\nonumber\\
		g^{(3)}(\boldsymbol{x})&=\langle K^{m}_{q_{2}j}M^{(1)}_{q_{1}}+ K^{m}_{ji}M^{(2)}_{q_{1}q_{2},i}\rangle \frac{\partial^{3}\Upsilon^{M}}{\partial x_{q_{1}} \partial x_{q_{2}}\partial x_{j}}+\\
		&+\langle K_{ji}^{m}\tilde{M}^{(2,1)}_{pq_{1},i} -\alpha^{m}_{iq_{2}}{N}^{(2)}_{ipq_{1}j,q_{2}}-\alpha_{ij}^{m}{N}^{(1)}_{ipq_{1}}\rangle\frac{\partial^{2} \dot{U}^{M}_{p}}{\partial x_{q_{1}}\partial x_{j}}+\nonumber \\
		&+\langle K_{ji}^{m}{M}^{(2,1)}_{,i}-p^{m}M^{(1)}_{j} -\alpha^{m}_{ij}\tilde{N}^{(1)}_{i}-\alpha_{iq_{1}}^{m}\tilde{N}^{(2)}_{ij,q_{1}}\rangle\frac{\partial \dot{\Upsilon}^{M}}{\partial x_{j}}+\nonumber\\
		&+\langle K_{ji}^{m}{M}^{(2,2)}_{,i}-p^{(m,0)}M^{(1)}_{j} -\alpha^{m}_{ij}\tilde{N}^{(1,1)}_{i}-\alpha_{iq_{1}}^{m}\tilde{N}^{(2,1)}_{ij,q_{1}}\rangle\frac{\partial \ddot{\Upsilon}^{M}}{\partial x_{j}}-\langle\alpha_{ij}^{m}\tilde{N}^{(2,2)}_{ip,j}\rangle \dddot{U}^{M}_{p}.\nonumber
	\end{align}
\end{subequations}
The solutions of the differential problems \eqref{eqn1:ep1}-\eqref{eqn2:ep1} are
\begin{subequations}
	\begin{align}
		\label{eqn:solfif1p1}
		u^{(3)}_{h}(\boldsymbol{x},\boldsymbol{\xi},t) &= N^{(3)}_{hpq_{1}q_{2}q_{3}}(\boldsymbol{\xi})\frac{\partial^{3}U^{M}_{p}}{\partial x_{q_{1}}\partial x_{q_{2}}\partial x_{q_{3}}}+\tilde{N}^{(3)}_{hq_{1}q_{2}}(\boldsymbol{\xi})\frac{\partial^{2} \Upsilon^{M}}{\partial x_{q_{1}}\partial x_{q_{2}}}+{\tilde{N}}^{(3,1)}_{hq_{1}q_{2}}(\boldsymbol{\xi})\frac{\partial^{2} \dot{\Upsilon}^{M}}{\partial x_{q_{1}}\partial x_{q_{2}}}+ \nonumber \\
		&+N^{(3,2)}_{hpq_{1}}(\boldsymbol{\xi})\frac{\partial \ddot{U}^{M}_{p} }{\partial x_{q_{1}}}+\tilde{\tilde{N}}_{h}^{(3,2)}(\boldsymbol{\xi})\ddot{\Upsilon}^{M}+\tilde{\tilde{N}}_{h}^{(3,3)}(\boldsymbol{\xi})\dddot{\Upsilon}^{M},\\
		\upsilon^{(3)}(\boldsymbol{x},\boldsymbol{\xi},t) &= M^{(3)}_{q_{1}q_{2}q_{3}}(\boldsymbol{\xi})\frac{\partial^{3}\Upsilon^{M}}{\partial x_{q_{1}}\partial x_{q_{2}}\partial x_{q_{3}}}+\tilde{M}^{(3,1)}_{pq_{1}q_{2}}(\boldsymbol{\xi})\frac{\partial^{2} \dot{U}^{M}_{p}}{\partial x_{q_{1}}\partial x_{q_{2}}}+M^{(3,1)}_{q_{1}}(\boldsymbol{\xi})\frac{\partial \dot{\Upsilon}^{M}}{\partial x_{q_{1}}}+ \nonumber \\
		&+M^{(3,2)}_{q_{1}}(\boldsymbol{\xi})\frac{\partial \ddot{\Upsilon}^{M}}{\partial x_{q_{1}}}+\tilde{M}^{(3,3)}_{p}(\boldsymbol{\xi})\dddot{U}^{M}_{p}  \label{eqn:solfif1p2}.	
	\end{align}
\end{subequations} 
The differential problems at the order $\varepsilon^{2}$ are 
\begin{subequations}
	\begin{align}
		&\Big({C}^{m}_{ijhk}\Big ( \frac{\partial {u}^{(3)}_{h}}{\partial x_{k}}+{u}^{(4)}_{h,k} \Big) \Big)_{,j}+\frac{\partial}{\partial x_{j}} \Big({C}^{m}_{ijhk}\Big(\frac{\partial{u}^{(2)}_{h}}{\partial x_{k}}+{u}^{(3)}_{h,k}\Big)\Big)-(\alpha_{ij}^{m}\upsilon^{(3)}+\alpha_{ij}^{(m,1)}\dot{\upsilon}^{(3)})_{,j}+\label{eqn1:ep2}\\
		&-\frac{\partial}{\partial x_{j}}(\alpha_{ij}^{m}\upsilon^{(2)}+\alpha_{ij}^{(m,1)}\dot{\upsilon}^{(2)})-\rho^{m}\ddot{u}^{(2)}_{i}=f^{(4)}_{i}(\boldsymbol{x}),\nonumber\\
		&\Big( {K}^{m}_{ij}\Big ( \frac{\partial {\upsilon}^{(3)}}{\partial x_{j}}+{\upsilon}^{(4)}_{,j} \Big) \Big)_{,i}+\frac{\partial}{\partial x_{i}} \Big({K}^{m}_{ij}\Big(\frac{\partial{\upsilon}^{(2)}}{\partial x_{j}}+{\upsilon}^{(3)}_{,j}\Big)\Big)-p^{m}\dot{\upsilon}^{(2)}-p^{(m,0)}\ddot{\upsilon}^{(2)}-\alpha_{ij}^{m}\Big(\frac{\partial \dot{u}^{(2)}_{i}}{\partial x_{j}}+\dot{u}^{(3)}_{i,j}\Big)=g^{(4)}(\boldsymbol{x}),\label{eqn2:ep2}
	\end{align}
\end{subequations}
with the interface conditions
\begin{subequations}
	\begin{align}
		&\Big[\Big[u^{(4)}_{h}\Big]\Big]\Big\vert_{\boldsymbol{\xi} \in \Sigma_{1}}=0 \quad \Big [\Big[ \Big ( C^{m}_{ijhk}\Big (\frac{\partial u^{(3)}_{h}}{\partial x_{k}}+u^{(4)}_{h,k}\Big)-(\alpha_{ij}^{m}\upsilon^{(3)}+\alpha_{ij}^{(m,1)}\dot{\upsilon}^{(3)}) \Big )n_{j}\Big ]\Big]\Big \vert_{\boldsymbol{\xi} \in \Sigma_{1}}=0,\\
		&\Big[\Big[\upsilon^{(4)}\Big]\Big]\Big\vert_{\boldsymbol{\xi} \in \Sigma_{1}}=0 \quad \Big [\Big[ \Big ( K^{m}_{ij}\Big (\frac{\partial \upsilon^{(3)}}{\partial x_{j}}+\upsilon^{(4)}_{,j}\Big) \Big )n_{i}\Big ]\Big]\Big \vert_{\boldsymbol{\xi} \in \Sigma_{1}}=0.
	\end{align}
\end{subequations}
Substituting the solutions (27a)-(27b) at the order $\varepsilon^{0}$ and the ones \eqref{eqn:solfif1p1}-\eqref{eqn:solfif1p2} at the order $\varepsilon$ into the equations \eqref{eqn1:ep2}-\eqref{eqn2:ep2} and considering the solvability conditions in the class of $\mathcal{Q}-$periodic functions yield
\begin{subequations}
	\begin{align}
		f^{(4)}_{i}(\boldsymbol{x})&= \langle C^{m}_{ikhj}N^{(3)}_{hpq_{1}q_{2}q_{3},j}+C^{m}_{iq_{3}hk}N^{(2)}_{hpq_{1}q_{2}}\rangle\frac{\partial^{4}U^{M}_{p}}{\partial x_{q_{1}} \partial x_{q_{2}}\partial x_{q_{3}}\partial x_{k}}+ \label{f4}\\
		&+\langle C^{m}_{iq_{2}hk}\tilde{N}^{(2)}_{hq_{1}} +C^{m}_{ikhj}{\tilde{N}}^{(3)}_{hq_{1}q_{2},j}-\alpha_{ik}^{m}M^{(2)}_{q_{1}q_{2}}\rangle\frac{\partial^{3} {\Upsilon}^{M}}{\partial x_{q_{1}} \partial x_{q_{2}}\partial x_{k}}+\nonumber \\		
		&+\langle C^{m}_{iq_{2}hk}\tilde{N}^{(2,1)}_{hq_{1}}+ C^{m}_{ikhj}{\tilde{N}}^{(3,1)}_{hq_{1}q_{2},j}-\alpha_{ik}^{(m,1)}M^{(2)}_{q_{1}q_{2}}\rangle\frac{\partial^{3} \dot{\Upsilon}^{M}}{\partial x_{q_{1}} \partial x_{q_{2}}\partial x_{k}}+\nonumber\\
		&+\langle C^{m}_{iq_{1}hk}\tilde{N}^{(2,2)}_{hp}+C^{m}_{ikhj}N^{(3,2)}_{hpq_{1},j}-\rho^{m}N^{(2)}_{ipq_{1}k}-\alpha_{ik}^{(m,1)}\tilde{M}^{(2,1)}_{pq_{1}}\rangle\frac{\partial^{2} \ddot{U}^{M}_{p}}{\partial x_{q_{1}}\partial x_{k}}+\nonumber \\
		&-\langle \alpha^{m}_{ik}M^{(2,1)} \rangle \frac{\partial \dot{\Upsilon}^{M}}{\partial x_{k}}+\langle C^{m}_{ikhj}{\tilde{\tilde{N}}}^{(3,2)}_{h,j}-\rho^{m}\tilde{N}^{(2)}_{ik}-\alpha^{m}_{ik}M^{(2,2)}-\alpha^{(m,1)}_{ik}M^{(2,1)} \rangle \frac{\partial \ddot{\Upsilon}^{M}}{\partial x_{k}}+\nonumber \\
		&+\langle C^{m}_{ikhj}{\tilde{\tilde{N}}}^{(3,3)}_{h,j}-\rho^{m}\tilde{N}^{(2,1)}_{ik}-\alpha^{(m,1)}_{ik}M^{(2,2)}\rangle \frac{\partial \dddot{\Upsilon}^{M}}{\partial x_{k}}+\nonumber\\
		&-\langle \rho^{m}N^{(2,2)}_{ip} \rangle \ddddot{U}_{p}^{M}-\langle \alpha_{ik}^{(m,1)}\tilde{M}^{(2,1)}_{pq_{1}}\rangle\frac{\partial^{2} \dot{U}^{M}_{p}}{\partial x_{q_{1}}\partial x_{k}},\nonumber\\
		g^{(4)}(\boldsymbol{x})&=\langle K^{m}_{q_{3}j}M^{(2)}_{q_{1}q_{2}}+ K^{m}_{ji}M^{(3)}_{q_{1}q_{2}q_{3},i}\rangle \frac{\partial^{4}\Upsilon^{M}}{\partial x_{q_{1}} \partial x_{q_{2}} \partial x_{q_{3}}\partial x_{j}}-\langle p^{m}\tilde{M}^{(2,1)}_{pj} \rangle \frac{\partial  \ddot{U}^{M}_{p}}{\partial x_{q_{1}}}+\label{g4} \\
		&+\langle K_{q_{2}j}^{m}\tilde{M}^{(2,1)}_{pq_{1}}+K^{m}_{ji}\tilde{M}^{(3,1)}_{q_{1}q_{2},i} -\alpha_{iq_{3}}^{m}{N}^{(3)}_{ipq_{1}q_{2}j,q_{3}}-\alpha_{ij}^{m}{N}^{(2)}_{ipq_{1}q_{2}}\rangle\frac{\partial^{3} \dot{U}^{M}_{p}}{\partial x_{q_{1}}\partial x_{q_{2}}\partial x_{j}}+ \nonumber \\
		&+\langle K_{q_{1}j}^{m}{M}^{(2,1)}+K_{ji}^{m}{M}^{(3,1)}_{q_{1},i}-p^{m}M^{(2)}_{q_{1}j} -\alpha^{m}_{iq_{2}}\tilde{N}^{(3)}_{iq_{1}j,q_{2}}-\alpha_{ij}^{m}\tilde{N}^{(2)}_{iq_{1}}\rangle\frac{\partial^{2} \dot{\Upsilon}^{M}}{\partial x_{q_{1}} \partial x_{j}}+\nonumber\\
		&+\langle K_{q_{1}j}^{m}{M}^{(2,2)}+K_{ji}^{m}{M}^{(3,2)}_{q_{1},i}-p^{(m,0)}M^{(2)}_{q_{1}j}-\alpha^{m}_{ij}\tilde{N}^{(2,1)}_{iq_{1}}-\alpha^{m}_{iq_{2}}\tilde{N}^{(3,1)}_{iq_{1}j,q_{2}} \rangle\frac{\partial^{2} \ddot{\Upsilon}^{M}}{\partial x_{q_{1}} \partial x_{j}}+\nonumber\\
		&+\langle K_{ji}^{m}\tilde{M}^{(3,3)}_{p,i}-p^{(m,0)}\tilde{M}^{(2,1)}_{pj}-\alpha^{m}_{ij}N^{(2,2)}_{ip}-\alpha_{iq_{1}}^{m}N^{(3,2)}_{ipj,q_{1}}\rangle\frac{\partial \dddot{U}^{M}_{p}}{\partial x_{j}}-\langle p^{m}M^{(2,1)}\rangle \ddot{\Upsilon}^{M}+\nonumber\\
		&-\langle p^{m}M^{(2,2)}+p^{(m,0)}M^{(2,1)}+\alpha_{ij}^{m}\tilde{\tilde{N}}^{(3,2)}_{i,j}\rangle \dddot{\Upsilon}^{M}-\langle p^{(m,0)}M^{(2,2)}+\alpha_{ij}^{m}\tilde{\tilde{N}}^{(3,3)}_{i,j}\rangle \ddddot{\Upsilon}^{M}.\nonumber
	\end{align}
\end{subequations}
The solutions of the problems \eqref{eqn1:ep2}-\eqref{eqn2:ep2} are
\begin{subequations}
	\begin{align}
		\label{eqn:solfif1p3}
		u^{(4)}_{h}(\boldsymbol{x},\boldsymbol{\xi},t)& = N^{(4)}_{hpq_{1}q_{2}q_{3}q_{4}}(\boldsymbol{\xi})\frac{\partial^{4}U^{M}_{p}}{\partial x_{q_{1}}\partial x_{q_{2}}\partial x_{q_{3}}\partial x_{q_{4}}}+\tilde{N}^{(4)}_{hq_{1}q_{2}q_{3}}(\boldsymbol{\xi})\frac{\partial^{3} \Upsilon^{M}}{\partial x_{q_{1}}\partial x_{q_{2}}\partial x_{q_{3}}}+{\tilde{N}}^{(4,1)}_{hq_{1}q_{2}q_{3}}(\boldsymbol{\xi})\frac{\partial^{3} \dot{\Upsilon}^{M}}{\partial x_{q_{1}}\partial x_{q_{2}} \partial x_{q_{3}}}+\nonumber\\
		&+N^{(4,2)}_{hpq_{1}q_{2}}(\boldsymbol{\xi})\frac{\partial^{2} \ddot{U}^{M}_{p} }{\partial x_{q_{1}} \partial x_{q_{2}}}+N^{(4,1)}_{hpq_{1}q_{2}}(\boldsymbol{\xi})\frac{\partial^{2} \dot{U}^{M}_{p} }{\partial x_{q_{1}} \partial x_{q_{2}}}+\nonumber\\ &+\tilde{\tilde{N}}_{hq_{1}}^{(4,1)}(\boldsymbol{\xi})\frac{\partial\dot{\Upsilon}^{M}}{\partial x_{q_{1}} }+\tilde{\tilde{N}}_{hq_{1}}^{(4,2)}(\boldsymbol{\xi})\frac{\partial\ddot{\Upsilon}^{M}}{\partial x_{q_{1}} }+\tilde{\tilde{N}}_{hq_{1}}^{(4,3)}(\boldsymbol{\xi})\frac{\partial\dddot{\Upsilon}^{M}}{\partial x_{q_{1}} }+N^{(4,4)}_{hp}(\boldsymbol{\xi})\ddddot{U}^{M}_{p},\\
		\upsilon^{(4)}(\boldsymbol{x},\boldsymbol{\xi},t) &= M^{(4)}_{q_{1}q_{2}q_{3}q_{4}}(\boldsymbol{\xi})\frac{\partial^{4}\Upsilon^{M}}{\partial x_{q_{1}}\partial x_{q_{2}}\partial x_{q_{3}}\partial x_{q_{4}}}+\tilde{M}^{(4,1)}_{pq_{1}q_{2}q_{3}}(\boldsymbol{\xi})\frac{\partial^{3} \dot{U}^{M}_{p}}{\partial x_{q_{1}}\partial x_{q_{2}}\partial x_{q_{3}}}+M^{(4,1)}_{q_{1}q_{2}}(\boldsymbol{\xi})\frac{\partial^{2} \dot{\Upsilon}^{M}}{\partial x_{q_{1}}\partial x_{q_{2}}}+\nonumber\\
		&+\tilde{\tilde{M}}^{(4,2)}_{q_{1}q_{2}}(\boldsymbol{\xi})\frac{\partial^{2} \ddot{\Upsilon}^{M}}{\partial x_{q_{1}}\partial x_{q_{2}}}+\tilde{M}^{(4,3)}_{pq_{1}}(\boldsymbol{\xi})\frac{\partial \dddot{U}^{M}_{p}}{\partial x_{q_{1}}}+ \tilde{M}^{(4,2)}_{pq_{1}}(\boldsymbol{\xi})\frac{\partial \ddot{U}^{M}_{p}}{\partial x_{q_{1}}}+\nonumber\\
		&+M^{(4,2)}(\boldsymbol{\xi})\ddot{\Upsilon}^{M}+M^{(4,3)}(\boldsymbol{\xi})\dddot{\Upsilon}^{M}+M^{(4,4)}(\boldsymbol{\xi})\ddddot{\Upsilon}^{M}\label{eqn:solfif1p4}.	
	\end{align}
\end{subequations} 
The differential problems at the order $\varepsilon^{3}$ are 
\begin{subequations}
	\begin{align}
		&\Big( {C}^{m}_{ijhk}\Big ( \frac{\partial {u}^{(4)}_{h}}{\partial x_{k}}+{u}^{(5)}_{h,k} \Big) \Big)_{,j}+\frac{\partial}{\partial x_{j}} \Big({C}^{m}_{ijhk}\Big(\frac{\partial{u}^{(3)}_{h}}{\partial x_{k}}+{u}^{(4)}_{h,k}\Big)\Big)-(\alpha_{ij}^{m}\upsilon^{(4)}+\alpha_{ij}^{(m,1)}\dot{\upsilon}^{(4)})_{,j}+\label{eqn1:ep3}\\
		&-\frac{\partial}{\partial x_{j}}(\alpha_{ij}^{m}\upsilon^{(3)}+\alpha_{ij}^{(m,1)}\dot{\upsilon}^{(3)})-\rho^{m}\ddot{u}^{(3)}_{i}=f^{(5)}_{i}(\boldsymbol{x}),\nonumber\\
		&\Big( {K}^{m}_{ij}\Big ( \frac{\partial {\upsilon}^{(4)}}{\partial x_{j}}+{\upsilon}^{(5)}_{,j} \Big) \Big)_{,i}+\frac{\partial}{\partial x_{i}} \Big({K}^{m}_{ij}\Big(\frac{\partial{\upsilon}^{(3)}}{\partial x_{j}}+{\upsilon}^{(4)}_{,j}\Big)\Big)-p^{m}\dot{\upsilon}^{(3)}-p^{(m,0)}\ddot{\upsilon}^{(3)}-\alpha_{ij}^{m}\Big(\frac{\partial \dot{u}^{(3)}_{i}}{\partial x_{j}}+\dot{u}^{(4)}_{i,j}\Big)=g^{(5)}(\boldsymbol{x}),\label{eqn2:ep3}
	\end{align}
\end{subequations}
with the interface conditions
\begin{subequations}
	\begin{align}
		&\Big[\Big[u^{(5)}_{h}\Big]\Big]\Big\vert_{\boldsymbol{\xi} \in \Sigma_{1}}=0 \quad \Big [\Big[ \Big ( C^{m}_{ijhk}\Big (\frac{\partial u^{(4)}_{h}}{\partial x_{k}}+u^{(5)}_{h,k}\Big)-(\alpha_{ij}^{m}\upsilon^{(4)}+\alpha_{ij}^{(m,1)}\dot{\upsilon}^{(4)}) \Big )n_{j}\Big ]\Big]\Big \vert_{\boldsymbol{\xi} \in \Sigma_{1}}=0,\\
		&\Big[\Big[\upsilon^{(5)}\Big]\Big]\Big\vert_{\boldsymbol{\xi} \in \Sigma_{1}}=0 \quad \Big [\Big[ \Big ( K^{m}_{ij}\Big (\frac{\partial \upsilon^{(4)}}{\partial x_{j}}+\upsilon^{(5)}_{,j}\Big) \Big )n_{i}\Big ]\Big]\Big \vert_{\boldsymbol{\xi} \in \Sigma_{1}}=0.
	\end{align}
\end{subequations}
Substituting the solutions \eqref{eqn:solfif1p1}-\eqref{eqn:solfif1p2} at the order $\varepsilon$ and the ones \eqref{eqn:solfif1p3}-\eqref{eqn:solfif1p4} at the order $\varepsilon^{2}$ into the equations \eqref{eqn1:ep3}-\eqref{eqn2:ep3} and taking into account	 the solvability conditions in the class of $\mathcal{Q}-$periodic functions entail 
\begin{subequations}
	\begin{align}
		f^{(5)}_{i}(\boldsymbol{x})&= \langle C^{m}_{ikhj}N^{(4)}_{hpq_{1}q_{2}q_{3}q_{4},j}+C^{m}_{iq_{4}hk}N^{(3)}_{hpq_{1}q_{2}q_{3}}\rangle\frac{\partial^{5}U^{M}_{p}}{\partial x_{q_{1}} \partial x_{q_{2}}\partial x_{q_{3}}\partial x_{q_{4}}\partial x_{k}}+ \label{f5}\\
		&+\langle C^{m}_{iq_{3}hk}\tilde{N}^{(3)}_{hq_{1}q_{2}} +C^{m}_{ikhj}{\tilde{N}}^{(4)}_{hq_{1}q_{2}q_{3},j}-\alpha_{ik}^{m}M^{(3)}_{q_{1}q_{2}q_{3}}\rangle\frac{\partial^{4} {\Upsilon}^{M}}{\partial x_{q_{1}} \partial x_{q_{2}}\partial x_{q_{3} }\partial x_{k}}+\nonumber \\		
		&+\langle C^{m}_{iq_{3}hk}\tilde{N}^{(3,1)}_{hq_{1}q_{2}}+ C^{m}_{ikhj}{\tilde{N}}^{(4,1)}_{hq_{1}q_{2}q_{3},j}-\alpha_{ik}^{(m,1)}M^{(3)}_{q_{1}q_{2}q_{3}}\rangle\frac{\partial^{4} \dot{\Upsilon}^{M}}{\partial x_{q_{1}} \partial x_{q_{2}}\partial x_{q_{3}}\partial x_{k}}+\nonumber\\
		&+\langle C^{m}_{iq_{2}hk}{N}^{(3,2)}_{hpq_{1}}+C^{m}_{ikhj}N^{(4,2)}_{hpq_{1}q_{2},j}-\rho^{m}N^{(3)}_{ipq_{1}q_{2}q_{3}}-\alpha_{ik}^{(m,1)}\tilde{M}^{(3,1)}_{pq_{1}q_{2}}\rangle\frac{\partial^{3} \ddot{U}^{M}_{p}}{\partial x_{q_{1}}\partial x_{q_{2}}\partial x_{k}}+\nonumber \\		
		&+\langle C^{m}_{ihkj}{N}^{(4,1)}_{hpq_{1}q_{2},j}-\alpha_{ik}^{(m,1)}\tilde{M}^{(3,1)}_{pq_{1}q_{2}}\rangle\frac{\partial^{3} \dot{U}^{M}_{p}}{\partial x_{q_{1}}\partial x_{q_{2}}\partial x_{k}}+\langle C^{m}_{ikhj}{\tilde{\tilde{N}}}^{(4,1)}_{hq_{1},j}-\alpha^{m}_{ik}M^{(3,1)}_{hq_{1}} \rangle \frac{\partial^{2} \dot{\Upsilon}^{M}}{\partial x_{q_{1}}\partial x_{k}}+\nonumber\\
		&+\langle C^{m}_{iq_{1}hk}{\tilde{\tilde{N}}}^{(3,2)}_{h}+ C^{m}_{ikhj}{\tilde{\tilde{N}}}^{(4,2)}_{hq_{1},j}-\rho^{m}\tilde{N}^{(3)}_{iq_{1}q_{2}}-\alpha^{m}_{ik}M^{(3,2)}_{hq_{1}}-\alpha^{(m,1)}_{ik}M^{(3,1)}_{hq_{1}} \rangle \frac{\partial^{2} \ddot{\Upsilon}^{M}}{\partial x_{q_{1}}\partial x_{k}}+\nonumber \\
		&+\langle C^{m}_{iq_{1}hk}{\tilde{\tilde{N}}}^{(3,3)}_{h}+ C^{m}_{ikhj}{\tilde{\tilde{N}}}^{(4,3)}_{hq_{1},j}-\rho^{m}\tilde{N}^{(3,1)}_{iq_{1}q_{2}}-\alpha^{(m,1)}_{ik}M^{(3,1)}_{q_{1}} \rangle \frac{\partial^{2} \dddot{\Upsilon}^{M}}{\partial x_{q_{1}}\partial x_{k}}+ \nonumber\\
		&+\langle C^{m}_{ikhj}N^{(4,4)}_{hp,j}- \alpha^{(m,1)}_{ik}\tilde{M}^{(3,3)}-\rho^{m}N^{(3,2)}_{ipk} \rangle \frac{\partial \ddddot{U}_{p}^{M}}{\partial x_{k}}-\langle \alpha^{m}_{ik}\tilde{M}^{(3,3)}_{p} \rangle \frac{\partial \dddot{U}_{p}^{M}}{\partial x_{k}}+\nonumber\\
		&-\langle \rho^{m}\tilde{\tilde{N}}^{(3,2)}_{i} \rangle \ddddot{\Upsilon}^{M}-\langle \rho^{m}\tilde{\tilde{N}}^{(3,3)}_{i} \rangle \frac{\partial^{5}{\Upsilon}^{M} }{\partial t^{5}},
		\nonumber\\
		g^{(5)}(\boldsymbol{x})&=\langle K^{m}_{q_{4}j}M^{(3)}_{q_{1}q_{2}q_{3}}+ K^{m}_{ji}M^{(4)}_{q_{1}q_{2}q_{3}q_{4},i}\rangle \frac{\partial^{5}\Upsilon^{M}}{\partial x_{q_{1}} \partial x_{q_{2}} \partial x_{q_{3}} \partial x_{q_{4}}\partial x_{j}}+\label{g5} \\
		&+\langle K_{q_{3}j}^{m}\tilde{M}^{(3,1)}_{pq_{1}q_{2}}+K^{m}_{ji}\tilde{M}^{(4,1)}_{q_{1}q_{2}q_{3},i} -\alpha_{ij}^{m}{N}^{(3)}_{ipq_{1}q_{2}q_{3}}-\alpha_{iq_{4}}^{m}{N}^{(4)}_{ipq_{1}q_{2}q_{3}j,q_{4}}\rangle\frac{\partial^{4} \dot{U}^{M}_{p}}{\partial x_{q_{1}}\partial x_{q_{2}}\partial x_{q_{3}}\partial x_{j}}+ \nonumber \\
		&+\langle K_{q_{2}j}^{m}{M}^{(3,1)}_{q_{1}}+K_{ji}^{m}{M}^{(4,1)}_{q_{1}q_{2},i}-p^{m}M^{(3)}_{q_{1}q_{2}j} -\alpha^{m}_{iq_{3}}\tilde{N}^{(4,1)}_{iq_{1}q_{2}j,q_{3}}-\alpha_{ij}^{m}\tilde{N}^{(3,1)}_{iq_{1}q_{2}}\rangle\frac{\partial^{3} \dot{\Upsilon}^{M}}{\partial x_{q_{1}} \partial x_{q_{2}} \partial x_{j}}+\nonumber\\
		&+\langle K_{q_{2}j}^{m}{M}^{(3,2)}_{q_{1}}+K_{ji}^{m}{M}^{(4,2)}_{q_{1}q_{2},i}-p^{(m,0)}M^{(3)}_{q_{1}q_{2}j} -\alpha^{m}_{iq_{3}}\tilde{N}^{(4)}_{iq_{1}q_{2}j,q_{3}}-\alpha_{ij}^{m}\tilde{N}^{(3,1)}_{iq_{1}q_{2}}\rangle\frac{\partial^{3} \ddot{\Upsilon}^{M}}{\partial x_{q_{1}} \partial x_{q_{2}} \partial x_{j}}+\nonumber\\
		&+\langle K_{q_{1}j}^{m}\tilde{M}^{(3,3)}_{p}+K_{ji}^{m}{M}^{(4,3)}_{pq_{1},i}-p^{(m,0)}\tilde{M}^{(3,1)}_{pq_{1}j} -\alpha_{ij}^{m}N^{(3,2)}_{ipq_{1}}-\alpha_{ij}^{m}{N}^{(4,2)}_{ipq_{1}k,j}\rangle \frac{\partial^{2}  \dddot{U}^{M}_{p}}{\partial x_{q_{1}}\partial x_{j}}+\nonumber\\
		&+\langle K_{ji}^{m}\tilde{M}^{(4,2)}_{pq_{1},i}-p^{m}\tilde{M}^{(3,1)}_{pq_{1}j}-\alpha_{ik}^{m}N^{(4,1)}_{ipq_{1}j,k} \rangle \frac{\partial^{2}  \ddot{U}^{M}_{p}}{\partial x_{q_{1}}\partial x_{j}}-\ddddot{U}^{M}_{p}\langle p^{m} \tilde{M}^{(3,3)}_{p} \rangle+\nonumber \\
		&- \frac{\partial^{5}{U}^{M}_{p}}{\partial t^{5}}\langle\alpha_{ij}^{m}N^{(4,4)}_{ip,j}+ p^{(m,0)} \tilde{M}^{(3,3)}_{p} \rangle+\nonumber\\
		&+\langle K_{ji}^{m}{M}^{(4,3)}-p^{m}{M}^{(3,2)}_{j}-p^{(m,0)}{M}^{(3,1)}_{j}-\alpha^{m}_{ij}\tilde{\tilde{N}}^{(3,2)}_{i}-\alpha_{ik}^{m}\tilde{\tilde{N}}^{(4,2)}_{ij,k}\rangle\frac{\partial \dddot{\Upsilon}^{M}}{\partial x_{j}}+\nonumber\\
		&+\langle p^{m}{M}^{(3,1)}_{j}-\alpha_{ik}^{m}{\tilde{N}}^{(4,1)}_{ij,k}\rangle\frac{\partial \ddot{\Upsilon}^{M}}{\partial x_{j}}+\nonumber\\
		&+\langle K_{ji}^{m}{M}^{(4,4)}-p^{(m,0)}{M}^{(3,2)}_{j}-\alpha^{m}_{ij}\tilde{\tilde{N}}^{(3,3)}_{i}-\alpha_{ik}^{m}\tilde{\tilde{N}}^{(4,3)}_{ij,k}\rangle\frac{\partial \ddddot{\Upsilon}^{M}}{\partial x_{j}}.\nonumber
	\end{align}
\end{subequations}
The solutions of the problems \eqref{eqn1:ep3}-\eqref{eqn2:ep3} are
\begin{subequations}
	\begin{align}
		\label{eqn:solfif1p5}
		u^{(5)}_{h}(\boldsymbol{x},\boldsymbol{\xi},t) &= N^{(5)}_{hpq_{1}q_{2}q_{3}q_{4}q_{5}}(\boldsymbol{\xi})\frac{\partial^{5}U^{M}_{p}}{\partial x_{q_{1}}\partial x_{q_{2}}\partial x_{q_{3}}\partial x_{q_{4}}\partial x_{q_{5}}}+\tilde{N}^{(5)}_{hq_{1}q_{2}q_{3}q_{4}}(\boldsymbol{\xi})\frac{\partial^{4} \Upsilon^{M}}{\partial x_{q_{1}}\partial x_{q_{2}}\partial x_{q_{3}}\partial x_{q_{4}}}+\\
		&+{\tilde{N}}^{(5,1)}_{hq_{1}q_{2}q_{3}q_{4}}(\boldsymbol{\xi})\frac{\partial^{4} \dot{\Upsilon}^{M}}{\partial x_{q_{1}}\partial x_{q_{2}} \partial x_{q_{3}}\partial x_{q_{4}}}+N^{(5,2)}_{hpq_{1}q_{2}q_{3}}(\boldsymbol{\xi})\frac{\partial^{3} \ddot{U}^{M}_{p} }{\partial x_{q_{1}} \partial x_{q_{2}}\partial x_{q_{3}}}+\nonumber \\
		&+\tilde{\tilde{N}}_{hq_{1}q_{2}}^{(5,1)}(\boldsymbol{\xi})\frac{\partial^{2}\dot{\Upsilon}^{M}}{\partial x_{q_{1}}\partial x_{q_{2}} }+\tilde{\tilde{N}}_{hq_{1}q_{2}}^{(5,2)}(\boldsymbol{\xi})\frac{\partial^{2}\ddot{\Upsilon}^{M}}{\partial x_{q_{1}}\partial x_{q_{2}} }+\tilde{\tilde{N}}_{hq_{1}q_{2}}^{(5,3)}(\boldsymbol{\xi})\frac{\partial^{2}\dddot{\Upsilon}^{M}}{\partial x_{q_{1}}\partial x_{q_{2}} }+\nonumber\\
		&N^{(5,4)}_{hpq_{1}}(\boldsymbol{\xi})\frac{\partial \ddddot{U}^{M}_{p}}{\partial x_{q_{1}}}+N^{(5,3)}_{hpq_{1}}(\boldsymbol{\xi})\frac{\partial \dddot{U}^{M}_{p}}{\partial x_{q_{1}}}+\nonumber\\
		&+\tilde{N}^{(5,4)}_{h}\ddddot{\Upsilon}^{M}+\tilde{N}^{(5,5)}_{h}\frac{\partial^{5} {\Upsilon}^{M} }{\partial t^{5}}+N^{(5,1)}_{hpq_{1}q_{2}q_{3}}(\boldsymbol{\xi})\frac{\partial^{3} \dot{U}^{M}_{p} }{\partial x_{q_{1}} \partial x_{q_{2}}\partial x_{q_{3}}},\nonumber \\
		\upsilon^{(5)}(\boldsymbol{x},\boldsymbol{\xi},t)& = M^{(5)}_{q_{1}q_{2}q_{3}q_{4}q_{5}}(\boldsymbol{\xi})\frac{\partial^{5}\Upsilon^{M}}{\partial x_{q_{1}}\partial x_{q_{2}}\partial x_{q_{3}}\partial x_{q_{4}}\partial x_{q_{5}}}+\tilde{M}^{(5,1)}_{pq_{1}q_{2}q_{3}q_{4}}(\boldsymbol{\xi})\frac{\partial^{3} \dot{U}^{M}_{p}}{\partial x_{q_{1}}\partial x_{q_{2}}\partial x_{q_{3}}\partial x_{q_{4}}}+\label{eqn:solfif1p6}\\
		&+M^{(5,1)}_{q_{1}q_{2}q_{3}}(\boldsymbol{\xi})\frac{\partial^{3} \dot{\Upsilon}^{M}}{\partial x_{q_{1}}\partial x_{q_{2}}\partial x_{q_{3}}}+M^{(5,2)}_{q_{1}q_{2}q_{3}}(\boldsymbol{\xi})\frac{\partial^{3} \ddot{\Upsilon}^{M}}{\partial x_{q_{1}}\partial x_{q_{2}}\partial x_{q_{3}}}+\tilde{M}^{(5,3)}_{pq_{1}q_{2}}(\boldsymbol{\xi})\frac{\partial^{2} \dddot{U}^{M}_{p}}{\partial x_{q_{1}}\partial x_{q_{2}}}+\nonumber\\
		&+\tilde{M}^{(5,2)}_{pq_{1}}(\boldsymbol{\xi})\frac{\partial^{2} \ddot{U}^{M}_{p}}{\partial x_{q_{1}}\partial x_{q_{2}}}+M^{(5,2)}_{q_{1}}(\boldsymbol{\xi})\frac{\partial \ddot{\Upsilon}^{M}}{\partial x_{q_{1}}}+M^{(4,3)}_{q_{1}}(\boldsymbol{\xi})\frac{\partial \dddot{\Upsilon}^{M}}{\partial x_{q_{1}}}+M^{(5,4)}_{q_{1}}(\boldsymbol{\xi})\frac{\partial \ddddot{\Upsilon}^{M}}{\partial x_{q_{1}}}+\nonumber\\
		&+\tilde{M}^{(5,4)}_{p}(\boldsymbol{\xi})\ddddot{U}^{M}_{p}+\tilde{M}^{(5,5)}_{p}(\boldsymbol{\xi})\frac{\partial^{5}U^{M}_{p}}{\partial t^{5}}\nonumber.
	\end{align}
\end{subequations} 
\section*{Section B \quad Setting matrices}
Section B displays $2 \times 2$ matrices appearing in Equation (59) of Section 4 of the main text. Indeed, they result to be 
\begin{subequations}
	\begin{align}
		\label{matrixformdue}
		\boldsymbol{\Gamma}^{(0,0)}&=
		\begin{bmatrix}
			\omega^{2}n^{(2,2)}_{ip} & 0\\
			0   & \begin{pmatrix}
				-\iota \omega (m^{(2,1)}+\\
				+\iota \omega m^{(2,2)})
			\end{pmatrix} 
		\end{bmatrix}, \quad
		\boldsymbol{\Gamma}^{(0,1)}=
		\begin{bmatrix}
			0 & \begin{pmatrix}
				-\omega^{2}(\tilde{\tilde{n}}^{(3,2)}_{i}+\\
				+\iota \omega \tilde{\tilde{n}}^{(3,3)}_{i})
			\end{pmatrix} \\
			-\iota\omega^{3}\tilde{m}^{(3,3)}_{p}   &  0\\
		\end{bmatrix},\\
		\boldsymbol{\Gamma}^{(0,2)}&=
		\begin{bmatrix}
			-\omega^{4}n^{(4,4)}_{ip} & 0 \\
			0  & \begin{pmatrix}
				\omega^{2}(\tilde{\tilde{m}}^{(4,2)}+\\
				+\iota \omega\tilde{\tilde{m}}^{(4,3)}+\\
				-\omega^{2}\tilde{\tilde{m}}^{(4,4)})
			\end{pmatrix}\\
		\end{bmatrix}, \quad \boldsymbol{\Gamma}^{(1,0)}=
		\begin{bmatrix}
			0 & \begin{pmatrix}
				-\iota (\iota \omega\tilde{n}^{(2,1)}_{iq_{1}}+\\
				+ \tilde{n}^{(2)}_{iq_{1}})v_{q_{1}} 
			\end{pmatrix}  \\
			\omega  \tilde{m}^{(2,1)}_{pq_{1}}v_{q_{1}}  & 0\\
		\end{bmatrix},
		\\
		\boldsymbol{\Gamma}^{(1,1)}&=
		\begin{bmatrix}
			\iota \omega^{2}n^{(3,2)}_{iq_{1}}v_{q_{1}}	 & 	0  \\
			0 &\begin{pmatrix}
				- \omega (\iota \omega  m^{(3,2)}_{q_{1}}+\\
				+ m^{(3,1)}_{q_{1}})v_{q_{1}}
			\end{pmatrix} \\
		\end{bmatrix}, \quad \boldsymbol{\Gamma}^{(1,2)}	=
		\begin{bmatrix}
			0	 &\begin{pmatrix}
				\omega (\tilde{n}^{(4,1)}_{iq_{1}}+\\
				+ \omega^{2}  \tilde{\tilde{n}}^{(4,3)}_{iq_{1}}+\\
				+ \iota \omega  \tilde{\tilde{n}}^{(4,2)}_{iq_{1}})v_{q_{1}}
			\end{pmatrix}  \\
			\begin{pmatrix}
				- \omega(\iota \omega   m^{(3,2)}_{q_{1}}+\\
				+  m^{(3,1)}_{q_{1}})v_{q_{1}}
			\end{pmatrix} & 0\\
		\end{bmatrix},
		\\
		\boldsymbol{\Gamma}^{(2,0)}	&=
		\begin{bmatrix}
			-n^{(2)}_{iq_{1} p q_{2}}v_{q_{1}}v_{q_{2}}	 & 	0\\
			0& -m^{(2)}_{q_{1} q_{2}}v_{q_{1}}v_{q_{2}}\\
		\end{bmatrix}, \quad
		\boldsymbol{\Gamma}^{(2,1)}=
		\begin{bmatrix}
			0	 & \begin{pmatrix}
				-(\tilde{n}^{(3)}_{i q_{1} q_{2}}+\\
				+\iota \omega \tilde{n}^{(3,1)}_{i q_{1} q_{2}})v_{q_{1}}v_{q_{2}}
			\end{pmatrix}	 \\
			-\iota \omega \tilde{m}^{(3,1)}_{p q_{1} q_{2}}v_{q_{1}}v_{q_{2}} & 0\\
		\end{bmatrix},\\
		\boldsymbol{\Gamma}^{(2,2)}&=
		\begin{bmatrix}
			\begin{pmatrix}
				(\iota \omega n^{(4,1)}_{iq_{1} p q_{2}}+\\
				-\omega^{2} n^{(4,2)}_{iq_{1} p q_{2}})v_{q_{1} }v_{q_{2}}
			\end{pmatrix}	 & 	0 \\
			0& \begin{pmatrix}
				-\iota \omega (m^{(4,1)}_{q_{1} q_{2}}+\\
				+\iota \omega m^{(4,2)}_{q_{1} q_{2}})v_{q_{1} }v_{q_{2}}
			\end{pmatrix}\\
		\end{bmatrix}, \quad
		\boldsymbol{\Gamma}^{(3,1)}=
		\begin{bmatrix}
			\begin{pmatrix}
				-\iota n^{(3)}_{ir_{1}r_{2} p q_{1}}\\
				v_{r_{1}}v_{r_{2}}v_{q_{1}}
			\end{pmatrix}	& 0\\
			0& \begin{pmatrix}
				-\iota m^{(3)}_{r_{1} r_{2} q_{1}}\\
				v_{r_{1}}v_{r_{2}}v_{q_{1}}
			\end{pmatrix}
		\end{bmatrix},\\
		\boldsymbol{\Gamma}^{(3,2)}&=
		\begin{bmatrix}
			0& \begin{pmatrix}(-\iota n^{(4)}_{ir_{1}r_{2} p q_{1}}+\\
				+\omega\tilde{n}^{(4,1)}_{ir_{1}r_{2}q_{1}})v_{r_{1}}v_{r_{2}}v_{q_{1}}\end{pmatrix}\\
			\begin{pmatrix}
				-\iota m^{(3)}_{r_{1}r_{2}q_{1}}\\
				v_{r_{1}}v_{r_{2}}v_{q_{1}}
			\end{pmatrix}& 0
		\end{bmatrix}, \quad	
		\boldsymbol{\Gamma}^{(4,2)}=
		\begin{bmatrix}
			\begin{pmatrix}
				n^{(4)}_{i q_{1}q_{2} pr_{1}r_{2}}\\
				v_{q_{1}}v_{q_{2}}v_{r_{1}}v_{r_{2}}
			\end{pmatrix}& 0\\
			0& \begin{pmatrix}
				m^{(4)}_{q_{1}q_{2} r_{1}r_{2}}\\
				v_{q_{1}}v_{q_{2}}v_{r_{1}}v_{r_{2}}
			\end{pmatrix}
		\end{bmatrix}.	
	\end{align}
\end{subequations}
\section*{Section C \quad Deriving invariants coefficients} 
\subsection*{Section C.1}
The invariants coefficients, referred to the characteristic polynomial (67) of Subsection  4.1 in the main text, are derived via the Faddeev-LeVerrier recursive formula  \cite{helmberg1993faddeev} as follows 
\begin{align}
	II_{4}(\omega)&=1,\quad II_{3}(\omega)=-\mathrm{tr}[\boldsymbol{S}(\omega)], \quad II_{2}(\omega)=-\tfrac{1}{2}\mathrm{tr}[\boldsymbol{S}(\omega)^{2}]+\tfrac{1}{2}(\mathrm{tr}[\boldsymbol{S}(\omega)])^{2},&\\
	II_{1}(\omega)&=-\tfrac{1}{3}\mathrm{tr}[\boldsymbol{S}(\omega)^{3}]+\tfrac{1}{2}\mathrm{tr}[\boldsymbol{S}(\omega)]\mathrm{tr}[\boldsymbol{S}(\omega)^{2}]-\tfrac{1}{6}(\mathrm{tr}[\boldsymbol{S}(\omega)])^{3}, &\nonumber\\
	II_{0}(\omega)&=-\tfrac{1}{4}\mathrm{tr}[\boldsymbol{S}(\omega)^{4}]-\tfrac{1}{4}(\mathrm{tr}[\boldsymbol{S}(\omega)])^{2}\mathrm{tr}[\boldsymbol{S}(\omega)^{2}]+\tfrac{1}{3}\mathrm{tr}[\boldsymbol{S}(\omega)]\mathrm{tr}[\boldsymbol{S}(\omega)^{3}]+\tfrac{1}{8}(\mathrm{tr}[\boldsymbol{S}(\omega)^{2}])^{2}+\tfrac{1}{24}(\mathrm{tr}[\boldsymbol{S}(\omega)])^{4}.\nonumber
\end{align} 
\subsection*{Section C.2}
The invariants coefficients, related to the characteristic polynomial (76) of Subsection  4.2 in the main text, depend on the microscopic lenght $\varepsilon$ and its powers. Indeed, by means of the Faddeev-LeVerrier recursive formula, they are determined as follows
\begin{align}
	\label{INV}
	I_{8}(\omega,\varepsilon)&=1,\quad I_{7}(\omega,\varepsilon)=-\mathrm{tr}[\boldsymbol{D}(\omega)], \quad I_{6}(\omega,\varepsilon)=-\tfrac{1}{2}\mathrm{tr}[\boldsymbol{D}(\omega)^{2}]+\tfrac{1}{2}(\mathrm{tr}[\boldsymbol{D}(\omega)])^{2},&\\
	I_{5}(\omega,\varepsilon)&=-\tfrac{1}{3}\mathrm{tr}[\boldsymbol{D}(\omega)^{3}]+\tfrac{1}{2}\mathrm{tr}[\boldsymbol{D}(\omega)]\mathrm{tr}[\boldsymbol{D}(\omega)^{2}]-\tfrac{1}{6}(\mathrm{tr}[\boldsymbol{D}(\omega)])^{3}, \nonumber\\
	I_{4}(\omega,\varepsilon)&=-\tfrac{1}{4}\mathrm{tr}[\boldsymbol{D}(\omega)^{4}]-\tfrac{1}{4}(\mathrm{tr}[\boldsymbol{D}(\omega)])^{2}\mathrm{tr}[\boldsymbol{D}(\omega)^{2}]+\tfrac{1}{3}\mathrm{tr}[\boldsymbol{D}(\omega)]\mathrm{tr}[\boldsymbol{D}(\omega)^{3}]+\tfrac{1}{8}(\mathrm{tr}[\boldsymbol{D}(\omega)^{2}])^{2}+\tfrac{1}{24}(\mathrm{tr}[\boldsymbol{D}(\omega)])^{4},\nonumber\\
	I_{3}(\omega,\varepsilon)&=-\tfrac{1}{5}\mathrm{tr}[\boldsymbol{D}(\omega)^{5}]+\tfrac{1}{4}\mathrm{tr}[\boldsymbol{D}(\omega)^{4}]\mathrm{tr}[\boldsymbol{D}(\omega)]+\tfrac{1}{12}(\mathrm{tr}[\boldsymbol{D}(\omega)])^{3}\mathrm{tr}[\boldsymbol{D}(\omega)^{2}]+\tfrac{1}{5}\mathrm{tr}[\boldsymbol{D}(\omega)^{3}]\mathrm{tr}[\boldsymbol{D}(\omega)^{2}]+\nonumber \\
	&-\tfrac{1}{8}(\mathrm{tr}[\boldsymbol{D}(\omega)^{2}])^{2}\mathrm{tr}[\boldsymbol{D}(\omega)]-\tfrac{1}{6}(\mathrm{tr}[\boldsymbol{D}(\omega)])^{2}\mathrm{tr}[\boldsymbol{D}(\omega)^{3}]-\tfrac{1}{120}(\mathrm{tr}[\boldsymbol{D}(\omega)])^{5},& \nonumber\\
	I_{2}(\omega,\varepsilon)&=-\tfrac{1}{6} \mathrm{tr}[\boldsymbol{D}(\omega)^{6}]+ \tfrac{1}{5} \mathrm{tr}[\boldsymbol{D}(\omega)] \mathrm{tr}[\boldsymbol{D}(\omega)^{5}]+ \tfrac{1}{8} \mathrm{tr}[\boldsymbol{D}(\omega)^{2}]\mathrm{tr}[\boldsymbol{D}(\omega)^{4}]-\tfrac{1}{8}\mathrm{tr}[\boldsymbol{D}(\omega)^{4}](\mathrm{tr} [\boldsymbol{D}(\omega)])^{2}+\nonumber\\
	&+\tfrac{1}{18}(\mathrm{tr}[\boldsymbol{D}(\omega)^{3}])^{2}-\tfrac{1}{6}\mathrm{tr}[\boldsymbol{D}(\omega)]\mathrm{tr}[\boldsymbol{D}(\omega)^{2}]\mathrm{tr}[\boldsymbol{D}(\omega)^{3}]+\tfrac{1}{18}\mathrm{tr}[\boldsymbol{D}(\omega)^{3}](\mathrm{tr}[\boldsymbol{D}(\omega)])^{3}-\tfrac{1}{48}(\mathrm{tr}[\boldsymbol{D}(\omega)^{2}])^{3}+\nonumber\\
	&+\tfrac{1}{16}(\mathrm{tr}[\boldsymbol{D}(\omega)^{2}])^{2}(\mathrm{tr}[\boldsymbol{D}(\omega)])^{2}-\tfrac{1}{48}(\mathrm{tr}[\boldsymbol{D}(\omega)])^{4}\mathrm{tr}[\boldsymbol{D}(\omega)^{2}]+\tfrac{1}{720}(\mathrm{tr}[\boldsymbol{D}(\omega)])^{6},\nonumber\\
	I_{1}(\omega,\varepsilon)&=-\tfrac{1}{7} \mathrm{tr}[\boldsymbol{D}(\omega)^{7}]+ \tfrac{1}{6} \mathrm{tr}[\boldsymbol{D}(\omega)] \mathrm{tr}[\boldsymbol{D}(\omega)^{6}]+ \tfrac{1}{10} \mathrm{tr}[\boldsymbol{D}(\omega)^{2}]\mathrm{tr}[\boldsymbol{D}(\omega)^{5}]-\tfrac{1}{10}\mathrm{tr}[\boldsymbol{D}(\omega)^{5}](\mathrm{tr} [\boldsymbol{D}(\omega)])^{2}+\nonumber\\
	&+ \tfrac{1}{12} \mathrm{tr}[\boldsymbol{D}(\omega)^{3}] \mathrm{tr}[\boldsymbol{D}(\omega)^{4}]- \tfrac{1}{8}\mathrm{tr}[\boldsymbol{D}(\omega)] \mathrm{tr}[\boldsymbol{D}(\omega)^{2}]\mathrm{tr}[\boldsymbol{D}(\omega)^{4}]+\tfrac{1}{24}\mathrm{tr}[\boldsymbol{D}(\omega)^{3}](\mathrm{tr} [\boldsymbol{D}(\omega)^{2}])^{2}+\nonumber\\
	&+ \tfrac{1}{12} \mathrm{tr}[\boldsymbol{D}(\omega)^{3}] \mathrm{tr}[\boldsymbol{D}(\omega)^{2}](\mathrm{tr}[\boldsymbol{D}(\omega)])^{2}- \tfrac{1}{72}(\mathrm{tr}[\boldsymbol{D}(\omega)])^{4} \mathrm{tr}[\boldsymbol{D}(\omega)^{3}]+\tfrac{1}{48}\mathrm{tr}[\boldsymbol{D}(\omega)](\mathrm{tr} [\boldsymbol{D}(\omega)^{2}])^{3}+\nonumber\\
	&-\tfrac{1}{48}(\mathrm{tr}[\boldsymbol{D}(\omega)])^{3}(\mathrm{tr} [\boldsymbol{D}(\omega)^{2}])^{2}+\tfrac{1}{240}(\mathrm{tr}[\boldsymbol{D}(\omega)])^{5}\mathrm{tr} [\boldsymbol{D}(\omega)^{2}]+\tfrac{1}{5040}(\mathrm{tr}[\boldsymbol{D}(\omega)])^{7},\nonumber\\
	I_{0}(\omega,\varepsilon)&=-\tfrac{1}{8} \mathrm{tr}[\boldsymbol{D}(\omega)^{8}]+ \tfrac{1}{7} \mathrm{tr}[\boldsymbol{D}(\omega)] \mathrm{tr}[\boldsymbol{D}(\omega)^{7}]+ \tfrac{1}{12} \mathrm{tr}[\boldsymbol{D}(\omega)^{2}]\mathrm{tr}[\boldsymbol{D}(\omega)^{6}]-\tfrac{1}{12}\mathrm{tr}[\boldsymbol{D}(\omega)^{6}](\mathrm{tr} [\boldsymbol{D}(\omega)])^{2}+\nonumber\\
	&+ \tfrac{1}{15} \mathrm{tr}[\boldsymbol{D}(\omega)^{3}] \mathrm{tr}[\boldsymbol{D}(\omega)^{5}]- \tfrac{1}{10}\mathrm{tr}[\boldsymbol{D}(\omega)] \mathrm{tr}[\boldsymbol{D}(\omega)^{2}]\mathrm{tr}[\boldsymbol{D}(\omega)^{5}]+\tfrac{1}{30}\mathrm{tr}[\boldsymbol{D}(\omega)^{5}](\mathrm{tr} [\boldsymbol{D}(\omega)^{2}])^{3}+\nonumber\\
	&- \tfrac{1}{12} \mathrm{tr}[\boldsymbol{D}(\omega)^{3}] \mathrm{tr}[\boldsymbol{D}(\omega)^{3}]\mathrm{tr}[\boldsymbol{D}(\omega)]- \tfrac{1}{32}(\mathrm{tr}[\boldsymbol{D}(\omega)^{2}])^{2} \mathrm{tr}[\boldsymbol{D}(\omega)^{4}]+\tfrac{1}{16}\mathrm{tr}[\boldsymbol{D}(\omega)^{2}](\mathrm{tr} [\boldsymbol{D}(\omega)])^{2}\mathrm{tr}[\boldsymbol{D}(\omega)^{4}]+\nonumber\\
	&-\tfrac{1}{96}(\mathrm{tr}[\boldsymbol{D}(\omega)])^{4}\mathrm{tr} [\boldsymbol{D}(\omega)^{4}]-\tfrac{1}{36}(\mathrm{tr}[\boldsymbol{D}(\omega)]^{3})^{2}(\mathrm{tr} [\boldsymbol{D}(\omega)])^{2}+\tfrac{1}{24}(\mathrm{tr}[\boldsymbol{D}(\omega)]^{2})^{2}\mathrm{tr}[\boldsymbol{D}(\omega)^{3}]\mathrm{tr}[\boldsymbol{D}(\omega)]+\nonumber\\
	&-\tfrac{1}{36}(\mathrm{tr}[\boldsymbol{D}(\omega)])^{3}\mathrm{tr} [\boldsymbol{D}(\omega)^{2}]\mathrm{tr} [\boldsymbol{D}(\omega)^{3}]+\tfrac{1}{360}(\mathrm{tr}[\boldsymbol{D}(\omega)])^{5}\mathrm{tr} [\boldsymbol{D}(\omega)^{3}]+\tfrac{1}{384}(\mathrm{tr}[\boldsymbol{D}(\omega)]^{2})^{4}+\nonumber\\
	&-\tfrac{1}{96}(\mathrm{tr}[\boldsymbol{D}(\omega)])^{2}(\mathrm{tr} [\boldsymbol{D}(\omega)^{2}])^{3}+\tfrac{1}{192}(\mathrm{tr}[\boldsymbol{D}(\omega)^{2}])^{2}(\mathrm{tr} [\boldsymbol{D}(\omega)])^{4}-\tfrac{1}{1440}(\mathrm{tr}[\boldsymbol{D}(\omega)])^{6}\mathrm{tr}[\boldsymbol{D}(\omega)^{2}]+\nonumber\\
	&+\tfrac{1}{32}(\mathrm{tr}[\boldsymbol{D}(\omega)^{4}])^{2}+\tfrac{1}{40320}(\mathrm{tr}[\boldsymbol{D}(\omega)])^{8}.\nonumber
\end{align} 
Bearing in mind that $\boldsymbol{D}=\boldsymbol{D}^{(0)}+\varepsilon \boldsymbol{D}^{(1)}+\varepsilon^{2}\boldsymbol{D}^{(2)}$ and leveraging the trinomial theorem, it is possible to manipulate the power of the $n^{th}$ matrix involved in equation (15), resulting in $(\boldsymbol{D}^{(0)}+ \boldsymbol{D}^{(1)}+\boldsymbol{D}^{(2)})^{n}=\sum_{h=0}^{n}\sum_{k=0}^{n-h}\binom{n}{h,k}(\boldsymbol{D}^{(0)})^{n-h-k}(\boldsymbol{D}^{(1)})^{k}(\boldsymbol{D}^{(2)})^{h}$. Relying of the linearity of the trace $\mathrm{tr}$, the invariants \eqref{INV} can be transformed as
\begin{subequations}
	\begin{align}
		&I_{8}(\omega,\varepsilon)=1, \quad I_{7}(\omega,\varepsilon)=- \mathrm{tr}[\boldsymbol{D}^{(0)}]-\varepsilon \mathrm{tr}[\boldsymbol{D}^{(1)}]-\varepsilon^{2}\mathrm{tr}[\boldsymbol{D}^{(2)}],\\
		&I_{6}(\omega,\varepsilon)=-\tfrac{1}{2}\sum_{h=0}^{2}\sum_{k=0}^{2-h}\binom{2}{h,k}\mathrm{tr}[(\boldsymbol{D}^{(0)})^{2-h-k}]\varepsilon^{k}\mathrm{tr}[(\boldsymbol{D}^{(1)})^{k}]\varepsilon^{2h}\mathrm{tr}[(\boldsymbol{D}^{(2)})^{h}]+\nonumber\\
		&+\tfrac{1}{2}\sum_{h=0}^{2}\sum_{k=0}^{2-h}\binom{2}{h,k}(\mathrm{tr}[\boldsymbol{D}^{(0)}])^{2-h-k}\varepsilon^{k}(\mathrm{tr}[(\boldsymbol{D}^{(1)})])^{k}\varepsilon^{2h}(\mathrm{tr}[\boldsymbol{D}^{(2)}])^{h},\\
		&I_{5}(\omega,\varepsilon)=-\tfrac{1}{3}\sum_{h=0}^{3}\sum_{k=0}^{3-h}\binom{3}{h,k}\mathrm{tr}[(\boldsymbol{D}^{(0)})^{3-h-k}]\varepsilon^{k}\mathrm{tr}[(\boldsymbol{D}^{(1)})^{k}]\varepsilon^{2h}\mathrm{tr}[(\boldsymbol{D}^{(2)})^{h}]+\nonumber\\	
		&+\tfrac{1}{2}\Big(\mathrm{tr}[\boldsymbol{D}^{(0)}]+\varepsilon \mathrm{tr}[\boldsymbol{D}^{(1)}]+\varepsilon^{2}\mathrm{tr}[\boldsymbol{D}^{(2)}]\Big)\sum_{h=0}^{2}\sum_{k=0}^{2-h}\binom{2}{h,k}\mathrm{tr}[(\boldsymbol{D}^{(0)})^{2-h-k}]\varepsilon^{k}\mathrm{tr}[(\boldsymbol{D}^{(1)})^{k}]\varepsilon^{2h}\mathrm{tr}[(\boldsymbol{D}^{(2)})^{h}]+\nonumber\\
		&-\tfrac{1}{6}\sum_{h=0}^{3}\sum_{k=0}^{3-h}\binom{3}{h,k}(\mathrm{tr}[\boldsymbol{D}^{(0)}])^{3-h-k}\varepsilon^{k}(\mathrm{tr}[(\boldsymbol{D}^{(1)})])^{k}\varepsilon^{2h}(\mathrm{tr}[\boldsymbol{D}^{(2)}])^{h},\\
		&I_{4}(\omega,\varepsilon)=-\tfrac{1}{4}\sum_{h=0}^{4}\sum_{k=0}^{4-h}\binom{4}{h,k}\mathrm{tr}[(\boldsymbol{D}^{(0)})^{4-h-k}]\varepsilon^{k}\mathrm{tr}[(\boldsymbol{D}^{(1)})^{k}]\varepsilon^{2h}\mathrm{tr}[(\boldsymbol{D}^{(2)})^{h}]+\nonumber\\
		&-\tfrac{1}{4}\sum_{h=0}^{2}\sum_{k=0}^{2-h}\binom{2}{h,k}(\mathrm{tr}[\boldsymbol{D}^{(0)}])^{2-h-k}\varepsilon^{k}(\mathrm{tr}[(\boldsymbol{D}^{(1)})])^{k}\varepsilon^{2h}(\mathrm{tr}[\boldsymbol{D}^{(2)}])^{h}\cdot \nonumber\\
		&\cdot\sum_{h=0}^{2}\sum_{k=0}^{2-h}\binom{2}{h,k}\mathrm{tr}[(\boldsymbol{D}^{(0)})^{2-h-k}]\varepsilon^{k}\mathrm{tr}[(\boldsymbol{D}^{(1)})^{k}]\varepsilon^{2h}\mathrm{tr}[(\boldsymbol{D}^{(2)})^{h}]+\nonumber\\
		&+\tfrac{1}{3}\Big(\mathrm{tr}[\boldsymbol{D}^{(0)}]+\varepsilon \mathrm{tr}[\boldsymbol{D}^{(1)}]+\varepsilon^{2}\mathrm{tr}[\boldsymbol{D}^{(2)}]\Big)\sum_{h=0}^{3}\sum_{k=0}^{3-h}\binom{3}{h,k}\mathrm{tr}[(\boldsymbol{D}^{(0)})^{3-h-k}]\varepsilon^{k}\mathrm{tr}[(\boldsymbol{D}^{(1)})^{k}]\varepsilon^{2h}\mathrm{tr}[(\boldsymbol{D}^{(2)})^{h}]+\nonumber\\
		&+\tfrac{1}{8}\Big(\sum_{h=0}^{2}\sum_{k=0}^{2-h}\binom{2}{h,k}\mathrm{tr}[(\boldsymbol{D}^{(0)})^{2-h-k}]\varepsilon^{k}\mathrm{tr}[(\boldsymbol{D}^{(1)})^{k}]\varepsilon^{2h}\mathrm{tr}[(\boldsymbol{D}^{(2)})^{h}]\Big)^{2}+\nonumber\\ &+\tfrac{1}{24}\sum_{h=0}^{4}\sum_{k=0}^{4-h}\binom{4}{h,k}(\mathrm{tr}[\boldsymbol{D}^{(0)}])^{4-h-k}\varepsilon^{k}(\mathrm{tr}[(\boldsymbol{D}^{(1)})])^{k}\varepsilon^{2h}(\mathrm{tr}[\boldsymbol{D}^{(2)}])^{h},\\
		&I_{3}(\omega,\varepsilon)=-\tfrac{1}{5}\sum_{h=0}^{5}\sum_{k=0}^{5-h}\binom{5}{h,k}\mathrm{tr}[(\boldsymbol{D}^{(0)})^{5-h-k}]\varepsilon^{k}\mathrm{tr}[(\boldsymbol{D}^{(1)})^{k}]\varepsilon^{2h}\mathrm{tr}[(\boldsymbol{D}^{(2)})^{h}]+\nonumber\\
		&+\tfrac{1}{4}\sum_{h=0}^{4}\sum_{k=0}^{4-h}\binom{4}{h,k}\mathrm{tr}[(\boldsymbol{D}^{(0)})^{4-h-k}]\varepsilon^{k}\mathrm{tr}[(\boldsymbol{D}^{(1)})^{k}]\varepsilon^{2h}\mathrm{tr}[(\boldsymbol{D}^{(2)})^{h}]\Big(\mathrm{tr}[\boldsymbol{D}^{(0)}]+\varepsilon \mathrm{tr}[\boldsymbol{D}^{(1)}]+\varepsilon^{2}\mathrm{tr}[\boldsymbol{D}^{(2)}]\Big)+\nonumber\\
		&+\tfrac{1}{12}\sum_{h=0}^{3}\sum_{k=0}^{3-h}\binom{3}{h,k}(\mathrm{tr}[\boldsymbol{D}^{(0)}])^{3-h-k}\varepsilon^{k}(\mathrm{tr}[(\boldsymbol{D}^{(1)})])^{k}\varepsilon^{2h}(\mathrm{tr}[\boldsymbol{D}^{(2)}])^{h}\cdot\nonumber\\
		&\cdot\sum_{h=0}^{2}\sum_{k=0}^{2-h}\binom{2}{h,k}\mathrm{tr}[(\boldsymbol{D}^{(0)})^{2-h-k}]\varepsilon^{k}\mathrm{tr}[(\boldsymbol{D}^{(1)})^{k}]\varepsilon^{2h}\mathrm{tr}[(\boldsymbol{D}^{(2)})^{h}]+\nonumber\\
		&+\tfrac{1}{6}\sum_{h=0}^{3}\sum_{k=0}^{3-h}\binom{3}{h,k}\mathrm{tr}[(\boldsymbol{D}^{(0)})^{3-h-k}]\varepsilon^{k}\mathrm{tr}[(\boldsymbol{D}^{(1)})^{k}]\varepsilon^{2h}\mathrm{tr}[(\boldsymbol{D}^{(2)})^{h}]\cdot\nonumber\\
		&\cdot\sum_{h=0}^{2}\sum_{k=0}^{2-h}\binom{2}{h,k}\mathrm{tr}[(\boldsymbol{D}^{(0)})^{2-h-k}]\varepsilon^{k}\mathrm{tr}[(\boldsymbol{D}^{(1)})^{k}]\varepsilon^{2h}\mathrm{tr}[(\boldsymbol{D}^{(2)})^{h}]+\nonumber \\
		&-\tfrac{1}{8}\Big(\sum_{h=0}^{2}\sum_{k=0}^{2-h}\binom{2}{h,k}\mathrm{tr}[(\boldsymbol{D}^{(0)})^{2-h-k}]\varepsilon^{k}\mathrm{tr}[(\boldsymbol{D}^{(1)})^{k}]\varepsilon^{2h}\mathrm{tr}[(\boldsymbol{D}^{(2)})^{h}]\Big)^{2} \Big(\mathrm{tr}[\boldsymbol{D}^{(0)}]+\varepsilon \mathrm{tr}[\boldsymbol{D}^{(1)}]+\varepsilon^{2}\mathrm{tr}[\boldsymbol{D}^{(2)}]\Big)+\nonumber\\
		&-\tfrac{1}{6}\sum_{h=0}^{2}\sum_{k=0}^{2-h}\binom{2}{h,k}(\mathrm{tr}[\boldsymbol{D}^{(0)}])^{2-h-k}\varepsilon^{k}(\mathrm{tr}[(\boldsymbol{D}^{(1)})])^{k}\varepsilon^{2h}(\mathrm{tr}[\boldsymbol{D}^{(2)}])^{h}\cdot\nonumber\\
		&\cdot\sum_{h=0}^{3}\sum_{k=0}^{3-h}\binom{3}{h,k}\mathrm{tr}[(\boldsymbol{D}^{(0)})^{3-h-k}]\varepsilon^{k}\mathrm{tr}[(\boldsymbol{D}^{(1)})^{k}]\varepsilon^{2h}\mathrm{tr}[(\boldsymbol{D}^{(2)})^{h}]+\nonumber\\
		&-\tfrac{1}{120}\sum_{h=0}^{5}\sum_{k=0}^{5-h}\binom{5}{h,k}(\mathrm{tr}[\boldsymbol{D}^{(0)}])^{5-h-k}\varepsilon^{k}(\mathrm{tr}[(\boldsymbol{D}^{(1)})])^{k}\varepsilon^{2h}(\mathrm{tr}[\boldsymbol{D}^{(2)}])^{h},\\
		&I_{2}(\omega,\varepsilon)=-\tfrac{1}{6}\sum_{h=0}^{6}\sum_{k=0}^{6-h}\binom{6}{h,k}\mathrm{tr}[(\boldsymbol{D}^{(0)})^{6-h-k}]\varepsilon^{k}\mathrm{tr}[(\boldsymbol{D}^{(1)})^{k}]\varepsilon^{2h}\mathrm{tr}[(\boldsymbol{D}^{(2)})^{h}]+\nonumber\\
		&+\tfrac{1}{5}\sum_{h=0}^{5}\sum_{k=0}^{5-h}\binom{5}{h,k}\mathrm{tr}[(\boldsymbol{D}^{(0)})^{5-h-k}]\varepsilon^{k}\mathrm{tr}[(\boldsymbol{D}^{(1)})^{k}]\varepsilon^{2h}\mathrm{tr}[(\boldsymbol{D}^{(2)})^{h}]\cdot\nonumber\\
		&\cdot\mathrm{tr}[\boldsymbol{D}^{(0)}]+\varepsilon \mathrm{tr}[\boldsymbol{D}^{(1)}]+\varepsilon^{2}\mathrm{tr}[\boldsymbol{D}^{(2)}]+\tfrac{1}{8}\sum_{h=0}^{2}\sum_{k=0}^{2-h}\binom{2}{h,k}\mathrm{tr}[(\boldsymbol{D}^{(0)})^{2-h-k}]\varepsilon^{k}\mathrm{tr}[(\boldsymbol{D}^{(1)})^{k}]\varepsilon^{2h}\mathrm{tr}[(\boldsymbol{D}^{(2)})^{h}]\cdot\nonumber\\
		&\sum_{h=0}^{4}\sum_{k=0}^{4-h}\binom{4}{h,k}\mathrm{tr}[(\boldsymbol{D}^{(0)})^{4-h-k}]\varepsilon^{k}\mathrm{tr}[(\boldsymbol{D}^{(1)})^{k}]\varepsilon^{2h}\mathrm{tr}[(\boldsymbol{D}^{(2)})^{h}]+\nonumber\\
		&+\tfrac{1}{18}\Big(\sum_{h=0}^{3}\sum_{k=0}^{3-h}\binom{3}{h,k}\mathrm{tr}[(\boldsymbol{D}^{(0)})^{3-h-k}]\varepsilon^{k}\mathrm{tr}[(\boldsymbol{D}^{(1)})^{k}]\varepsilon^{2h}\mathrm{tr}[(\boldsymbol{D}^{(2)})^{h}]\Big)^{2}+\nonumber\\ &-\tfrac{1}{8}\sum_{h=0}^{2}\sum_{k=0}^{2-h}\binom{2}{h,k}(\mathrm{tr}[\boldsymbol{D}^{(0)}])^{2-h-k}\varepsilon^{k}(\mathrm{tr}[(\boldsymbol{D}^{(1)})])^{k}\varepsilon^{2h}(\mathrm{tr}[\boldsymbol{D}^{(2)}])^{h}\cdot\nonumber\\
		&\cdot\sum_{h=0}^{4}\sum_{k=0}^{4-h}\binom{4}{h,k}\mathrm{tr}[(\boldsymbol{D}^{(0)})^{4-h-k}]\varepsilon^{k}\mathrm{tr}[(\boldsymbol{D}^{(1)})^{k}]\varepsilon^{2h}\mathrm{tr}[(\boldsymbol{D}^{(2)})^{h}]+\nonumber\\
		&-\tfrac{1}{6}\Big(\mathrm{tr}[\boldsymbol{D}^{(0)}]+\varepsilon \mathrm{tr}[\boldsymbol{D}^{(1)}]+\varepsilon^{2}\mathrm{tr}[\boldsymbol{D}^{(2)}]\Big)\sum_{h=0}^{2}\sum_{k=0}^{2-h}\binom{2}{h,k}\mathrm{tr}[(\boldsymbol{D}^{(0)})^{2-h-k}]\varepsilon^{k}\mathrm{tr}[(\boldsymbol{D}^{(1)})^{k}]\varepsilon^{2h}\mathrm{tr}[(\boldsymbol{D}^{(2)})^{h}]\cdot\nonumber\\
		&\cdot\sum_{h=0}^{3}\sum_{k=0}^{3-h}\binom{3}{h,k}\mathrm{tr}[(\boldsymbol{D}^{(0)})^{3-h-k}]\varepsilon^{k}\mathrm{tr}[(\boldsymbol{D}^{(1)})^{k}]\varepsilon^{2h}\mathrm{tr}[(\boldsymbol{D}^{(2)})^{h}]+\nonumber\\
		&-\tfrac{1}{48}\Big(\sum_{h=0}^{2}\sum_{k=0}^{2-h}\binom{2}{h,k}\mathrm{tr}[(\boldsymbol{D}^{(0)})^{2-h-k}]\varepsilon^{k}\mathrm{tr}[(\boldsymbol{D}^{(1)})^{k}]\varepsilon^{2h}\mathrm{tr}[(\boldsymbol{D}^{(2)})^{h}]\Big)^{3}+\nonumber\\ &+\tfrac{1}{18}\sum_{h=0}^{3}\sum_{k=0}^{3-h}\binom{3}{h,k}\mathrm{tr}[(\boldsymbol{D}^{(0)})^{3-h-k}]\varepsilon^{k}\mathrm{tr}[(\boldsymbol{D}^{(1)})^{k}]\varepsilon^{2h}\mathrm{tr}[(\boldsymbol{D}^{(2)})^{h}]\cdot\nonumber\\
		&\cdot\sum_{h=0}^{3}\sum_{k=0}^{3-h}\binom{3}{h,k}(\mathrm{tr}[\boldsymbol{D}^{(0)}])^{3-h-k}\varepsilon^{k}(\mathrm{tr}[(\boldsymbol{D}^{(1)})])^{k}\varepsilon^{2h}(\mathrm{tr}[\boldsymbol{D}^{(2)}])^{h}+\nonumber\\
		&+\tfrac{1}{16}\Big(\sum_{h=0}^{2}\sum_{k=0}^{2-h}\binom{2}{h,k}\mathrm{tr}[(\boldsymbol{D}^{(0)})^{2-h-k}]\varepsilon^{k}\mathrm{tr}[(\boldsymbol{D}^{(1)})^{k}]\varepsilon^{2h}\mathrm{tr}[(\boldsymbol{D}^{(2)})^{h}]\Big)^{2}\cdot\nonumber\\
		&\cdot\sum_{h=0}^{2}\sum_{k=0}^{2-h}\binom{2}{h,k}(\mathrm{tr}[\boldsymbol{D}^{(0)}])^{2-h-k}\varepsilon^{k}(\mathrm{tr}[(\boldsymbol{D}^{(1)})])^{k}\varepsilon^{2h}(\mathrm{tr}[\boldsymbol{D}^{(2)}])^{h}\nonumber\\
		&-\tfrac{1}{48}\sum_{h=0}^{2}\sum_{k=0}^{2-h}\binom{2}{h,k}\mathrm{tr}[(\boldsymbol{D}^{(0)})^{2-h-k}]\varepsilon^{k}\mathrm{tr}[(\boldsymbol{D}^{(1)})^{k}]\varepsilon^{2h}\mathrm{tr}[(\boldsymbol{D}^{(2)})^{h}]\cdot\nonumber\\
		&\cdot \sum_{h=0}^{4}\sum_{k=0}^{4-h}\binom{4}{h,k}(\mathrm{tr}[\boldsymbol{D}^{(0)}])^{4-h-k}\varepsilon^{k}(\mathrm{tr}[(\boldsymbol{D}^{(1)})])^{k}\varepsilon^{2h}(\mathrm{tr}[\boldsymbol{D}^{(2)}])^{h}+\nonumber\\
		&+\tfrac{1}{720}\sum_{k=0}^{6-h}\binom{6}{h,k}(\mathrm{tr}[\boldsymbol{D}^{(0)}])^{6-h-k}\varepsilon^{k}(\mathrm{tr}[(\boldsymbol{D}^{(1)})])^{k}\varepsilon^{2h}(\mathrm{tr}[\boldsymbol{D}^{(2)}])^{h},\\
		&I_{1}(\omega,\varepsilon)=-\tfrac{1}{7}\sum_{h=0}^{7}\sum_{k=0}^{7-h}\binom{7}{h,k}\mathrm{tr}[(\boldsymbol{D}^{(0)})^{7-h-k}]\varepsilon^{k}\mathrm{tr}[(\boldsymbol{D}^{(1)})^{k}]\varepsilon^{2h}\mathrm{tr}[(\boldsymbol{D}^{(2)})^{h}]+\nonumber\\
		&+\tfrac{1}{6}\sum_{h=0}^{6}\sum_{k=0}^{6-h}\binom{6}{h,k}\mathrm{tr}[(\boldsymbol{D}^{(0)})^{6-h-k}]\varepsilon^{k}\mathrm{tr}[(\boldsymbol{D}^{(1)})^{k}]\varepsilon^{2h}\mathrm{tr}[(\boldsymbol{D}^{(2)})^{h}]\Big(\mathrm{tr}(\boldsymbol{D}^{(0)})+\varepsilon \mathrm{tr}(\boldsymbol{D}^{(1)})+\varepsilon^{2}\mathrm{tr}(\boldsymbol{D}^{(2)})\Big)+\nonumber\\
		&-\tfrac{1}{5040}\sum_{h=0}^{7}\sum_{k=0}^{7-h}\binom{7}{h,k}(\mathrm{tr}[\boldsymbol{D}^{(0)}])^{7-h-k}\varepsilon^{k}(\mathrm{tr}[(\boldsymbol{D}^{(1)})])^{k}\varepsilon^{2h}(\mathrm{tr}[\boldsymbol{D}^{(2)}])^{h}+\nonumber\\
		&+\tfrac{1}{10}\sum_{h=0}^{2}\sum_{k=0}^{2-h}\binom{2}{h,k}\mathrm{tr}[(\boldsymbol{D}^{(0)})^{2-h-k}]\varepsilon^{k}\mathrm{tr}[(\boldsymbol{D}^{(1)})^{k}]\varepsilon^{2h}\mathrm{tr}[(\boldsymbol{D}^{(2)})^{h}]\cdot\nonumber\\
		&\cdot\sum_{h=0}^{5}\sum_{k=0}^{5-h}\binom{5}{h,k}\mathrm{tr}[(\boldsymbol{D}^{(0)})^{5-h-k}]\varepsilon^{k}\mathrm{tr}[(\boldsymbol{D}^{(1)})^{k}]\varepsilon^{2h}\mathrm{tr}[(\boldsymbol{D}^{(2)})^{h}]+\nonumber\\
		&+\tfrac{1}{12}\sum_{h=0}^{3}\sum_{k=0}^{3-h}\binom{3}{h,k}\mathrm{tr}[(\boldsymbol{D}^{(0)})^{3-h-k}]\varepsilon^{k}\mathrm{tr}[(\boldsymbol{D}^{(1)})^{k}]\varepsilon^{2h}\mathrm{tr}[(\boldsymbol{D}^{(2)})^{h}]\cdot\nonumber\\
		&\cdot\sum_{h=0}^{4}\sum_{k=0}^{4-h}\binom{4}{h,k}\mathrm{tr}[(\boldsymbol{D}^{(0)})^{4-h-k}]\varepsilon^{k}\mathrm{tr}[(\boldsymbol{D}^{(1)})^{k}]\varepsilon^{2h}\mathrm{tr}[(\boldsymbol{D}^{(2)})^{h}]+\nonumber\\
		&+\tfrac{1}{24}\sum_{h=0}^{4}\sum_{k=0}^{4-h}\binom{4}{h,k}\mathrm{tr}[(\boldsymbol{D}^{(0)})^{4-h-k}]\varepsilon^{k}\mathrm{tr}[(\boldsymbol{D}^{(1)})^{k}]\varepsilon^{2h}\mathrm{tr}[(\boldsymbol{D}^{(2)})^{h}]\cdot\nonumber\\
		&\cdot\sum_{h=0}^{3}\sum_{k=0}^{3-h}\binom{3}{h,k}(\mathrm{tr}[\boldsymbol{D}^{(0)}])^{3-h-k}\varepsilon^{k}(\mathrm{tr}[(\boldsymbol{D}^{(1)})])^{k}\varepsilon^{2h}(\mathrm{tr}[\boldsymbol{D}^{(2)}])^{h}+\nonumber\\
		&-\tfrac{1}{18}\mathrm{tr}[\boldsymbol{D}^{(0)}]+\varepsilon \mathrm{tr}[\boldsymbol{D}^{(1)}]+\varepsilon^{2}\mathrm{tr}[\boldsymbol{D}^{(2)}]\Big(\sum_{h=0}^{3}\sum_{k=0}^{3-h}\binom{3}{h,k}\mathrm{tr}[(\boldsymbol{D}^{(0)})^{3-h-k}]\varepsilon^{k}\mathrm{tr}[(\boldsymbol{D}^{(1)})^{k}]\varepsilon^{2h}\mathrm{tr}[(\boldsymbol{D}^{(2)})^{h}]\Big)^{2}+\nonumber\\
		&-\tfrac{1}{24}\sum_{h=0}^{3}\sum_{k=0}^{3-h}\binom{3}{h,k}\mathrm{tr}[(\boldsymbol{D}^{(0)})^{3-h-k}]\varepsilon^{k}\mathrm{tr}[(\boldsymbol{D}^{(1)})^{k}]\varepsilon^{2h}\mathrm{tr}[(\boldsymbol{D}^{(2)})^{h}]\cdot\nonumber\\
		&\cdot\Big(\sum_{h=0}^{2}\sum_{k=0}^{2-h}\binom{2}{h,k}\mathrm{tr}[(\boldsymbol{D}^{(0)})^{2-h-k}]\varepsilon^{k}\mathrm{tr}[(\boldsymbol{D}^{(1)})^{k}]\varepsilon^{2h}\mathrm{tr}[(\boldsymbol{D}^{(2)})^{h}]\Big)^{2}+\nonumber\\
		&+\tfrac{1}{12}\sum_{h=0}^{2}\sum_{k=0}^{2-h}\binom{2}{h,k}\mathrm{tr}[(\boldsymbol{D}^{(0)})^{2-h-k}]\varepsilon^{k}\mathrm{tr}[(\boldsymbol{D}^{(1)})^{k}]\varepsilon^{2h}\mathrm{tr}[(\boldsymbol{D}^{(2)})^{h}]\cdot \nonumber\\
		&\cdot\sum_{h=0}^{2}\sum_{k=0}^{2-h}\binom{2}{h,k}(\mathrm{tr}[\boldsymbol{D}^{(0)}])^{2-h-k}\varepsilon^{k}(\mathrm{tr}[(\boldsymbol{D}^{(1)})])^{k}\varepsilon^{2h}(\mathrm{tr}[\boldsymbol{D}^{(2)}])^{h}\cdot\nonumber\\
		&\cdot \sum_{h=0}^{3}\sum_{k=0}^{3-h}\binom{3}{h,k}\mathrm{tr}[(\boldsymbol{D}^{(0)})^{3-h-k}]\varepsilon^{k}\mathrm{tr}[(\boldsymbol{D}^{(1)})^{k}]\varepsilon^{2h}\mathrm{tr}[(\boldsymbol{D}^{(2)})^{h}]+\nonumber\\
		&-\tfrac{1}{72}\sum_{h=0}^{3}\sum_{k=0}^{3-h}\binom{3}{h,k}\mathrm{tr}[(\boldsymbol{D}^{(0)})^{3-h-k}]\varepsilon^{k}\mathrm{tr}[(\boldsymbol{D}^{(1)})^{k}]\varepsilon^{2h}\mathrm{tr}[(\boldsymbol{D}^{(2)})^{h}]\cdot\nonumber\\
		&\cdot\sum_{h=0}^{4}\sum_{k=0}^{4-h}\binom{4}{h,k}(\mathrm{tr}[\boldsymbol{D}^{(0)}])^{4-h-k}\varepsilon^{k}(\mathrm{tr}[(\boldsymbol{D}^{(1)})])^{k}\varepsilon^{2h}(\mathrm{tr}[\boldsymbol{D}^{(2)}])^{h}+\nonumber\\
		&+\tfrac{1}{48}\Big(\sum_{h=0}^{2}\sum_{k=0}^{2-h}\binom{2}{h,k}\mathrm{tr}[(\boldsymbol{D}^{(0)})^{2-h-k}]\varepsilon^{k}\mathrm{tr}[(\boldsymbol{D}^{(1)})^{k}]\varepsilon^{2h}\mathrm{tr}[(\boldsymbol{D}^{(2)})^{h}]\Big)^{3}\Big(\mathrm{tr}[\boldsymbol{D}^{(0)}]+\varepsilon \mathrm{tr}[\boldsymbol{D}^{(1)}]+\varepsilon^{2}\mathrm{tr}[\boldsymbol{D}^{(2)}]\Big)+\nonumber\\
		&-\tfrac{1}{48}\Big(\sum_{h=0}^{2}\sum_{k=0}^{2-h}\binom{2}{h,k}\mathrm{tr}[(\boldsymbol{D}^{(0)})^{2-h-k}]\varepsilon^{k}\mathrm{tr}[(\boldsymbol{D}^{(1)})^{k}]\varepsilon^{2h}\mathrm{tr}[(\boldsymbol{D}^{(2)})^{h}]\Big)^{2}\cdot\nonumber\\
		&\cdot\sum_{h=0}^{3}\sum_{k=0}^{3-h}\binom{3}{h,k}(\mathrm{tr}[\boldsymbol{D}^{(0)}])^{3-h-k}\varepsilon^{k}(\mathrm{tr}[(\boldsymbol{D}^{(1)})])^{k}\varepsilon^{2h}(\mathrm{tr}[\boldsymbol{D}^{(2)}])^{h}+\nonumber\\
		&+\tfrac{1}{240}\sum_{h=0}^{2}\sum_{k=0}^{2-h}\binom{2}{h,k}\mathrm{tr}[(\boldsymbol{D}^{(0)})^{2-h-k}]\varepsilon^{k}\mathrm{tr}[(\boldsymbol{D}^{(1)})^{k}]\varepsilon^{2h}\mathrm{tr}[(\boldsymbol{D}^{(2)})^{h}]\cdot\nonumber\\
		&\cdot\sum_{h=0}^{5}\sum_{k=0}^{5-h}\binom{5}{h,k}(\mathrm{tr}[\boldsymbol{D}^{(0)}])^{5-h-k}\varepsilon^{k}(\mathrm{tr}[(\boldsymbol{D}^{(1)})])^{k}\varepsilon^{2h}(\mathrm{tr}[\boldsymbol{D}^{(2)}])^{h}+\nonumber\\
		&-\tfrac{1}{8}\sum_{h=0}^{2}\sum_{k=0}^{2-h}\binom{2}{h,k}\mathrm{tr}[(\boldsymbol{D}^{(0)})^{2-h-k}]\varepsilon^{k}\mathrm{tr}[(\boldsymbol{D}^{(1)})^{k}]\varepsilon^{2h}\mathrm{tr}[(\boldsymbol{D}^{(2)})^{h}]\Big(\mathrm{tr}[\boldsymbol{D}^{(0)}]+\varepsilon \mathrm{tr}[\boldsymbol{D}^{(1)}]+\varepsilon^{2} \mathrm{tr}[\boldsymbol{D}^{(2)}]\Big)\cdot\nonumber\\
		&\cdot\sum_{h=0}^{4}\sum_{k=0}^{4-h}\binom{4}{h,k}\mathrm{tr}[(\boldsymbol{D}^{(0)})^{4-h-k}]\varepsilon^{k}\mathrm{tr}[(\boldsymbol{D}^{(1)})^{k}]\varepsilon^{2h}\mathrm{tr}[(\boldsymbol{D}^{(2)})^{h}]+\nonumber\\
		&-\tfrac{1}{10}\sum_{h=0}^{2}\sum_{k=0}^{2-h}\binom{2}{h,k}(\mathrm{tr}[\boldsymbol{D}^{(0)}])^{2-h-k}\varepsilon^{k}(\mathrm{tr}[(\boldsymbol{D}^{(1)})])^{k}\varepsilon^{2h}(\mathrm{tr}[\boldsymbol{D}^{(2)}])^{h}\cdot\nonumber\\
		&\cdot \sum_{h=0}^{5}\sum_{k=0}^{5-h}\binom{5}{h,k}\mathrm{tr}[(\boldsymbol{D}^{(0)})^{5-h-k}]\varepsilon^{k}\mathrm{tr}[(\boldsymbol{D}^{(1)})^{k}]\varepsilon^{2h}\mathrm{tr}[(\boldsymbol{D}^{(2)})^{h}],\\
		&I_{0}(\omega,\varepsilon)=-\tfrac{1}{8}\sum_{h=0}^{8}\sum_{k=0}^{8-h}\binom{8}{h,k}\mathrm{tr}[(\boldsymbol{D}^{(0)})^{8-h-k}]\varepsilon^{k}\mathrm{tr}[(\boldsymbol{D}^{(1)})^{k}]\varepsilon^{2h}\mathrm{tr}[(\boldsymbol{D}^{(2)})^{h}]+\nonumber\\
		&+\tfrac{1}{7}\Big(\mathrm{tr}[\boldsymbol{D}^{(0)}]+\varepsilon \mathrm{tr}[\boldsymbol{D}^{(1)}]+\varepsilon^{2}\mathrm{tr}[\boldsymbol{D}^{(2)}]\Big)\sum_{h=0}^{7}\sum_{k=0}^{7-h}\binom{7}{h,k}\mathrm{tr}[(\boldsymbol{D}^{(0)})^{7-h-k}]\varepsilon^{k}\mathrm{tr}[(\boldsymbol{D}^{(1)})^{k}]\varepsilon^{2h}\mathrm{tr}[(\boldsymbol{D}^{(2)})^{h}]+\nonumber\\
		&+\tfrac{1}{12}\sum_{h=0}^{2}\sum_{k=0}^{2-h}\binom{2}{h,k}\mathrm{tr}[(\boldsymbol{D}^{(0)})^{2-h-k}]\varepsilon^{k}\mathrm{tr}[(\boldsymbol{D}^{(1)})^{k}]\varepsilon^{2h}\mathrm{tr}[(\boldsymbol{D}^{(2)})^{h}]\cdot \nonumber\\
		&\cdot\sum_{h=0}^{6}\sum_{k=0}^{6-h}\binom{6}{h,k}\mathrm{tr}[(\boldsymbol{D}^{(0)})^{6-h-k}]\varepsilon^{k}\mathrm{tr}[(\boldsymbol{D}^{(1)})^{k}]\varepsilon^{2h}\mathrm{tr}[(\boldsymbol{D}^{(2)})^{h}]+\nonumber\\
		&-\tfrac{1}{12}\sum_{h=0}^{2}\sum_{k=0}^{2-h}\binom{2}{h,k}(\mathrm{tr}[\boldsymbol{D}^{(0)}])^{2-h-k}\varepsilon^{k}(\mathrm{tr}[(\boldsymbol{D}^{(1)})])^{k}\varepsilon^{2h}(\mathrm{tr}[\boldsymbol{D}^{(2)}])^{h}\cdot\nonumber\\
		&\cdot\sum_{h=0}^{6}\sum_{k=0}^{6-h}\binom{6}{h,k}\mathrm{tr}[(\boldsymbol{D}^{(0)})^{6-h-k}]\varepsilon^{k}\mathrm{tr}[(\boldsymbol{D}^{(1)})^{k}]\varepsilon^{2h}\mathrm{tr}[(\boldsymbol{D}^{(2)})^{h}]+\nonumber\\
		&+\tfrac{1}{15}\sum_{h=0}^{3}\sum_{k=0}^{3-h}\binom{3}{h,k}\mathrm{tr}[(\boldsymbol{D}^{(0)})^{3-h-k}]\varepsilon^{k}\mathrm{tr}[(\boldsymbol{D}^{(1)})^{k}]\varepsilon^{2h}\mathrm{tr}[(\boldsymbol{D}^{(2)})^{h}]\cdot\nonumber\\
		&\cdot\sum_{h=0}^{5}\sum_{k=0}^{5-h}\binom{5}{h,k}\mathrm{tr}[(\boldsymbol{D}^{(0)})^{5-h-k}]\varepsilon^{k}\mathrm{tr}[(\boldsymbol{D}^{(1)})^{k}]\varepsilon^{2h}\mathrm{tr}[(\boldsymbol{D}^{(2)})^{h}]+\nonumber\\
		&-\tfrac{1}{10}\sum_{h=0}^{2}\sum_{k=0}^{2-h}\binom{2}{h,k}\mathrm{tr}[(\boldsymbol{D}^{(0)})^{2-h-k}]\varepsilon^{k}\mathrm{tr}[(\boldsymbol{D}^{(1)})^{k}]\varepsilon^{2h}\mathrm{tr}[(\boldsymbol{D}^{(2)})^{h}]\Big(\mathrm{tr}[\boldsymbol{D}^{(0)}]+\varepsilon \mathrm{tr}[\boldsymbol{D}^{(1)}]+\varepsilon^{2} \mathrm{tr}[\boldsymbol{D}^{(2)}]\Big)\cdot\nonumber\\
		&\cdot\sum_{h=0}^{5}\sum_{k=0}^{5-h}\binom{5}{h,k}\mathrm{tr}[(\boldsymbol{D}^{(0)})^{5-h-k}]\varepsilon^{k}\mathrm{tr}[(\boldsymbol{D}^{(1)})^{k}]\varepsilon^{2h}\mathrm{tr}[(\boldsymbol{D}^{(2)})^{h}]+\nonumber\\
		&+\tfrac{1}{30}\sum_{h=0}^{5}\sum_{k=0}^{5-h}\binom{5}{h,k}\mathrm{tr}[(\boldsymbol{D}^{(0)})^{5-h-k}]\varepsilon^{k}\mathrm{tr}[(\boldsymbol{D}^{(1)})^{k}]\varepsilon^{2h}\mathrm{tr}[(\boldsymbol{D}^{(2)})^{h}]\cdot\nonumber\\
		&\cdot\sum_{h=0}^{3}\sum_{k=0}^{3-h}\binom{3}{h,k}(\mathrm{tr}[\boldsymbol{D}^{(0)}])^{3-h-k}\varepsilon^{k}(\mathrm{tr}[(\boldsymbol{D}^{(1)})])^{k}\varepsilon^{2h}(\mathrm{tr}[\boldsymbol{D}^{(2)}])^{h}+\nonumber\\
		&-\tfrac{1}{12}\sum_{h=0}^{3}\sum_{k=0}^{3-h}\binom{3}{h,k}\mathrm{tr}[(\boldsymbol{D}^{(0)})^{3-h-k}]\varepsilon^{k}\mathrm{tr}[(\boldsymbol{D}^{(1)})^{k}]\varepsilon^{2h}\mathrm{tr}[(\boldsymbol{D}^{(2)})^{h}]\Big(\mathrm{tr}[\boldsymbol{D}^{(0)}]+\varepsilon \mathrm{tr}[\boldsymbol{D}^{(1)}]+\varepsilon^{2} \mathrm{tr}[\boldsymbol{D}^{(2)}]\Big)\cdot\nonumber\\
		&\cdot\sum_{h=0}^{4}\sum_{k=0}^{4-h}\binom{4}{h,k}\mathrm{tr}[(\boldsymbol{D}^{(0)})^{4-h-k}]\varepsilon^{k}\mathrm{tr}[(\boldsymbol{D}^{(1)})^{k}]\varepsilon^{2h}\mathrm{tr}[(\boldsymbol{D}^{(2)})^{h}]+\nonumber\\
		&-\tfrac{1}{32}\sum_{h=0}^{4}\sum_{k=0}^{4-h}\binom{4}{h,k}\mathrm{tr}[(\boldsymbol{D}^{(0)})^{4-h-k}]\varepsilon^{k}\mathrm{tr}[(\boldsymbol{D}^{(1)})^{k}]\varepsilon^{2h}\mathrm{tr}[(\boldsymbol{D}^{(2)})^{h}]\cdot\nonumber\\
		&\cdot\Big(\sum_{h=0}^{2}\sum_{k=0}^{2-h}\binom{2}{h,k}\mathrm{tr}[(\boldsymbol{D}^{(0)})^{2-h-k}]\varepsilon^{k}\mathrm{tr}[(\boldsymbol{D}^{(1)})^{k}]\varepsilon^{2h}\mathrm{tr}[(\boldsymbol{D}^{(2)})^{h}]\Big)^{2}+\nonumber\\
		&+\tfrac{1}{16}\sum_{h=0}^{2}\sum_{k=0}^{2-h}\binom{2}{h,k}\mathrm{tr}[(\boldsymbol{D}^{(0)})^{2-h-k}]\varepsilon^{k}\mathrm{tr}[(\boldsymbol{D}^{(1)})^{k}]\varepsilon^{2h}\mathrm{tr}[(\boldsymbol{D}^{(2)})^{h}]\cdot \nonumber\\
		&\cdot\sum_{h=0}^{2}\sum_{k=0}^{2-h}\binom{2}{h,k}(\mathrm{tr}[\boldsymbol{D}^{(0)}])^{2-h-k}\varepsilon^{k}(\mathrm{tr}[(\boldsymbol{D}^{(1)})])^{k}\varepsilon^{2h}(\mathrm{tr}[\boldsymbol{D}^{(2)}])^{h}\cdot\nonumber\\
		&\cdot \sum_{h=0}^{4}\sum_{k=0}^{4-h}\binom{4}{h,k}\mathrm{tr}[(\boldsymbol{D}^{(0)})^{4-h-k}]\varepsilon^{k}\mathrm{tr}[(\boldsymbol{D}^{(1)})^{k}]\varepsilon^{2h}\mathrm{tr}[(\boldsymbol{D}^{(2)})^{h}]+\nonumber\\
		&-\tfrac{1}{96}\sum_{h=0}^{4}\sum_{k=0}^{4-h}\binom{4}{h,k}\mathrm{tr}[(\boldsymbol{D}^{(0)})^{4-h-k}]\varepsilon^{k}\mathrm{tr}[(\boldsymbol{D}^{(1)})^{k}]\varepsilon^{2h}\mathrm{tr}[(\boldsymbol{D}^{(2)})^{h}]\cdot\nonumber\\
		&\cdot\sum_{h=0}^{4}\sum_{k=0}^{4-h}\binom{4}{h,k}(\mathrm{tr}[\boldsymbol{D}^{(0)}])^{4-h-k}\varepsilon^{k}(\mathrm{tr}[(\boldsymbol{D}^{(1)})])^{k}\varepsilon^{2h}(\mathrm{tr}[\boldsymbol{D}^{(2)}])^{h}+\nonumber\\
		&-\tfrac{1}{36}\sum_{h=0}^{2}\sum_{k=0}^{2-h}\binom{2}{h,k}\mathrm{tr}[(\boldsymbol{D}^{(0)})^{2-h-k}]\varepsilon^{k}\mathrm{tr}[(\boldsymbol{D}^{(1)})^{k}]\varepsilon^{2h}\mathrm{tr}[(\boldsymbol{D}^{(2)})^{h}]\cdot\nonumber\\
		&\cdot\Big(\sum_{h=0}^{3}\sum_{k=0}^{3-h}\binom{3}{h,k}\mathrm{tr}[(\boldsymbol{D}^{(0)})^{3-h-k}]\varepsilon^{k}\mathrm{tr}[(\boldsymbol{D}^{(1)})^{k}]\varepsilon^{2h}\mathrm{tr}[(\boldsymbol{D}^{(2)})^{h}]\Big)^{2}+\nonumber\\
		&+\tfrac{1}{36}\sum_{h=0}^{2}\sum_{k=0}^{2-h}\binom{2}{h,k}(\mathrm{tr}[\boldsymbol{D}^{(0)}])^{2-h-k}\varepsilon^{k}(\mathrm{tr}[(\boldsymbol{D}^{(1)})])^{k}\varepsilon^{2h}(\mathrm{tr}[\boldsymbol{D}^{(2)}])^{h}\cdot\nonumber\\
		&\cdot\Big(\sum_{h=0}^{3}\sum_{k=0}^{3-h}\binom{3}{h,k}\mathrm{tr}[(\boldsymbol{D}^{(0)})^{3-h-k}]\varepsilon^{k}\mathrm{tr}[(\boldsymbol{D}^{(1)})^{k}]\varepsilon^{2h}\mathrm{tr}[(\boldsymbol{D}^{(2)})^{h}]\Big)^{2}+\nonumber\\
		&+\tfrac{1}{24}\sum_{h=0}^{3}\sum_{k=0}^{3-h}\binom{3}{h,k}\mathrm{tr}[(\boldsymbol{D}^{(0)})^{3-h-k}]\varepsilon^{k}\mathrm{tr}[(\boldsymbol{D}^{(1)})^{k}]\varepsilon^{2h}\mathrm{tr}[(\boldsymbol{D}^{(2)})^{h}]\Big(\mathrm{tr}[\boldsymbol{D}^{(0)}]+\varepsilon \mathrm{tr}[\boldsymbol{D}^{(1)}]+\varepsilon^{2} \mathrm{tr}[\boldsymbol{D}^{(2)}]\Big)\cdot\nonumber\\
		&\cdot\Big(\sum_{h=0}^{2}\sum_{k=0}^{2-h}\binom{2}{h,k}\mathrm{tr}[(\boldsymbol{D}^{(0)})^{2-h-k}]\varepsilon^{k}\mathrm{tr}[(\boldsymbol{D}^{(1)})^{k}]\varepsilon^{2h}\mathrm{tr}[(\boldsymbol{D}^{(2)})^{h}]\Big)^{2}+\nonumber\\
		&-\tfrac{1}{36}\sum_{h=0}^{2}\sum_{k=0}^{2-h}\binom{2}{h,k}\mathrm{tr}[(\boldsymbol{D}^{(0)})^{2-h-k}]\varepsilon^{k}\mathrm{tr}[(\boldsymbol{D}^{(1)})^{k}]\varepsilon^{2h}\mathrm{tr}[(\boldsymbol{D}^{(2)})^{h}]\cdot \nonumber\\
		&\cdot\sum_{h=0}^{3}\sum_{k=0}^{3-h}\binom{3}{h,k}(\mathrm{tr}[\boldsymbol{D}^{(0)}])^{3-h-k}\varepsilon^{k}(\mathrm{tr}[(\boldsymbol{D}^{(1)})])^{k}\varepsilon^{2h}(\mathrm{tr}[\boldsymbol{D}^{(2)}])^{h}\cdot\nonumber\\
		&\cdot \sum_{h=0}^{3}\sum_{k=0}^{3-h}\binom{3}{h,k}\mathrm{tr}[(\boldsymbol{D}^{(0)})^{3-h-k}]\varepsilon^{k}\mathrm{tr}[(\boldsymbol{D}^{(1)})^{k}]\varepsilon^{2h}\mathrm{tr}[(\boldsymbol{D}^{(2)})^{h}]+\nonumber\\
		&+\tfrac{1}{360}\sum_{h=0}^{3}\sum_{k=0}^{3-h}\binom{3}{h,k}\mathrm{tr}[(\boldsymbol{D}^{(0)})^{3-h-k}]\varepsilon^{k}\mathrm{tr}[(\boldsymbol{D}^{(1)})^{k}]\varepsilon^{2h}\mathrm{tr}[(\boldsymbol{D}^{(2)})^{h}]\cdot\nonumber\\
		&\cdot\sum_{h=0}^{5}\sum_{k=0}^{5-h}\binom{5}{h,k}(\mathrm{tr}[\boldsymbol{D}^{(0)}])^{5-h-k}\varepsilon^{k}(\mathrm{tr}[(\boldsymbol{D}^{(1)})])^{k}\varepsilon^{2h}(\mathrm{tr}[\boldsymbol{D}^{(2)}])^{h}+\nonumber\\
		&+\tfrac{1}{384}(\sum_{h=0}^{2}\sum_{k=0}^{2-h}\binom{2}{h,k}\mathrm{tr}[(\boldsymbol{D}^{(0)})^{2-h-k}]\varepsilon^{k}\mathrm{tr}[(\boldsymbol{D}^{(1)})^{k}]\varepsilon^{2h}\mathrm{tr}[(\boldsymbol{D}^{(2)})^{h}])^{4}+\nonumber\\
		&-\tfrac{1}{96}\Big(\sum_{h=0}^{2}\sum_{k=0}^{2-h}\binom{2}{h,k}\mathrm{tr}[(\boldsymbol{D}^{(0)})^{2-h-k}]\varepsilon^{k}\mathrm{tr}[(\boldsymbol{D}^{(1)})^{k}]\varepsilon^{2h}\mathrm{tr}[(\boldsymbol{D}^{(2)})^{h}]\Big)^{3}\cdot\nonumber\\
		&\cdot\sum_{h=0}^{2}\sum_{k=0}^{2-h}\binom{2}{h,k}(\mathrm{tr}[\boldsymbol{D}^{(0)}])^{2-h-k}\varepsilon^{k}(\mathrm{tr}[(\boldsymbol{D}^{(1)})])^{k}\varepsilon^{2h}(\mathrm{tr}[\boldsymbol{D}^{(2)}])^{h}+\nonumber\\
		&+\tfrac{1}{192}\Big(\sum_{h=0}^{2}\sum_{k=0}^{2-h}\binom{2}{h,k}\mathrm{tr}[(\boldsymbol{D}^{(0)})^{2-h-k}]\varepsilon^{k}\mathrm{tr}[(\boldsymbol{D}^{(1)})^{k}]\varepsilon^{2h}\mathrm{tr}[(\boldsymbol{D}^{(2)})^{h}]\Big)^{2}\cdot\nonumber\\
		&\cdot\sum_{h=0}^{4}\sum_{k=0}^{4-h}\binom{4}{h,k}(\mathrm{tr}[\boldsymbol{D}^{(0)}])^{4-h-k}\varepsilon^{k}(\mathrm{tr}[(\boldsymbol{D}^{(1)})])^{k}\varepsilon^{2h}(\mathrm{tr}[\boldsymbol{D}^{(2)}])^{h}+\nonumber\\
		&-\tfrac{1}{1442}\sum_{h=0}^{6}\sum_{k=0}^{6-h}\binom{6}{h,k}(\mathrm{tr}[\boldsymbol{D}^{(0)}])^{6-h-k}\varepsilon^{k}(\mathrm{tr}[(\boldsymbol{D}^{(1)})])^{k}\varepsilon^{2h}(\mathrm{tr}[\boldsymbol{D}^{(2)}])^{h}\cdot\nonumber\\
		&\cdot\sum_{h=0}^{2}\sum_{k=0}^{2-h}\binom{2}{h,k}\mathrm{tr}[(\boldsymbol{D}^{(0)})^{2-h-k}]\varepsilon^{k}\mathrm{tr}[(\boldsymbol{D}^{(1)})^{k}]\varepsilon^{2h}\mathrm{tr}[(\boldsymbol{D}^{(2)})^{h}]\nonumber+\\
		&+\tfrac{1}{32}\Big(\sum_{h=0}^{4}\sum_{k=0}^{4-h}\binom{4}{h,k}\mathrm{tr}[(\boldsymbol{D}^{(0)})^{4-h-k}]\varepsilon^{k}\mathrm{tr}[(\boldsymbol{D}^{(1)})^{k}]\varepsilon^{2h}\mathrm{tr}[(\boldsymbol{D}^{(2)})^{h}]\Big)^{2}+\nonumber\\
		&+\tfrac{1}{40320}\sum_{h=0}^{8}\sum_{k=0}^{8-h}\binom{8}{h,k}(\mathrm{tr}[\boldsymbol{D}^{(0)}])^{8-h-k}\varepsilon^{k}(\mathrm{tr}[(\boldsymbol{D}^{(1)})])^{k}\varepsilon^{2h}(\mathrm{tr}[\boldsymbol{D}^{(2)}])^{h}.
	\end{align}
\end{subequations}
\subsection*{Section C.3}
The invariants coefficients, referred to the characteristic polynomial (81) of Subsection 5.1 in the main text, are derived via the Faddeev-LeVerrier recursive formula  \cite{helmberg1993faddeev} as follows 
\begin{subequations}
\begin{align}
	III_{6}(\omega)&=1,\quad III_{5}(\omega)=-\mathrm{tr}[\boldsymbol{T}(\omega)], \quad III_{4}(\omega)=-\tfrac{1}{2}\mathrm{tr}[\boldsymbol{T}(\omega)^{2}]+\tfrac{1}{2}(\mathrm{tr}[\boldsymbol{T}(\omega)])^{2},&\\
	III_{3}(\omega)&=-\tfrac{1}{3}\mathrm{tr}[\boldsymbol{T}(\omega)^{3}]+\tfrac{1}{2}\mathrm{tr}[\boldsymbol{T}(\omega)]\mathrm{tr}[\boldsymbol{T}(\omega)^{2}]-\tfrac{1}{6}(\mathrm{tr}[\boldsymbol{T}(\omega)])^{3}.
\end{align}
\end{subequations} 
\section*{Section D \quad Perturbation functions of first order}
Section D provides details on the first-order perturbation functions associated with the example discussed in Section 5 of the main text.
\subsection*{Section D.1 \quad Perturbation function of first order $N_{hpq}^{(1)}$}
Let $\mathcal{N}$ be a layered domain obtained as a $d_{2}$-periodic arrangement of two layers with thickness $s_{1}$ and $s_{2}$ (here $d_{2}=s_{1}+s_{2}$ and $\eta=s_{1}/s_{2}$ are defined).
The phases are supposed to be homogeneous and orthotropic, with
orthotropic axis coincident with the layering direction $\boldsymbol{e}_{1}$. The micro-fluctuation functions $N_{hpq}^{(1)_{i}}$ are analytically obtained by solving the first cell problem (29a) in the main text. The superscript $i=\{1,2\}$ stands for the phase 1 and the phase 2 and they are formulated as
\begin{flalign}
	&N_{211}^{(1)_1}=-{\frac { \left( { {{C}_{1122}^{1}}}-{ {{C}_{1122}^{2}}} \right) { \xi_2}}{{
				{{C}_{2222}^{2}}}\,\eta+{ {{C}_{2222}^{1}}}}}
	, \quad 	N_{211}^{(1)_2}={\frac {\eta \, \left( { {{C}_{1122}^{1}}}-{ {{C}_{1122}^{2}}}
			\right) { \xi_2}}{{ {{C}_{2222}^{2}}}\, \eta +{ 
				{{C}_{2222}^{1}}}}},&	 
\end{flalign}
\begin{flalign}
	&N_{222}^{(1)_1}=-{\frac { \left( { {{C}_{2222}^{1}}}-{ {{C}_{2222}^{2}}} \right) { \xi_2}}{{
				{{C}_{2222}^{2}}}\,\eta+{ {{C}_{2222}^{1}}}}}
	, \quad 	N_{222}^{(1)_2}={\frac {\eta \, \left( { {{C}_{2222}^{1}}}-{ {{C}_{2222}^{2}}}
			\right) { \xi_2}}{{ {{C}_{2222}^{2}}}\, \eta +{ 
				{{C}_{2222}^{1}}}}},&	 
\end{flalign}
\begin{flalign}
	&N_{112}^{(1)_1}=N_{121}^{(1)_1}=-{\frac { \left( { {{C}_{1212}^{1}}}-{ {{C}_{1212}^{2}}} \right) { \xi_2}}{{
				{{C}_{1212}^{2}}}\,\eta+{ {{C}_{1212}^{1}}}}}
	, \quad 	N_{112}^{(1)_2}=N_{121}^{(1,0)_2}={\frac {\eta \, \left( { {{C}_{1212}^{1}}}-{ {{C}_{1212}^{2}}}
			\right) { \xi_2}}{{ {{C}_{1212}^{2}}}\, \eta +{ 
				{{C}_{1212}^{1}}}}}.&	 
\end{flalign}
Such functions depend on the fast variable $\boldsymbol{\xi}$, since the microstructure enjoys the simmetry property. In the following it is assumed that the coordinate $\xi_{2}$ is centered in both layers.
\subsection*{Section D.2 \quad Perturbation function of first order $\tilde{N}_{h}^{(1)}$}
The micro-fluctuation functions $\tilde{N}_{h}^{(1)_{i}}$ are analytically obtained by solving the cell problem (29b) in the main text as
\begin{flalign}
	&\tilde{N}_{2}^{(1)_1}={\frac { \left( { {{\alpha}_{22}^{1}}}-{ {{\alpha}_{22}^{2}}} \right) { \xi_2}}{{
				{{C}_{2222}^{2}}}\,\eta+{ {{C}_{2222}^{1}}}}}
	, \quad 	\tilde{N}_{2}^{(1)_2}=-{\frac {\eta \, \left( { {{\alpha}_{22}^{1}}}-{ {{\alpha}_{22}^{2}}}
			\right) { \xi_2}}{{ {{C}_{2222}^{2}}}\, \eta +{ 
				{{C}_{2222}^{1}}}}}.&	 
\end{flalign}
\subsection*{Section D.3 \quad Perturbation function of first order $\tilde{N}_{h}^{(1,1)}$}
The micro-fluctuation functions $\tilde{N}_{h}^{(1,1)_{i}}$ are analytically obtained by solving the cell problem (29c) in the main text as
\begin{flalign}
	&\tilde{N}_{2}^{(1,1)_1}={\frac { \left( { {{\alpha}_{22}^{(1,1)}}}-{ {{\alpha}_{22}^{(2,1)}}} \right) { \xi_2}}{{
				{{C}_{2222}^{2}}}\,\eta+{ {{C}_{2222}^{1}}}}}
	, \quad 	\tilde{N}_{2}^{(1,1)_2}=-{\frac {\eta \, \left( { {{\alpha}_{22}^{(1,1)}}}-{ {{\alpha}_{22}^{(2,1)}}}
			\right) { \xi_2}}{{ {{C}_{2222}^{2}}}\, \eta +{ 
				{{C}_{2222}^{1}}}}}.&	 
\end{flalign}
\subsection*{Section D.4 \quad Perturbation function of first order $M_{q}^{(1)}$}
The micro-fluctuation functions $M_{q}^{(1)_{i}}$ are analytically obtained by solving the cell problem (29d) in the main text as
\begin{flalign}
	&M_{2}^{(1)_1}=-{\frac { \left( { {{K}_{22}^{1}}}-{ {{K}_{22}^{2}}} \right) { \xi_2}}{{
				{{K}_{22}^{2}}}\,\eta+{ {{K}_{22}^{1}}}}}
	, \quad 	M_{2}^{(1)_2}={\frac {\eta \, \left( { {{K}_{22}^{1}}}-{ {{K}_{22}^{2}}}
			\right) { \xi_2}}{{ {{K}_{22}^{2}}}\, \eta +{ 
				{{K}_{22}^{1}}}}}.&	 
\end{flalign}
\section*{Section E \quad Perturbation functions of second order}
Section E provides details on the second-order perturbation functions associated with the example discussed in Section 5 of the main text.
\subsection*{Section E.1 \quad Perturbation functions of second order $N_{hpqr}^{(2)}$}
The perturbation functions $N_{hpqr}^{(2)_{i}}$ (with $i=\{1,2\}$), deriving from the cell problem (31a) of the main text, are:	
\begin{flalign}
	&N_{1111}^{(2)_1}=A_{1111}^{2} (\xi_{2})^2+A_{1111}^{0}, \quad N_{1111}^{(2)_2}=B_{1111}^{2} \xi_{2}^2+B_{1111}^{0},&
\end{flalign}	
\begin{flalign}
	&N_{2211}^{(2)_1}=A_{2211}^{2} (\xi_{2})^2+A_{2211}^{0}, \quad N_{2211}^{(2)_2}=B_{2211}^{2} \xi_{2}^2+B_{2211}^{0},&
\end{flalign}
\begin{flalign}
	&N_{2222}^{(2)_1}=A_{2222}^{2} (\xi_{2})^2+A_{2222}^{0}, \quad N_{2222}^{(2)_2}=B_{2222}^{2} \xi_{2}^2+B_{2222}^{0},&
\end{flalign}
\begin{flalign}
	&N_{1122}^{(2)_1}=A_{1122}^{2} (\xi_{2})^2+A_{1122}^{0}, \quad N_{1122}^{(2)_2}=B_{1122}^{2} \xi_{2}^2+B_{1122}^{0}.&
\end{flalign}
The constants $A_{1111}^{2}$, $A_{1111}^{0}$, $B_{1111}^{2}$, $B_{1111}^{0}$, $A_{2211}^{2}$, $A_{2211}^{0}$, $B_{2211}^{2}$, $B_{2211}^{0}$, $A_{2222}^{2}$, $A_{2222}^{0}$, $B_{2222}^{2}$, $B_{2222}^{0}$, $A_{1122}^{2}$,\\ \\ $A_{1122}^{0}$, $B_{1122}^{2}$ and $B_{1122}^{0}$ are determined as follows
\begin{flalign}
	&A_{1111}^{2}=-\frac{1}{2}\,{\frac {A_{1111}^{2,0}+\eta A_{1111}^{2,1} }{ \left( \eta+1 \right)  \left( { C_{2222}^{{2}}}\,\eta+ C_{2222}^{{1}} \right) { C_{1212}^{{1}}}}},&
\end{flalign}	             
\begin{flalign}
	&A_{1111}^{0}=\frac{1}{24}\,{\frac {\eta\, \left( { A_{1111}^{0,3}}\,({\eta})^{3}+{ A_{1111}^{0,2}}
			\,({\eta})^{2}+{ A_{1111}^{0,1}}\,\eta+{ A_{1111}^{0,0}} \right) }{ \left( 
			\eta+1 \right) ^{4} \left( { C_{2222}^{{2}}}\,\eta+{ C_{2222}^{{1}}}
			\right) { C_{1212}^{{2}}}\,{ C_{1212}^{{1}}}}},&
\end{flalign}

\begin{flalign}
	&B_{1111}^{2}=\frac{1}{2}\,{\frac {\eta B_{1111}^{2,1}+(\eta)^2 B_{1111}^{2,2} }{ \left( \eta+1 \right)  \left( { C_{2222}^{{2}}}\,\eta+ C_{2222}^{{1}} \right) { C_{1212}^{{2}}}}},&
\end{flalign}
\begin{flalign}
	&B_{1111}^{0}=\frac{1}{24}\,{\frac {\eta\, \left( { -2 A_{1111}^{0,3}}\,({\eta})^{3}+{ B_{1111}^{0,2}}
			\,({\eta})^{2}+{ B_{1111}^{0,1}}\,\eta-{\frac{A_{1111}^{0,0}}{2}} \right) }{ \left( 
			\eta+1 \right) ^{4} \left( { C_{2222}^{{2}}}\,\eta+{ C_{2222}^{{1}}}
			\right) { C_{1212}^{{1}}}\,{ C_{1212}^{{2}}}}},&
\end{flalign}
\begin{flalign}
	&A_{2211}^{2}=\frac{1}{2}\,{\frac {{C_{1122}^{{1}}}\, \left( { C_{1212}^{{1}}}-{ C_{1212}^{{2}}}
			\right) }{ \left( { C_{1212}^{{2}}}\,\eta+{ C_{1212}^{{1}}} \right) { 
				C_{2222}^{{1}}}}},&
\end{flalign}
\begin{flalign}
	&A_{2211}^{0}=-\frac{1}{24}\,{\frac { \left( { C_{1212}^{{1}}}-{ C_{1212}^{{2}}} \right) \eta\,
			\left( { C_{1122}^{{1}}}\,({\eta})^{2}{ C_{2222}^{{2}}}+3\,{ C_{1122}^{{1}}}\,
			\eta\,{ C_{2222}^{{2}}}+2\,{ C_{1122}^{{2}}}\,{ C_{2222}^{{1}}} \right) }{
			\left( \eta+1 \right) ^{3} \left( { C_{1212}^{{2}}}\,\eta+{ C_{1212}^{{1}}}
			\right) { C_{2222}^{{2}}}\,{ C_{2222}^{{1}}}}},&
\end{flalign}	
\begin{flalign}
	&B_{2211}^{2}=-\frac{\eta  C_{1122}^{2}  C_{2222}^{{1}}}{ C_{2222}^{{2}} C_{1122}^{1}}A_{2211}^{2},&
\end{flalign}	
\begin{flalign}
	&B_{2211}^{0}=\frac{1}{24}\,{\frac { \left( 2\,{ C_{1122}^{{1}}}\,({\eta})^{2}{ C_{2222}^{{2}}}+3\,
			\eta\,{ C_{1122}^{{2}}}\,{ C_{2222}^{{1}}}+{ C_{1122}^{{2}}}\,{ C_{2222}^{{1}}}
			\right)  \left( { C_{1212}^{{1}}}-{ C_{1212}^{{2}}} \right) \eta}{ \left( 
			\eta+1 \right) ^{3} \left( { C_{1212}^{{2}}}\,\eta+{ C_{1212}^{{1}}}
			\right) { C_{2222}^{{2}}}\,{ C_{2222}^{{1}}}}},&
\end{flalign}
\begin{flalign}
	&A_{2222}^{2}={\frac {{ C_{2222}^{{1}}}-{ C_{2222}^{{2}}}}{2\,{ C_{2222}^{{2}}}\,\eta+2\,{
				C_{2222}^{{1}}}}}, \quad A_{2222}^{0}=-\frac{1}{24}\,{\frac { \left( { C_{2222}^{{1}}}-{ C_{2222}^{{2}}} \right) \eta\,
			\left( \eta+2 \right) }{ \left( { C_{2222}^{{2}}}\,\eta+{ C_{2222}^{{1}}}
			\right)  \left( \eta+1 \right) ^{2}}},&
\end{flalign}
\begin{flalign}
	&B_{2222}^{2}=-\eta A_{2222}^{2}, \quad B_{2222}^{0}=-\frac{2\eta+1}{\eta+2}A_{2222}^{0},&
\end{flalign}
\begin{flalign}
	&A_{1122}^{2}={\frac {{ \hat{C}_{1212}^{{1}}}-{ C_{1212}^{{2}}}}{2\,{ C_{1212}^{{2}}}\,\eta+2\,{
				C_{1212}^{{1}}}}}, \quad A_{1122}^{0}=-\frac{1}{24}\,{\frac { \left( { C_{1212}^{{1}}}-{ C_{1212}^{{2}}} \right) \eta\,
			\left( \eta+2 \right) }{ \left( { C_{1212}^{{2}}}\,\eta+{ C_{1212}^{{1}}}
			\right)  \left( \eta+1 \right) ^{2}}},&
\end{flalign}
\begin{flalign}
	&B_{1122}^{2}=-\eta A_{1122}^{2}, \quad B_{1122}^{0}=-\frac{2\eta+1}{\eta+2}A_{1122}^{0},&
\end{flalign}
where the costants $A_{1111}^{2,0}$, $A_{1111}^{2,1}$, $A_{1111}^{0,3}$, $A_{1111}^{0,2}$, $A_{1111}^{0,1}$, $A_{1111}^{0,0}$, $B_{1111}^{2,1}$, $B_{1111}^{2,2}$, $B_{2222}^{0,2}$ and $B_{2222}^{0,1}$ are not reported here for sake of simplicity but they can be found in \cite{del2019characterization}.
\subsection*{Section E.2 \quad Perturbation functions $\tilde{N}_{hq}^{(2)}$ and $\tilde{N}_{hq}^{(2,1)}$}
The perturbation functions $\tilde{N}_{hq}^{(2)_{i}}$ (with $i=\{1,2\}$), obtained by performing the cell problem (31b) in the main text, are:
\begin{flalign}
	\label{CPC}
	&\tilde{N}_{11}^{(2)_1}=M_{11}^{2} (\xi_{2})^2+M_{11}^{0}, \quad \tilde{N}_{11}^{(2)_2}=R_{11}^{2} \xi_{2}^2+R_{11}^{0},&
\end{flalign}
\begin{flalign}
	\label{CPC2}
	&\tilde{N}_{22}^{(2)_1}=M_{22}^{2} (\xi_{2})^2+M_{22}^{0}, \quad \tilde{N}_{22}^{(2)_2}=R_{22}^{2} \xi_{2}^2+R_{22}^{0},&
\end{flalign}
where the constants $M_{11}^{2}$, $M_{11}^{0}$, $R_{11}^{2}$, $R_{11}^{0}$, $M_{22}^{2}$, $M_{22}^{0}$, $R_{22}^{2}$ and $R_{22}^{0}$ are expressed as
\begin{flalign}
	&M_{11}^{2}=\frac{1}{2}\frac{M_{11}^{2,1}\eta+M_{11}^{2,0}}{C^{1}_{1212}(\eta+1)(C^{2}_{2222}\eta+C^{1}_{2222})},\quad M_{11}^{0}=-\frac{1}{24}\frac{\eta(M_{11}^{0,3}\eta^{3}+M_{11}^{0,2}\eta^{2}+M_{11}^{0,1}\eta+M_{11}^{0,0})}{C^{2}_{1212}C^{1}_{1212}(\eta+1)^{4}(C^{2}_{2222}\eta+C^{1}_{2222})},&
\end{flalign}  
\begin{flalign}
	&R_{11}^{2}=-\frac{1}{2}\frac{\eta(R_{11}^{2,1}\eta+R_{11}^{2,0})}{C^{2}_{1212}(\eta+1)(C^{2}_{2222}\eta+C^{1}_{2222})},\quad R_{11}^{0}=\frac{1}{12}\frac{\eta(R_{11}^{0,3}(\eta)^{3}+R_{11}^{0,2}(\eta)^{2}+R_{11}^{0,1}\eta+R_{11}^{0,0})}{C^{2}_{1212}C^{1}_{1212}(\eta+1)^{4}(C^{2}_{2222}\eta+C^{1}_{2222})},&
\end{flalign}
\begin{flalign}
	&M_{22}^{2}=-\frac{1}{2}\frac{M_{22}^{2,2}(\eta)^{2}+M_{22}^{2,1}\eta+M_{22}^{2,0}}{C^{1}_{2222}(\eta+1)(C^{2}_{2222}\eta+C^{1}_{2222})(\alpha^{2}_{22}\eta+\alpha^{1}_{22})},&
\end{flalign}
\begin{flalign}
	&M_{22}^{0}=-\frac{1}{24}\frac{\eta(M_{22}^{0,4}\eta^{4}+M_{22}^{0,3}(\eta)^{3}+M_{22}^{0,2}(\eta)^{2}+M_{22}^{0,1}\eta+M_{22}^{0,0})}{(C^{2}_{2222}\eta+C^{1}_{2222})C^{2}_{2222}C^{1}_{2222}(\eta+1)^{4}(\alpha^{2}_{22}\eta+\alpha^{1}_{22})},&
\end{flalign}
\begin{flalign}
	&R_{22}^{2}=\frac{1}{2}\frac{\eta(R_{22}^{2,2}(\eta)^{2}+R_{22}^{2,1}\eta+R_{22}^{2,0})}{C^{2}_{2222}(\eta+1)(C^{2}_{2222}\eta+C^{1}_{2222})(\alpha^{2}_{22}\eta+\alpha^{1}_{22})},&
\end{flalign}
\begin{flalign}	
	&R_{22}^{0}=-\frac{1}{12}\frac{\eta(R_{22}^{0,4}(\eta)^{4}+R_{22}^{0,3}(\eta)^{3}+R_{22}^{0,2}(\eta)^{2}+R_{22}^{0,1}\eta+R_{22}^{0,0})}{(C^{2}_{2222}\eta+C^{1}_{2222})C^{2}_{2222}C^{1}_{2222}(\eta+1)^{4}(\alpha^{2}_{22}\eta+\alpha^{1}_{22})},&
\end{flalign}  
where the constants $M_{11}^{2,1}$, $M_{11}^{2,0}$, $M_{11}^{0,3}$, $M_{11}^{0,2}$, $M_{11}^{0,1}$, $M_{11}^{0,0}$, $R_{11}^{2,1}$, $R_{11}^{2,0}$, $R_{11}^{0,3}$, $R_{11}^{0,2}$, $R_{11}^{0,1}$, $R_{11}^{0,0}$, $M_{22}^{2,2}$, $M_{22}^{2,1}$, $M_{22}^{2,0}$, $M_{22}^{0,4}$, $M_{22}^{0,3}$, $M_{22}^{0,2}$, $M_{22}^{0,1}$ , $M_{22}^{0,0}$, $R_{22}^{2,2}$, $R_{22}^{2,1}$, $R_{22}^{2,0}$, $R_{22}^{0,4}$, $R_{22}^{0,3}$, $R_{22}^{0,2}$, $R_{22}^{0,1}$ and $R_{22}^{0,0}$ depend on the components of the tensor $\boldsymbol{\alpha}^{(m)}$. Meanwhile, the perturbation function $\tilde{N}_{hq}^{(2,1)_{i}}$, derived from the cell problem (31c) of the main text, has the same structure as \eqref{CPC}-\eqref{CPC2}, but its coefficients rely on the components of the tensor $\boldsymbol{\alpha}^{(m,1)}$. 
\subsection*{Section E.3 \quad Perturbation functions $N_{hp}^{(2,2)}$}
The perturbation functions $N_{hp}^{(2,2)_{i}}$ (with $i=\{1,2\}$), obtained by performing the cell problem (31d) of the main text, are:
\begin{flalign}
	&N_{11}^{(2,2)_1}=A_{11}^{2} (\xi_{2})^2+A_{11}^{0}, \quad N_{11}^{(2,2)_2}=B_{11}^{2} (\xi_{2})^2+B_{11}^{0},&
\end{flalign}		
\begin{flalign}
	\label{zia}
	&N_{22}^{(2,2)_1}=A_{22}^{2} (\xi_{2})^2+A_{22}^{0}, \quad N_{22}^{(2,2)_2}=B_{22}^{2} (\xi_{2})^2+B_{22}^{0},&
\end{flalign}
where the constants $A_{11}^{2}$, $A_{11}^{0}$, $B_{11}^{2}$, $B_{11}^{0}$, $A_{22}^{2}$, $A_{22}^{0}$, $B_{22}^{2}$ and $B_{22}^{0}$ are     
\begin{flalign}
	&A_{11}^{2}=\frac{1}{2}\,{\frac { \left( { \rho_{1}}-{ \rho_{2}} \right) { }}{
			\left( \eta +1 \right) { C_{1212}^{{1}}}}}, \quad A_{11}^{0}=-\frac{1}{24}\,{\frac { \left( { \rho_{1}}-{ \rho_{2}} \right) { }
			\, \eta \, \left( { C_{1212}^{{2}}}\,({\eta })^{2}+3\,{
				C_{1212}^{{2}}}\, \eta +2\,{ C_{1212}^{{1}}} \right) }{{
				C_{1212}^{{2}}}\, \left( \eta +1 \right) ^{4}{ C_{1212}^{{1}}}}},&	
\end{flalign}
\begin{flalign}
	&B_{11}^{2}=-\frac{1}{2}\,{\frac { \left( { \rho_{1}}-{ \rho_{2}} \right) \eta { }}{
			\left( \eta +1 \right) { C_{1212}^{{2}}}}}, \quad B_{11}^{0}=\frac{1}{24}\,{\frac { \left( { \rho_{1}}-{ \rho_{2}} \right) { }
			\, \eta \, \left( { 2C_{1212}^{{2}}}\,({\eta })^{2}+3\,{
				C_{1212}^{{1}}}\, \eta +{ C_{1212}^{{1}}} \right) }{{
				C_{1212}^{{2}}}\, \left( \eta +1 \right) ^{4}{ C_{1212}^{{1}}}}},&	
\end{flalign}
\begin{flalign}
	&A_{22}^{2}=\frac{1}{2}\,{\frac { \left( { \rho_{1}}-{ \rho_{2}} \right) { }}{
			\left( \eta +1 \right) { C_{2222}^{{1}}}}}, \quad A_{22}^{0}=-\frac{1}{24}\,{\frac { \left( { \rho_{1}}-{ \rho_{2}} \right) { }
			\, \eta \, \left( { C_{2222}^{{2}}}\,({\eta })^{2}+3\,{
				C_{2222}^{{2}}}\, \eta +2\,{ C_{2222}^{{1}}} \right) }{{
				C_{2222}^{{2}}}\, \left( \eta +1 \right) ^{4}{ C_{2222}^{{1}}}}},&
\end{flalign}
\begin{flalign}
	&B_{22}^{2}=-\frac{1}{2}\,{\frac { \left( { \rho_{1}}-{ \rho_{2}} \right) \eta { }}{
			\left( \eta +1 \right) { C_{2222}^{{2}}}}}, \quad B_{22}^{0}=\frac{1}{24}\,{\frac { \left( { \rho_{1}}-{ \rho_{2}} \right) { }
			\, \eta \, \left( { 2C_{2222}^{{2}}}\,({\eta })^{2}+3\,{
				C_{2222}^{{1}}}\, \eta +{ C_{2222}^{{1}}} \right) }{{
				C_{2222}^{{2}}}\, \left( \eta +1 \right) ^{4}{ C_{2222}^{{1}}}}}.&	
\end{flalign}
\subsection*{Section E.4 \quad Perturbation functions $\tilde{M}_{pq}^{(2,1)}$}
The perturbation functions $\tilde{M}_{pq}^{(2,1)}$ (with $i=\{1,2\}$), obtained by performing the cell problem (33b) of the main text, are:
\begin{flalign}
	&\tilde{M}_{11}^{(2,1)_1}=E_{11}^{2} (\xi_{2})^2+E_{11}^{0}, \quad \tilde{M}_{11}^{(2,1)_2}=D_{11}^{2} (\xi_{2})^2+D_{11}^{0},&
\end{flalign}		
\begin{flalign}
	\label{tildem}
	&\tilde{M}_{22}^{(2,1)_1}=E_{22}^{2} (\xi_{2})^2+E_{22}^{0}, \quad \tilde{M}_{22}^{(2,1)_2}=D_{22}^{2} (\xi_{2})^2+D_{22}^{0},&
\end{flalign}
where the constants $E_{11}^{2}$, $E_{11}^{0}$, $D_{11}^{2}$, $D_{11}^{0}$, $E_{22}^{2}$, $E_{22}^{0}$, $D_{22}^{2}$ and $D_{22}^{0}$ are
\begin{flalign}
	&E_{11}^{2}=-\frac{1}{2}\frac{(E_{11}^{2,1}\eta+E_{11}^{2,0})}{K^{1}_{22}(\eta+1)(C^{2}_{2222}\eta+C^{1}_{2222})},\quad E_{11}^{0}=\frac{1}{24}\frac{\eta(E_{11}^{0,3}(\eta)^{3}+E_{11}^{0,2}(\eta)^{2}+E_{11}^{0,1}\eta+E_{11}^{0,0})}{K^{2}_{22}K^{1}_{22}(\eta+1)^{4}(C^{2}_{2222}\eta+C^{1}_{2222})},&
\end{flalign}
\begin{flalign}
	&D_{11}^{2}=\frac{1}{2}\frac{\eta(D_{11}^{2,1}\eta+D_{11}^{2,0})}{K^{2}_{22}(\eta+1)(C^{2}_{2222}\eta+C^{1}_{2222})},\quad
	D_{11}^{0}=-\frac{1}{12}\frac{\eta(D_{11}^{0,3}(\eta)^{3}+D_{11}^{0,2}(\eta)^{2}+D_{11}^{0,1}\eta+D_{11}^{0,0})}{K^{2}_{22}K^{1}_{22}(\eta+1)^{4}(C^{2}_{2222}\eta+C^{1}_{2222})},&
\end{flalign}
\begin{flalign}
	&E_{22}^{2}=-\frac{1}{2}\frac{C^{1}_{2222}\alpha^{2}_{22}-C^{2}_{2222}\alpha^{1}_{22}}{K^{1}_{22}(C^{2}_{2222}\eta+C^{1}_{2222})},\quad E_{22}^{0}=\frac{1}{24}\frac{\eta(E_{22}^{0,2}(\eta)^{2}+E_{22}^{0,1}\eta+E_{22}^{0,0})}{K^{1}_{22}K^{2}_{22}(\eta+1)^{3}(C^{2}_{2222}\eta+C^{1}_{2222})},&
\end{flalign}	
\begin{flalign}
	&D_{22}^{2}=\frac{1}{2}\frac{\eta(C^{1}_{2222}\alpha^{2}_{22}-C^{2}_{2222}\alpha^{1}_{22})}{K^{2}_{22}(C^{2}_{2222}\eta+C^{1}_{2222})},\quad D_{22}^{0}=-\frac{1}{24}\frac{\eta(D_{22}^{0,2}(\eta)^{2}+D_{22}^{0,1}\eta+D_{22}^{0,0})}{K^{1}_{22}K^{2}_{22}(\eta+1)^{3}(C^{2}_{2222}\eta+C^{1}_{2222})}.& 
\end{flalign}
The coefficients $E_{11}^{2,1}$, $E_{11}^{2,0}$, $E_{11}^{0,3}$, $E_{11}^{0,2}$, $E_{11}^{0,1}$, $E_{11}^{0,0}$, $D_{11}^{2,1}$, $D_{11}^{2,0}$, $D_{11}^{0,3}$, $D_{11}^{0,2}$, $D_{11}^{0,1}$, $D_{11}^{0,0}$, $E_{22}^{0,2}$, $E_{22}^{0,1}$, $E_{22}^{0,0}$, $D_{22}^{0,2}$, $D_{22}^{0,1}$ and $D_{22}^{0,0}$ are not reported here for sake of simplicity.
\subsection*{Section E.5 \quad Perturbation functions $M^{(2,1)}$ and $M^{(2,2)}$ }
The perturbation functions $M^{(2,1)}$ (with $i=\{1,2\}$), derived from the cell problem (33c), are:
\begin{flalign}
	\label{FPM21}
	&M^{(2,1)_1}=E^{2} (\xi_{2})^2+E^{0}, \quad M^{(2,1)_2}=D^{2} (\xi_{2})^2+D^{0},&
\end{flalign}	
where the constants $E^{2}$, $E^{0}$, $D^{2}$ and $D^{0}$ are 
\begin{flalign}
	&E^{2}=\frac{1}{2}\frac{E^{2,1}\eta+E^{2,0}}{K^{1}_{22}(\eta+1)(C^{2}_{2222}\eta+C^{1}_{2222})},\quad E^{0}=-\frac{1}{24}\frac{\eta(E^{0,3}(\eta)^{3}+E^{0,2}(\eta)^{2}+E^{0,1}\eta+E^{0,0})}{K^{2}_{22}K^{1}_{22}(\eta+1)^{4}(C^{2}_{2222}\eta+C^{1}_{2222})},&
\end{flalign}
\begin{flalign}
	&D^{2}=-\frac{1}{2}\frac{\eta(D^{2,1}\eta+D^{2,0})}{K^{2}_{22}(\eta+1)(C^{2}_{2222}\eta+C^{1}_{2222})},\quad D^{0}=\frac{1}{12}\frac{\eta(D^{0,3}(\eta)^{3}+D^{0,2}(\eta)^{2}+D^{0,1}\eta+D^{0,0})}{K^{2}_{22}K^{1}_{22}(\eta+1)^{4}(C^{2}_{2222}\eta+C^{1}_{2222})}.&
\end{flalign}
The coefficients $E^{2,1}$, $E^{2,0}$, $E^{0,3}$, $E^{0,2}$, $E^{0,1}$, $E^{0,0}$, $D^{2,1}$, $D^{2,0}$, $D^{0,3}$, $D^{0,2}$, $D^{0,1}$ and $D^{0,0}$ are not reported here for sake of simplicity. Moreover, they depend, among the others, on $p^{m}$. On the other hand, the perturbation function $M^{(2,2)}$, derived from the cell problem (33d), has the same structure as \eqref{FPM21}, though its coefficients rely on $p^{(m,0)}$. 
\subsection*{Section E.6 \quad Perturbation functions $M^{(2)}_{q}$}  
The perturbation functions $M^{(2)}_{q}$ (with $i=\{1,2\}$), derived from the cell problem (33a) of the main text, are:  
\begin{flalign}
	\label{FPM22}
	&M^{(2)_1}_{22}=L^{2}_{22} (\xi_{2})^2+L^{0}_{22}, \quad M^{(2)_2}_{22}=T^{2}_{22} (\xi_{2})^2+T^{0}_{22},&
\end{flalign}	
where the constants $L^{2}_{22}$, $L^{0}_{22}$, $T^{2}_{22}$ and $T^{0}_{22}$ are 
\begin{flalign}
	&L^{2}_{22}=\frac{1}{2}\frac{K_{22}^{1}-K_{22}^{2}}{(\eta K_{22}^{2}-K_{22}^{1})}, \quad L^{0}_{22}=-\frac{\eta(\eta+2)(K_{22}^{1}-K_{22}^{2})}{24(\eta+1)^{2}(\eta K_{22}^{2}-K_{22}^{1})},&
\end{flalign} 
\begin{flalign}
	&T^{2}_{22}=-\frac{1}{2}\frac{\eta(K_{22}^{1}-K_{22}^{2})}{(\eta K_{22}^{2}-K_{22}^{1})}, \quad T^{0}_{22}=\frac{\eta(2\eta+1)(K_{22}^{1}-K_{22}^{2})}{24(\eta+1)^{2}(\eta K_{22}^{2}-K_{22}^{1})}.&
\end{flalign} 
\section*{Acknowledgement}
The authors gratefully acknowledge financial support from National Group of Mathematical Physics, Italy (GNFM-IN$\delta$AM),  from University of Chieti-Pescara project Search for Excellence Ud’A 2019 and from University of Trento, project UNMASKED 2020.
	\bibliographystyle{myplainnat}
\bibliography{hom}